\documentclass[prd,showpacs,altaffilletter,twocolumn,
superscriptaddress,floatfix,nofootinbib,preprintnumbers]{revtex4-1} 

\pdfoutput=1
\usepackage{amsfonts,amsmath,graphicx,bm} 
\usepackage{float}
\usepackage{dcolumn}
\usepackage[pdftex,
       colorlinks=true,
       urlcolor=blue,       
       linkcolor=red,       
       pdfpagemode=None,
       bookmarksopen=true]{hyperref}
\usepackage[capitalise]{cleveref}
\usepackage{ifpdf}
\usepackage{multirow}
\usepackage[colorlinks=true]{hyperref}
\usepackage{url}
\usepackage{slashed}

\newcommand{\glasgow}{SUPA, School of Physics and Astronomy,
  University of Glasgow, Glasgow, G12 8QQ, UK}

\def\today{\number\day\space\ifcase\month\or
January\or February\or March\or April\or May\or June\or
July\or August\or September\or October\or November\or December\fi
\space\number\year}
\newcount\mins \newcount\hours
\def\now{\hours=\time \mins=\time
	\divide\hours by60 \multiply\hours by60 \advance\mins by-\hours
	\divide\hours by60 
	\number\hours:\ifnum\mins<10 0\fi\number\mins }

\newcommand{\RDstar}{0.273(15)}       
\newcommand{\RDstarIMP}{0.342(6)}     
\newcommand{\RDsstar}{0.266(9)}      
\newcommand{\RDsstarIMP}{0.340(3)}    

\newcommand{\RDstarexp}{0.2482(20)}           
\newcommand{\RDstarIMPexp}{0.3372(23)}        
\newcommand{\RDsstarexp}{ 0.2459(34)}         
\newcommand{\RDsstarIMPexp}{0.3358(21)}       

\newcommand{\vcb}{39.03(56)_\mathrm{exp}(67)_\mathrm{latt}\times 10^{-3}}     

\newcommand{\totalVCB}{$V_{cb}=42.9(0.5)_\mathrm{exp}(2.2)_\mathrm{latt}\times 10^{-3}$}    
\newcommand{\totalVCBpdg}{$V_{cb}=43.4(0.9)_\mathrm{exp}(2.2)_\mathrm{latt}\times 10^{-3}$} 

\newcommand{\VCBMU}{38.75(96)\times 10^{-3}}      
\newcommand{\VCBE}{39.26(91)\times 10^{-3}}       
\newcommand{\RATIOVCBEMU}{1.013(17)}              

\newcommand{\AFBE}{0.270(33)}             
\newcommand{\AFBMU}{0.266(34)}            
\newcommand{\DELTAAFBMUE}{-0.0036(10)}    

\newcommand{\RDstaroverRDsstar}{1.028(50)}      
\newcommand{\dFdw}{-0.97(15)}                   
\newcommand{\dFdws}{-0.94(11)}                  
\newcommand{\Gammaetot}{ 1.13(12)\times 10^{-11}~\mathrm{GeV}}    
\newcommand{\Gammamutot}{ 1.13(11)\times 10^{-11}~\mathrm{GeV}}   
\newcommand{\Gammatautot}{  3.10(17)\times 10^{-12}~\mathrm{GeV}}

\newcommand{\Gammaetots}{ 1.201(63)\times 10^{-11}~\mathrm{GeV}} 
\newcommand{\Gammamutots}{ 1.197(63)\times 10^{-11}~\mathrm{GeV}}
\newcommand{\Gammatautots}{ 3.20(10)\times 10^{-12}~\mathrm{GeV}}

\newcommand{\Alambdatauratio}{ 1.040(29)}  
\newcommand{\FLepratio}{ 0.942(46)}        
\newcommand{\AFBratio}{ 1.19(23)}          

\newcommand{\Alambdatau}{  0.547(19)}      
\newcommand{\FLep}{  0.395(24)}            
\newcommand{\AFB }{ 0.100(25)}             

\newcommand{\Alambdataus }{ 0.5331(91)}    
\newcommand{\FLeps }{ 0.420(12) }          
\newcommand{\AFBs }{ 0.084(12)}            

\begin{document}

\title{$B \rightarrow D^*$ vector, axial-vector and tensor form factors for the full $q^2$ range from lattice QCD}

\author{Judd \surname{Harrison}}
\email[]{judd.harrison@glasgow.ac.uk}
\affiliation{\glasgow}

\author{Christine~T.~H.~\surname{Davies}} 
\email[]{christine.davies@glasgow.ac.uk}
\affiliation{\glasgow}

\collaboration{HPQCD Collaboration}
\email[]{http://www.physics.gla.ac.uk/HPQCD}


\begin{abstract}
We compute the complete set of SM and tensor $B_{(s)}\to D_{(s)}^*\ell\bar{\nu}$ semileptonic form factors across the full kinematic range of the decay using second generation MILC $n_f=2+1+1$ HISQ gluon field configurations and HISQ valence quarks, with the \textit{heavy-HISQ} method. Lattice spacings range from $0.09\mathrm{fm}$ to $0.044\mathrm{fm}$ with pion masses from $\approx 300\mathrm{MeV}$ down to the physical value and heavy quark masses ranging between $\approx 1.5 m_c$ and $4.1 m_c \approx 0.9 m_b$; currents are normalised nonperturbatively. Using the recent untagged $B\to D^*\ell\bar{\nu}_\ell$ data from Belle and $B_s\to D_s^*\mu\bar{\nu}_\mu$ from LHCb together with our form factors we determine a model independent value of $V_{cb}=\vcb$, in agreement with previous exclusive determinations and in tension with the most recent inclusive result at the level of $3.6\sigma$. We also observe a $\approx 1\sigma$ tension between the shape of the differential decay rates computed using our form factors and those measured by Belle. We compute a purely theoretical Standard Model value for the ratio of semitauonic and semimuonic decay rates, $R(D^*)=\RDstar$, which we find to be closer to the recent Belle measurement and HFLAV average than theory predictions using fits to experimental differential rate data for $B\to D^*\ell\bar{\nu}_\ell$. Determining $V_{cb}$ from our form factors and the experimental total rate for $B\to D^*\ell\nu$ also gives a value in agreement with inclusive results. We also compute the longitudinal polarisation fraction for the semitauonic mode, $F_L^{D^*}=\FLep$, which is in tension at the level of $2.2\sigma$ with the recent Belle measurement. Our calculation combines $B\to D^*$ and $B_s\to D_s^*$ lattice results, and we provide an update which supersedes our previous lattice computation of the $B_s\to D_s^*$ form factors. We also give the chiral perturbation theory needed to analyse the tensor form factors.
\end{abstract}

\maketitle



\section{Introduction}
\label{sec:intro}

Semileptonic and leptonic decays of mesons allow for many high precision tests of the Standard Model~(SM) description of the weak interaction. For example, in the SM the Cabibbo-Kobayashi-Maskawa~(CKM) matrix, which encodes the couplings of flavor-changing quark currents with the SM $W$-bosons, is unitary. Determinations of the CKM matrix elements using the weak decays of mesons~\cite{Chakraborty:2021qav,Bazavov:2017lyh} allow us to check if the unitarity constraints are satisfied. Currently those coming from the first row and column, which describe the couplings with up and down quarks, are in tension with unitarity at the level of $3\sigma$~\cite{pdg20}. 

The CKM matrix element $V_{cb}$, governing the strength of the quark level $b\to c\ell\bar{\nu}_\ell$ transition, can be determined most precisely either from inclusive semileptonic $B$ decays, where all charmed final states are included, or from exclusive semileptonic decays to a specific charmed meson. The inclusive determination of $V_{cb}$, which uses the operator product expansion~\cite{Gambino:2015ima} to express the nonperturbative physics in terms of matrix elements of local operators with $B$ mesons, gives $|V_{cb}|=42.16(51)\times 10^{-3}$~\cite{Bordone:2021oof}. 

Until very recently the exclusive determination only used experimental data for $B\to D$ and $B \to D^*$. This data has typically been extrapolated to the zero recoil point, where the $D^{(*)}$ meson is at rest, before being compared to lattice determinations~\cite{Bailey:2014tva,Harrison:2017fmw} of the single form factor relevant at this point. Recently $B_s\to D_s^{(*)}$ experimental data from LHCb was used together with HPQCD's early calculation of the $B_s\to D_s$ form factors~\cite{EuanBsDs} (as well as the $B_s\to D_s^*$ form factor at zero recoil~\cite{EuanBsDsstar}) to provide a complementary determination of $V_{cb}$. Averaging $B_{(s)}\to D_{(s)}^{(*)}$ results gives a value of $|V_{cb}|=38.90(53)\times 10^{-3}$~\cite{HFLAV:2022pwe}, in tension at the level of $4.4\sigma$ with the most recent inclusive result. This determination is most sensitive to $B\to D^*$ data, which is much more precise than existing $B_s\to D_s^{(*)}$ data, and is preferred over $B\to D$ owing to the kinematic factors appearing in the differential rate, which allow for more data to be collected near zero recoil and thus for a more precise extrapolation to this point. Note that while lattice form factors for $B\to D$ are available away from zero recoil~\cite{Lattice:2015rga}, extrapolation of experimental data to zero recoil is still used in order to straightforwardly average experimental results~\cite{HFLAV:2022pwe}.

The extrapolation of experimental data to the zero recoil point has typically been done using either the Caprini, Lellouch and Neubert~(CLN) parameterisation scheme~\cite{Caprini:1997mu}, or the Boyd, Grinstein and Lebed~(BGL) parameterisation scheme~\cite{Boyd:1997kz}. The CLN scheme imposes strong unitarity constraints based on heavy quark symmetry, and uses heavy quark effective theory~(HQET) to reduce the number of independent parameters. This results in a highly constrained fit with only a single parameter able to modify the shape of the form factors. This approach has been widely criticised as underestimating residual uncertainties~\cite{Bigi:2017jbd,Grinstein:2017nlq,Bigi:2016mdz}, and theoretical analyses of the 2017 Belle dataset~\cite{Belle:2017rcc} indicated that CLN was not well suited to describe the data~\cite{Bigi:2017njr,Bordone:2019vic}. 
The BGL scheme is more general, imposing unitarity bounds based on analyticity~\cite{Boyd:1997kz}. Early analyses of the 2017 Belle dataset indicated that the use of BGL, as opposed to CLN, would go some way to resolving the tension between inclusive and exclusive decays~\cite{Bigi:2017njr,Grinstein:2017nlq,Jaiswal:2017rve}. However, analysis of the more recent larger untagged dataset from Belle~\cite{Belle:2018ezy} instead finds very similar central values and uncertainties for $V_{cb}$ using BGL and CLN schemes, both in similar tension with the inclusive result at the same level as previous exclusive results.

Recent advances in lattice QCD have allowed for the calculation of pseudoscalar to vector form factors for $b$-quark decays across the full kinematic range of the decays, with HPQCD producing the first calculations for $B_c\to J/\psi$~\cite{Harrison:2020gvo} and $B_s\to D_s^*$~\cite{Harrison:2021tol}, related to $B\to D^*$ by the exchange of the light spectator quark with a charm or strange quark respectively. These calculations used highly improved staggered quarks~(HISQ)~\cite{PhysRevD.75.054502} for all quarks, and were carried out using the $n_\mathrm{f}=2+1+1$ second generation MILC gauge configurations including up/down, strange and charm HISQ quarks in the sea. In order to extract form factors for mesons including a physically massive $b$ quark the so called \textit{heavy-HISQ} method was used. This framework involves using a heavy quark, $h$, in place of the $b$, and varying the mass of $h$ from close to the charm quark mass all the way up to the physical $b$ quark mass. By using multiple lattices with different lattice spacings this procedure allows us to map out discretisation effects and the physical dependence on the $h$ quark mass in the quantities of interest and to extract precise values at the physical point where the $h$ quark mass is equal to that of the $b$. The determination of the full set of $B_s\to D_s^*$ form factors allowed for a model-independent determination of $V_{cb}$~\cite{Harrison:2021tol}, using recent experimental results from LHCb~\cite{Aaij:2020hsi}.

The Fermilab Lattice and MILC Collaborations have recently also published first results for $B\to D^*$ form factors away from zero recoil~\cite{FermilabLattice:2021cdg}, with lattice data exting across $\approx 1/3$ the full kinematic range of the decay, using the Fermilab action~\cite{El-Khadra:1996wdx} for $b$ and $c$ quarks and using gluon field configurations with $n_f=2+1$ flavours of asqtad sea quarks. They found, using the recent untagged data from Belle~\cite{Belle:2018ezy} and synthetic data from BaBar~\cite{BaBar:2019vpl}, $|V_{cb}|=38.40(66)_\mathrm{th}(34)_\mathrm{exp}\times 10^{-3}$, in tension at the level of $\approx 4\sigma$ with the most recent inclusive determinations, and confirming the persistent tension currently seen in global averages~\cite{HFLAV:2022pwe}. The JLQCD Collaboration has also presented preliminary results for the $B\to D^*$ form factors~\cite{Kaneko:2019vkx,Kaneko:2021tlw}. Note that these lattice results have been used in combination with unitarity constraints via the `Dispersive Matrix' method to extend these form factors across the kinematic range. Those studies found values of $V_{cb}$ closer to the inclusive result~\cite{Martinelli:2021onb}.

Semileptonic decays of mesons also allow us to search directly for violations of the universality of the SM coupling between leptons and $W$-bosons, as might result from new physics~(NP) beyond the Standard Model. The most common method by which this is done is to construct ratios of branching fractions to final states with different leptons. This results in the cancellation of the CKM matrix element factors, as well as a substantial cancellation of correlated uncertainties entering through the form factors. The ratio relevant for $B\to D^*$ is 
\begin{equation}\label{RDstardef}
R(D^*)=\frac{\Gamma(B\to D^*\tau\bar{\nu}_\tau)}{\Gamma(B\to D^*\mu\bar{\nu}_\mu)}.
\end{equation}
The most precise theoretical determinations of $R(D^*)$ in the SM use fits to experimental data for $B\to D^*\mu\bar{\nu}_\mu$, together with the assumption that NP can only appear in the semitauonic mode, to pin down the 3 form factors needed for the light lepton case$~(\ell=e,\mu)$. Until recently, the remaining pseudoscalar form factor relevant for the case of the heavy $\tau$ lepton was determined using HQET inputs~\cite{Gambino:2019sif,Bordone:2019vic,Jaiswal:2017rve}. This approach results in a very precise theory prediction for $R(D^*)=0.254(5)$~\cite{HFLAV:2022pwe} in tension with the most recent experimental average, $R^\mathrm{HFLAV}(D^*)=0.295(14)$~\cite{HFLAV:2022pwe}, at the level of $2.7\sigma$. This tension increases to $\approx 3\sigma$ if $R(D)$ is included. However, more recent measurements from the BaBar, Belle and LHCb collaborations are closer to the SM prediction~\cite{PhysRevD.88.072012,Belle:2019rba,LHCb:2017rln}. 

Recently, the Fermilab-MILC collaboration presented a lattice-only determination of $R(D^*)$ as well as a determination using a joint fit to lattice and experimental data~\cite{FermilabLattice:2021cdg}, resulting in values of $R(D^*)=0.265(13)$ and $R(D^*)=0.2483(13)$ respectively. The difference between these results, while only at the $1\sigma$ level, is surprising and makes clear the desirability of additional precise lattice-only determinations of $R(D^*)$, as well as direct comparisons of the shape of the differential rate between theory and experiment, where some tension was also seen in~\cite{FermilabLattice:2021cdg}.

The ratio of~\cref{RDstardef} was also computed using lattice QCD for $B_s \to D_s^*\ell\bar{\nu}$ and $B_c\to J/\psi$ in~\cite{Harrison:2021tol} and~\cite{Harrison:2020gvo} respectively. The former is of particular interest as the value computed there, $R(D_s^*)=0.2490(69)$, is in agreement with the theory prediction for $R(D^*)$ using experimental data as input. The form factors for $B \to D^*\ell\bar{\nu}$ and $B_s \to D_s^*\ell\bar{\nu}$ are related by the change of spectator quark from up/down to strange, and the corresponding $SU(3)_\mathrm{flav}$ symmetry breaking effects are expected to be small, at the level of $\approx 1\%$~\cite{Harrison:2017fmw}. As such, a simultaneous analysis of $B_s\to D_s^*$ and $B\to D^*$ is desirable in order to investigate the differences between the results presented in~\cite{Harrison:2021tol} and those in~\cite{FermilabLattice:2021cdg}. 

In addition to $R(D^*)$, there are other observables, such as the $\tau$ lepton polarisation asymmetry, the forward-backward asymmetry and the $D^*$ longitudinal polarisation fraction. These are expected to be sensitive to NP~\cite{Becirevic:2019tpx} and theoretical predictions for these would be valuable for future measurements. They also provide further tests of $SU(3)_\mathrm{flav}$ breaking, which is expected to be small~\cite{Bordone:2019guc} as for $R(D^*)$. The Belle collaboration has recently measured both the lepton polarisation asymmetry~\cite{Belle:2017ilt} and the $D^*$ longitudinal polarisation fraction~\cite{Belle:2019ewo}, both of which may be computed directly on the lattice without the need for inputs such as $V_{cb}$.


Until now, lattice calculations of form factors for pseudoscalar to vector decays have focused exclusively on those form factors needed to describe the decay within the SM. Specifically, these are the two axial-vector form factors, one vector form factor and one pseudoscalar form factor. However, assuming left handed neutrinos, there are two additional dimension-6, parity-conserving four-fermion operators which can appear in the effective Hamiltonion whose matrix elements between $B$ and $D^*$ states are nonzero. These are the tensor operators:
\begin{align}
(\bar{c}\sigma_{\mu\nu}b)&(\bar{\ell}_R\sigma^{\mu\nu} \nu_L)\nonumber\\
(\bar{c}\sigma_{\mu\nu}\gamma_5b)&(\bar{\ell}_R\sigma^{\mu\nu} \nu_L).
\end{align}
The form factors for pseudoscalar to vector decays for the quark currents $\bar{c}\sigma_{\mu\nu}b$ and $\bar{c}\sigma_{\mu\nu}\gamma_5b$ have not previously been computed on the lattice, though the single form factor for the related $\bar{s}\sigma_{\mu\nu}b$ was computed for the rare decay $B_c\to D_s\ell^+\ell^-(\nu\bar{\nu})$ in~\cite{Cooper:2021bkt}, as well as for $B\to K$~\cite{Parrott:2022dnu}, using the \textit{heavy-HISQ} method together with renormalisation factors matching the lattice tensor currents to those in the continuum $\overline{\mathrm{MS}}$ scheme, computed in~\cite{Hatton:2020vzp} using an intermediate RI-SMOM scheme.

In this work, we build on previous \textit{heavy-HISQ} calculations of pseudoscalar to vector decays and compute both the SM and tensor form factors for $B\to D^*$. We also compute the SM and tensor form factors for $B_s\to D_s^*$, which we analyse simultaneously in order to better map out the dependence of the form factors on the spectator quark mass and in order to study $SU(3)_\mathrm{flav}$ breaking effects between the two. We then give values for $|V_{cb}|$, $R(D^*)$ and other observables. 

The remaining sections are organised as follows:
\begin{itemize}
\item In \cref{sec:theory} we detail the theoretical framework relevant for semileptonic $B \to D^*$ decays, including the effective Hamiltonian, definitions of form factors and helicity amplitudes and expressions for the differential decay rate.
\item \cref{lattcalc} contains the details of our lattice calculation, including our correlator fitting procedure, current renormalisation and how form factors are extracted from correlator fit results.
\item In \cref{Results} we give the results of our lattice calculation and describe our chiral-continuum fit procedure including the heavy quark mass dependence. We give our results for the SM and tensor form factors and demonstrate the stability of our results to changes in correlator fits and changes to our chiral-continuum fit procedure.
\item In \cref{Discussion} we use our form factors to compute observables including $R(D^*)$. We compare our results to the recent measurement by Belle and determine a value of $V_{cb}$.
\item Finally, in \cref{Conclusion} we summarise our findings and suggest directions for future investigations.
\item In \cref{fulldiff} we compute expressions for the full differential decay rate including all operators relevant for NP. In~\cref{ntbinning} we discuss our approach to binning correlator data. In~\cref{lattdat} we give the numerical results for the form factors on each ensemble, extracted from fits to correlation functions. In~\cref{chilogs} we compute the next-to-leading order chiral logarithms, needed for the chiral-continuum extrapolation of the tensor form factors, using heavy-meson rooted staggered chiral perturbation theory. In~\cref{comptoprev} we compare the updated $B_s\to D_s^*$ form factor results of this work to those in~\cite{Harrison:2021tol}.

\end{itemize}

\section{Theoretical Background}
\label{sec:theory}

\begin{figure}
\includegraphics[scale=0.4]{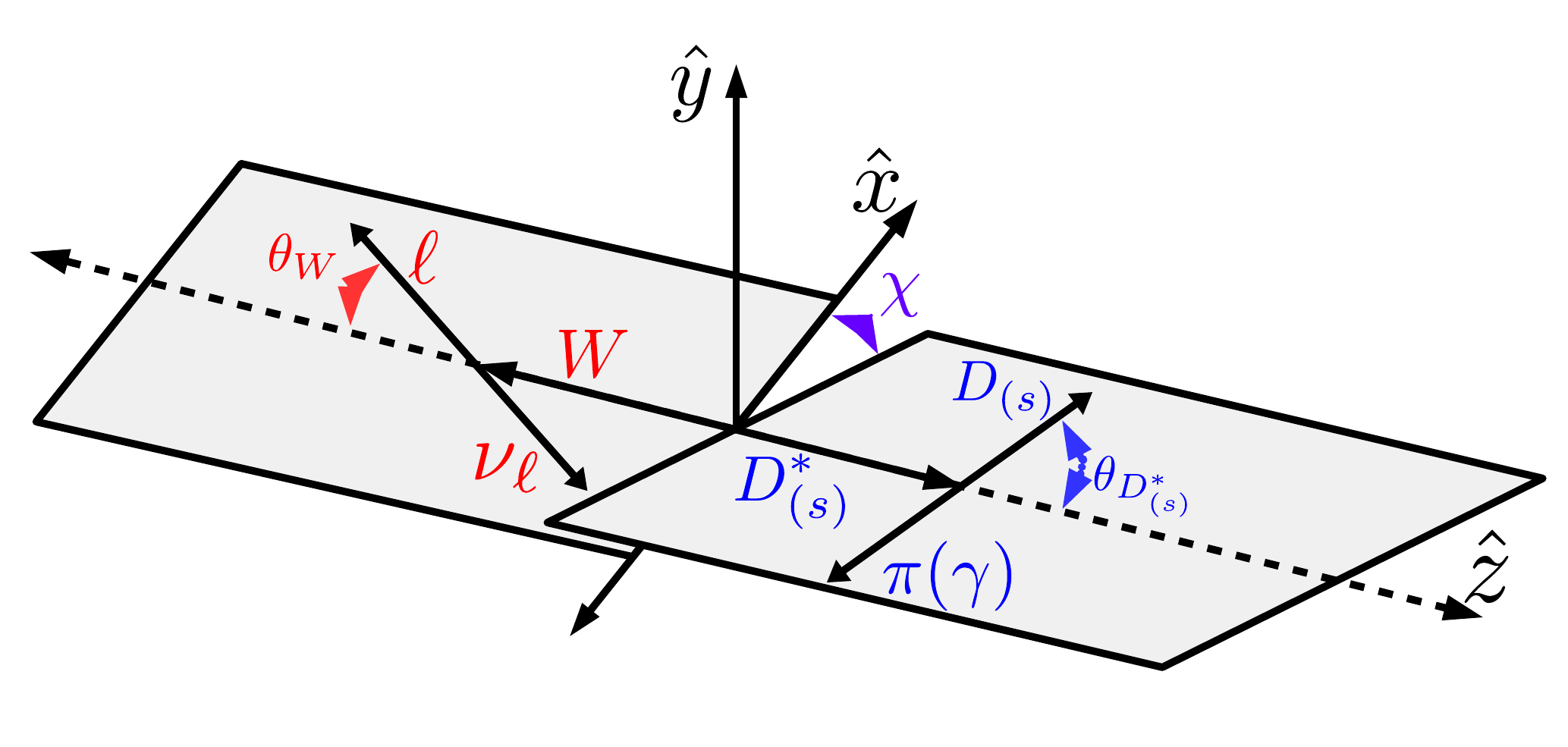}
\caption{\label{BsDsstarangles}Conventions for the angular variables entering the differential decay rate.}
\end{figure}


The effective Hamiltonian relevant for semileptonic $b\to c$ decays is, assuming left-handed neutrinos, 
\begin{align}\label{EffectiveHamiltonian}
\mathcal{H}_\mathrm{eff}=\sqrt{2}G_{F}V_{cb}\Big[&g_V\bar{c}\gamma_\mu b \bar{\ell}_L\gamma^\mu \nu_L+g_A\bar{c}\gamma_\mu \gamma_5b \bar{\ell}_L\gamma^\mu \nu_L\nonumber\\
+&g_S\bar{c}b \bar{\ell}_R \nu_L\nonumber\\
+&g_P\bar{c}\gamma_5b \bar{\ell}_R \nu_L\nonumber\\
+&g_T \bar{c}\sigma_{\mu\nu}b\bar{\ell}_R\sigma^{\mu\nu} \nu_L\nonumber\\
+&g_{T5} \bar{c}\sigma_{\mu\nu}\gamma^5 b\bar{\ell}_R\sigma^{\mu\nu} \nu_L + \mathrm{h.c.}\Big]
\end{align}
where $\sigma_{\mu\nu}=i/2[\gamma_\mu,\gamma_\nu]$ and $g_X$ are potentially complex coefficients. In the SM $g_T=g_{T5}=g_P=g_S=0$ and $g_V=-g_A=1$. 

The differential decay rate to $D\pi\ell\bar{\nu}_\ell$ is, taking the $D^*$ as a narrow resonance,
\begin{align}\label{diffrateM}
\frac{d\Gamma}{dq^2d\mathrm{cos}(\theta_{D^*})d\mathrm{cos}(\theta_W)d\chi}=N(q^2)\sum_{\lambda_\ell}\Big|\sum_{\lambda_{D^*}} \mathcal{M}^{\lambda_{D^*}\lambda_\ell}\Big|^2
\end{align}
where $N(q^2)$ is an overall kinematic factor
\begin{align}
N(q^2)=\frac{3G_F^2|V_{cb}\eta_{EW}|^2}{8(4\pi)^4}\frac{k (q^2-m_\ell^2)^2}{q^2M_B^2}\mathcal{B}(D^*\to D\pi)
\end{align}
and the angular variables are defined in~\cref{BsDsstarangles}. 

The right hand side of~\cref{diffrateM} is conventionally expressed in terms of helicity amplitudes, which are related to the form factors that parameterise the nonperturbative matrix elements of the quark currents in~\cref{EffectiveHamiltonian}. The form factors, $h_Y$, for $B\to D^*$ are defined in the HQET basis as~\cite{Bernlochner:2017jka}:
\begin{align}\label{formfactors}
\langle D^*|\bar{c} b|\overline{B}\rangle             =&~ 0,\nonumber\\
\langle D^*|\bar{c}\gamma^5 b|\overline{B}\rangle             =& -\sqrt{M_BM_{D^*}}(\epsilon^*\cdot v)h_P,\nonumber\\
\langle D^*|\bar{c}\gamma^\mu b|\overline{B}\rangle           =&~ i\sqrt{M_BM_{D^*}}\varepsilon^{\mu\nu\alpha\beta}\epsilon^{*}_\nu v^\prime_\alpha v_\beta h_V,\nonumber\\
\langle D^*|\bar{c} \gamma^\mu \gamma^5 b|\overline{B}\rangle =& ~\sqrt{M_BM_{D^*}}\big[ h_{A_1}(w+1)\epsilon^{*\mu} \nonumber\\
&-h_{A_2}(\epsilon^*\cdot v)v^\mu-h_{A_3}(\epsilon^*\cdot v)v^{\prime\mu} \big],\nonumber\\
\langle D^*|\bar{c}\sigma^{\mu\nu} b|\overline{B}\rangle      =& -\sqrt{M_BM_{D^*}}\varepsilon^{\mu\nu\alpha\beta}\big[ h_{T_1}\epsilon^*_\alpha(v+v^\prime)_\beta \nonumber\\
&+h_{T_2}\epsilon^*_\alpha(v-v^\prime)_\beta +h_{T_3}(\epsilon^*\cdot v)v_\alpha v^\prime_\beta \big],
\end{align}
where $w=v^\prime\cdot v$ and $v^\prime$ and $v$ are the four velocities of the $D^*$ and $B$ respectively. Note that the matrix element of $\bar{c}\sigma_{\mu\nu}\gamma^5 b$ is related to that of $\bar{c}\sigma_{\mu\nu} b$, since $\sigma_{\mu\nu}\gamma^5=\frac{i}{2}\epsilon_{\mu\nu\alpha\beta}\sigma^{\alpha\beta}$. The tensor current is renormalised in the SM, and so the tensor form factors depend on the renormalisation scale which for $b$ decays is typically taken as $\mu=m_b^\mathrm{pole}$. 

In terms of these form factors, the nonzero helicity amplitudes for the (axial-)vector currents are
\begin{align}\label{smhelicityampls}
H_\pm=&-g_A^*h_{A_1} \sqrt{M_BM_{D^*}}(1+w)\nonumber\\
&\mp g_V^*h_V \sqrt{M_BM_{D^*}(w^2-1)}\\
H_0=& -g_A^*M_B(1+w)\sqrt{\frac{M_BM_{D^*}}{q^2}}\nonumber\\
&\times\big[ h_{A_1}(w-r)- (w-1)[h_{A_3}+r h_{A_2}] \big]\\
H_t=& -g_A^* M_B\sqrt{\frac{M_BM_{D^*}(w^2-1)}{q^2}}\nonumber\\
&\times\big[ h_{A_1}(1+w) -h_{A_2}(1-wr)-h_{A_3}(w-r) \big]
\end{align}
where $r=M_{D^*}/M_B$. Note that the complex conjugates of the coefficients $g_X$ appear in the conjugate mode ${B}^0\to D^{*-}\ell^+\nu$ for the general complex $g_X$ appearing in~\cref{EffectiveHamiltonian}. Expressions for the tensor helicity amplitudes are given in~\cref{fulldiff}. 

The squared matrix element entering the differential rate may be written as
\begin{align}\label{eq:helicitytoMsq}
\sum_{\lambda_\ell}\Big|\sum_{\lambda_{D^*}} \mathcal{M}^{\lambda_{D^*}\lambda_\ell}\Big|^2=\sum_i k_i(\theta_W,\theta_{D^*},\chi) \mathcal{H}_i.
\end{align}
The combinations $k_i$ and $\mathcal{H}_i$ are given in~\cref{tab:diffterms} for the charge conjugate mode, ${B}^0\to D^{*-}\ell^+\nu$, for the case where only $g_A$ and $g_V$ are nonzero. We have checked that this expression matches that given in~\cite{Belle:2018ezy}. Note that it also agrees with the expression for the $\ell^+\nu$ final state given in~\cite{Cohen:2018vhw}, though there one must also take $H_+\leftrightarrow H_-$ for the conjugate hadronic current.
\begin{table}
\centering
\caption{The helicity amplitude combinations and coefficients for them that 
appear in~\cref{eq:helicitytoMsq} for ${B}^0\to D^{*-}\ell^+\nu$ when only $g_A$ and $g_V$ in~\cref{EffectiveHamiltonian} are nonzero. \label{tab:diffterms}}
\begin{tabular}{ c c | c  }
\hline
$i$ & $\mathcal{H}_i$ & $k_i(\theta_W,\theta_{D^*},\chi)$\\
\hline
1 & $|H_+(q^2)|^2$ & $(1-\cos(\theta_W))^2(\sin^2(\theta_{D_s^*}))$\\
2 & $|H_-(q^2)|^2$ & $(1+\cos(\theta_W))^2(\sin^2(\theta_{D_s^*}))$\\
3 & $|H_0|^2$& $4\sin^2(\theta_W)\cos^2(\theta_{D_s^*})$\\
4 & $\mathrm{Re}(H_+H_0^*)$&$-2\sin(\theta_W)\sin(2\theta_{D_s^*})\cos(\chi)(1-\cos(\theta_W))$\\
5 & $\mathrm{Re}(H_-H_0^*)$&$ 2\sin(\theta_W)\sin(2\theta_{D_s^*})\cos(\chi)(1+\cos(\theta_W))$\\
6 & $\mathrm{Re}(H_+H_-^*)$& $-2\sin^2(\theta_W)\sin^2(\theta_{D_s^*})\cos(2\chi)$\\
7 & $\frac{m_\ell^2}{q^2}|H_+(q^2)|^2$ & $\sin^2(\theta_W)\sin^2(\theta_{D_s^*})$\\
8 & $\frac{m_\ell^2}{q^2}|H_-(q^2)|^2$ & $\sin^2(\theta_W)\sin^2(\theta_{D_s^*})$\\
9 & $\frac{m_\ell^2}{q^2}|H_0|^2$& $4\cos^2(\theta_W)\cos^2(\theta_{D_s^*})$\\
10 & $\frac{m_\ell^2}{q^2}|H_t(q^2)|^2$ & $4\cos^2(\theta_{D_s^*})$\\
11 & $\frac{m_\ell^2}{q^2}\mathrm{Re}(H_+H_0^*)$&$-2\sin(\theta_W)\sin(2\theta_{D_s^*})\cos(\chi)\cos(\theta_W)$\\
12 & $\frac{m_\ell^2}{q^2}\mathrm{Re}(H_-H_0^*)$&$-2\sin(\theta_W)\sin(2\theta_{D_s^*})\cos(\chi)\cos(\theta_W)$\\
13 & $\frac{m_\ell^2}{q^2}\mathrm{Re}(H_+H_-^*)$& $2\sin^2(\theta_W)\sin^2(\theta_{D_s^*})\cos(2\chi)$\\
14 & $\frac{m_\ell^2}{q^2}\mathrm{Re}(H_tH_0^*)$& $-8\cos^2(\theta_{D_s^*})\cos(\theta_W)$\\
15 & $\frac{m_\ell^2}{q^2}\mathrm{Re}(H_+H_t^*)$& $4\sin(\theta_W)\sin(2\theta_{D_s^*})\cos(\chi)$\\
16 & $\frac{m_\ell^2}{q^2}\mathrm{Re}(H_-H_t^*)$& $4\sin(\theta_W)\sin(2\theta_{D_s^*})\cos(\chi)$
\end{tabular}
\end{table}

The construction of the full differential rate including tensor, axial-tensor and pseudoscalar currents is described in~\cref{fulldiff}, together with the combinations $k_i$ and $\mathcal{H}_i$. The explicit coefficients for the full and partially integrated differential rate are also provided as a supplementary python script for the general case of complex $g_X$.

\section{Lattice Calculation}\label{lattcalc}

Our lattice QCD calculation of the $B\to D^*$ form factors follows broadly the same heavy-HISQ approach as those presented in~\cite{Harrison:2021tol,Harrison:2020nrv} for the related $B_s\to D_s^*$ and $B_c\to J/\psi$ form factors respectively. We use a range of masses for a heavy quark, $h$, between the charm and physical bottom quark mass. The heavy-light pseudoscalar meson, which we will refer to as $H$, is at rest on the lattice. We give momentum to the charm quark using twisted boundary conditions so that the $D^*$ covers the range of physical momenta for $H\to D^*$ decay. We use the HISQ action~\cite{PhysRevD.75.054502} for all valence quarks and use the second generation $N_f=2+1+1$ MILC ensembles of gauge configurations, which include equal mass~$(m_u=m_d)$ HISQ light quarks in the sea, as well as physically tuned strange and charm sea quarks~\cite{PhysRevD.82.074501,PhysRevD.87.054505}. We include ensembles with a range of lattice spacings from $0.09~\mathrm{fm}$ down to $0.045~\mathrm{fm}$ and a range of light quark masses. On the finest ensemble with $0.045~\mathrm{fm}$ we are able to reach very close to the physical bottom quark mass for $h$. The details of these ensembles are given in~\cref{tab:gaugeinfo}. Note that compared to~\cite{Harrison:2021tol,Harrison:2020nrv} we include an additional ensemble, set 5, with $w_0/a=3.0170(23)$~\cite{Hatton:2020qhk}, which we refer to as `physical superfine'. This additional ensemble is important, along with `physical fine' lattices, for resolving the logarithmic dependence of the form factors on the pion mass~\cite{Randall:1993qg} arising from the proximity of the $D^*$ to the $D^*\to D\pi$ threshold. The heavy quark masses used, together with the valence charm and strange quark masses~(for the $B_s\to D_s^*$ case), are given in~\cref{valmasses}. We use valence light quarks with masses equal to the sea light quark masses in~\cref{tab:gaugeinfo}. 

\begin{figure}
\includegraphics[scale=0.4]{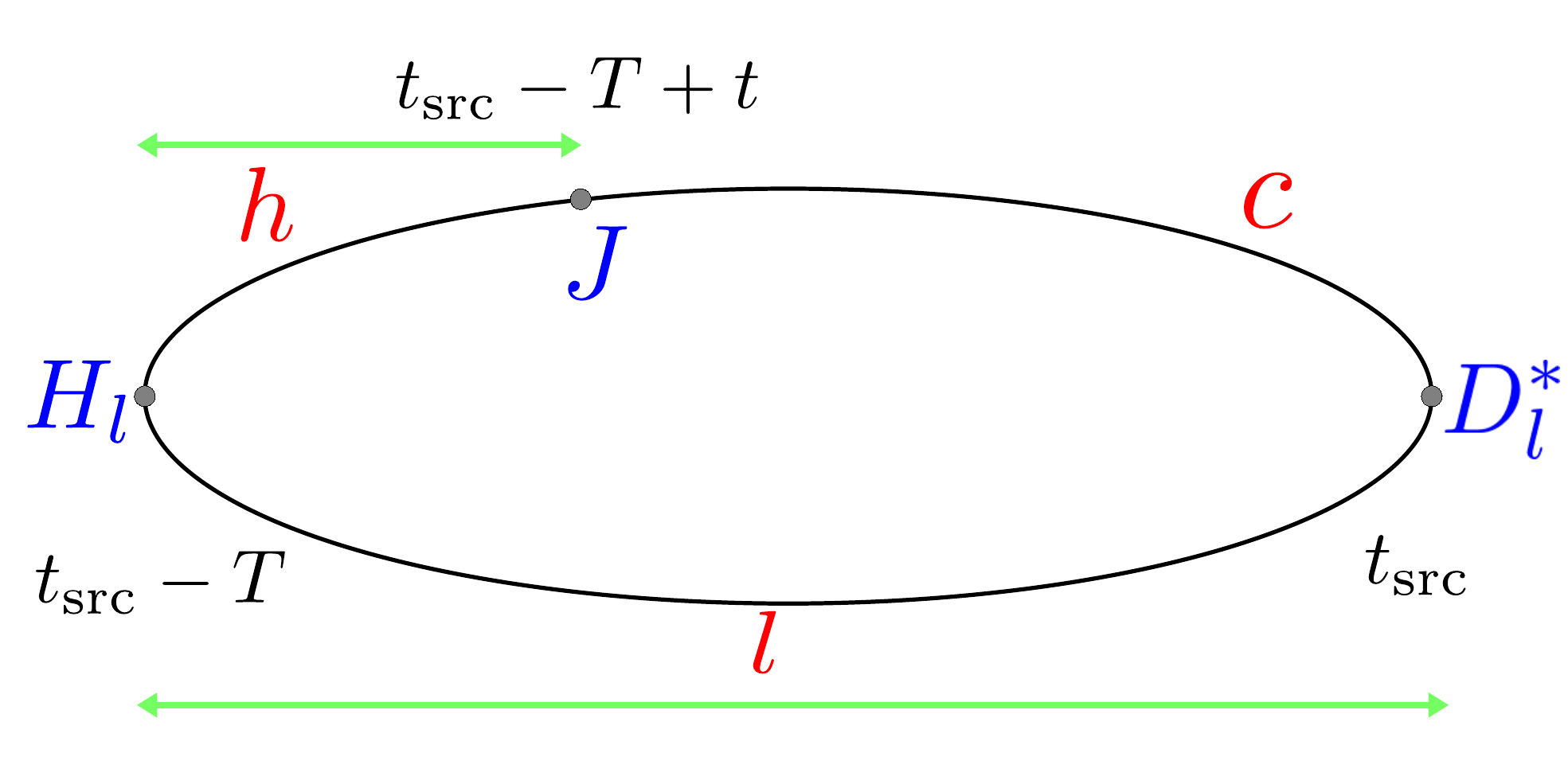}
\caption{\label{3ptfig}Arrangement of propagators in the three point function; we refer to $c$ as the `active' charm quark, $h$ as the `extended' heavy quark and $l$ as the `spectator' light/strange quark. $J$ represents the insertion of either a vector, axial-vector or tensor current and $H_l$ and $D_l^*$ represent the insertion of the corresponding meson interpolating operators.}
\end{figure}

On the lattice, we compute 2-point and 3-point correlation functions of meson interpolating operators and currents in order to extract matrix elements, amplitudes and energies. Note that in our lattice calculation the correlation functions are constructed from staggered spin-taste operators~\cite{PhysRevD.75.054502}. In this section, for notational simplicity, we write the correlation functions in terms of the equivalent continuum operators built from Dirac fermions. For a general current operator $\bar{c}\Gamma h$, the 2-point and 3-point correlation functions take the form
\begin{align}
C_\mathrm{2pt}^{D^*_l}(t,0) =& \langle  0|\bar{l}\gamma^\nu c(t) \left(\bar{l}\gamma^\nu c(0)\right)^\dagger| 0 \rangle,\nonumber\\
C_\mathrm{2pt}^{H_l}(t,0) =& \langle  0|\left(\bar{h}\gamma^5 l(t)\right)^\dagger\bar{h}\gamma^5 l(0) | 0 \rangle,\nonumber\\
C_\mathrm{3pt}(T,t,0) =& \langle  0|\bar{l}\gamma^\nu c(T) ~ \bar{c}\Gamma h(t) ~ \bar{h}\gamma^5 l(0)| 0 \rangle. \label{threepointcorr}
\end{align}
We compute correlation functions for both $l=u/d$ and $l=s$, and we will distinguish the mesons with $l=s$ with a subscript~$s$. We use random wall sources at time $t_\mathrm{src}$ for the light and charm quark propagators in order to improve statistics, as well as for the heavy quarks entering the 2-point functions, and we use twisted boundary conditions~\cite{Sachrajda:2004mi, Guadagnoli:2005be} to give momentum to the charm quark. We use the light quark propagator at time $t_\mathrm{src}-T$ to construct the source for the heavy quark propagator needed for the 3-point correlation functions. Finally, this heavy quark propagator is tied together with the charm propagator at time $t_\mathrm{src}-T+t$ to form the 3-point correlation function. The arrangement of quark propagators entering the 3-point functions is shown in~\cref{3ptfig}. We compute 3-point correlation functions using multiple values of $T$ in order to resolve the $T$ dependence of the correlation functions. The values of $T$ used on each ensemble, together with the twists used to give momentum to the charm quarks, are given in~\cref{twists}. Note that the twists differ slightly from those used in~\cite{Harrison:2021tol}.
\begin{table}
\caption{Details of the gauge field configurations used in our calculation \cite{PhysRevD.87.054505,PhysRevD.82.074501}. We use the Wilson flow parameter~\cite{Borsanyi:2012zs}, $w_0$, to fix 
the lattice spacing given in column 2. The physical value of $w_0$ was determined in \cite{PhysRevD.88.074504} to be 0.1715(9)fm and the values of $w_0/a$, 
which are used together with $w_0$ to compute $a$, were taken from \cite{PhysRevD.96.034516,PhysRevD.91.054508,EuanBsDs}. Set 1 with $w_0/a=1.9006(20)$ is referred to as `fine', set 2 with $w_0/a=2.896(6)$ as `superfine', set 3 with $w_0/a=3.892(12)$ as `ultrafine' and set 4 with $w_0/a=1.9518(7)$ as `physical fine'. Note that compared to~\cite{Harrison:2021tol,Harrison:2020nrv} we include an additional ensemble, set 5, with $w_0/a=3.0170(23)$~\cite{Hatton:2020qhk}, which we refer to as `physical superfine', that includes physical light quarks. $n_\mathrm{cfg}$ and $n_\mathrm{t}$ give the number of configurations and the number of time sources respectively. $am_{l0}$, $am_{s0}$ and $am_{c0}$ are the masses of the sea up/down, strange and charm quarks in lattice units. We also include the approximate mass of the Goldstone pion, computed in~\cite{Bazavov:2017lyh}.\label{tab:gaugeinfo}}
\begin{tabular}{c c c c c c c c c}\hline
 Set &$a$ & $N_x\times N_t$ &$am_{l0}$&$am_{s0}$& $am_{c0}$ & $M_\pi$ &$n_\mathrm{cfg}\times n_\mathrm{t}$ \\ 
  & $(\mathrm{fm})$&  &&&  & $(\mathrm{MeV})$ & \\ \hline
1 & $0.0902$   & $32\times 96 $    &$0.0074$ &$0.037$  & $0.440$ & $316$ & $1000\times 16$\\
2 & $0.0592$   & $48\times 144  $    &$0.0048$ &$0.024$  & $0.286$ & $329$ & $500\times 4$\\
3 & $0.0441$   &$ 64\times 192  $    &$0.00316$ &$0.0158$  & $0.188$ & $315$ & $375\times 4$\\
4 & $0.0879$  &$ 64\times 96  $    &$0.0012$ &$0.0363$  & $0.432$ & $129$ & $600\times 8$\\
5 & $0.0568$   &$ 96\times 192  $    &$0.0008$ &$0.0219$  & $0.2585$ & $135$ & $100\times 4$\\\hline
\end{tabular}
\end{table}

\begin{table}
\centering
\caption{Details of the strange, charm and heavy valence masses. \label{valmasses}}
\begin{tabular}{c c c c }\hline
 Set &$am_h^\mathrm{val}$ & $am_s^\mathrm{val} $ & $am_c^\mathrm{val} $\\ \hline
1 & $0.65,0.725,0.8$&   $0.0376$     & $0.449$  \\
2 & $0.427,0.525,0.65,0.8$ &  $0.0234$   & $0.274$   \\
3 & $0.5,0.65,0.8$&   $0.0165$    & $0.194$  \\
4 & $0.65,0.725,0.8$ &   $0.036$   & $0.433$ \\
5 & $0.427,0.525,0.65,0.8$&  $0.0165$    & $0.2585$   \\ \hline
\end{tabular}
\end{table}

\begin{table}
\centering
\caption{Values of twists, $\theta$, together with values of $T$ used in the three point functions in~\cref{threepointcorr}. Note that we use a momentum direction $\vec{p}'=(k,k,0)$ with $ak=\theta\pi/N_x$.\label{twists}}
\begin{tabular}{c c c }\hline
Set &$\theta$ & $T/a$\\ \hline
1&0.0, 0.3859, 0.7718, 1.1577, 1.5436, 1.9295& 14,17,20\\
2&0.0, 0.8464, 1.6929, 2.5393, 3.3857, 4.2322& 22,25,28\\
3&0.0, 1.2596, 2.5192, 3.7788, 5.0384, 6.2981& 31,36,41\\
4&0.0, 0.7672, 1.5343, 2.3015, 3.0687, 3.8358& 14,17,20\\
5&0.0, 1.6929, 3.3857, 5.0786, 6.7715, 8.4643& 22,25,28\\\hline
\end{tabular}
\end{table}

\subsection{Correlator Fits}\label{corrfitting}
We fit the correlation functions in~\cref{threepointcorr} to exponentials, including time-oscillating terms as is typical when using staggered quarks~\cite{PhysRevD.75.054502,Parrott:2022dnu,Harrison:2020gvo,Harrison:2021tol,Cooper:2020wnj}:
\begin{align}\label{twopointfit}
{C}_\mathrm{2pt}^{D_s^*}(t,0) =&\sum_{i}\Big((A^i_n)^2{e^{-tE^{i}_n}}\nonumber+(-1)^{t}(A^i_o)^2{e^{-tE^i_o}}\Big),\\
{C}_\mathrm{2pt}^{H_s}(t,0) =&\sum_{i}\Big((B^i_n)^2{e^{-tM^{i}_n}}+(-1)^{t}(B^i_o)^2{e^{-tM^i_o}}\Big)
\end{align}
and
\begin{align}\label{threepointfit}
{C}_\mathrm{3pt}(T,t,0) &=\sum_{i,j}\Big({  A^i_n B^j_n J^{ij}_{nn} e^{-(T-t)E^{i}_n - tM^{j}_n} }\nonumber\\
+&{(-1)^{T+t}}  A^i_o B^j_n J^{ij}_{on} e^{-(T-t)E^i_o - tM^{j}_n} \nonumber\\
+&{(-1)^{t}}  A^i_n B^j_o J^{ij}_{no} e^{-(T-t)E^{i}_n - tM^j_o} \nonumber\\
+&{(-1)^{T}}  A^i_o B^j_o J^{ij}_{oo} e^{-(T-t)E^i_o - tM^j_o} \Big) .
\end{align}
Here $i$ and $j$ are integers corresponding to on-shell particle states of increasing energies, $A^i$ and $B^i$ are the amplitudes (together with relativistic normalisation factors) of the $D_{(s)}^*$ and $H_{(s)}$ operators respectively and $E_{i}$ and $M_{i}$ are their energies and masses. The time-oscillating terms, with subscript~`o' are a consequence of the use of staggered quarks; since our interpolating operators are only projected onto definite spatial momentum they may couple to `time doubled' states. The subscript $n$ indicates non-oscillating states. $J^{ij}_{nn}$ is then related to the matrix element of the current $\bar{c}\Gamma h(t)$ in~\cref{threepointcorr} between the non-oscillating states labelled $i$ and $j$. 
The ground state parameters are related to matrix elements as:
\begin{align}
A^0_n=&\frac{N_{D_{(s)}^*} }{\sqrt{2E_{D_{(s)}^*}}}\left(1+\frac{{\vec{p}}_{\nu}^{~\prime 2}}{M_{D_{(s)}^*}^2}\right)^{1/2},\nonumber\\
B^0_n=&\frac{N_{H_{(s)}}}{\sqrt{2M_{H_{(s)}}}}
\end{align}
where
\begin{align}
\langle 0 |\bar{l}\gamma^\nu c|D^*_{(s)}(p',\lambda)\rangle &= N_{D^*_{(s)}}\epsilon^\nu(p',\lambda),\nonumber\\
\langle {H_{(s)}}(p) |\bar{h}\gamma^5 c|0)\rangle &= N_{H_{(s)}}.
\end{align}
\begin{equation}
J^{00}_{nn(\nu,\Gamma)} = \sum_{\lambda}\frac{\epsilon^\nu(p',\lambda) \langle  D_{(s)}^*(p',\lambda ) |\bar{c}\Gamma h |H_{(s)}\rangle}{\sqrt{2E_{D_{(s)}^*}2M_{H_{(s)}}\left(1+{\vec{p}}_{\nu}^{~\prime 2}/M_{D_{(s)}^*}^2\right)}}\label{relnorm}
\end{equation}
where ${\vec{p}}~'_{\nu}$ is the $\nu$ component of the $D_{(s)}^*$ spatial momentum, with $\nu$ corresponding to the choice of polarisation in~\cref{threepointcorr}, with current $\overline{c}\Gamma h$.

\subsection{Extracting Form Factors}
\label{ExtractingFormFactors}
In order to extract the form factors from our correlator fits, we must use appropriate combinations of $D^*_{(s)}$ momentum, $p^\prime$, four-vector component, $\nu$, and current Dirac matrix, $\Gamma$, when computing correlation functions. These combinations must produce matrix elements corresponding to linearly independent combinations of form factors. In order to isolate $h_V$ and $h_{A_{1,2,3}}$ on each ensemble for each combination of $q^2$ and $am_h$ we use the same combinations of $\nu$ and $\Gamma$ as described in~\cite{Harrison:2020gvo,Harrison:2021tol}. We give the relation of these matrix elements to the form factors below. We work with the $H$ at rest throughout.

\subsubsection{Vector and Axial-Vector Form Factors}
As in~\cite{Harrison:2020gvo,Harrison:2021tol} we define $\Phi_\nu$, corresponding to the denominator in the right hand side of~\cref{relnorm}
\begin{align}
\Phi_\nu={\sqrt{2E_{D^*}2M_{H}\left(1+{\vec{p}}_{\nu}^{~\prime 2}/M_{D^*}^2\right)}}.
\end{align}
With this definition, together with~\cref{formfactors} and the completeness relation for the $D^*$ polarisation vectors $\sum_\lambda \epsilon^\mu\epsilon^{\nu*} = -g^{\mu\nu}+v^{'\mu}v^{'\nu}$, we have for the combinations listed in~\cref{spintastetableSM}
\begin{align}\label{SMmatels}
\tilde{J}^{00}_{nn(1,\gamma^{3})} =&ih_V \tilde{k},\nonumber\\
\tilde{J}^{00}_{nn(1,\gamma^{5})} =&\frac{M_{H}w\tilde{k} }{m_h+m_c}\Big(h_{A_1}(w+1)  \nonumber\\
&-h_{A_2}(1-rw) -h_{A_3}(w-r)\Big),\nonumber\\
\tilde{J}^{00}_{nn(3,\gamma^{3}\gamma^{5})} =&(1+w)h_{A_1},\nonumber\\   
\tilde{J}^{00}_{nn(1,\gamma^{1}\gamma^{5})}=&h_{A_1}(w+1)(1+\tilde{k}^2) - wh_{A_3}\tilde{k}^2,
\end{align}
where we have defined the reduced combination $\tilde{J}^{00}_{nn(\nu,\Gamma)}={J}^{00}_{nn(\nu,\Gamma)}\Phi_\nu/\sqrt{M_{H_{(s)}}M_{D_{(s)}^*}}$, $\tilde{k}=k/M_{D^*}$ and where $\vec{p}_{D^*}=\vec{p}'=(k,k,0)$.
Note that when converting between form factors and matrix elements, we use the masses obtained from the local spin-taste operators for the $D^*_{(s)}$ and $H_{(s)}$. Discretisation effects resulting from this choice only enter at the level of the taste splittings, which for heavy-light mesons using HISQ quarks are very small~\cite{PhysRevD.75.054502,PhysRevD.87.054505}, and will be consistently included in our chiral-continuum extrapolation along with other discretisation effects.
\subsubsection{Tensor Form Factors}
We now proceed to detail the procedure adopted here for isolating the tensor form factors.
For the tensor current the sum over $D^*$ polarisations in~\cref{relnorm} gives
\begin{align}
\tilde{J}^{00}_{nn(\kappa,\sigma^{\mu\nu})}=&\varepsilon^{\mu\nu\alpha\beta}\big[ h_{T_1}(\delta^\kappa_\alpha-v'^\kappa v'_\alpha)(v+v^\prime)_\beta \nonumber\\
&+h_{T_2}(\delta^\kappa_\alpha-v'^\kappa v'_\alpha)(v-v^\prime)_\beta \nonumber\\
&+h_{T_3}(v^\kappa-wv'^\kappa)v_\alpha v^\prime_\beta \big]·
\end{align}
We choose combinations of Lorentz indices for the tensor current and $D_{(s)}^*$ operator, $\mu\nu=12$ and $\kappa=3$, $\mu\nu=14$ and $\kappa=3$,  $\mu\nu=23$ and $\kappa=1$. These choices give
\begin{align}\label{tensormatels}
\tilde{J}^{00}_{nn(3,\sigma^{12})} =& h_{T_1}(1+w) +h_{T_2}(1-w),\nonumber\\
\tilde{J}^{00}_{nn(3,\sigma^{14})} =&\tilde{k}\big( h_{T_1} -h_{T_2} \big),\nonumber\\
\tilde{J}^{00}_{nn(1,\sigma^{23})} =~&h_{T_1}(1+w + \tilde{k}^2) \nonumber\\
~+&h_{T_2}(1-w + \tilde{k}^2) \nonumber\\
~-&h_{T_3}w\tilde{k}^2.
\end{align}

\subsubsection{Spin-Taste Operators}
We implement the meson interpolator and current operators as staggered spin-tastes operators. The combinations of spin-taste operators we use are given in \cref{spintastetableSM,spintastetableTENSOR}. These have been chosen so that the current operator is the local one for which the renormalisation factors were computed.
\begin{table}
\centering
\caption{Spin-taste operators used to isolate the SM form factors, $h_{V,A_{1,2,3}}$. The first column is the operator used for the $H_{(s)}$, the second for the $D_{(s)}^*$ and the third column is the operator used at the current. \label{spintastetableSM}}
\begin{tabular}{c | c c c }\hline
 &$\mathcal{O}_{H_{(s)}}$ & $\mathcal{O}_{D_{(s)}^*}$ & $\mathcal{O}_J$   \\
\hline
$\tilde{J}^{00}_{nn(1,\gamma^{3})}$ & $\gamma_0\gamma_5\otimes \gamma_0\gamma_5$ & $\gamma_1\otimes \gamma_1\gamma_2$ & $\gamma_3\otimes \gamma_3$  \\
$\tilde{J}^{00}_{nn(1,\gamma^{5})}$& $\gamma_5\otimes \gamma_5$ & $\gamma_1\otimes 1$ & $\gamma_5\otimes \gamma_5$  \\
$\tilde{J}^{00}_{nn(3,\gamma^{3}\gamma^{5})}$& $\gamma_5\otimes \gamma_5$ & $\gamma_3\otimes \gamma_3$ & $\gamma_3\gamma_5\otimes \gamma_3\gamma_5$ \\
$\tilde{J}^{00}_{nn(1,\gamma^{1}\gamma^{5})}$& $\gamma_5\otimes \gamma_5$ & $\gamma_1\otimes \gamma_1$ & $\gamma_1\gamma_5\otimes \gamma_1\gamma_5$ \\ \hline    
\end{tabular}
\end{table}                                
\begin{table}
\centering
\caption{Spin-taste operators used to isolate the tensor form factors $h_{T_{1,2,3}}$. The first column is the operator used for the $H_{(s)}$, the second for the $D_{(s)}^*$ and the third column is the operator used at the current. \label{spintastetableTENSOR}}
\begin{tabular}{c | c c c }\hline
 &$\mathcal{O}_{H_s}$ & $\mathcal{O}_{D_s^*}$ & $\mathcal{O}_J$   \\
\hline
$\tilde{J}^{00}_{nn(3,\sigma^{12})}$ & $\gamma_0\gamma_5\otimes\gamma_0 \gamma_5$ & $\gamma_3\otimes \gamma_3$ & $\gamma_1\gamma_2\otimes \gamma_1\gamma_2$  \\
$\tilde{J}^{00}_{nn(3,\sigma^{14})}$ & $\gamma_5\otimes \gamma_5$ & $\gamma_3\otimes \gamma_2\gamma_3$ & $\gamma_0\gamma_1\otimes \gamma_0\gamma_1$ \\
$\tilde{J}^{00}_{nn(1,\sigma^{23})}$ & $\gamma_0\gamma_5\otimes\gamma_0 \gamma_5$ & $\gamma_1\otimes \gamma_1$ & $\gamma_2\gamma_3\otimes \gamma_2\gamma_3$  \\ \hline    
\end{tabular}
\end{table}     

\subsection{Current Renormalisation}
The lattice currents used require renormalisation factors to match them to the continuum operators, and for the tensor current, we match to the $\overline{\mathrm{MS}}$ scheme. The axial-vector and vector current pieces, $Z_A$ and $Z_V$ respectively, are given in~\cref{Zfactors}. These were computed in~\cite{EuanBsDsstar} and~\cite{EuanBsDs} for sets 1, 2, 3 and 4. On set 5, we use the values from set 2, adding a conservative $1.0\%$ uncertainty motivated by the observed maximum change between sets 1 and 4 for a somewhat smaller difference in lattice spacings. The $Z^A$ and $Z^V$ values for $am_h=0.725$ on set 1 and $am_h=0.65$ on set 4 were obtained by interpolation from the other values for those sets, using the largest uncertainty of the other factors on that set. The tensor renormalisation factors, $Z_T$, were computed using an intermediate RI-SMOM scheme in~\cite{Hatton:2020vzp}. We use the factors computed at an intermediate scale of $\mu=2~\mathrm{GeV}$, and then run to $\mu=4.8~\mathrm{GeV}\sim m_b^\mathrm{pole}$, with the condensate correction applied. Since we are only interested in the $m_h=m_b$ point, we use $Z_T(4.8~\mathrm{GeV})$ on each lattice, rather than trying to estimate a value of $m_h^\mathrm{pole}$ to run $Z_T$ to for each $m_h$. The values of $Z_T$ are given in~\cref{ZTfactors}. Note that the tensor renormalisation factors were defined in the limit that the valence masses in lattice units are taken to zero, and as such are independent of $am_h$.

\begin{table}
\centering
\caption{ $Z$ factors from \cite{EuanBsDsstar} and \cite{EuanBsDs} for the axial-vector and vector operators used in this work, together with the discretisation corrections. $Z^A$ and $Z^V$ values for $am_h=0.725$ on set 1 and $am_h=0.65$ on set 4 were obtained by interpolation from the other values for those sets. The total renormalisation factor is given by $Z^{A(V)}Z^\text{disc}$. \label{Zfactors}}
\begin{tabular}{ c c c c c }
\hline
Set & $am_h$ & $Z^A$& $Z^V$& $Z^\text{disc}$ \\\hline
1 & 0.65 & 1.03740(58)& 1.0254(35) & 0.99635\\
 & 0.725 & 1.04030(58)& 1.0309(35) & 0.99491 \\
 & 0.8 & 1.04367(56)& 1.0372(32) & 0.99306 \\
\hline
2 & 0.427 & 1.0141(12)& 1.0025(31) & 0.99931 \\
 & 0.525 & 1.0172(12)& 1.0059(33) & 0.99859\\
 & 0.65 & 1.0214(12)& 1.0116(37) & 0.99697 \\
 & 0.8 & 1.0275(12)& 1.0204(46) & 0.99367 \\
\hline
3 & 0.5 & 1.00896(44)& 1.0029(38) & 0.99889 \\
 & 0.65 & 1.01363(49)& 1.0081(43) & 0.99704 \\
 & 0.8 & 1.01968(55)& 1.0150(49) & 0.99375 \\
\hline
4 & 0.65 & 1.03717(47)& 1.0229(29) & 0.99645 \\
 & 0.725 & 1.04037(47)& 1.0285(29) & 0.995 \\
 & 0.8 & 1.04390(39)& 1.0348(29) & 0.99315 \\
\hline
5 & 0.427 & 1.014(10)& 1.002(10) & 0.99931 \\
 & 0.525 & 1.017(10)& 1.006(11) & 0.99859\\
 & 0.65 & 1.021(10)& 1.012(11) & 0.99697 \\
 & 0.8 & 1.028(10)& 1.020(11) & 0.99367 \\
\hline
\end{tabular}
\end{table}

\begin{table}
\centering
\caption{ $Z_T(\mu=4.8~\mathrm{GeV})$ factors from~\cite{Hatton:2020vzp} for the tensor operators used in this work.\label{ZTfactors}}
\begin{tabular}{ c c }
\hline
Set & $Z_T$ \\\hline
1&1.0029(43)\\
2&1.0342(43)\\
3&1.0476(42)\\
4&1.0029(43)\\
5&1.0342(43)\\\hline
\end{tabular}
\end{table}

\subsection{Correlator Fit Parameters}
We perform correlator fits of our lattice data to~\cref{twopointfit,threepointfit} using the \textbf{corrfitter} python package \cite{corrfitter}.
Our fits are done to all correlation functions simultaneously.

The prior values and uncertainties of the fit parameters that we use here are very similar to those used in~\cite{Harrison:2021tol}, with only small differences in the heuristic forms chosen for the $m_h$ dependance of $M_{H_{(s)}}$ and $M_{D_{(s)}^*}$. For ground-state priors we take $E^{D^*_{(s)}}_0=\sqrt{M_{D_{(s)}^*}^2+2k^2}\times 1(0.3)~\mathrm{GeV}$ and $M^{H_{(s)}}_0=(M^{H_s}_\mathrm{max}+m_h-0.8)\times 1(0.3)~\mathrm{GeV}$. Here we use $M_{D_s^*}=M_{D^*}+m_s$, where $m_s$ is the mass in GeV of the valence strange quark given in~\cref{valmasses}. For $M^{H_s}_\mathrm{max}$ we use the value of $M_{H_s}$ from \cite{EuanBsDs} corresponding to the largest value of $am_h=0.8$. Note that our priors for $H$ and $H_s$ masses have the same central value and uncertainty, and we use separate priors with equal central values and uncertainties for the energies and amplitudes of meson operators in different taste multiplets. Our priors for the lowest oscillating state energies, as well as amplitudes, are given in~\cref{priortable}. For the matrix elements, $J^{ij}_{n(o)n(o)}$, we take priors $0(1)$ for all except those proportional to $ak$. For these, we first divide by $ak$ before fitting, since $ak$ is known exactly from the twists~(\cref{twists}). We increase the uncertainty on the corresponding priors for the oscillating state matrix elements $J^{ij}_{no}$, $J^{ij}_{oo}$, and $J^{ij}_{oo}$ by a factor of 4 relative to $J^{ij}_{nn}$ to account for this rescaling, and take priors of $0(4)$.

\begin{table}
\centering
\caption{Correlator fit priors. We take $\Delta E^{(o)}_i=\Lambda_\mathrm{QCD}\times 1.0(0.75)$ where $\Delta E^{(o)}_i = E^{(o)}_{i+1}-E^{(o)}_{i},~i\geq 0$ and here for our correlator fits we take $\Lambda_\mathrm{QCD}=0.75~\mathrm{GeV}$. In the table we have defined $\Omega_{H_{(s)}}=M^{H_s}_\mathrm{max}+m_h-0.8$ and $\Omega_{D_{(s)}^*}=\sqrt{M_{D_{(s)}^*}^2+2k^2}$ following the relativistic dispersion relation.\label{priortable}}
\begin{tabular}{ c c c c c }\hline
 Prior & $D_{(s)}^*(k)$ & $H_s$\\\hline
$E_n^0/\mathrm{GeV}$	 	&$\Omega_{D_{(s)}^*}\times 1.0(0.3)$ 	&$\Omega_{H_{(s)}}\times 1(0.3)$ 	\\
$E^0_o/\mathrm{GeV}$		&$\Omega_{D_{(s)}^*}\times 1.2(0.5)$	&$\Omega_{H_{(s)}}\times 1.2(0.5)$		\\
$A(B)^{n(o)}_i$			&$0.1(5.0)$			&$0.1(5.0)$\\\hline
\end{tabular}
\end{table}

In order to fit our data simultaneously, it is necessary to implement an SVD cut~(see Appendix D of~\cite{Dowdall:2019bea}). The size of the SVD cut used on each lattice was chosen based on the values used in~\cite{Harrison:2021tol}, though note that by omitting the highly correlated $\eta_c$ and $\eta_h$ correlator data, as well as by only partially binning over time sources as discussed in~\cref{ntbinning}, we are able to use smaller SVD cuts, resulting in more stable fits. We also omit correlator data points close to the source and sink operators that contain significant excited state contamination. These data points are not included when computing correlations, further helping to improve resolution of the covariance matrix for the correlator data and reducing the size of the required SVD cut. The number of data points excluded from close to the source and sink operators are given in~\cref{fitparams}, together with the number of exponentials included in~\cref{twopointfit,threepointfit}. \cref{fitparams} also includes the value of $\chi^2/\mathrm{dof}$ estimated using prior and SVD noise as in~\cite{corrfitter}, following~\cite{Harrison:2020gvo,Harrison:2021tol}. In~\cref{stabilitytests} we investigate the effect of using different combinations the fit parameters in~\cref{fitparams}. We find that our results are very stable to changes in $\Delta T$ and the choice of SVD cut.

\begin{table}
\centering
\caption{Details of fit parameters. $\Delta T$ indicates the number of data points at the extremities of correlation functions not included in the fit, and $n_\mathrm{exp}$ is the number of non-oscillating and time-oscillating exponentials included in our correlator fits to~\cref{twopointfit,threepointfit}. \label{fitparams} $\chi^2/\mathrm{dof}$ is estimated by introducing SVD and prior noise as in~\cite{corrfitter}. We use the fit parameters in bold for our subsequent analysis. $\delta$ is a label for the other fits that we will use later in~\cref{stabilitytests} to investigate the sensitivity of our final results to these parameters.}
\begin{tabular}{ c c c c c c c c }\hline
Set & $n_\mathrm{exp}$ & $\Delta T_\mathrm{3pt}$ & $\Delta T_\mathrm{2pt}^{D_{(s)}^*}$& $\Delta T^{H_{(s)}}_\mathrm{2pt}$ & SVD cut & $\chi^2/\mathrm{dof}$ & $\delta$ \\ \hline
1  & \textbf{3} & \textbf{2} & \textbf{4} & \textbf{4} & \textbf{0.005} & \textbf{1.02} & 0\\
   & {3} & {3} & {6} & {6} & {0.005} & {0.99} & 1\\
   & {3} & {2} & {4} & {4} & {0.001} & {1.04} & 2\\\hline
2  & \textbf{3} & \textbf{4} & \textbf{9} & \textbf{9} & \textbf{0.005} & \textbf{1.01}& 0\\
   & {3} & {4} & {9} & {9} & {0.001} & {1.05} & 1\\
   & {3} & {4} & {8} & {8} & {0.005} & {1.04} & 2\\\hline
3  & \textbf{3} & \textbf{6} & \textbf{12} & \textbf{12} & \textbf{0.001} & \textbf{1.01}& 0\\
   & {3} & {5} & {11} & {11} & {0.001} & {1.02} & 1\\
   & {3} & {6} & {12} & {12} & {0.0005} & {1.07} & 2\\\hline
4  & \textbf{3} &  \textbf{2} & \textbf{4} & \textbf{4} & \textbf{0.01} & \textbf{1.02}& 0\\
   & {3} & {2} & {5} & {5} & {0.01} & {1.03} & 1\\
   & {3} & {2} & {4} & {4} & {0.005} & {1.02} & 2\\\hline
5  & \textbf{3} &  \textbf{5} & \textbf{10} & \textbf{10} & \textbf{0.001} & \textbf{1.1}& 0\\
   & {3} & {5} & {10} & {10} & {0.005} & {1.1} & 1\\
   & {3} & {4} & {8} & {8} & {0.001} & {1.1} & 2\\\hline
\end{tabular}
\end{table}

\section{Results}
\label{Results}

In this section we give the numerical results from the correlator fits described in~\cref{corrfitting}. We then describe our extrapolation of the form factors to the physical-continuum point. We demonstrate that our physical-continuum form factors are insensitive to reasonable changes to our fitting and extrapolation procedure, then we provide a breakdown of the sources of uncertainty entering the form factors across the kinematic range of the decay.

\subsection{Correlator Fit Results}
\begin{table}
\centering
\caption{$D^*$ masses for the local spin-taste operator $\gamma_1\otimes \gamma_1$ and $1-$link operators $\gamma_1\otimes 1$ and $\gamma_1\otimes \gamma_1\gamma_2$ used in our calculation, see~\cref{spintastetableSM,spintastetableTENSOR}. \label{charmMasses1}}
\begin{tabular}{ c c c c c }
\hline
& $aM_{D^*}$ & & &\\\hline
Set & $\gamma_1\otimes \gamma_1$&  $\gamma_1\otimes 1$ & $\gamma_1\otimes \gamma_1\gamma_2$  \\\hline
1&0.9289(26)	&0.9292(31)	&0.9277(34)	\\
\hline
2&0.6110(25)	&0.6110(36)	&0.6108(37)	\\
\hline
3&0.4556(14)	&0.4536(21)	&0.4551(18)	\\
\hline
4&0.8949(42)	&0.8954(53)	&0.8953(49)	\\
\hline
5&0.5829(49)	&0.5823(73)	&0.5790(75)	\\
\hline
\end{tabular}
\end{table}

\begin{table}
\centering
\caption{$D_s^*$ masses for the local spin-taste operator $\gamma_1\otimes \gamma_1$ and $1-$link operators $\gamma_1\otimes 1$ and $\gamma_1\otimes \gamma_1\gamma_2$ used in our calculation, see~\cref{spintastetableSM,spintastetableTENSOR}. \label{charmMasses2}}
\begin{tabular}{ c c c c c }
\hline
& $aM_{D_s^*}$ & & &\\\hline
Set & $\gamma_1\otimes \gamma_1$&  $\gamma_1\otimes 1$ & $\gamma_1\otimes \gamma_1\gamma_2$  \\\hline
1&0.96499(76)	&0.9649(11)	&0.9644(13)	\\
\hline
2&0.6349(12)	&0.6348(15)	&0.6346(16)	\\
\hline
3&0.47183(68)	&0.47155(85)	&0.47202(75)	\\
\hline
4&0.93970(62)	&0.93952(91)	&0.93964(93)	\\
\hline
5&0.6075(12)	&0.6084(13)	&0.6078(13)	\\
\hline
\end{tabular}
\end{table}

\begin{table}
\centering
\caption{$H_{(s)}$ masses for the local spin-taste operators $\gamma_5\otimes \gamma_5$ and $\gamma_0\gamma_5\otimes \gamma_0\gamma_5$ that we use in our calculation, see~\cref{spintastetableSM,spintastetableTENSOR}. \label{Massesheavy}}
\begin{tabular}{ c c | c c | c c }
\hline
&  & $\gamma_5\otimes \gamma_5$& & $\gamma_0\gamma_5\otimes \gamma_0\gamma_5$ & \\\hline
Set & $am_h$ & $aM_{H}$& $aM_{H_s}$& $aM_{H}$& $aM_{H_s}$\\\hline
1&	0.65	&1.08972(80)	&1.12504(26)	&1.0894(13)	&1.12556(46)	\\
&	0.725	&1.16928(88)	&1.20424(28)	&1.1689(14)	&1.20472(48)	\\
&	0.8	&1.24659(95)	&1.28127(29)	&1.2462(15)	&1.28172(50)	\\
\hline
2&	0.427	&0.7510(13)	&0.77418(55)	&0.7499(22)	&0.77410(70)	\\
&	0.525	&0.8617(15)	&0.88450(62)	&0.8607(24)	&0.88452(76)	\\
&	0.65	&0.9969(18)	&1.01962(71)	&0.9962(26)	&1.01976(84)	\\
&	0.8	&1.1516(21)	&1.17452(81)	&1.1513(29)	&1.17477(93)	\\
\hline
3&	0.5	&0.78656(92)	&0.80250(31)	&0.7863(13)	&0.80249(41)	\\
&	0.65	&0.9482(11)	&0.96383(38)	&0.9479(15)	&0.96372(50)	\\
&	0.8	&1.1021(14)	&1.11777(46)	&1.1019(17)	&1.11761(59)	\\
\hline
4&	0.65	&1.0783(15)	&1.12007(22)	&1.0768(23)	&1.12053(40)	\\
&	0.725	&1.1578(16)	&1.19923(23)	&1.1564(24)	&1.19966(43)	\\
&	0.8	&1.2352(17)	&1.27624(25)	&1.2338(26)	&1.27663(46)	\\
\hline
5&	0.427	&0.7440(22)	&0.76937(35)	&0.7419(44)	&0.76975(49)	\\
&	0.525	&0.8548(27)	&0.87950(44)	&0.8527(48)	&0.87994(58)	\\
&	0.65	&0.9902(32)	&1.01437(57)	&0.9882(52)	&1.01489(69)	\\
&	0.8	&1.1452(37)	&1.16898(74)	&1.1435(56)	&1.16958(82)	\\

\hline
\end{tabular}
\end{table}

The ground state $D^*$ and $D^*_s$ masses resulting from our correlator fits are given in~\cref{charmMasses1,charmMasses2}, where we see some changes compared to~\cite{Harrison:2021tol} on set 3 on the order of~$\approx 1.5\sigma$. Such changes are not surprising, and are a result of the exclusion of highly correlated $\eta_c$ data, as well as the inclusion of additonal $D_s^*$ polarisations and $D^*$ data, together with the improved resolution of the covariance matrix as discussed in~\cref{ntbinning}. We note that our fit results for the $D_s^*$ masses on set 3 are in good agreement with those given in~\cite{EuanBsDsstar}, which included a much smaller set of correlators and so had better resolution of the data covariance matrix. The $H_{(s)}$ masses are given in~\cref{Massesheavy} in lattice units, where we see good agreement with those in~\cite{Harrison:2021tol,EuanBsDsstar}.

\begin{figure*}
\centering
\includegraphics[scale=0.075]{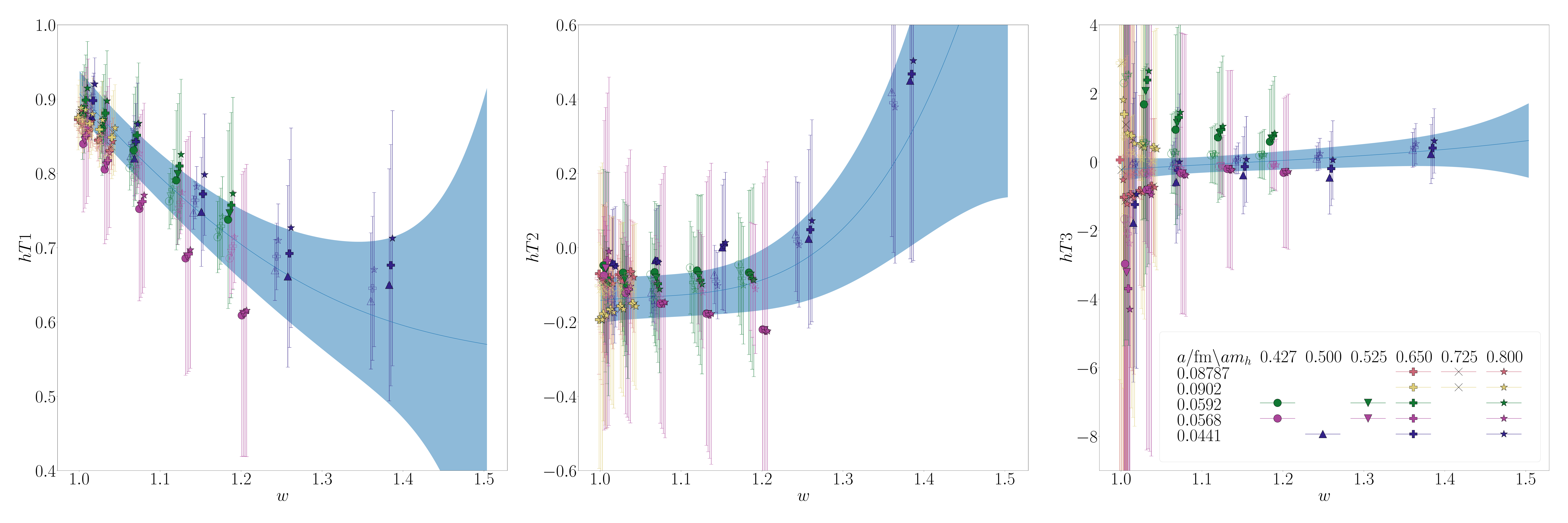}
\caption{\label{TFFqsq} The points show our lattice QCD results for each tensor
form factor as given in~\cref{l_T_set1,l_T_set2,l_T_set3,l_T_set4,l_T_set5} for $B\to D^*$ (filled points) and~\cref{s_T_set1,s_T_set2,s_T_set3,s_T_set4,s_T_set5} for $B_s\to D_s^*$ (empty points)  as a function of the recoil parameter, $w$. 
The legend gives the mapping between symbol colour and shape and the 
set of gluon field configurations used, as given by the lattice 
spacing, and the heavy quark mass in lattice units (see~\cref{tab:gaugeinfo,valmasses}). 
The blue curve with error band is the result of our fit in the continuum 
limit and with the physical $b$ quark mass. Note that we include the data points for both the $B_s\to D_s^*$ and $B\to D^*$ form factors, and that for clarity data points at fixed $w$ for  different values of $m_h$ are offset a small amount. 
}
\end{figure*}

The full set of numerical results for the SM form factors for $B\to D^*$ are given in~\cref{l_SM_set1,l_SM_set2,l_SM_set3,l_SM_set4,l_SM_set5} and in~\cref{s_SM_set1,s_SM_set2,s_SM_set3,s_SM_set4,s_SM_set5} for $B_s\to D_s^*$ in~\cref{lattdat}. There, the tensor form factors for $B\to D^*$ are also given in~\cref{l_T_set1,l_T_set2,l_T_set3,l_T_set4,l_T_set5} and in~\cref{s_T_set1,s_T_set2,s_T_set3,s_T_set4,s_T_set5} for $B_s\to D_s^*$. Note that $h_{T_3}$ is particularly noisy, owing to the factor of $k^2$ appearing in~\cref{tensormatels}. These data points are shown in~\cref{SMFFqsq,TFFqsq}, where we also show the $B\to D^*$ form factors extrapolated to the physical-continuum point.

\subsection{Physical-Continuum Extrapolation}
\label{physextrap}

\begin{figure*}
\centering
\includegraphics[scale=0.11]{SMFFstogetherw}
\caption{\label{SMFFqsq} The points show our lattice QCD results for each SM
form factor as given in~\cref{l_SM_set1,l_SM_set2,l_SM_set3,l_SM_set4,l_SM_set5} for $B\to D^*$ (filled points) and~\cref{s_SM_set1,s_SM_set2,s_SM_set3,s_SM_set4,s_SM_set5} for $B_s\to D_s^*$ (empty points) as a function of the recoil parameter, $w$. 
The legend gives the mapping between symbol colour and shape and the 
set of gluon field configurations used, as given by the lattice 
spacing, and the heavy quark mass in lattice units (see~\cref{tab:gaugeinfo,valmasses}). 
The blue curve with error band is the result of our fit in the continuum 
limit and with the physical $b$ quark mass for $B\to D^*$. Note that we include the data points for both the $B_s\to D_s^*$ and $B\to D^*$ form factors, and that for clarity data points at fixed $w$ for  different values of $m_h$ are offset a small amount.
}
\end{figure*}

In order to determine the physical-continuum form factors we must fit our lattice form factor data to an appropriate function describing its kinematic and physical $m_h$ dependence, as well as discretisation effects and quark mass mistuning effects. At the physical-continuum point with $m_h=m_b$, the BGL parameterisation is often used to describe the kinematic dependence of the form factors in the helicity basis, with the BGL coefficients guaranteed to be between $-1$ and $1$ by unitarity constraints. However, the BGL parameterisation (see \cref{sec:bglparams} for details) depends on the masses of several mesons containing a $b$ quark, as well as susceptibilities which also depend on the $b$ quark mass and are computed perturbatively. This makes it impractical for our purposes to use it here, where we require our fit function to describe the $m_h$ dependence of our form factors.

Instead we use a more straightforward power series in $(w-1)$, $(\Lambda_\mathrm{QCD}/m_h)$ and $\delta_{m_q}$ to parameterise the continuum HQET form factors. Using a power series in $(w-1)$ to describe the kinematic dependence of the form factors allows us to describe the physical $m_h$ dependence away from the point $m_h=m_b$ as modifications to the coefficients. These appear as multiplicative corrections, in powers of $(\Lambda_\mathrm{QCD}/m_h)$ motivated by HQET. However, we must be careful to choose prior widths for our coefficients that do not overly constrain the shape of the form factors. In order to set our priors for the physical-continuum coefficient of each power of $(w-1)$, we make use of the physical-continuum BGL expansion~\cite{Boyd:1997kz} at $m_h=m_b$ where the masses and susceptibilities are well known. We can then compute each physical-continuum $(w-1)$ coefficient in terms of the physical-continuum BGL expansion coefficients, and use priors for the physical-continuum BGL coefficients directly, choosing prior widths motivated by the unitarity bounds.

To compute the physical-continuum $(w-1)$ coefficients of the HQET form factors, we start with the physical-continuum BGL parameterisation of the helicity basis form factors at the $m_h=m_b$ point, which we then convert to the HQET basis. The BGL parameterisation is given in terms of $z$, mapped from $q^2$ (see~\cref{bgldefinition}). We set $t_0=q^2_\mathrm{max}$ in this mapping and then expand $z$, the Blaschke factors $P(z)$, and outer functions $\phi(z)$ appearing in the BGL expansion in powers of $(w-1)$.
This provides a linear map, which we call $M^{\mathrm{BGL}\to\mathrm{HQET}}_{nm,YX}$, from the physical-continuum BGL coefficients for the helicity basis form factors, to each physical-continuum $(w-1)^n$ coefficient for the HQET form factors. Note that since the BGL expansion describes the form factors in the helicity basis, we must explicitly impose the kinematical constraints $F_1(w=1)=M_B(1-r)f(w=1)$ and $F_2(w_\mathrm{max})=(1+r)/(M_B^2(1+w_\mathrm{max})(1-r)r) F_1(w_\mathrm{max})$ in order to convert to the HQET basis consistently. This is done by fixing the zeroth order BGL coefficient of $F_1$ and $F_2$ in terms of the remaining coefficients such that the constraints are satisfied. We follow the conventions for masses and resonances entering the BGL expansion given in~\cite{Bigi:2017jbd}, athough we have checked that other choices do not significantly impact the mapping to $(w-1)$ coefficients. We use Gaussian priors for the BGL coefficients of $0\pm 5$, which are very conservative compared to the unitarity constraints which force them to be less than 1.

Since the $z$ expansion converges quickly owing to the small size of $z$, we include only up to $z^4$ in the $z$ expansion. When we look at the numerical values appearing in  $M^{\mathrm{BGL}\to\mathrm{HQET}}_{nm,YX}$ we see that some are substantially greater than 1. For instance, the coefficient of $(w-1)^5$ for $h_{A_1}$ includes a term $\approx -50 a_0^{f~\mathrm{BGL}}$, where $ a_0^{f~\mathrm{BGL}}$ is the leading ($z^0$) coefficient in the BGL expansion for the form factor $f$. For $a_0^{f~\mathrm{BGL}}\sim \mathcal{O}(1)$ this would give a contribution of $\mathcal{O}(1)$ to the form factor close to $w^\mathrm{max}$ where we have lattice data. In order to ensure that we do not bias our fit to small values of the BGL coefficients, it is therefore important that we go to sufficiently high order in $(w-1)$. We find that the $(w-1)^{10}$ coefficients for any of the HQET form factors give a maximum contribution of $\mathcal{O}(0.01)$ for $\mathcal{O}(1)$ BGL coefficients close to $w^\mathrm{max}$. This is an order of magnitude smaller than the uncertainties on our lattice data points in this region, and so we truncate the power series in $(w-1)$ at order 10.

The $(\Lambda_\mathrm{QCD}/m_h)$, and $\delta_{m_q}$ polynomial terms are then included as modifications to the continuum $(w-1)$ coefficients. Note that for the tensor form factors, since there is currently no equivalent BGL expansion available in the literature, we instead use Gaussian priors of $0\pm 20$ for each $(w-1)$ coefficient.

Additionally, our fit function must describe the pion mass dependence of our form factor data, including logarithms determined from staggered chiral perturbation theory~\cite{Chow:1993hr,Laiho:2005ue}. The staggered chiral logarithms for the SM form factors were given in~\cite{FermilabLattice:2021cdg}. Following the methods in~\cite{Chow:1993hr,Aubin:2003mg,Laiho:2005ue} we find that the staggered chiral logarithms for the tensor form factors for $B\to D^{*}$ are related straightforwardly to those for the SM form factors, with $\mathrm{logs}_{SU(3)}^{h_{T_1}^{(s)}}=\mathrm{logs}_{SU(3)}^{h_{A_1}^{(s)}}$, $\mathrm{logs}_{SU(3)}^{h_{T_3}^{(s)}}=\mathrm{logs}_{SU(3)}^{h_{A_2}^{(s)}}$ and $\mathrm{logs}_{SU(3)}^{h_{T_2}^{(s)}}=0$ to 1-loop. For completeness, we also compute the logarithms for $B\to D$ and find that $\mathrm{logs}_{SU(3)}^{f_T}=\mathrm{logs}_{SU(3)}^{f_+}$. Full expressions for $\mathrm{logs}_{SU(3)}^{Y^{(s)}}$ are given in~\cref{chilogs}. We also include polynomial terms in $(M_\pi/\Lambda_\chi)^2$ in our fit form, contained in $\delta_\chi^{(s)}$ in~\cref{eq:anfitform}, where we take $\Lambda_\chi=1~\mathrm{GeV}$. We then fit our $B\to D^*$ and $B_s\to D_s^*$ data together by taking $M_\pi\to M_K$, $M_K\to M_S$, swapping $M_U\leftrightarrow M_S$ in the taste-axial-vector and taste-vector hairpin terms and suitably modifying the flavour-neutral taste-singlet terms. We label the form factors and observables corresponding to $B_s\to D_s^*$ with a superscript `$s$', ${Y^{s}}$. We use the taste splittings determined in~\cite{PhysRevD.87.054505} together with the relations given in~\cite{Laiho:2005ue} for the flavour-neutral pion mass eigenstates. We assume that the taste splittings behave as $M_{\pi_\xi}^2-M_{\pi_5}^2= n_\xi a^2\delta_t$, where $n_A=1$, $n_T=2$, $n_V=3$ and $n_I=4$, and use the value for $\xi=A$ to determine $\delta_t$. Note that on set 3 we use the observation that $\delta_{t}\propto a^2$ to fix the taste splitting, $a^2\delta_t$, to be $0.31$ times that on set 2. We assume that the taste splittings are equal on sets 1 and 4, and on sets 2 and 5 respectively. We use the relation $\delta_A^\prime=\delta_V^\prime=-\delta_t$, which was found to be a good approximation for HISQ~\cite{PhysRevD.88.074504,Colquhoun:2015mfa}, to fix the hairpin coefficients.

\begin{figure}
\centering
\includegraphics[scale=0.225]{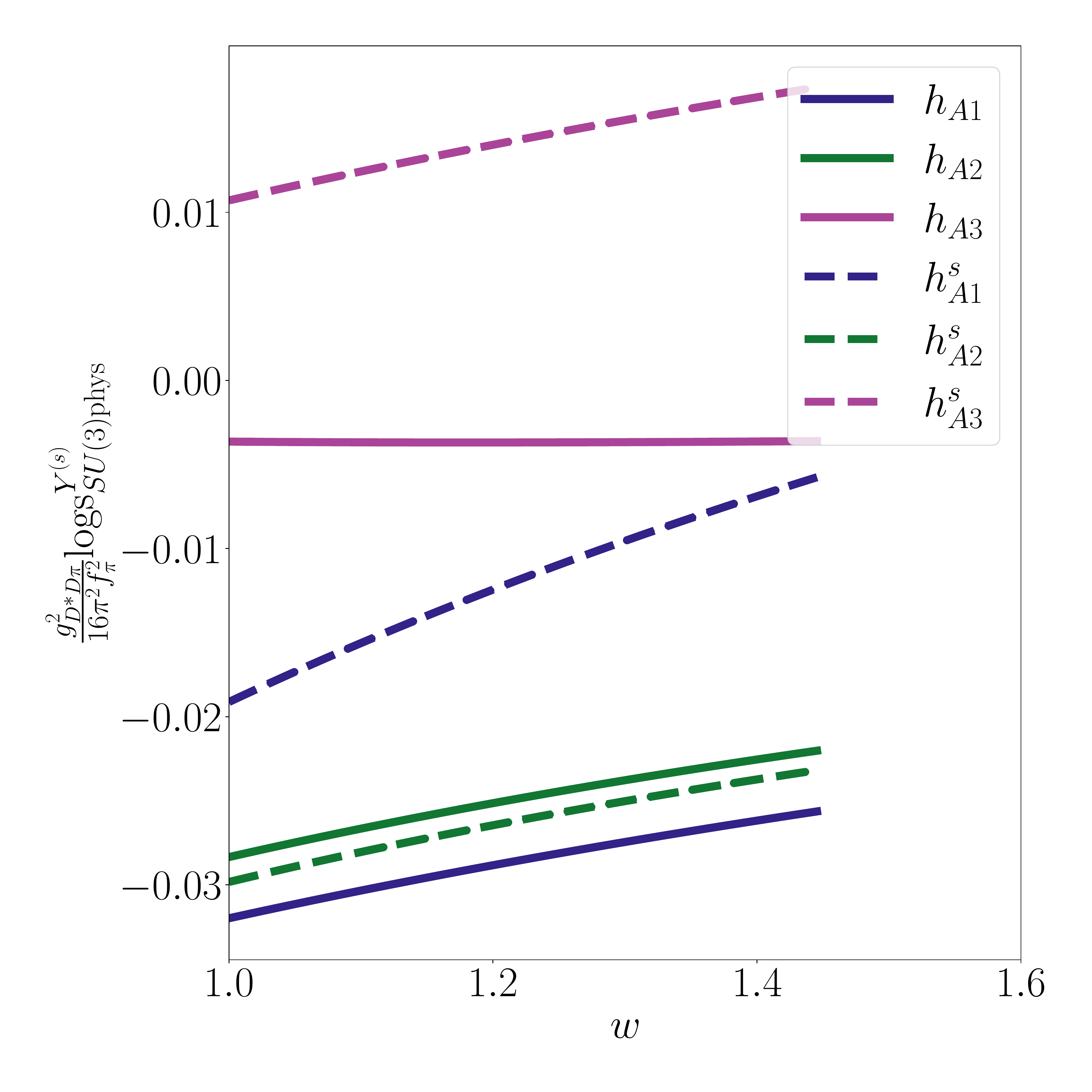}
\caption{\label{logsplot} The physical value of the logs in~\cref{fitfunctionequation}, $\frac{g_{D^*D\pi}^2}{16\pi^2f^2_\pi }\mathrm{logs}_{SU(3)}^{Y^{(s)}}$, plotted for $g_{D^*D\pi}=0.53$, illustrating the $w$ and chiral dependence of our fit function. It can be seen that the log term varies slowly with $w$ relative to the $(w-1)^n$ terms in our fit, and so we expect the subtraction of the physical $B\to D^*$ logs in~\cref{fitfunctionequation} to only slightly modify the coefficients of the $(w-1)^n$ terms.}
\end{figure}

Our fit function takes the explicit form 
\begin{align}
F^{Y^{(s)}}(w) = \sum_{n=0}^{10}& a^{Y^{(s)}}_n (w-1)^n \mathcal{N}^{Y^{(s)}}_n \nonumber\\
&+ \frac{g_{D^*D\pi}^2}{16\pi^2f^2_\pi }\left(\mathrm{logs}_{SU(3)}^{Y^{(s)}}-\mathrm{logs}_{SU(3)\mathrm{phys}}^{Y^{(s)}}\right)\label{fitfunctionequation},
\end{align}
where $g_{D^*D\pi}$ is the $D^*\to D\pi$ coupling, which is the same for $B\to D^*$ and $B_{s}\to D_{s}^*$ at the order to which we work in chiral perturbation theory. We take $f_\pi=130\mathrm{MeV}$ and neglect the uncertainty in $f_\pi$, since the uncertainty of the overall coefficient of the logs is dominated by that of $g_{D^*D\pi}$. Note that we subtract the physical point logarithms for $B\to D^*$ and $B_{s}\to D_{s}^*$ in each case, this ensures that at the physical point our fit function for $B\to D^*$ reduces to a polynomial in $(w-1)$. The physical chiral logs entering our fit function depend only mildly on $w$, as illustrated in~\cref{logsplot}, and so we expect the subtraction of the physical logs to only slightly modify the coefficients of the $(w-1)^n$ terms. We use the values of $M_\pi$ computed in~\cite{Bazavov:2017lyh} given in~\cref{tab:gaugeinfo} and $M_K$ computed in~\cite{Chakraborty:2021qav} for sets 1-4. On set~5 we determine $M_K=493~\mathrm{MeV}$ from independent correlator fits and take the physical values to be ${M_{\pi}^\mathrm{phys}}=139.6~\mathrm{MeV}$, ${M_{K}^\mathrm{phys}}=493.7~\mathrm{MeV}$ and ${M_{\eta}^\mathrm{phys}}=547.9~\mathrm{MeV}$.

The coefficients, $a^Y_n$, for each form factor take the form
\begin{align}
\label{eq:anfitform}
a^{Y^{(s)}}_n =& \alpha^{Y}_n\nonumber\\
\times \Big(1&+\sum_{j\neq 0}^3 b_n^{Y,j}\Delta_{h}^{(j)} +\delta_\chi^{(s)}\sum_{j=0}^3\tilde{b}_n^{Y^{(s)},j}\Delta_{h}^{(j)} \Big),
\end{align}
where
\begin{equation}
\delta_\chi^{(s)}=\left(\frac{M_{\pi(K)}}{\Lambda_\chi}\right)^2-\left(\frac{M_{\pi}^\mathrm{phys}}{\Lambda_\chi}\right)^2
\end{equation}
allows for up to $\approx 25\%$ difference between the $B\to D^*$ and $B_s\to D_s^*$ form factors. For $Y=h_{A_1},h_{A_2},h_{A_3},h_{V}$ the $(w-1)^n$ coefficient, $\alpha^{Y}_n$, is given by
\begin{align}
\alpha^{Y}_n = \sum_{\substack{m=0,\\X=f,F_1,F_2,g}}^4 M^{\mathrm{BGL}\to\mathrm{HQET}}_{nm,YX} a^{X,\mathrm{BGL}}_m
\end{align}
with $M$ the linear mapping from the continuum BGL $z$ expansion parameterisation to the expansion in powers of $(w-1)$ and $a^{X,\mathrm{BGL}}_m$ the BGL $z^m$ coefficient for form factor $X$.
Note that because of Luke's theorem~\cite{LUKE1990447} we set the coefficients $b_0^{h_{A_1},1}$ and $\tilde{b}_0^{h_{A_1},1}$, corresponding to the zero recoil continuum $\Lambda/m_h$ term, equal to zero. The $\Delta_{h}^{(j)}$ allow for the dependence on the heavy quark mass. Here, we use the $H_{s}$ mass as a proxy for the heavy quark mass. Note that $w_0$ and $w_0/a$, which are used to determine the lattice spacing on each set, are included as priors.
We use $\Delta_{h}^{(0)}=1$ and
\begin{equation}
\Delta_{h}^{(j\neq 0)}=\left(\frac{\Lambda}{M_{H_s}}\right)^j-\left(\frac{\Lambda}{M_{B_s}^\mathrm{phys}}\right)^j.
\end{equation}
We take the physical value of the $B_s$ mass to be $M_{B_s}=5.36688~\mathrm{GeV}$~\cite{pdg20} and we take $\Lambda = 0.5 ~\mathrm{GeV}$. 

The remainder of~\cref{fitfunctionequation}, $\mathcal{N}^{Y^{(s)}}_n$, takes into account the small mistuning of the valence and sea quark masses.
\begin{equation}
\mathcal{N}^{Y^{(s)}}_n = 1 + A^{Y^{(s)}}_n \delta_{m_c}^\mathrm{val}+ B^{Y^{(s)}}_n \delta_{m_c}^\mathrm{sea}+ C^{Y^{(s)}}_n \delta_{m_s}^\mathrm{val}+ D^{Y^{(s)}}_n \delta_{m_s}^\mathrm{sea}
\end{equation}
with
\begin{align}
\delta_{m_c}^\mathrm{val} = (am_c^\mathrm{val}-am_c^\mathrm{tuned})/am_c^\mathrm{tuned},\nonumber\\
\delta_{m_c}^\mathrm{sea} = (am_{c}^\mathrm{sea}-am_c^\mathrm{tuned})/am_c^\mathrm{tuned},\nonumber\\
\delta_{m_s}^\mathrm{val} = (am_{s}^\mathrm{val} - am_{s}^\mathrm{tuned})/(10am_{s}^\mathrm{tuned}),\nonumber\\
\delta_{m_{s}}^\mathrm{sea} = (am_{s}^\mathrm{sea} - am_{s}^\mathrm{tuned})/(10am_{s}^\mathrm{tuned}).\label{deltatermseq}
\end{align}
Note that $C^{Y}_n =0$ so that the valence strange quark mistuning term is only included for the $B_s\to D_s^*$ case. The tuned values of the quark masses are given by
\begin{equation}
am_c^\mathrm{tuned} = am_c^\mathrm{val}\frac{M_{D_s^*}^\mathrm{phys}}{M_{D_s^*}},
\end{equation}
and
\begin{equation}
am_s^\mathrm{tuned} = am_s^\mathrm{val}\left(\frac{M_{\eta_s}^\mathrm{phys}}{M_{\eta_s}}\right)^2
\end{equation}
$M_{D_s^*}$ on each set is given in lattice units in~\cref{charmMasses2} and we use the values of $M_{\eta_s}$ given in~\cite{EuanBsDsstar} 
which used the same values of $am_s^\mathrm{val}$. We determine $M_{\eta_s}$ on set 5 from independent correlator fits to be $0.19824(8)$. Since the $\eta_s$ masses are only used to determine the strange quark mistuning, and because they are very precise, we neglect their correlations with our other data. We take priors of $0(1)$ for each $b_n$ and $\tilde{b}_n$. We also use priors of $0.0(0.1)$ for $B^{Y^{(s)}}_n$, motivated by the results of the analysis 
of $m_c^\mathrm{sea}$ effects on $w_0$ in \cite{PhysRevD.91.054508}. 
We take priors of $0.0(0.5)$ for $D^{Y^{(s)}}_n$ for each form factor, since sea quark mistuning effects enter at 1-loop. We take a prior for $g_{D^*D\pi}$ of $0.53(8)$, following~\cite{FermilabLattice:2021cdg}.

Discretisation effects enter our lattice calculation at the level of matrix elements. It is therefore important to account for them at this level, rather than at the level of the form factors, where cancellations may cause them to be underestimated. To do this, we convert the continuum form factors given by~\cref{fitfunctionequation} to the matrix elements $J_\mathrm{phys}^{\nu,\Gamma{(s)}}\equiv{J}^{00}_{nn(\nu,\Gamma)}$ given in~\cref{SMmatels,tensormatels} and allow for discretisation effects in this quantity. We then perform the fit against the matrix elements directly, simultaneously for the different combinations listed in~\cref{spintastetableSM,spintastetableTENSOR}, including discretisation effects using the fit form
\begin{align}\label{cterms}
J_\mathrm{latt}^{\nu,\Gamma{(s)}} = J_\mathrm{phys}^{\nu,\Gamma{(s)}} + &\sum_{j,n=0}^3\sum_{k,l\neq 0}^3 c_n^{(\nu,\Gamma),jkl}\Delta_{h}^{(j)} (w-1)^n\nonumber\\
& \times\left(\frac{am_c^\mathrm{val}}{\pi}\right)^{2k}  \left(\frac{am_h^\mathrm{val}}{\pi}\right)^{2l}\nonumber\\
+ &\sum_{j,n=0}^3\sum_{k,l\neq 0}^3 \tilde{c}_n^{(\nu,\Gamma){(s)},jkl}\Delta_{h}^{(j)} (w-1)^n\nonumber\\
& \times\left(\frac{am_c^\mathrm{val}}{\pi}\right)^{2k}  \left(\frac{am_h^\mathrm{val}}{\pi}\right)^{2l}\delta_\chi^{(s)}.
\end{align}
We take priors of $0(1)$ for each $c_n$ and $\tilde{c}_n$, multiplying terms of order $\mathcal{O}(a^2)$ by $0.5$ in line with the tree level $a^2$ improvement of the HISQ action~\cite{PhysRevD.75.054502}. All of the remaining priors are taken as $0(1)$.

\subsection{Tests of the Stability of the Analysis}
\begin{figure*}
\includegraphics[scale=0.18]{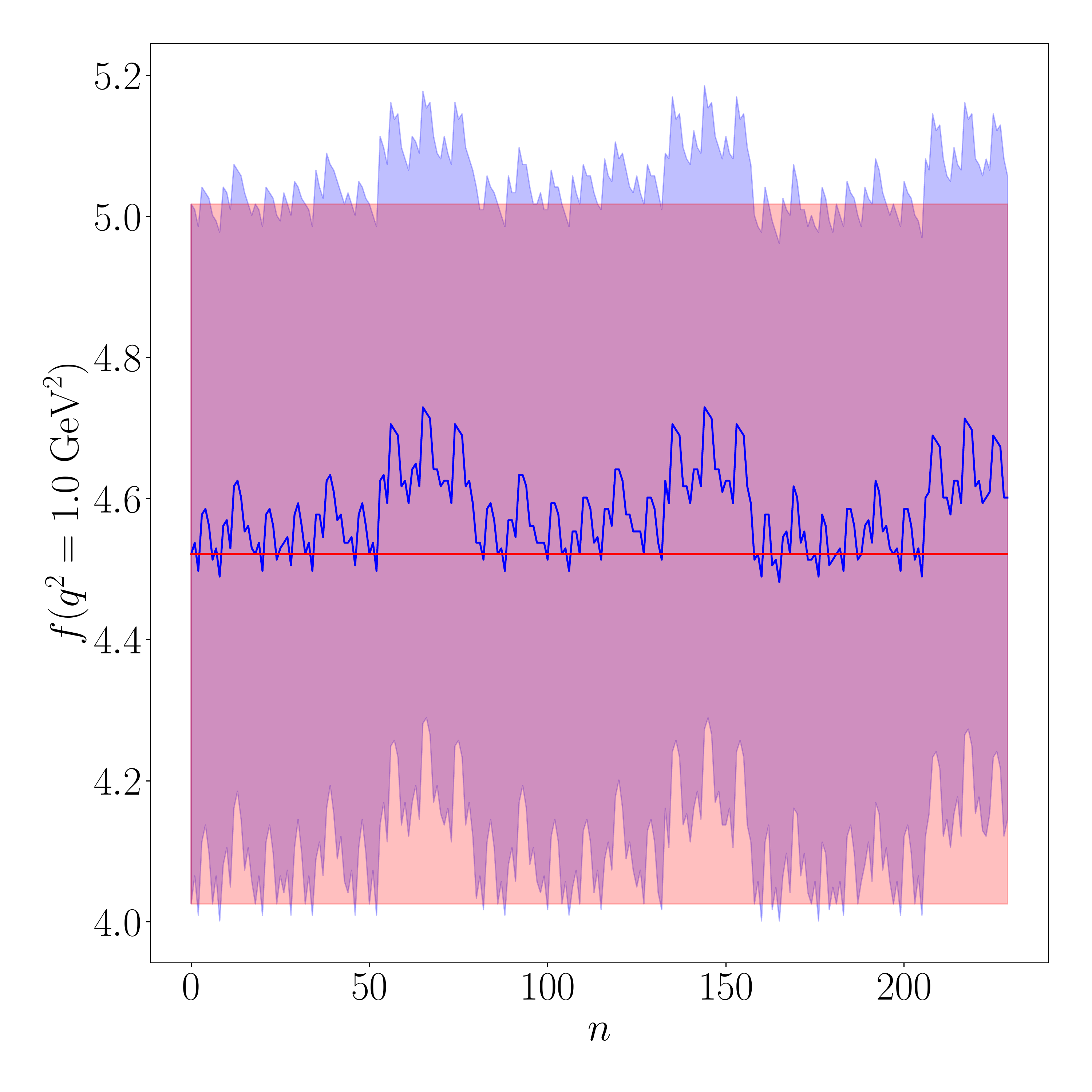}\includegraphics[scale=0.18]{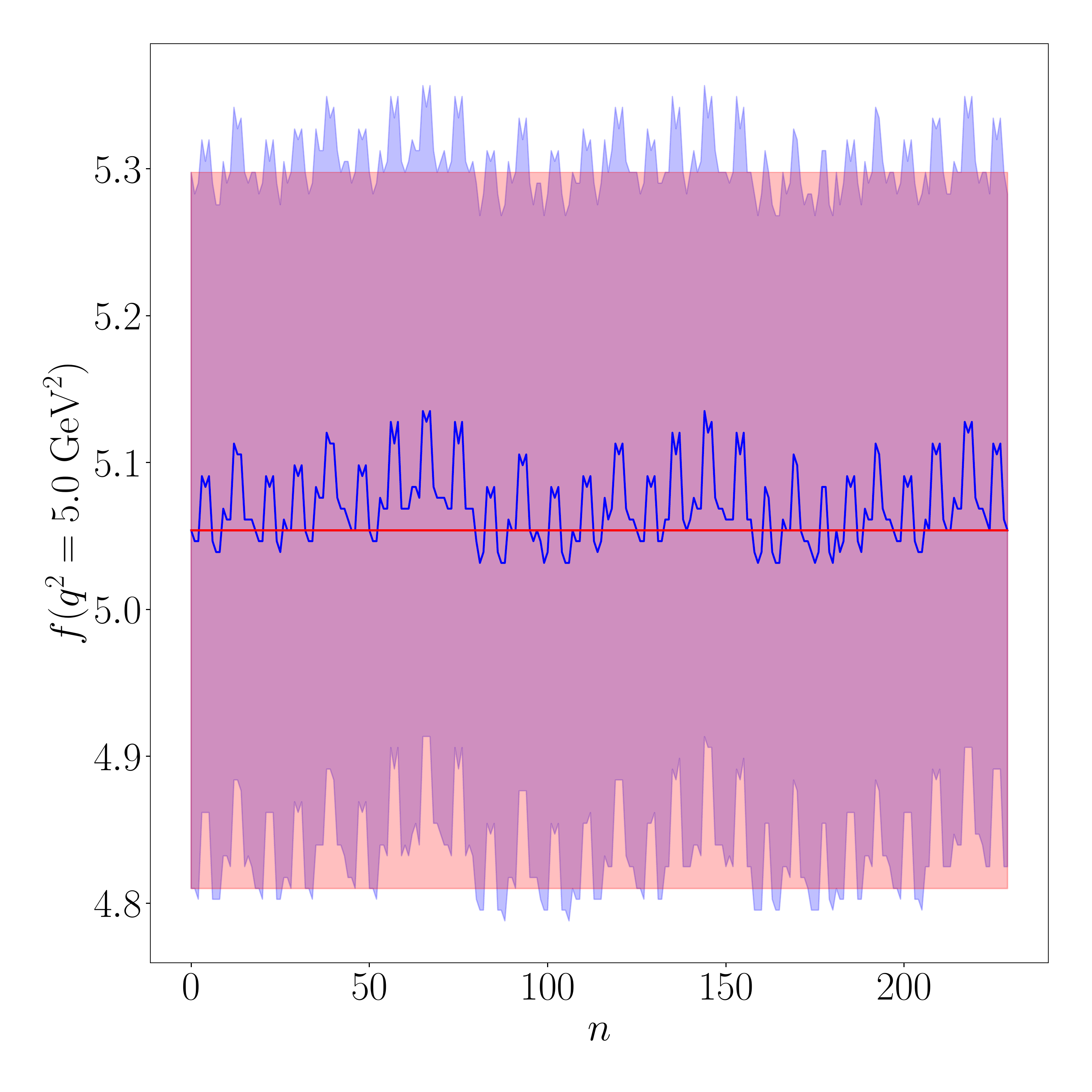}\includegraphics[scale=0.18]{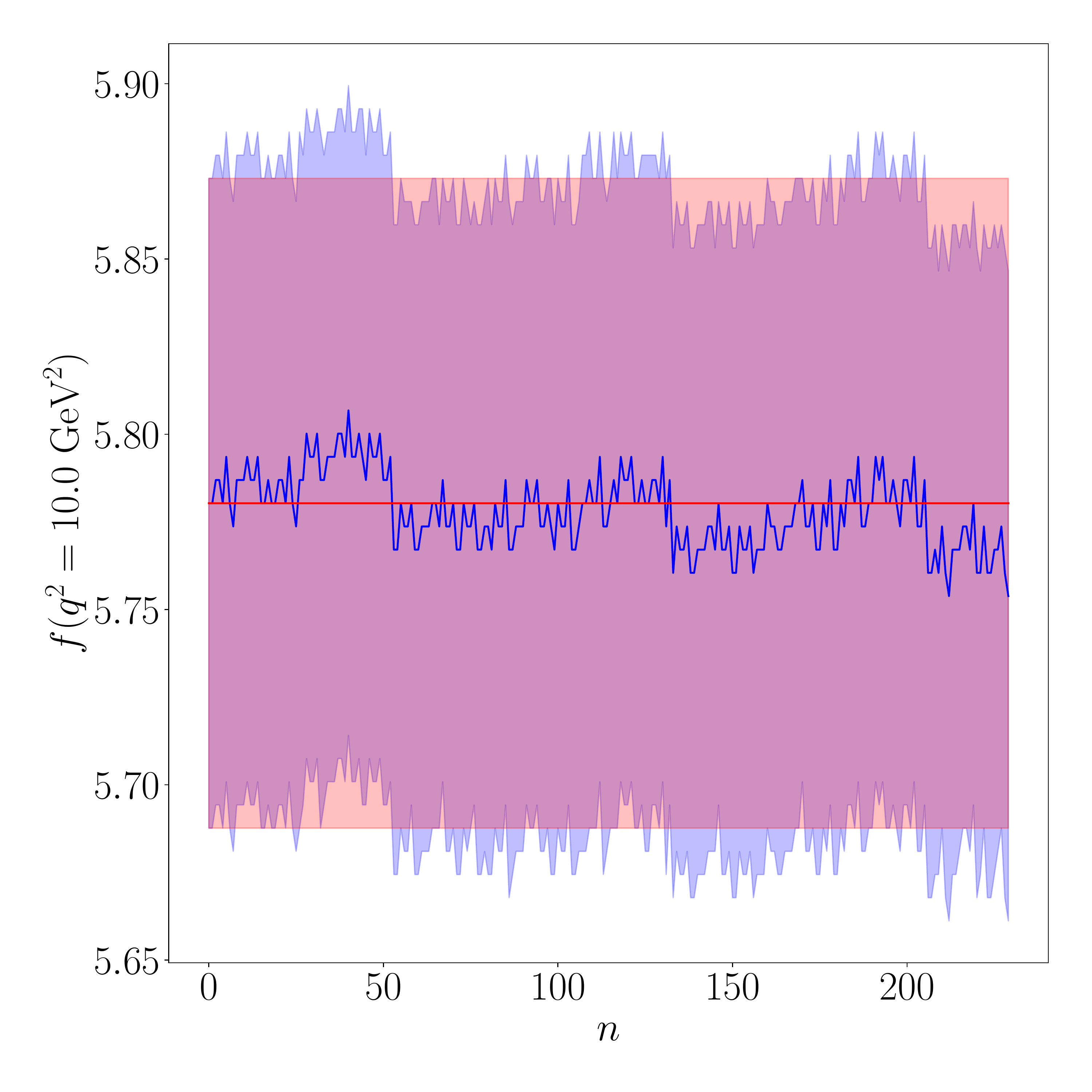}\\
\includegraphics[scale=0.18]{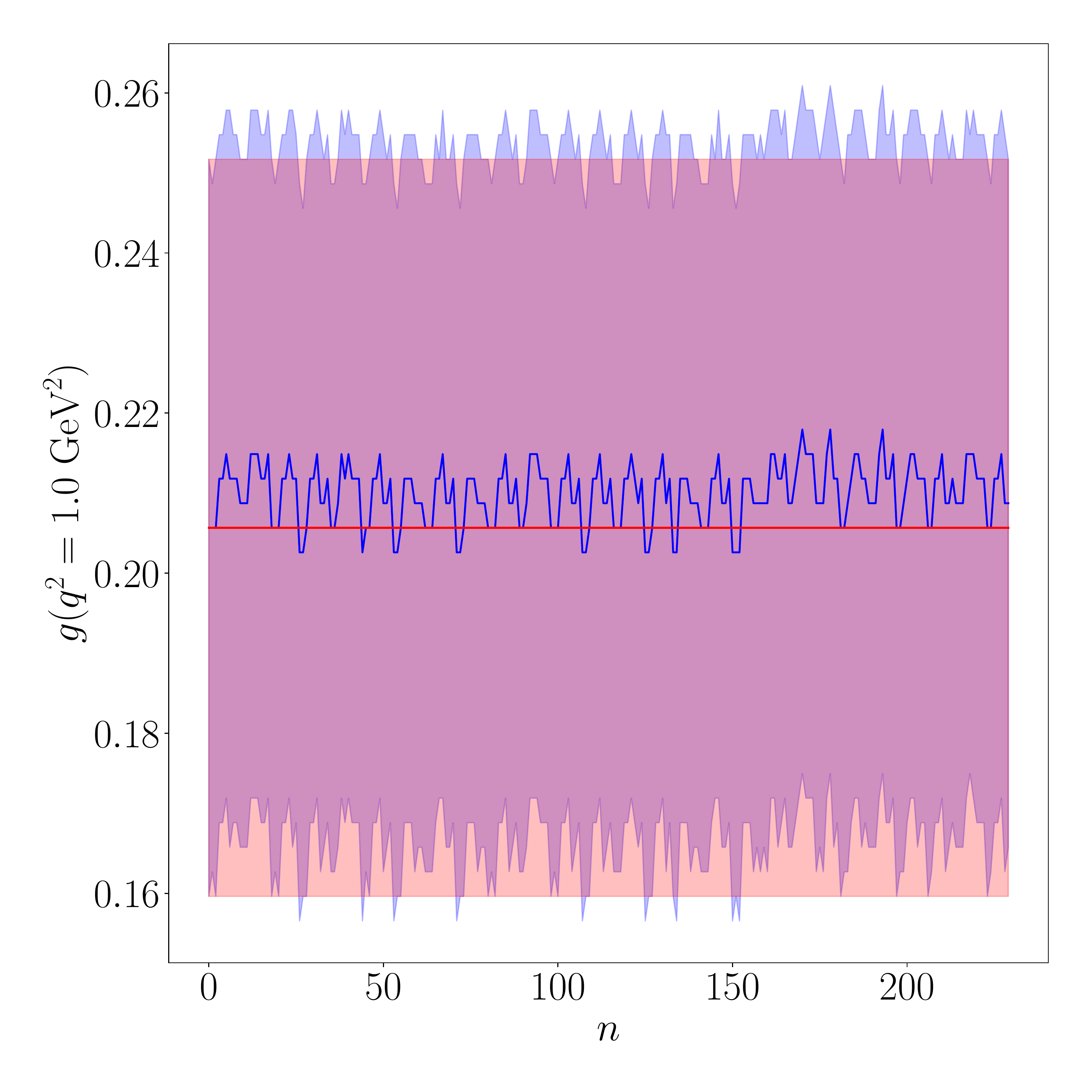} \includegraphics[scale=0.18]{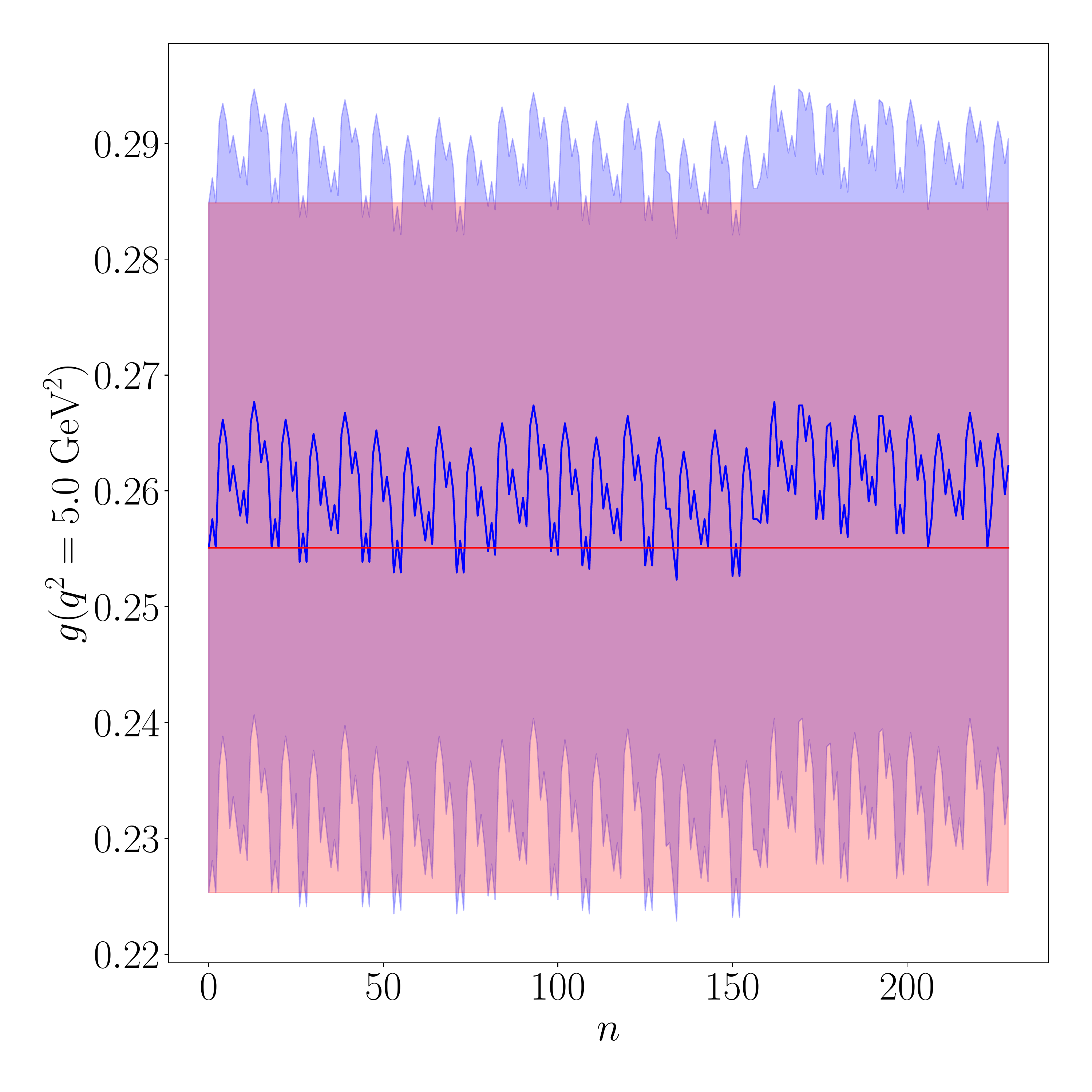} \includegraphics[scale=0.18]{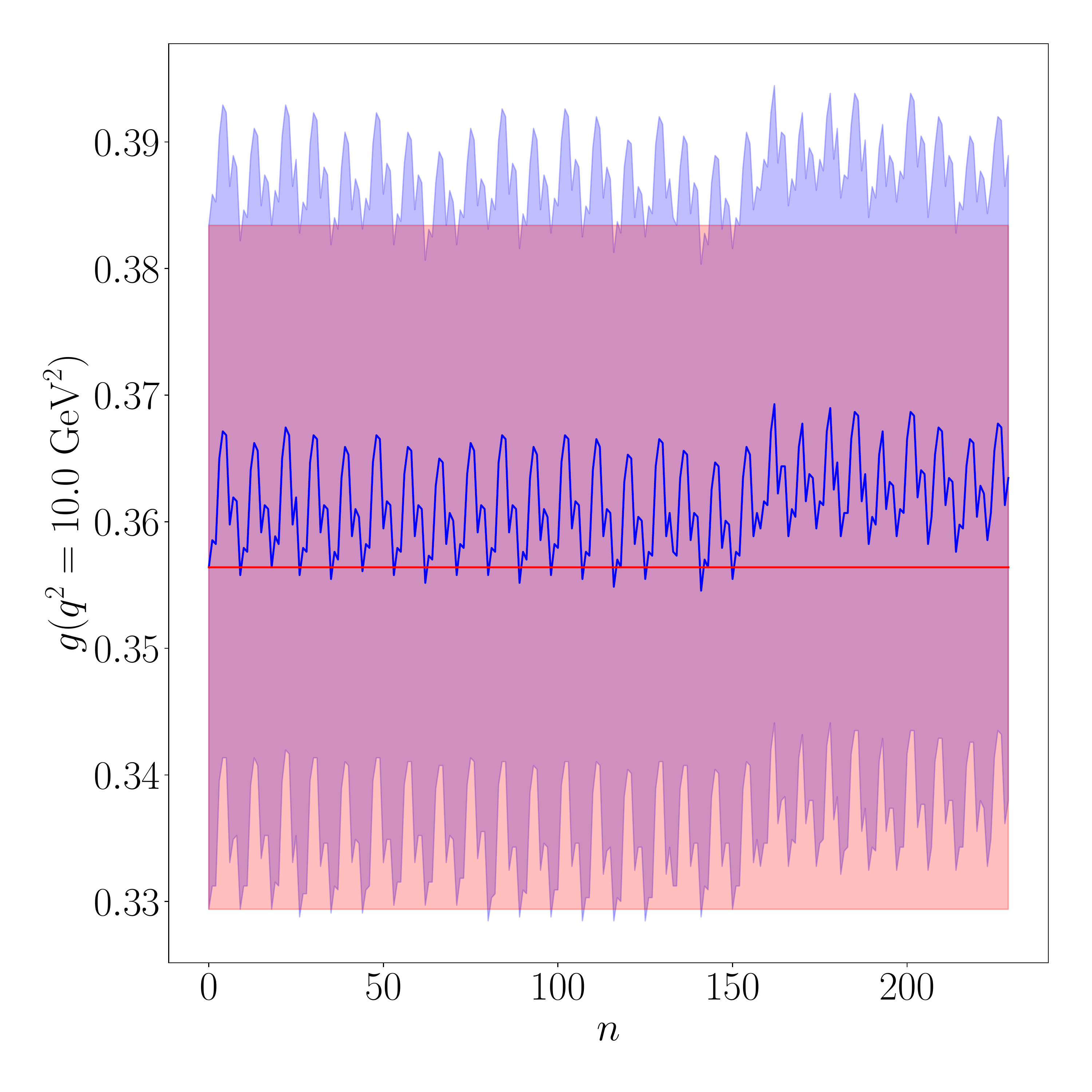}
\caption{\label{stabplothfg}Values of the form factors $f$ and $g$ for $B\to D^*$ evaluated at $q^2=1~\mathrm{GeV}^2$, $q^2=5~\mathrm{GeV}^2$ and $q^2=10~\mathrm{GeV}^2$ for different combinations of correlator fits using different parameters. The red line and error band corresponds to our chosen combination and the blue line and error band corresponds to the form factors resulting from different fit combinations. Here, $n=\delta_3 + 3\delta_2+9\delta_1+27\delta_4+81\delta_5$ where $\delta_i$ is the value of $\delta$ for set $i$ given in~\cref{fitparams}. We see that no combination of correlator fits results in a significant variation of $f$ or $g$.}
\end{figure*}

\begin{figure*}
\includegraphics[scale=0.25]{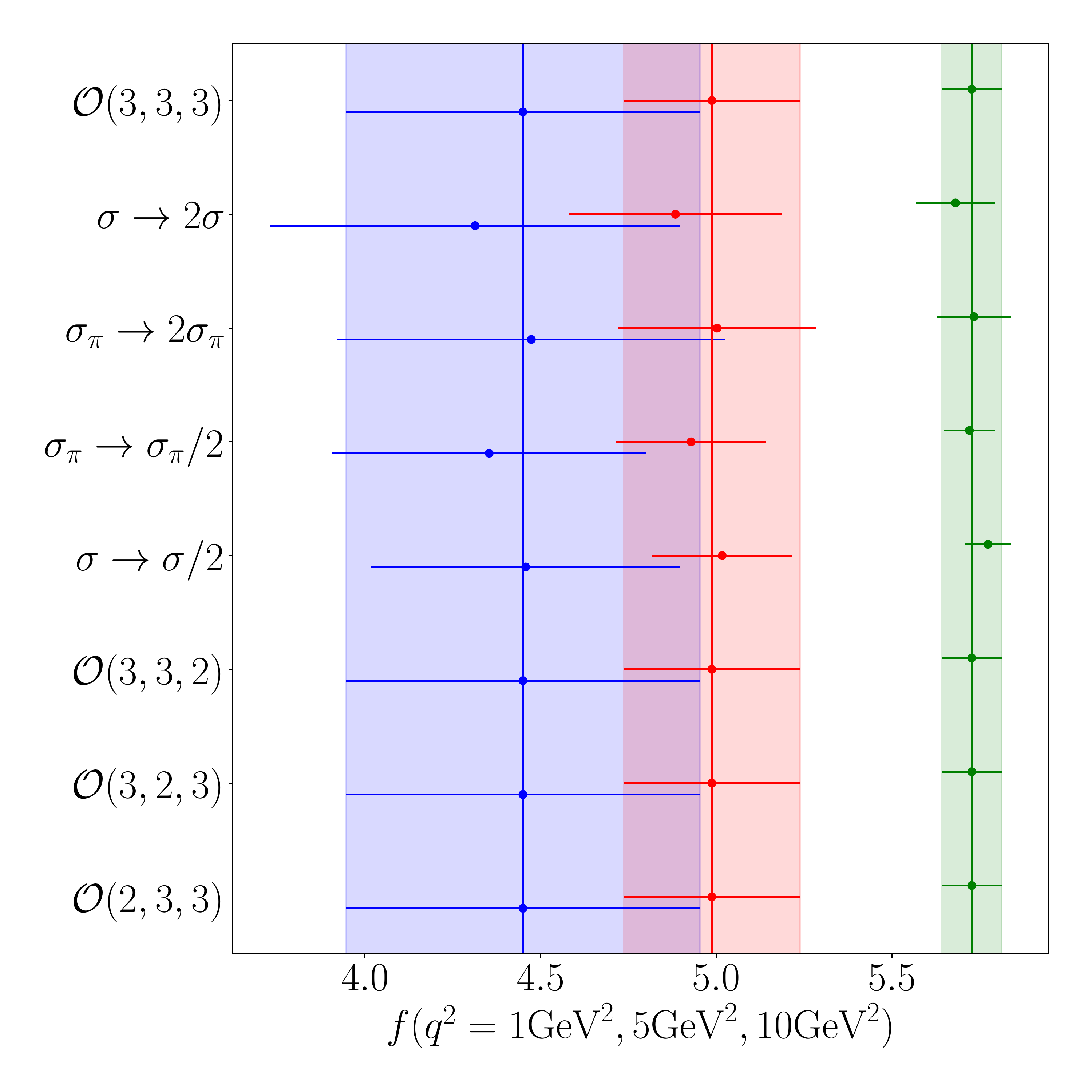}\includegraphics[scale=0.25]{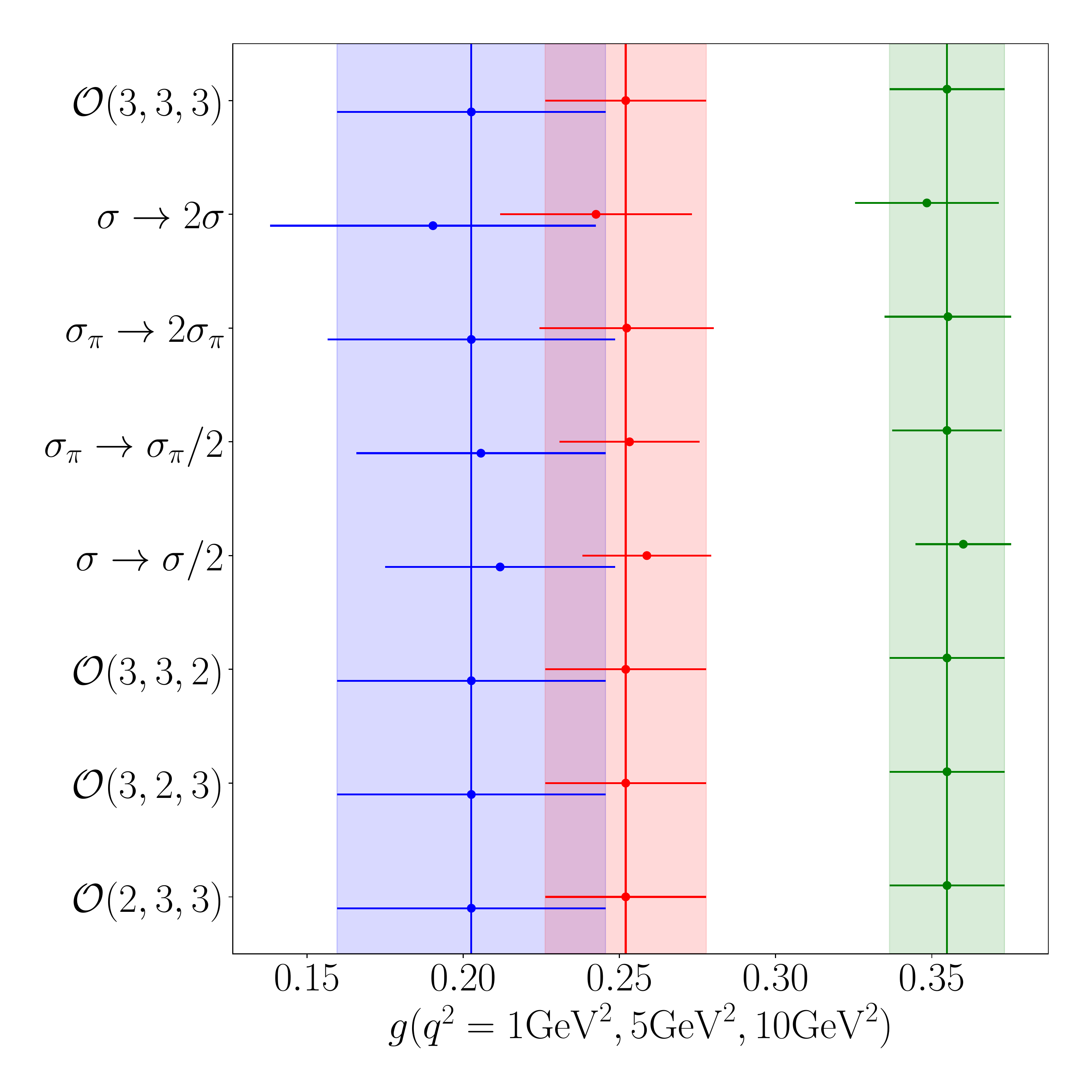}
\caption{\label{stabplothfgordervars}Values of the form factors $f$ and $g$ evaluated at $q^2=1~\mathrm{GeV}^2$, $q^2=5~\mathrm{GeV}^2$ and $q^2=10~\mathrm{GeV}^2$, corresponding to the blue, red and green points respectively, for different combinations of chiral-continuum extrapolation parameters. $\sigma\to \sigma/2$ indicates that we multiply the prior widths of $b_n^{j}$ and $c_n^{jkl}$ defined in~\cref{eq:anfitform,cterms} by 0.5 and $\sigma\to 2\sigma$ indicates that we multiply the prior widths of $b_n^{j}$ and $c_n^{jkl}$ defined in~\cref{eq:anfitform,cterms} by 2. $\sigma_\pi\to \sigma_\pi/2$ indicates that we multiply the prior widths of $\tilde{b}_n^{j}$, $\tilde{c}_n^{jkl}$ and $g_{D^*D\pi}$  by 0.5 in~\cref{fitfunctionequation,eq:anfitform,cterms} and $\sigma_\pi\to 2\sigma_\pi$ indicates that we multiply the prior widths of $\tilde{b}_n^{j}$, $\tilde{c}_n^{jkl}$ and $g_{D^*D\pi}$ by 2. $\mathcal{O}(n_{M_H},n_{am_c},n_{am_h})$ indicates the order to which we sum in $j,k,l$ respectively in~\cref{fitfunctionequation,eq:anfitform,cterms}.   }
\end{figure*}

\label{stabilitytests}
Here, we demonstrate that our physical-continuum results are insensitive to variations in both the parameters chosen when fitting correlator data, as well as the parameters entering the chiral-continuum extrapolation. First, we repeat the analysis described in~\cref{physextrap} using different combinations of the fits detailed in~\cref{fitparams}. In order to assess the sensitivity of our results, we follow~\cite{Harrison:2021tol} and compare the values of the form factors, evaluated at $q^2=1~\mathrm{GeV}^2$, $q^2=5~\mathrm{GeV}^2$ and $q^2=10~\mathrm{GeV}^2$. We perform this analysis in the physically important helicity basis, in which each form factor corresponds to a definite $D_{(s)}^*$ and $W$ polarisation. The SM form factors in this basis are defined via:
\begin{align}\label{helicitybasis}
g=&\frac{h_V}{M_{B_s}\sqrt{r}}\nonumber\\
f=&M_{B_s}\sqrt{r}(1+w)h_{A_1}\nonumber\\
F_1=&M_{B_s}^2\sqrt{r}(1+w)\left((w-r)h_{A_1}-(w-1)(rh_{A_2}+h_{A_3}) \right)\nonumber\\
F_2=&\frac{1}{\sqrt{r}}\left( (1+w)h_{A_1} + (rw-1)h_{A_2} + (r-w)h_{A_3}\right).
\end{align}
We also define definite helicity tensor form factors, related to the tensor helicity amplitudes given in~\cref{tensorhelicityamplitudes},
\begin{align}\label{helicitybasisT}
F_{T_1}&=(1+w)h_{T_1}+(w-1)h_{T_2}-h_{T_3}(w^2-1),\nonumber\\
F_{T_2}&=h_{T_1}(1-r)(1+w) -h_{T_2}(1+r)(w-1) ,\nonumber\\
F_{T_3}&=h_{T_1}(1+r)-h_{T_2}(1-r).
\end{align}
These are plotted in~\cref{stabplothfg} for $f$ and $g$ at $q^2=1~\mathrm{GeV}^2$, $q^2=5~\mathrm{GeV}^2$ and $q^2=10~\mathrm{GeV}^2$, with $n=\delta_3 + 3\delta_2+9\delta_1+27\delta_4+81\delta_5$ tracking the different fit parameters, where $\delta_i$ is the value of $\delta$ given in~\cref{fitparams}. In~\cref{stabplothfg} we see that no combination of alternative correlator fit parameters listed in~\cref{fitparams} results in a significant variation of $f$ or $g$ across the full kinematic range of the decay. Similar plots for the remaining form factors, including those for the tensor form factors, are given in the Supplementary Material~\cite{supplementarymat}, where we see that the other form factors are also stable to these variations.

We also investigate the effect of reducing the prior widths, as well as reducing the order summed to for each expansion parameter in~\cref{fitfunctionequation}. We evaluate the form factors, again at $q^2=1~\mathrm{GeV}^2$, $q^2=5~\mathrm{GeV}^2$ and $q^2=10~\mathrm{GeV}^2$, for different combinations of these chiral-continuum extrapolation parameters. We also investigate the effect of reducing the order to which we sum in $j,k,l$ in~\cref{fitfunctionequation,eq:anfitform,cterms}, as well as the effect of halving the prior widths of ${b}_n^{j}$ and $c_n^{jkl}$ defined in~\cref{eq:anfitform} and halving the prior widths of $\tilde{b}_n^{j}$, $\tilde{c}_n^{jkl}$ and $g_{D^*D\pi}$ in~\cref{fitfunctionequation,eq:anfitform}. The resulting form factors for each modification of chiral-continuum extrapolation procedure are plotted in~\cref{stabplothfgordervars} for the form factors $f$ and $g$, where we see that none of these changes to the extrapolation procedure result in a significant change to the form factors. Plots for the remaining $B\to D^*$ and $B_s\to D_s^*$ form factors are given in the Supplementary Material~\cite{supplementarymat}, where we see that none of our form factors are sensitive to these changes.

\subsection{Error Budget}
\begin{figure}
\includegraphics[scale=0.5]{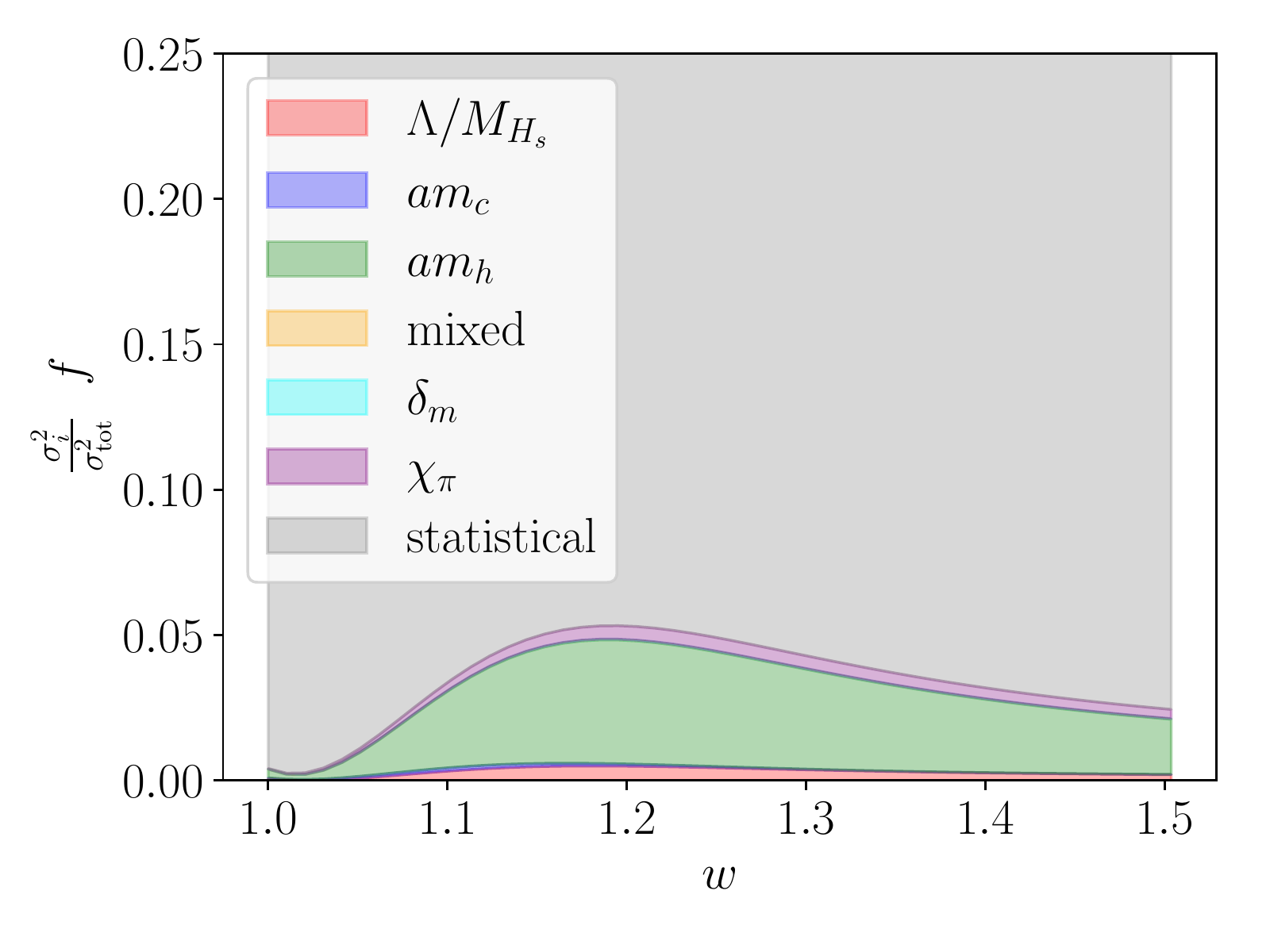}\\
\includegraphics[scale=0.5]{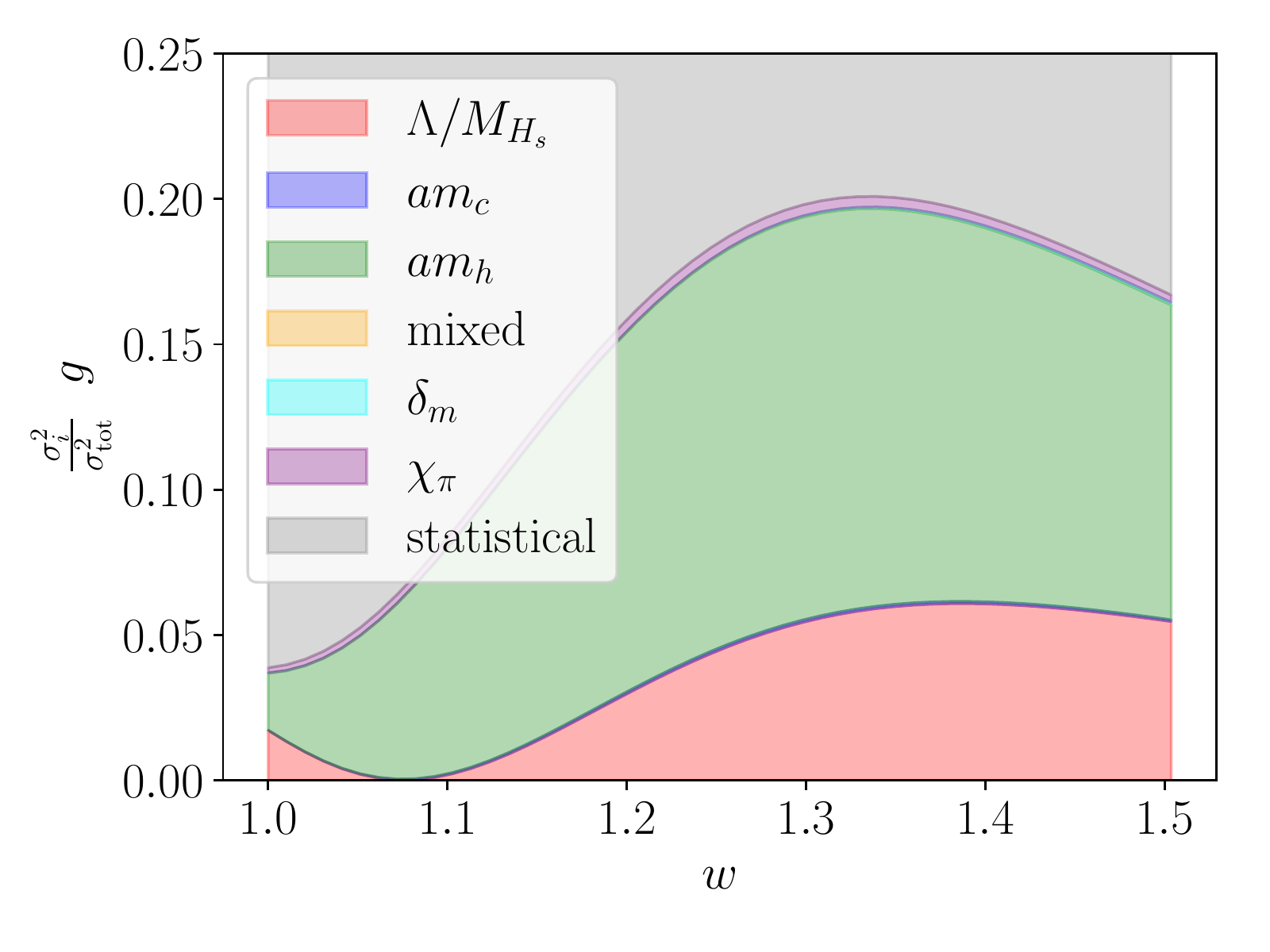}
\caption{\label{errorbandsfg}Plots showing the fractional contribution of each source of uncertainty to the total variance for the form factors $f$ and $g$ across the full kinematic range. The vertical axis is truncated at $0.25$ for clarity, with the remaining variance between $0.25$ and $1$ attributable to statistics.}
\end{figure}
In~\cref{errorbandsfg} we plot the fractional contribution of each source of uncertainty to the total variance for the form factors $f$ and $g$ across the full kinematic range of the decay. These are computed from the partial variance of the form factor at each $w$ with respect to the priors, and so the size of each band represents the extent to which the corresponding terms in the chiral-continuum fit are not constrained by the data. The band labelled $\chi_\pi$ corresponds to the priors $\tilde{c}_n^{jkl}$, $\tilde{b}_n^{j}$ and $g_{D^*D\pi}$, $\Lambda/M_h$ corresponds to the priors $b_n^{j\neq 0}$, $am_c$ to $c_n^{0~k\neq 0~0}$, $am_h$ to $c_n^{00~l\neq 0}$, $\delta_m$ to the priors entering $\mathcal{N}^Y_n$, and `mixed' corresponds to priors for ${b}_n^{j}$ $c_n^{jkl}$ where at least two of $j$, $k$ or $l$ are nonzero. `Statistical' corresponds to the uncertainty from our data. Unsurprisingly, we see that close to $w=1$ where we have data on all ensembles for all masses we have very good control over the discretisation, chiral and heavy-mass dependence, whereas towards the maximum value of $w$, corresponding to $q^2=0$, where we have less coverage with our data, we see that the uncertainty coming from unconstrained terms in our fit function is larger. For the SM form factors, we generally find that control over discretisation effects set by $am_h$, as well as control over the physical heavy mass dependence, are the dominant sources of uncertainty not constrained by the data. Plots for $F_1$ and $F_2$ as well as the tensor form factors in the helicity basis defined in~\cref{helicitybasisT}, are given in~\cref{errorbandsF1F2,errorbandsFT1FT2FT3} in~\cref{lattdat}, where we see a similar situation for $F_1$, $F_2$, $F_{T_2}$ and $F_{T_3}$. The uncertainty in $F_{T_1}$ is dominated by the unconstrained chiral dependence of the factor $h_{T_3}$, shown in~\cref{TFFqsq}. We also show plots for the $B_s\to D_s^*$ form factor uncertainties in~\cref{lattdat}, with similar behaviour to those for $B\to D^*$.

\subsection{BGL Form Factor Parameterisation}
\label{sec:bglparams}
\begin{table}
\centering
\caption{\label{bglfitparams}BGL fit parameters, defined in~\cref{bgldefinition}, for our $B\to D^*$ form factors. Here we also include the sums of squared coefficients, which we see are far from saturating the unitarity bounds in~\cref{unitaritybounds}.}
\begin{tabular}{ c c }\hline
$a_0^{g}$ & 0.0318(17) \\
$a_1^{g}$ & -0.128(95) \\
$a_2^{g}$ & 0.08(76) \\\hline
$a_0^{f}$ & 0.01212(17) \\
$a_1^{f}$ & 0.012(17) \\
$a_2^{f}$ & -0.18(45) \\\hline
$a_0^{F1}$ & 0.002029(29) \\
$a_1^{F1}$ & -0.0081(86) \\
$a_2^{F1}$ & -0.08(27) \\\hline
$a_0^{{F2}}$ & 0.0415(27) \\
$a_1^{{F2}}$ & -0.20(12) \\
$a_2^{{F2}}$ & -0.01(80) \\\hline
$\sum_i^3|a_i^{g}|^2$ & 0.02(13) \\
$\sum_i^3|a_i^{f}|^2+|a_i^{F_1}|^2$ & 0.04(17) \\
$\sum_i^3|a_i^{F_2}|^2$ & 0.042(48) 
\end{tabular}
\end{table}
For comparison to other lattice and experimental determinations, we fit synthetic data points generated at $w=1.025$, $w=1.225$, and $w=1.425$ for each $B\to D^*$ form factor in the helicity basis, defined in~\cref{helicitybasis}, using the BGL parameterisation~\cite{Boyd:1997kz}. The BGL parameterisation expresses the form factors as
\begin{align}\label{bgldefinition}
\mathcal{F}(t) = \frac{1}{P(z)\phi(z)}\sum_{n=0}^\infty a^\mathcal{F}_n z(t,t_0)^n.
\end{align}
Here we adopt the conventions for Blaschke factors $P(z)$, outer functions $\phi(z)$, $B_c^{(*)}$ resonances of~\cite{Bigi:2017jbd} which were also used in~\cite{FermilabLattice:2021cdg}. We include up to quadratic order in $z$, though we have confirmed that going to cubic order has only a very small effect on the resulting coefficients. We also enforce the condition $F_1(w=1)=M_B(1-r)f(w=1)$ by fixing $a^0_{F_1} =a^0_f(1-r)/\sqrt{2}(1+\sqrt{r})^2$. Note that here we take uniformly distributed priors between $-1$ and $1$ for each $a^\mathcal{F}_n$. Alhough we do not enforce the condition at $w_\mathrm{max}$, $F_2(w_\mathrm{max})=(1+r)/(M_B^2(1+w_\mathrm{max})(1-r)r) F_1(w_\mathrm{max})$, we find that our fit satisfies this condition to within $0.07\sigma$. The fit paramaters $a_i$ should satisfy the unitarity bounds given by
\begin{align}\label{unitaritybounds}
\sum_i^\infty|a_i^{g}|^2 &\le 1,\nonumber\\
\sum_i^\infty|a_i^{f}|^2+|a_i^{F_1}|^2&\nonumber\le 1,\\
\sum_i^\infty|a_i^{F_2}|^2&\le 1.
\end{align}
The results of this fit are given in~\cref{bglfitparams}, where we see that for the form factors $g$, $f$ and $F1$ we have a reasonably good agreement with~\cite{FermilabLattice:2021cdg}, and comparable uncertainties (c.f. Table 11 in that paper). However for $F_2$ we have significant tension at the level of $3\sigma$. \cref{bglfitparams} includes checks of the unitarity bounds, \cref{unitaritybounds}, which we find to be far from saturation for the number of coefficients we include.

\subsection{Reconstructing our Form Factors}
\label{reconstructingffs}

We have included in the Supplementary Material~\cite{supplementarymat} a Python script, \textbf{LOAD\_FIT.py}, that reads our physical-continuum HQET fit parameters~(see~\cref{fitfunctionequation}) and their correlations from the file \textbf{hpqcd\_BDstar.pydat}, in order to build the $B\to D^*$ and $B_s\to D_s^*$ form factors in the HQET basis. Note that the $B_s\to D_s^*$ form factors given here supersede those given in~\cite{Harrison:2021tol}. The script also performs checks against the values of the form factors at five equally spaced values of $q^2$, stored in \textbf{CHECKS.txt} and \textbf{CHECKS\_s.txt}. We also provide a file \textbf{synthetic\_data.pydat}, which may be loaded into Python using \textbf{gvar.gload}, that contains synthetic data points for the form factors in the HQET basis computed at 4 equally spaced values of $q^2=i\times q^2_\mathrm{max,(s)}/3, i \in [0,1,2,3]$ for the $B_{(s)}\to D_{(s)}^*$ form factors. These synthetic data points are also checked against those computed from our fit parameter text files in \textbf{LOAD\_FIT.py}. We have run these scripts using Python-3.10.6, using the packages numpy-1.21.5, scipy-1.8.0, gvar-11.10.1 and matplotlib-3.5.1.

\section{Discussion}
\label{Discussion}

\subsection{Comparison to Experiment, $|V_{cb}|$}\label{shapetolattbelle}
\begin{figure*}
\includegraphics[scale=0.09]{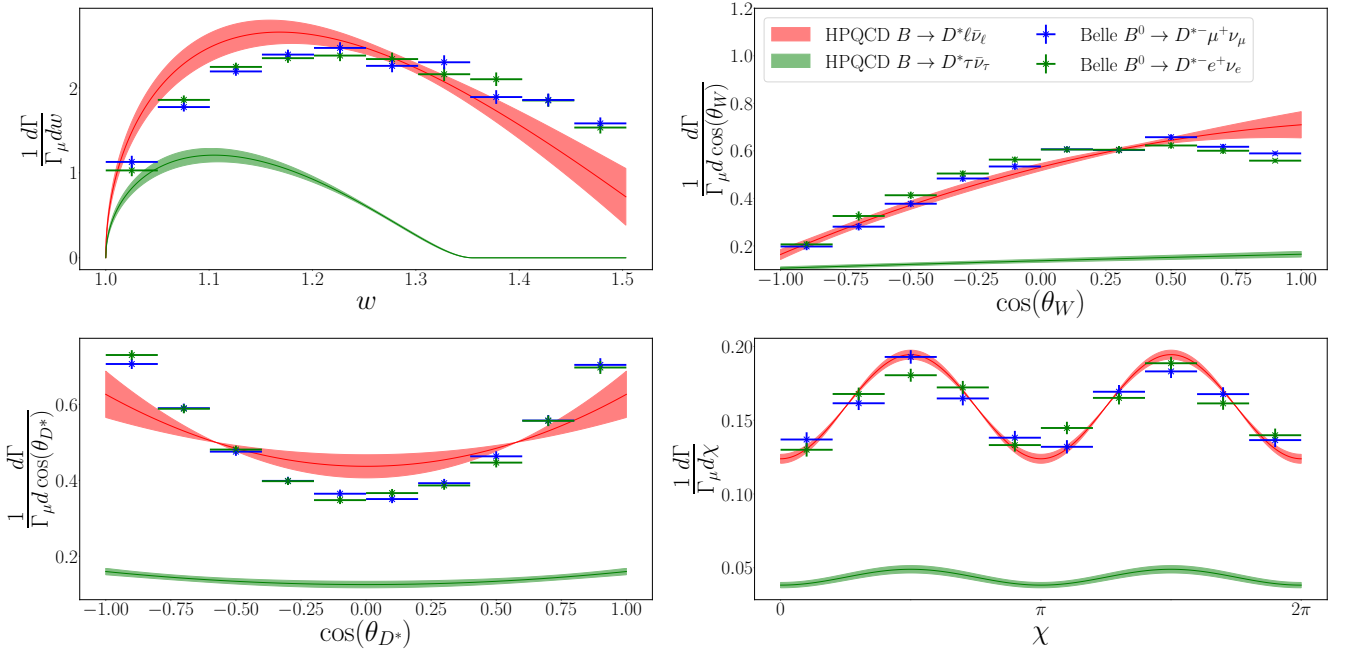}
\caption{\label{normplotsbellehpqcd}Our lattice-only normalised differential decay rates for $B\to D^*\ell\bar{\nu}$, with respect to the angular variables defined in~\cref{BsDsstarangles}, are shown as the red bands. We also include binned untagged data for $e/\mu$ from Belle~\cite{Belle:2018ezy}. Note the clear difference in shape, particularly for the differential rate with respect to $w$. Our tauonic differential decay rates are shown in green.}
\end{figure*}
\begin{figure}
\includegraphics[scale=0.105]{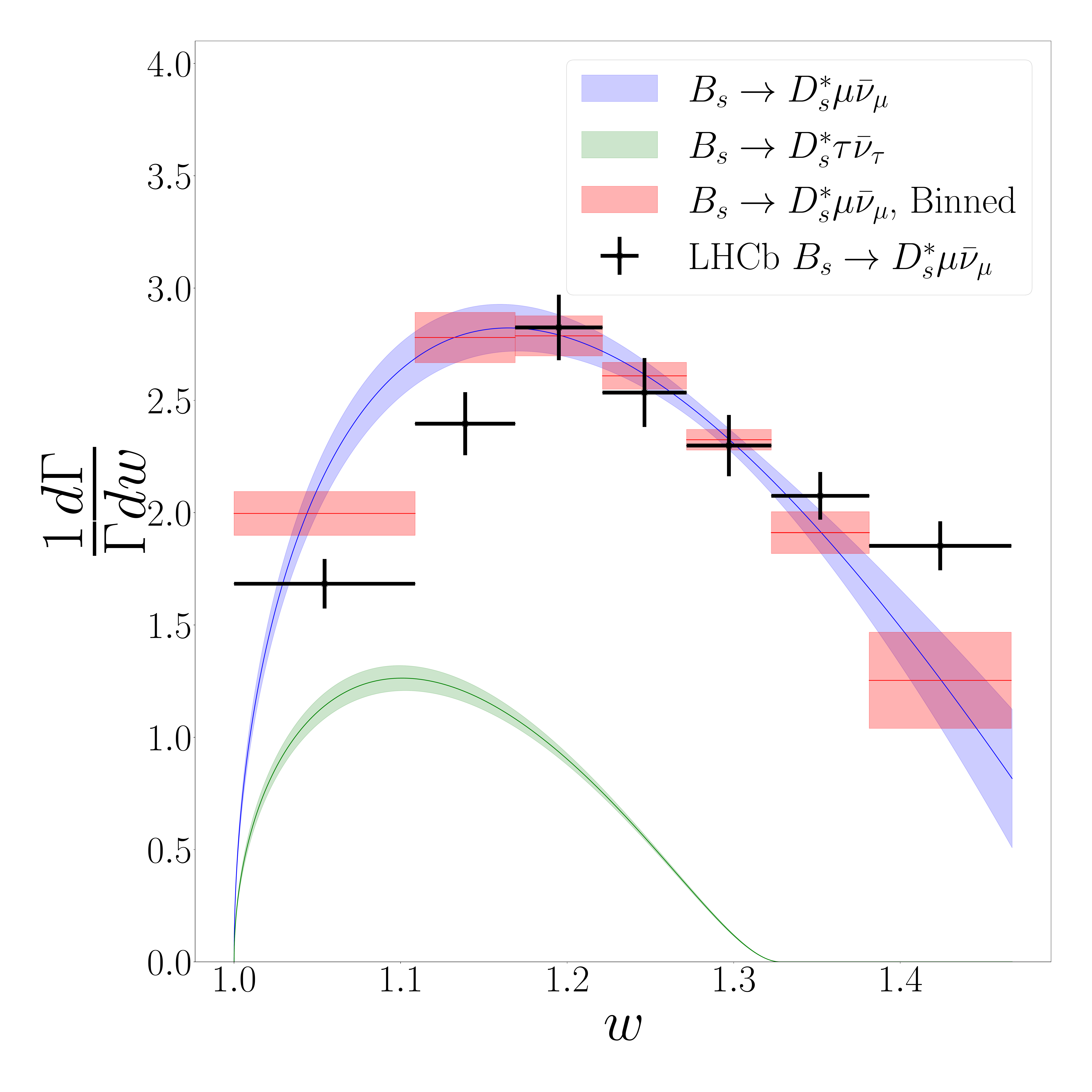}
\caption{\label{normplotsLHCbhpqcd}Our normalised differential decay rate for $B_s\to D_s^*\ell\bar{\nu}$ with respect to $w$ is shown as the blue band. We also include binned data from LHCb~\cite{Aaij:2020xjy}. Here, as for $B\to D^*$, we see a similar difference in shape between SM theory and experiment to that seen for Belle $B\to D^*$ data in~\cref{normplotsbellehpqcd}. The semitauonic mode is plotted as the green band.}
\end{figure}
\begin{figure}
\includegraphics[scale=0.105]{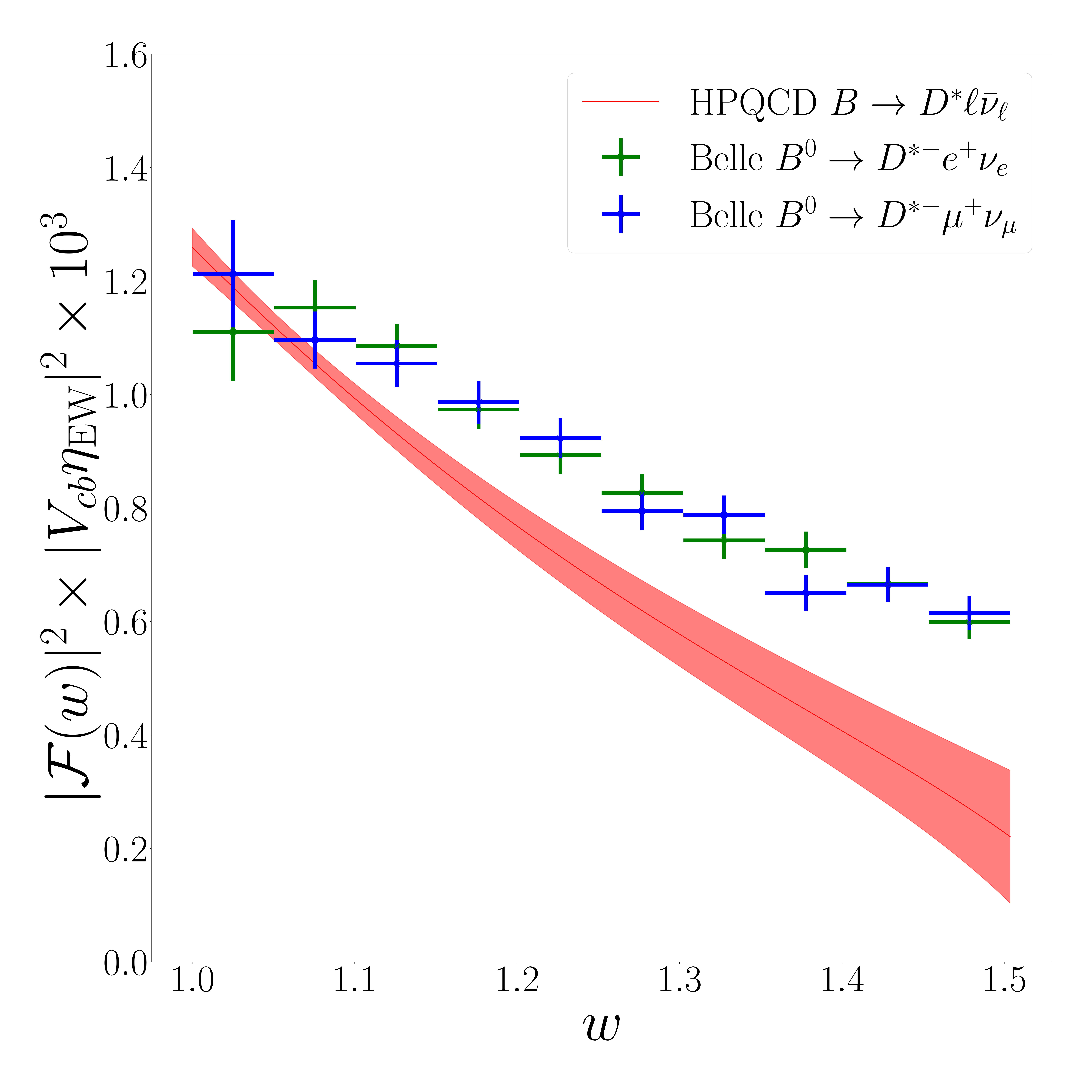}
\caption{\label{FsquareBelle}$|\mathcal{F}(w)\eta_\mathrm{EW}V_{cb}|^2$, defined via~\cref{Fdefeq}, plotted against $w$. Our lattice-only $|\mathcal{F}(w)|^2$ is multiplied by $V_{cb}$ extracted from the joint theory/experiment fit.}
\end{figure}
We can use our form factors together with the untagged data for $B\to D^*e^-\bar{\nu}_e$ and $B\to D^*\mu^-\bar{\nu}_\mu$ from Belle~\cite{Belle:2018ezy} in order to extract $|V_{cb}|$. We use our physical-continuum form factor parameters, given in the Supplementary Material~\cite{supplementarymat} as described in~\cref{reconstructingffs}, as priors to fit the experimental differential rate data from Belle, which has been binned in each of the variables $w$, $\theta_{D^*}$, $\theta_W$ and $\chi$ defined in~\cref{BsDsstarangles}. Note that throughout this section we assume no lepton flavour universality~(LFU) violation between the light $\ell=\mu$ and $\ell=e$ modes.

The covariance matrix for the Belle data does not include the zero eigenvalues expected from the fact that the bins for a given variable must sum to the same total. In order to remedy this issue we normalise the bins for each variable so that they sum to 1. This ensures that the covariance matrix contains the expected zero eigenmodes, which we then remove explicitly using an SVD cut. Following the observation in~\cite{FermilabLattice:2021cdg} that the experimental data used to extract $V_{cb}$ was dominated by the Belle dataset, we do not include any synthetic data points generated using fits from BaBar~\cite{BaBar:2019vpl}.

Once the fit to Belle data described above has been performed, a value of $|V_{cb}|$ can be read off by comparing the total number of events to $\Gamma/|V_{cb}\eta_{EW}|^2$ computed using the form factors resulting from the joint theory/experiment fit. We fit all four variables simultaneously, though we have checked that fitting the Belle data for any single variable on its own does not change the uncertainty in the resulting value of $|V_{cb}|$, exactly as one would expect from the fact that the sum of each set of 10 bins must be equal. In order to reconstruct the combined electron-muon $80\times 80$ covariance matrix we follow the procedure described in~\cite{Bobeth:2021lya} so that we may fit the $\ell=\mu$ and $\ell=e$ cases simultaneously, assuming no NP in either mode.

Since we have computed fully correlated form factors for both $B\to D^*$ and $B_s \to D_s^*$, it is possible for us to include data from LHCb~\cite{Aaij:2020xjy} for $B_s\to D^*_s$ in our fits. Even though this data is more limited, it can still inform the shape of the form factors. We include the LHCb $B_s\to D_s^*$ data in our fits in the same manner as the Belle data, integrating our differential decay rate over the bins used by LHCb and then including these in our $\chi^2$ minimisation.  However, since the available $B_s \to D_s^*$ experimental data is significantly less precise than that for $B \to D^*$, the inclusion of the LHCb data does not significantly change the central value or uncertainty of $|V_{cb}|$ determined in this way.

Our lattice-only normalised differential decay rates for $B\to D^*$ and $B_s\to D_s^*$ are shown in~\cref{normplotsbellehpqcd,normplotsLHCbhpqcd} respectively, together with the experimental data points for each bin. We see a difference in shape between our results and the binned data from Belle and LHCb. The fit to our results along with Belle and LHCb data gives $\chi^2/\mathrm{dof} = 0.95$ and $Q=0.55$. The visible disagreement in shape we see here is qualitatively similar to what was seen in~\cite{FermilabLattice:2021cdg}, where the authors observed a $2\sigma$ discrepancy across the full kinematic range of the decay after extrapolating their lattice results (covering $1 \le w \le 1.175$) to $w_\mathrm{max}$ using the BGL parameterisation. 

Using our fit to our lattice results along with the experimental data enables us to determine $|V_{cb}|$. We find
\begin{align}\label{eqn:vcb}
|V_{cb}|=\vcb
\end{align}
in good agreement with previous exclusive determinations~\cite{HFLAV:2022pwe}. Note that in determining $V_{cb}$ we take $|\eta_\mathrm{EW}|^2 = (1.00662)^2\times (1+\alpha_\mathrm{QED}\pi)$, with an additional Coulomb factor~\cite{Sirlin:1981ie,PhysRevD.41.1736} for the charged final states in the decay measured by Belle, $B^0\to D^{*-}\ell^+\bar{\nu}_\ell$, and neglect the uncertainty. 

For the purpose of comparison to other lattice QCD results from~\cite{FermilabLattice:2021cdg}, we plot $|\mathcal{F}(w)V_{cb}\eta_\mathrm{EW}|^2$ in~\cref{FsquareBelle}, where we use $V_{cb}$ extracted from our joint theory/experiment fit, \cref{eqn:vcb}, to multiply $|\mathcal{F}(w)|^2$ computed using only our form factors. $\mathcal{F}(w)$ is defined according to the equation for the differential rate with respect to $w$:
\begin{align}\label{Fdefeq}
&\frac{d\Gamma(B_{(s)}\to D^*_{(s)}\ell\bar{\nu}_\ell)}{dw} =\nonumber\\
& \frac{G_F^2}{48\pi^3}(M_{B_{(s)}}-M_{D^*_{(s)}})^2 M_{D^*_{(s)}}^3\sqrt{w^2-1}(w+1)^2\nonumber\\
&\times \left[1+\frac{4w}{w+1}\frac{M_{B_{(s)}}^2-2wM_{B_{(s)}}M_{D^*_{(s)}}+M_{D^*_{(s)}}^2}{(M_{B_{(s)}}-M_{D^*_{(s)}})^2} \right]\nonumber\\
&\times|\mathcal{F}^{(s)}(w)\eta_\mathrm{EW}V_{cb}|^2.
\end{align}
\cref{FsquareBelle} confirms the disagreement in shape of $|\mathcal{F}|^2$ seen in~\cite{FermilabLattice:2021cdg} between the SM and Belle data.

It has been emphasised that a precise determination of the slope of $\mathcal{F}$ at $w=1$ could significantly reduce the uncertainty in $V_{cb}$~\cite{Bigi:2017njr}. While it is preferable to extract $V_{cb}$ using lattice and experimental data across the full kinematic range, it is still interesting to examine the slope at $w=1$. We find, for $B\to D^*$ and $B_s\to D_s^*$,
\begin{align}
\frac{dF}{dw}\Big{|}_{w=1} &= \dFdw,\nonumber\\
\frac{dF^s}{dw}\Big{|}_{w=1} &= \dFdws.
\end{align}
The value for ${dF^s}/{dw}$ is in good agreement with the value of $-0.94(15)$ from our previous study~\cite{Harrison:2021tol}, and we find that the slope in both light and strange spectator cases agree well.

$V_{cb}$ may also be computed by combining the total decay rate from our lattice form factors and the Belle total rate without using the differential rate information. Doing this, we find \totalVCB, a larger value than that in~\cref{eqn:vcb} and in much better agreement with the inclusive value. This value may be understood from~\cref{FsquareBelle}, where it is apparent that our lattice results multiplied by $|V_{cb}|^2$ from~\cref{eqn:vcb} lie below the binned experimental data, and so give a greater value of $V_{cb}$ when only the total rate is considered. This approach discards information about the form factors contained in the shape of the experimentally measured differential rate, that otherwise constrains the form factors, and so results in a greater uncertainty. We may also use the experimental average branching fraction, $\mathcal{B}(B^+\to D^{*0}\ell^+\nu_\ell)$, and $B^+$ lifetime from the Particle Data Group~\cite{Workman:2022ynf} to calculate the total decay rate, which we can then combine with our lattice results to find \totalVCBpdg, consistent with the value above determined using the Belle total decay rate alone. Note that the total rate for $B^+\to D^{*0}\ell^+\nu_\ell$ does not include the additional Coulomb factor, $(1+\alpha_\mathrm{QED}\pi)$, required for $B^0\to D^{*-}\ell^+\nu_\ell$~\cite{Sirlin:1981ie,PhysRevD.41.1736}.

\subsection{$\Gamma$, $R(D^*_{(s)})$ and Angular Observables}\label{angularobsRdstarGamma}\begin{figure}
\includegraphics[scale=0.24]{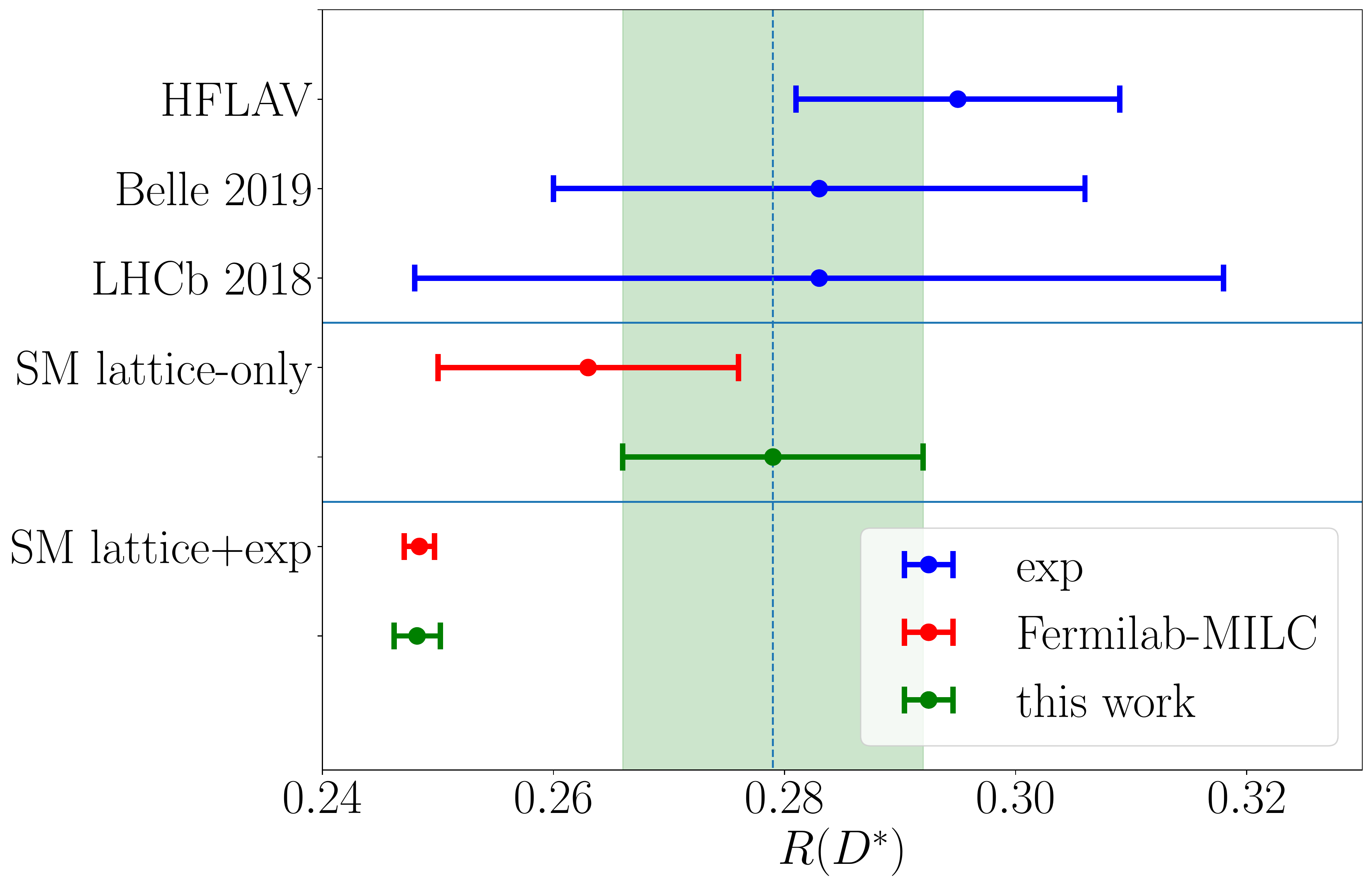}
\caption{\label{exp_plus_latt}`Lattice-only' and `lattice+experiment' values of $R(D^*)$. The results of this work are shown in green, while the recent results from the Fermilab-MILC collaboration~\cite{FermilabLattice:2021cdg} are shown in red. The inclusion of experimental data produces a similar downward shift in both cases. The two most recent experimental measurements of $R(D^*)$, from Belle~\cite{Belle:2019rba} and LHCb~\cite{LHCb:2017rln}, are also shown in blue, together with the HFLAV average value.}
\end{figure}

We can use our form factors to compute the total decay rates for the different processes, normalised by the combination $|V_{cb}\eta_\mathrm{EW}|^2$. We find:
\begin{align}\label{rates}
\Gamma\left(B\to D^*e\bar{\nu}_e\right)/|V_{cb}\eta_\mathrm{EW}|^2 		&= \Gammaetot, \nonumber \\
\Gamma\left(B\to D^*\mu\bar{\nu}_\mu\right)/|V_{cb}\eta_\mathrm{EW}|^2 		&= \Gammamutot, \nonumber \\
\Gamma\left(B\to D^*\tau\bar{\nu}_\tau\right)/|V_{cb}\eta_\mathrm{EW}|^2	&= \Gammatautot, \nonumber \\
\Gamma\left(B_s\to D_s^*e\bar{\nu}_e\right)/|V_{cb}\eta_\mathrm{EW}|^2 		&= \Gammaetots, \nonumber \\
\Gamma\left(B_s\to D_s^*\mu\bar{\nu}_\mu\right)/|V_{cb}\eta_\mathrm{EW}|^2 	&= \Gammamutots, \nonumber \\
\Gamma\left(B_s\to D_s^*\tau\bar{\nu}_\tau\right)/|V_{cb}\eta_\mathrm{EW}|^2	&= \Gammatautots. 
\end{align}
Note that the total decay rates for $B_s\to D_s^*$ are approximately $1\sigma$ lower than those computed by us previously in~\cite{Harrison:2021tol}. This is discussed further in~\cref{comptoprev}, where we compare our updated form factors for $B_s \to D_s^*$ to those in~\cite{Harrison:2021tol}.

We use our form factors to compute $R(D^*_{(s)})$, defined in~\cref{RDstardef}. We compute both a `lattice-only' value, using only our computed form factors, as well as a `lattice+experiment' value where we use the form factors resulting from our fits to lattice and experimental data in~\cref{shapetolattbelle}. These are given in~\cref{RDstartable}, together with the improved ratios in which the rates are integrated only between $q^2_\mathrm{max}$ and $m_\tau^2$,
\begin{equation}
R^\mathrm{imp}(D_{(s)}^*)=\frac{\int_{m_\tau^2}^{q^2_\mathrm{max}}dq^2 \frac{d\Gamma}{dq^2}({B}_{(s)}\rightarrow D_{(s)}^{*}\tau\bar{\nu}_\tau)}{\int_{m_\tau^2}^{q^2_\mathrm{max}}dq^2 \frac{d\Gamma}{dq^2}({B}_{(s)}\rightarrow D_{(s)}^{*}\mu\bar{\nu}_\mu)}.
\end{equation}
We see that the inclusion of experimental data shifts $R(D^*)$ downwards significantly and reduces the uncertainty. Our `lattice-only' $R(D^*)$ is shown in~\cref{exp_plus_latt}, together with the `lattice+experiment' value. In that figure we also plot the `lattice-only' and `lattice+experiment' values of $R(D^*)$ computed by the Fermilab-MILC collaboration~\cite{FermilabLattice:2021cdg}, where the inclusion of experimental data produces a similar downward shift. The two most recent experimental measurements of $R(D^*)$, from Belle~\cite{Belle:2019rba} and LHCb~\cite{LHCb:2017rln}, are also shown, together with the HFLAV average value.

\begin{table}
\centering
\caption{\label{RDstartable}$R(D^*_{(s)})$ and $R^\mathrm{imp}(D^*_{(s)})$ computed first using our form factors only, as well as computed using our form factors together with the joint fits to experimental data described in the text. Here we see that the inclusion of experimental data moves the values down by~$\approx 2\sigma$, and reduces their uncertainties.}
\begin{tabular}{ c | c c }
 & `lattice-only'   & `lattice+experiment' \\
\hline
$R(D^*)$   &  \RDstar   & \RDstarexp \\
$R(D^*_s)$ &  \RDsstar  & \RDsstarexp \\
$R^\mathrm{imp}(D^*)$   &  \RDstarIMP  & \RDstarIMPexp \\
$R^\mathrm{imp}(D^*_s)$ &  \RDsstarIMP  & \RDsstarIMPexp 
\end{tabular}
\end{table}

We may also use our form factors to compute observables related to the angular asymmetry of the decay. Here we compute the lepton polarisation asymmetry, $\mathcal{A}_{\lambda_\ell}$, the longitudinal polsarisation fraction, $F_{L}^{D_{(s)}^*}$, and the forward-backward asymmetry, $\mathcal{A}_{FB}$. These are defined as
\begin{align}
\label{angasymeq}
\mathcal{A}_{\lambda_\ell}(q^2) =& \frac{d\Gamma^{\lambda_\ell=-1/2}/dq^2-d\Gamma^{\lambda_\ell=+1/2}/dq^2}{d\Gamma/dq^2},\nonumber\\
F_{L}^{D_{(s)}^*}(q^2) =& \frac{d\Gamma^{\lambda_{D_{(s)}^*}=0}/dq^2}{d\Gamma/dq^2},\nonumber\\
\mathcal{A}_{FB}(q^2) =& -\frac{1}{d\Gamma/dq^2} \frac{2}{\pi}\int_0^\pi\frac{d\Gamma}{dq^2d\cos(\theta_W)}\cos(\theta_W)d\theta_W.
\end{align}
The integrated observables related to these quantities are defined as in~\cite{Harrison:2021tol} with the numerators and denominators integrated over $q^2$ independently. We find for $B\to D^*\tau\bar{\nu}_\tau$
\begin{align}\label{angobs}
\langle \mathcal{A}_{\lambda_\tau}\rangle=&\Alambdatau,\nonumber\\
\langle F_{L}^{D^*}\rangle=&\FLep,  \nonumber\\
\langle \mathcal{A}_\mathrm{FB}\rangle=&\AFB,
\end{align}
and for $B_s\to D_s^*\tau\bar{\nu}_\tau$
\begin{align}\label{angobs_s}
\langle \mathcal{A}^s_{\lambda_\tau}\rangle=&\Alambdataus,\nonumber\\
\langle F_{L}^{D^*_s}\rangle=&\FLeps, \nonumber\\
\langle \mathcal{A}^s_\mathrm{FB}\rangle=&\AFBs.
\end{align}
These values are in disagreement with expectations from the heavy quark expansion~(HQE)~\cite{Bordone:2019guc}. They are also in tension at the level of $2.2\sigma$ with the recent measurement of the $D^*$ longitudinal polarisation fraction by Belle~\cite{Belle:2019ewo}, $F_L^{D^*~\mathrm{Belle}}=0.60(8)_\mathrm{stat}(4)_\mathrm{sys}$. Our value of $\langle\mathcal{A}_{\lambda_\tau}\rangle=-P_\tau(D^*)$ is in good agreement with the measured value from Belle~\cite{Belle:2017ilt}, although there is a large statistical uncertainty on the experimental measurement: $P_\tau(D^*)=-0.38\pm 51_{(\mathrm{stat})}~^{+21}_{{-16}(\mathrm{syst})}$.

\subsection{$SU(3)_\mathrm{flav}$}\label{su3f}

\begin{figure}
\includegraphics[scale=0.25]{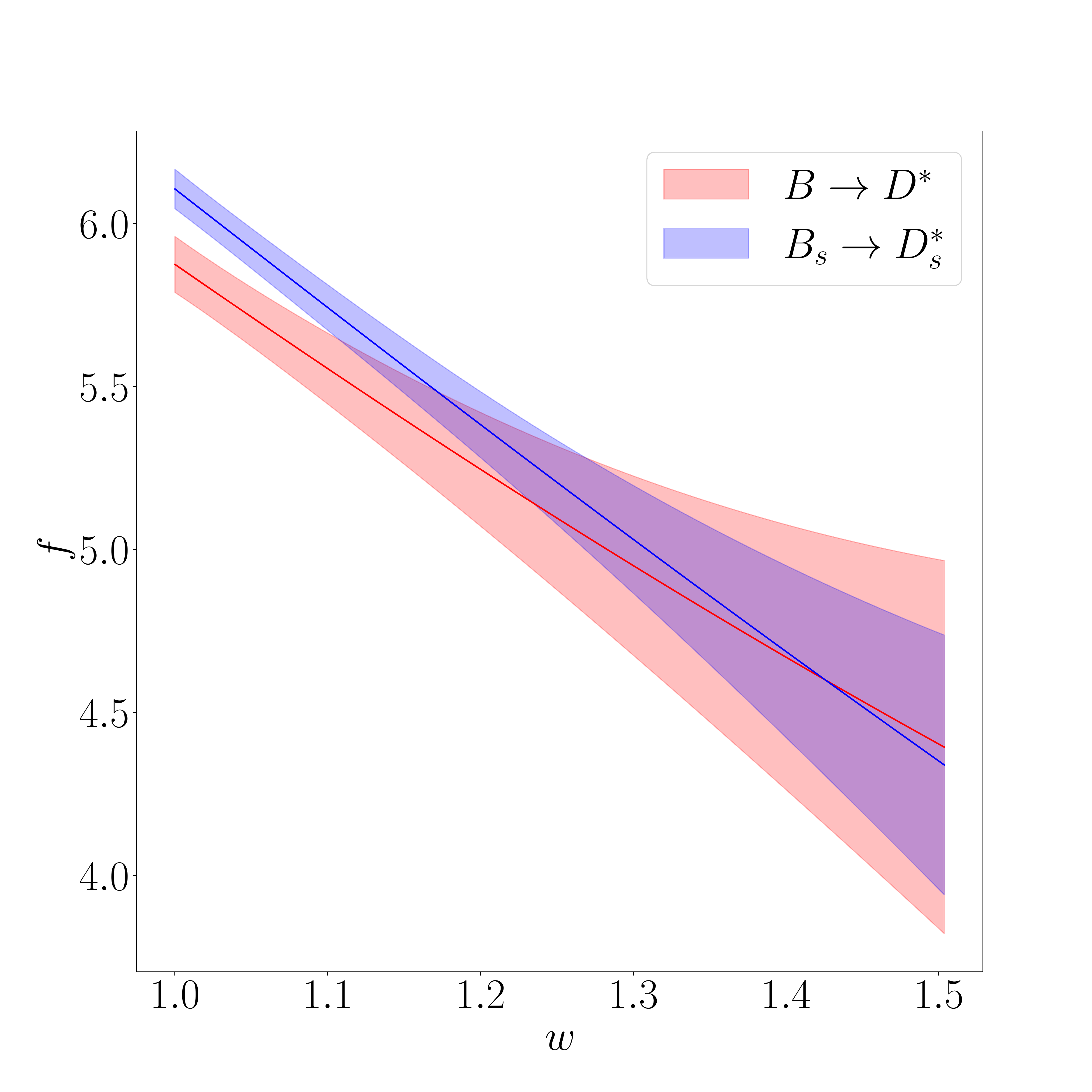}\\
\includegraphics[scale=0.25]{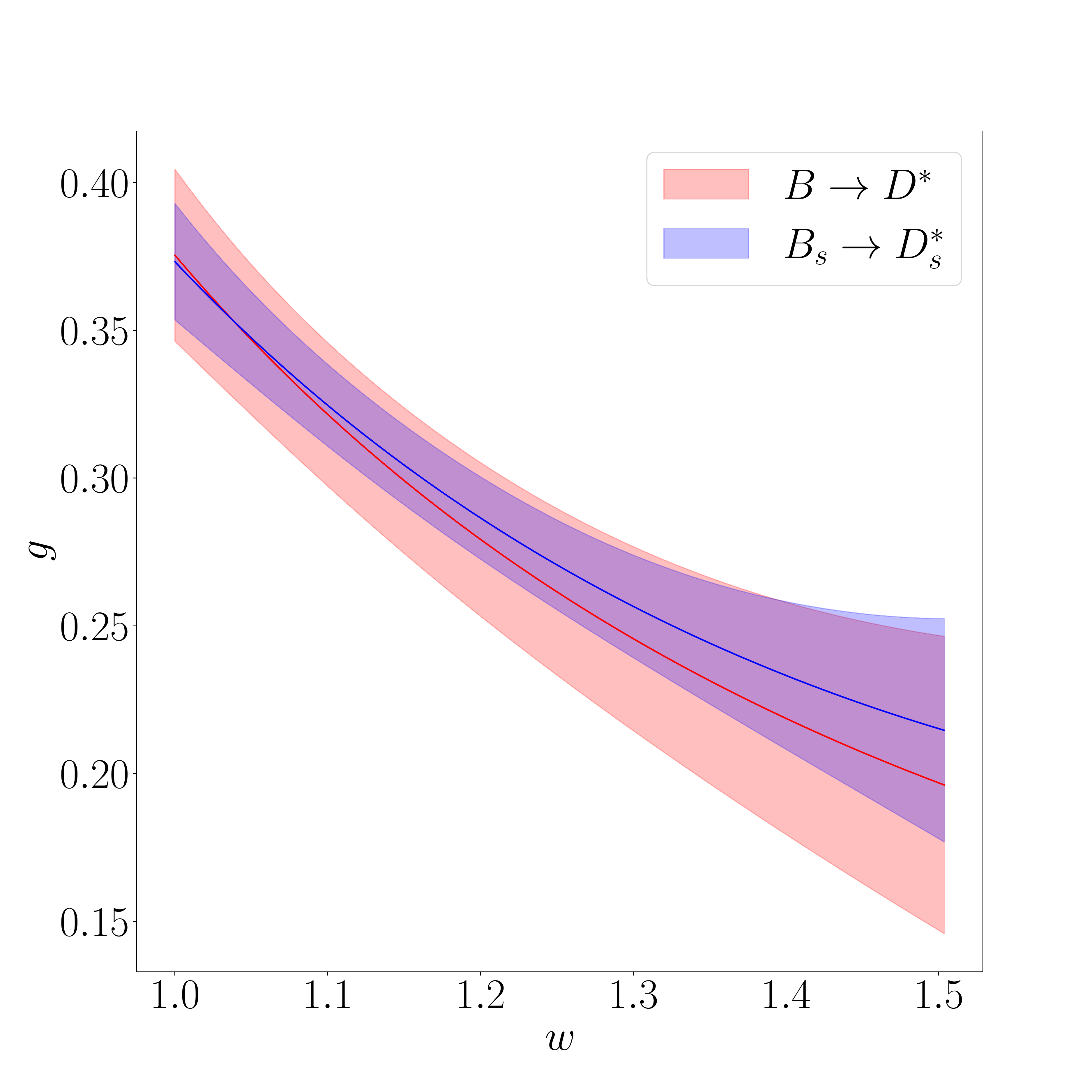}
\caption{\label{HPQCD-HQET_bsdss_comparison1}$B\to D^*$ and $B_s\to D_s^*$ helicity basis form factors $f$ and $g$, defined in~\cref{helicitybasis}.}
\end{figure}

\begin{figure}
\includegraphics[scale=0.25]{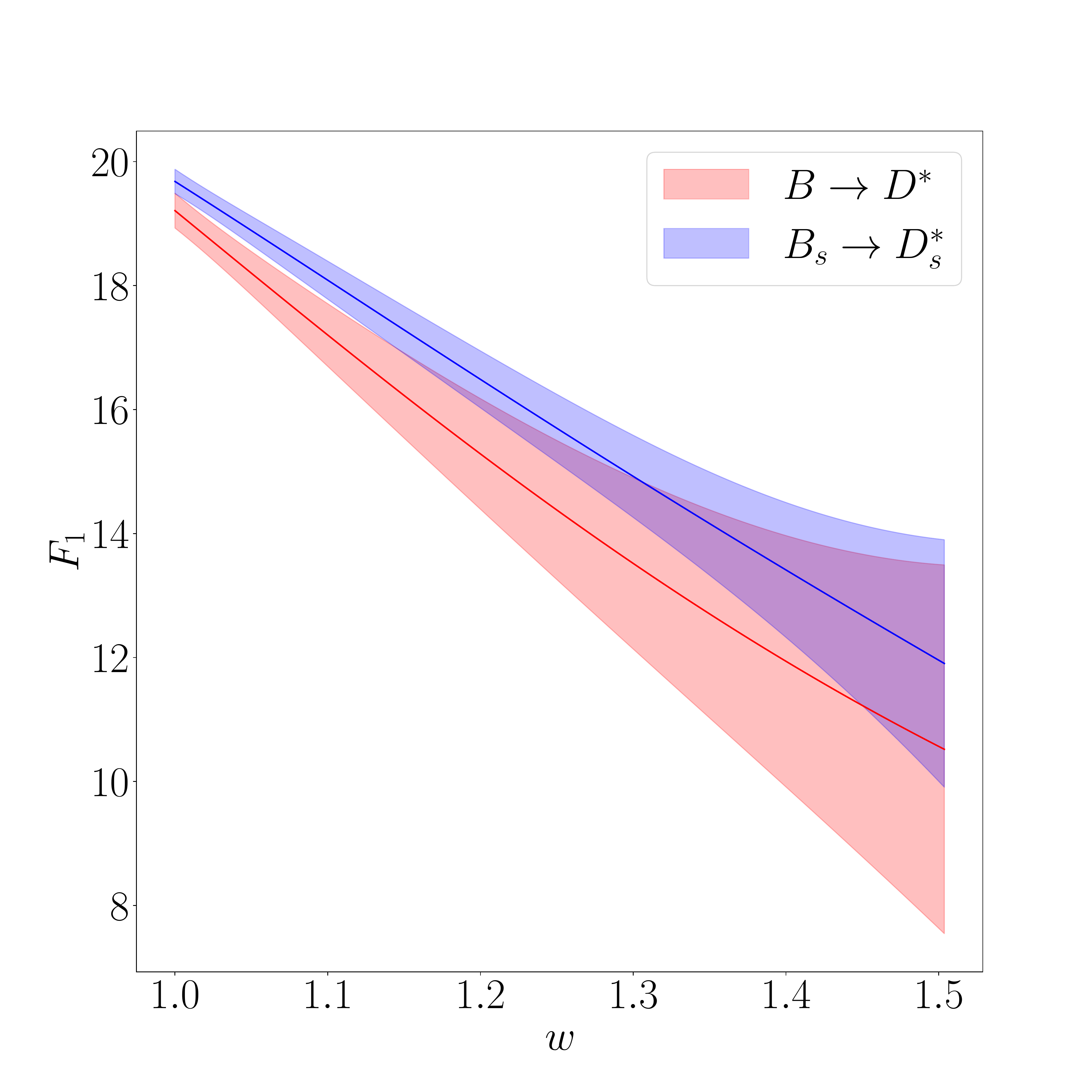}\\
\includegraphics[scale=0.25]{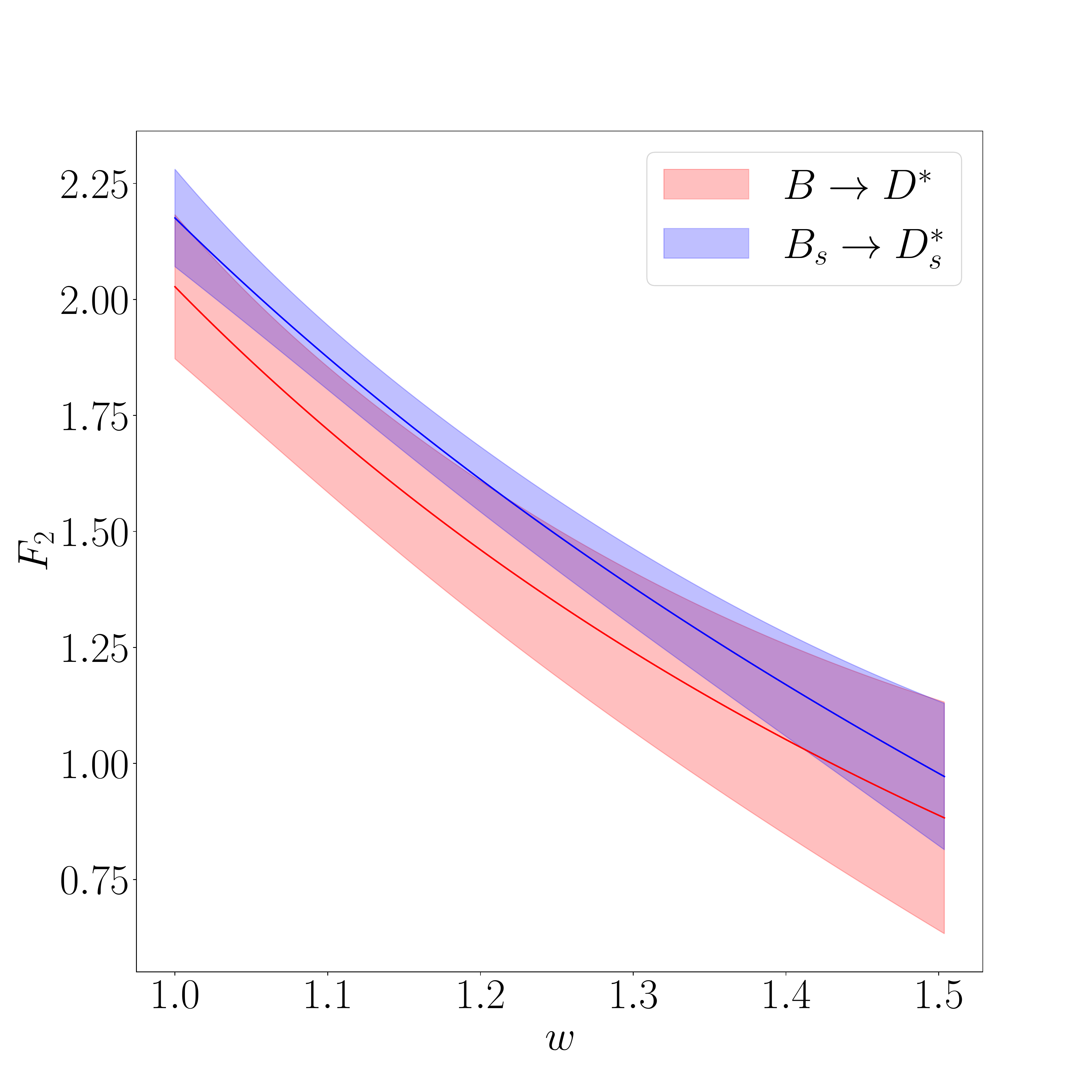}
\caption{\label{HPQCD-HQET_bsdss_comparison2}$B\to D^*$ and $B_s\to D_s^*$ helicity basis form factors $F_1$ and $F_2$, defined in~\cref{helicitybasis}.}
\end{figure}

$SU(3)_\mathrm{flav}$ symmetry breaking effects between $B\to D^*$ and $B_s\to D_s^*$ are expected to be small~\cite{Bordone:2019guc}. $R(D^*)$ and $R(D_s^*)$ are expected to differ by $\approx 1\%$. Here, using our lattice-only results, we find
\begin{equation}
R(D^*)/R(D_s^*) = \RDstaroverRDsstar.
\end{equation}
This result, together with our result for $R(J/\psi)$~\cite{Harrison:2020nrv}, implies the simple relation to increasing spectator quark mass $R(D^*)>R(D_s^*)>R(J/\psi)$. We also compute the ratios of the angular observables given in~\cref{angularobsRdstarGamma}. We find
\begin{align}
\langle \mathcal{A}_{\lambda_\tau}\rangle/\langle \mathcal{A}^s_{\lambda_\tau}\rangle=&\Alambdatauratio,\nonumber\\
\langle F_{L}^{D^*}\rangle/\langle F_{L}^{D_s^*}\rangle=&\FLepratio,  \nonumber\\
\langle \mathcal{A}_\mathrm{FB}\rangle/\langle \mathcal{A}^s_\mathrm{FB}\rangle=&\AFBratio.
\end{align}
These results are in slight tension with the HQE expectation of $\approx 1\%$ $SU(3)_\mathrm{flav}$ symmetry breaking, though this tension is not significant for our level of uncertainty. The SM form factors for $B\to D^*$ and $B_s\to D_s^*$ are plotted in the helicity basis in~\cref{HPQCD-HQET_bsdss_comparison1,HPQCD-HQET_bsdss_comparison2}, where we see $SU(3)_\mathrm{flav}$ symmetry breaking with differences ranges from $\approx 1\%$ for $f$ and $g$ up to $\approx 10\%$ for $F_2$. 

\subsection{Constraining New Physics in $B\to D^*\ell\bar{\nu}_\ell$}
\label{muNP}

The effective Hamiltonian~\cref{EffectiveHamiltonian} is most commonly expressed in terms of left and right handed fermions as
\begin{align}
\mathcal{H}_\mathrm{eff}=\frac{4G_{F}V_{cb}}{\sqrt{2}}\Big[&g_{V_L}\bar{c}_L\gamma_\mu b_L \bar{\ell}_L\gamma^\mu \nu_L+g_{V_R}\bar{c}_R\gamma_\mu b_R \bar{\ell}_L\gamma^\mu \nu_L\nonumber\\
+&g_{S_L}\bar{c}_Rb_L \bar{\ell}_R \nu_L\nonumber\\
+&g_{S_R}\bar{c}_L\gamma_5b_R \bar{\ell}_R \nu_L\nonumber\\
+&g_{T_L} \bar{c}_R\sigma_{\mu\nu}b_L\bar{\ell}_R\sigma^{\mu\nu} \nu_L + \mathrm{h.c.}\Big]
\end{align}
where $g_{T_L}=(g_{T}-g_{T5})/2$, $g_{V_L} = (g_V-g_A)/2$, $g_{V_R} = (g_V+g_A)/2$,  $g_{S_L} = (g_S-g_P)/2$ and $g_{S_R} = (g_S+g_P)/2$. Note that there is no $g_{T_R}$, since the corresponding current, $\bar{c}_L\sigma_{\mu\nu}b_R\bar{\ell}_R\sigma^{\mu\nu} \nu_L$, is identically zero. Here we have given the effective Hamiltonian for only a single flavour of lepton. Unlike in~\cref{shapetolattbelle}, we will now not assume LFU between the $\ell=\mu$ and $\ell=e$ modes and instead study each case separately. The couplings for each lepton flavour will be indicated by a superscript, as in $g^\ell_X$.

In~\cite{Jung:2018lfu} the authors give the patterns of couplings produced by different tree level models of NP. For the models they considered at most one of either the left or right handed vector coupling differed from its SM value, together with different nonzero combinations of the left and right handed scalar couplings and left handed tensor coupling. Throughout this subsection we compute constraints for different combinations of the various couplings and $V_{cb}$. Because $g_{V_L}$ may be absorbed into $V_{cb}$, it is sufficient in our case to fix $g_{V_L}=1$. For $g_{V_L}\neq 1$ one should take $g_X\to \tilde{g}_X=g_X/g_{V_L}$ and $V_{cb}\to \tilde{V}_{cb}=V_{cb}\times g_{V_L}$ in the constraints given below. In order to compute the constraints we fit the Belle data using our lattice FFs in the same manner as described in~\cref{shapetolattbelle}, for fixed numerical values of $g_X$. We fit the normalised binned differential data from Belle, and only include the total rate, $\Gamma$, as a single additional data point when we compute constraints including $V_{cb}$.

\subsubsection{Scalar Operators}
\begin{figure}
\includegraphics[scale=0.5]{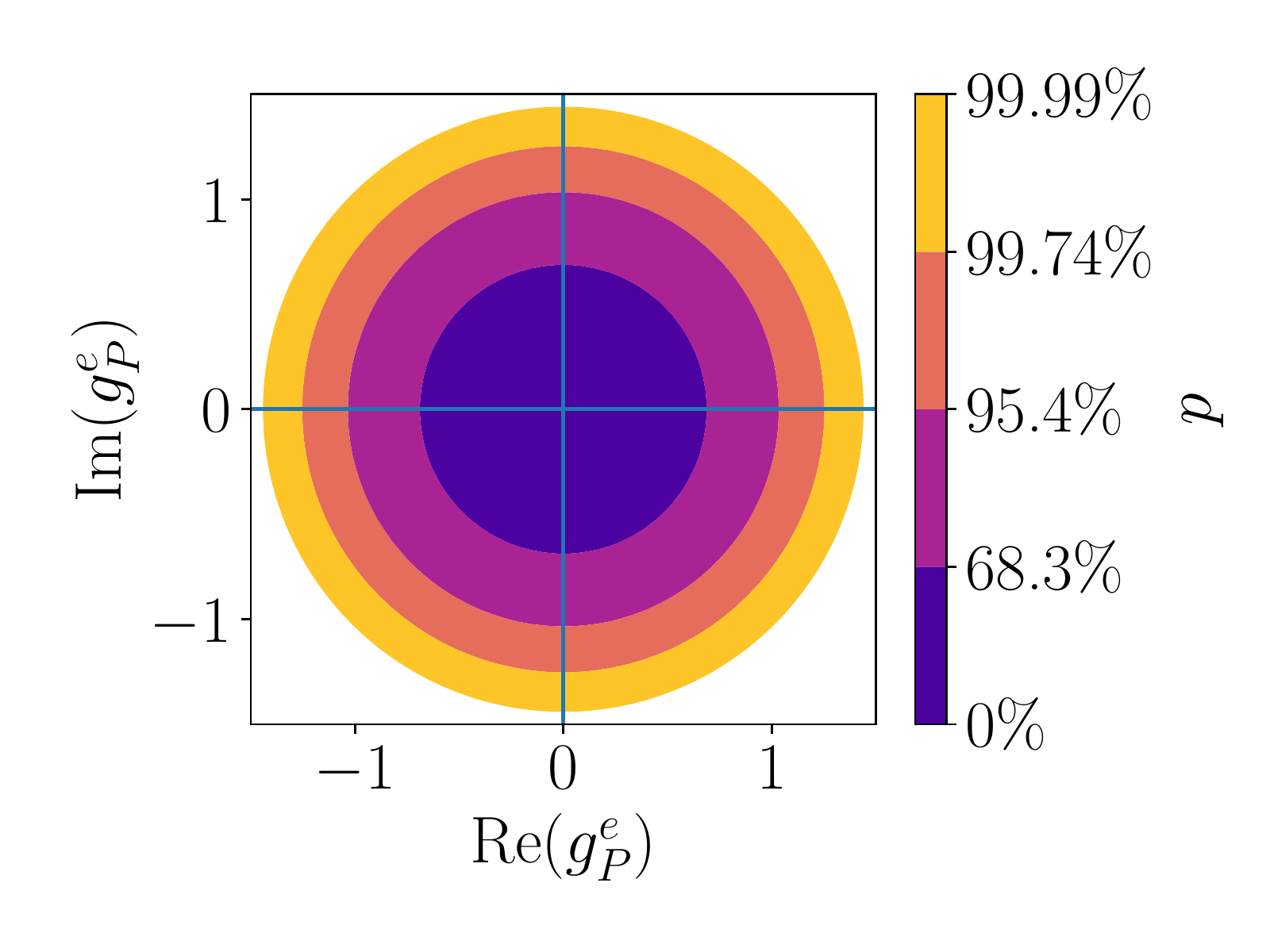}\\
\includegraphics[scale=0.5]{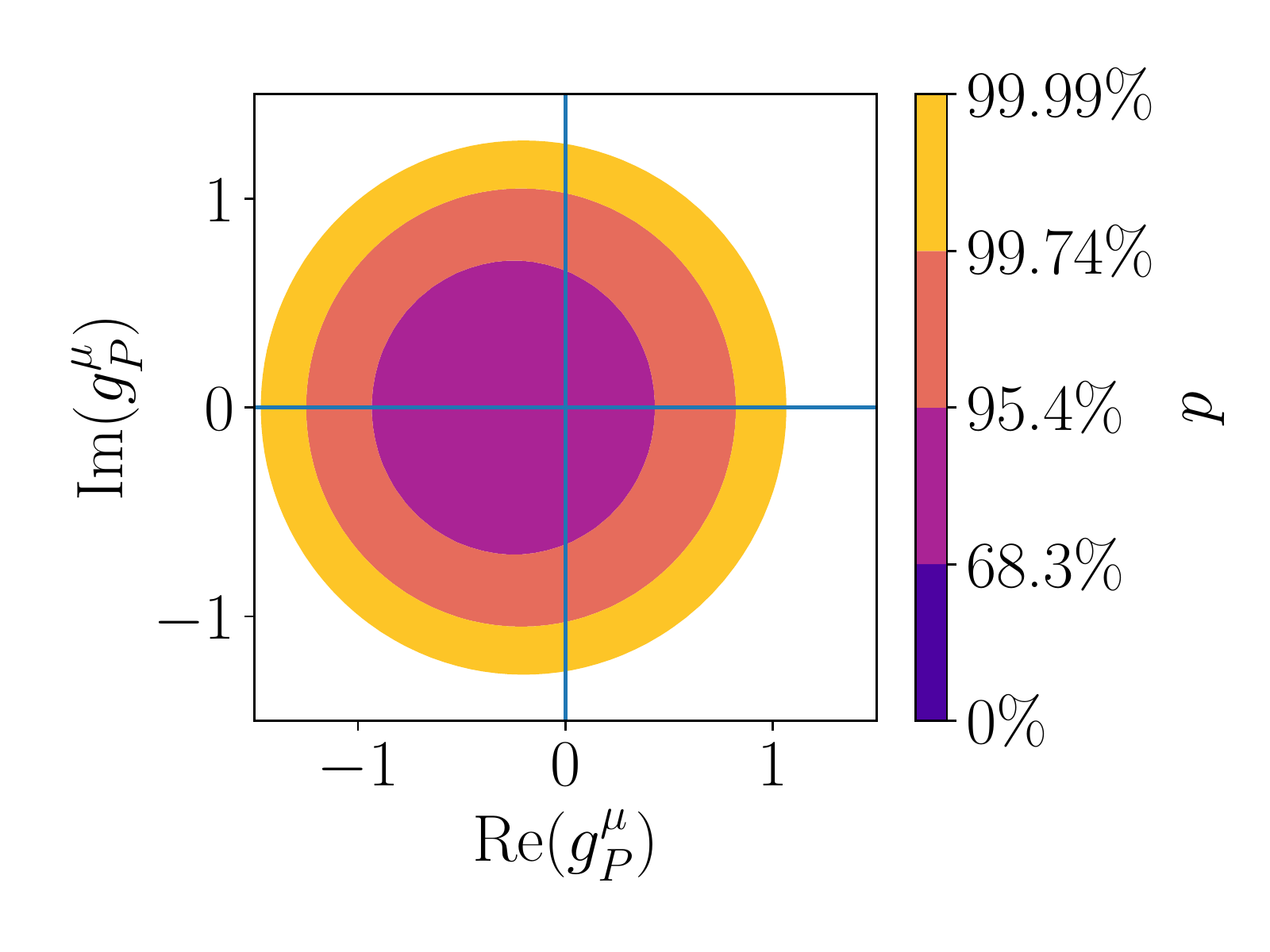}
\caption{\label{VcbRegP_sigma_plot} The top~(bottom) plot shows the constraints on $g_P^{e(\mu)}$ using our theory-only differential decay rate and the data for $B^0\to D^{*-} e^+(\mu^+){\nu}_{e(\mu)}$ from Belle~\cite{Belle:2018ezy} for different combinations of $g^{e(\mu)}_{P}$ and $V_{cb}$.  The different shaded regions correspond to the confidence level to which those values are excluded with intervals of $p=0\%,~68.3\%,~95.4\%,~99.7\%,~99.99\%$. The unshaded regions of the plot have $p>99.99\%$. The vertical and horizontal blue lines correspond to the SM value.}
\end{figure}

In~\cite{Jung:2018lfu} it was found that $B\to D^*$ produces only very weak constraints on the left handed scalar coupling. Indeed, using just $B\to D^*$ it is only possible to constrain the pseudoscalar combination $g_{S_L}-g_{S_R}=g_P$, with the QCD matrix element of the scalar current zero by parity as shown in~\cref{formfactors}. We find the constraints considering modifications to $g_P$ alone are very weak, as shown in~\cref{VcbRegP_sigma_plot} for both $\ell=e$ and $\ell=\mu$. Note that, since $B\to D\ell\nu$ provides complementary constraints for the scalar operators, fully correlated lattice results for both $B \to D^*$ and $B\to D$ SM and NP form factors would allow for the simultaneous constraint of all NP couplings.

\subsubsection{Tensor Operator}
\begin{figure}
\includegraphics[scale=0.5]{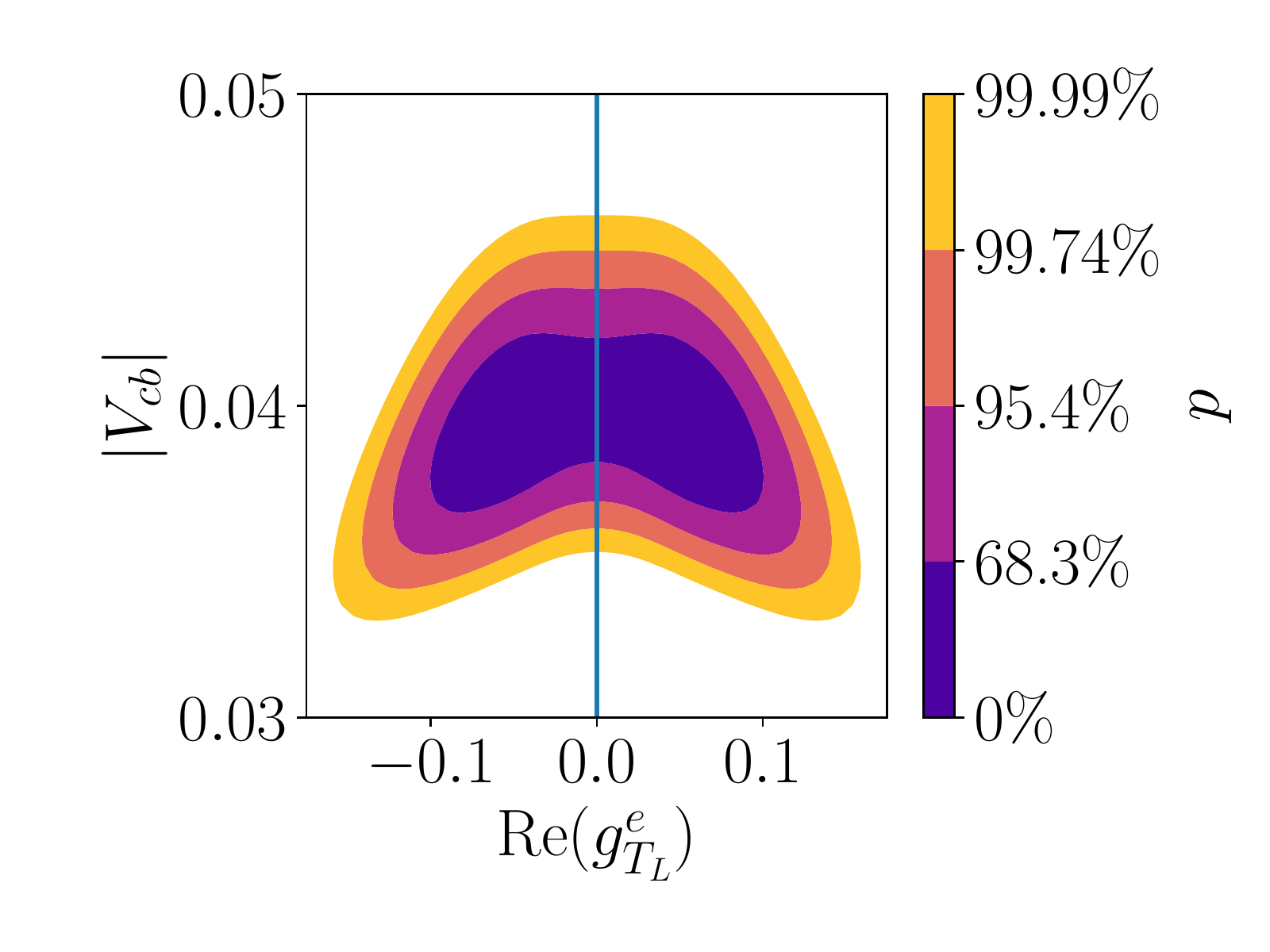}\\
\includegraphics[scale=0.5]{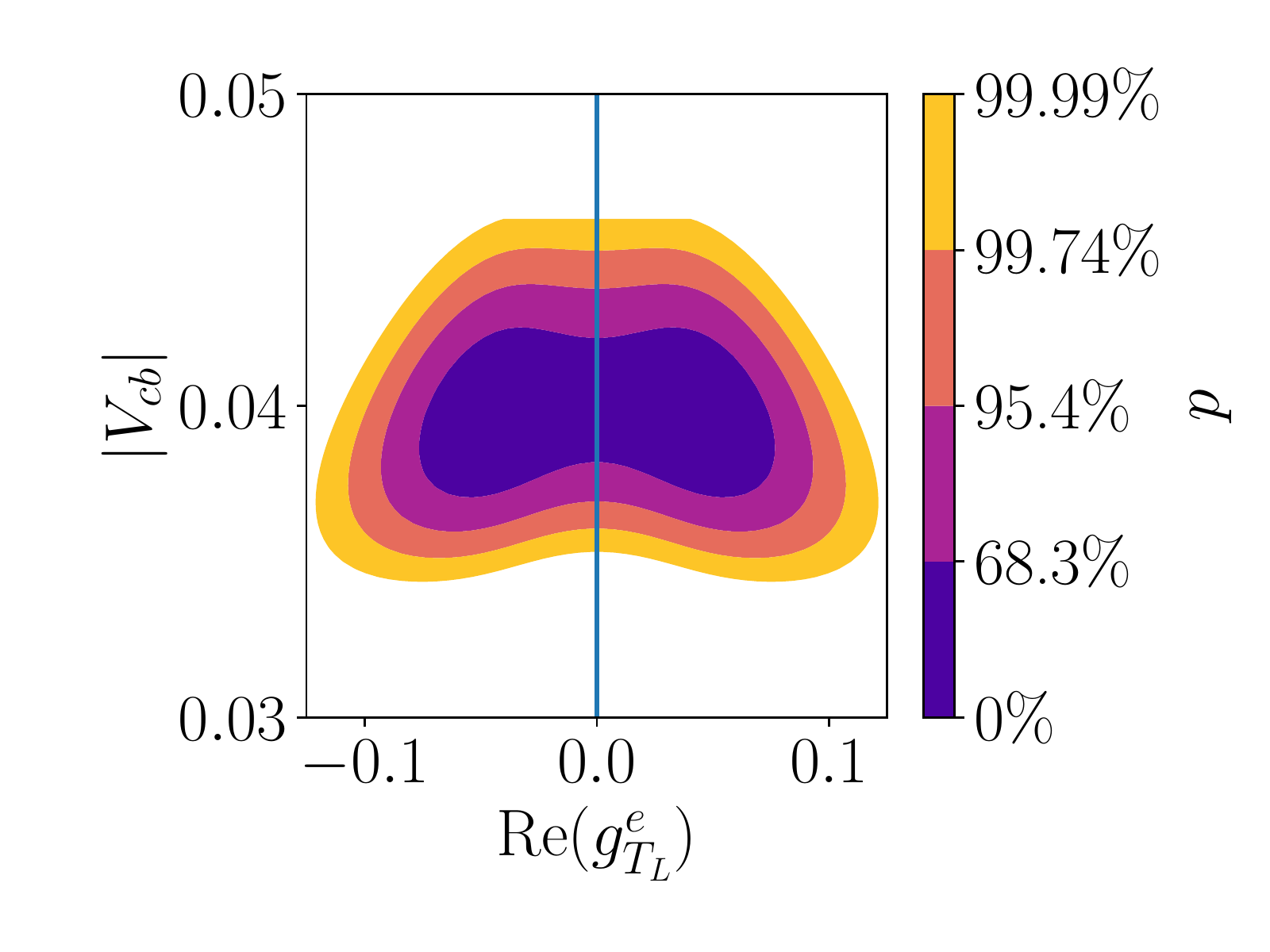}
\caption{\label{VcbRegT_L_e_sigma_plot} Tension between our theory-only differential decay rate and the data for $B^0\to D^{*-} e^+{\nu}_e$  from Belle~\cite{Belle:2018ezy} for different combinations of $\mathrm{Re}(g^e_{T_L})$ and $V_{cb}$. The top~(bottom) plot corresponds to the $g_P\approx +(-)4 g_{T_L}$ case described in the text. The different shaded regions correspond to the confidence level to which those values are excluded with intervals of $p=0\%,~68.3\%,~95.4\%,~99.7\%,~99.99\%$. The unshaded regions of the plot have $p>99.99\%$. The vertical blue line corresponds to the SM value. }
\end{figure}

\begin{figure}
\includegraphics[scale=0.5]{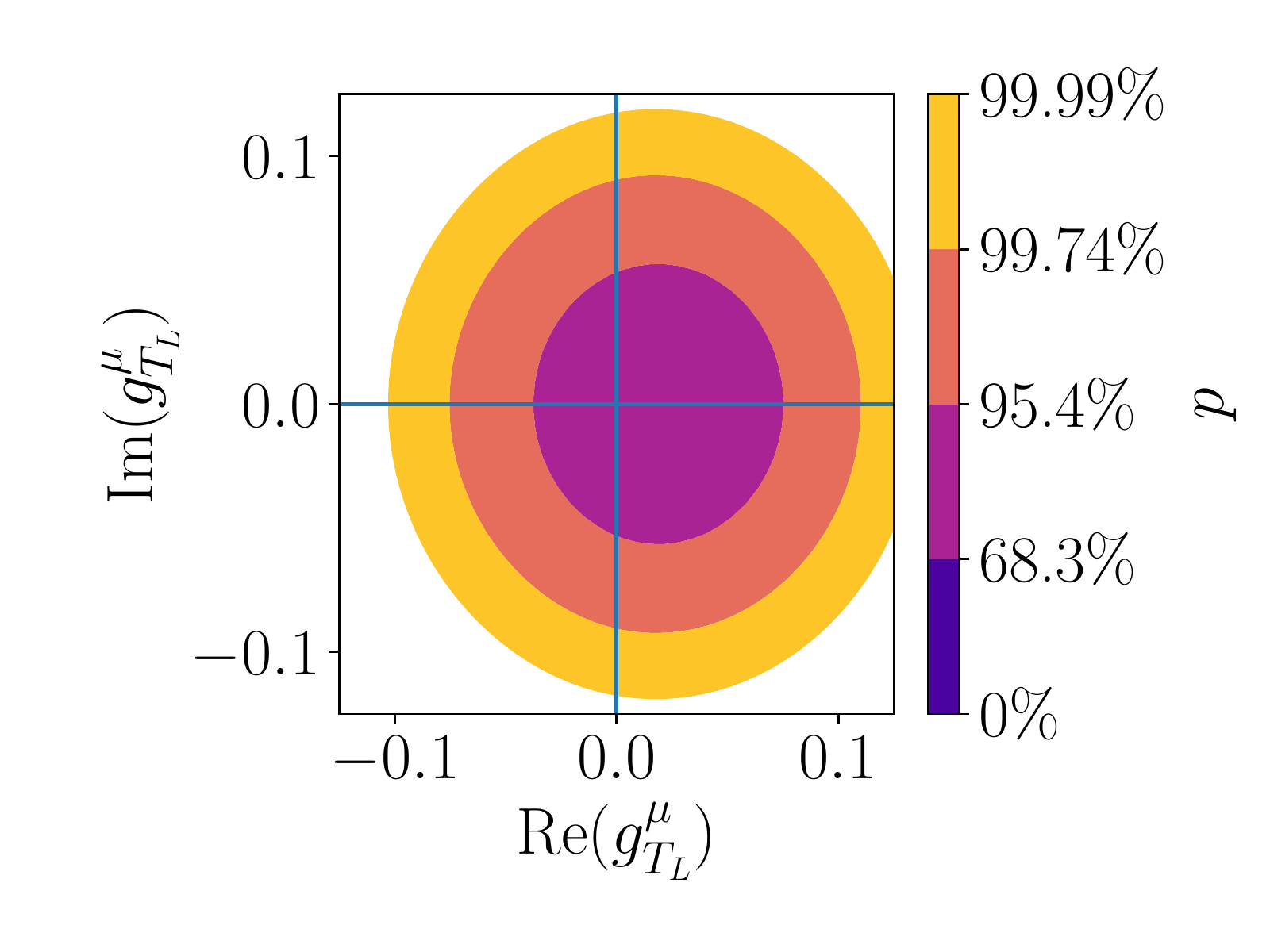}\\
\includegraphics[scale=0.5]{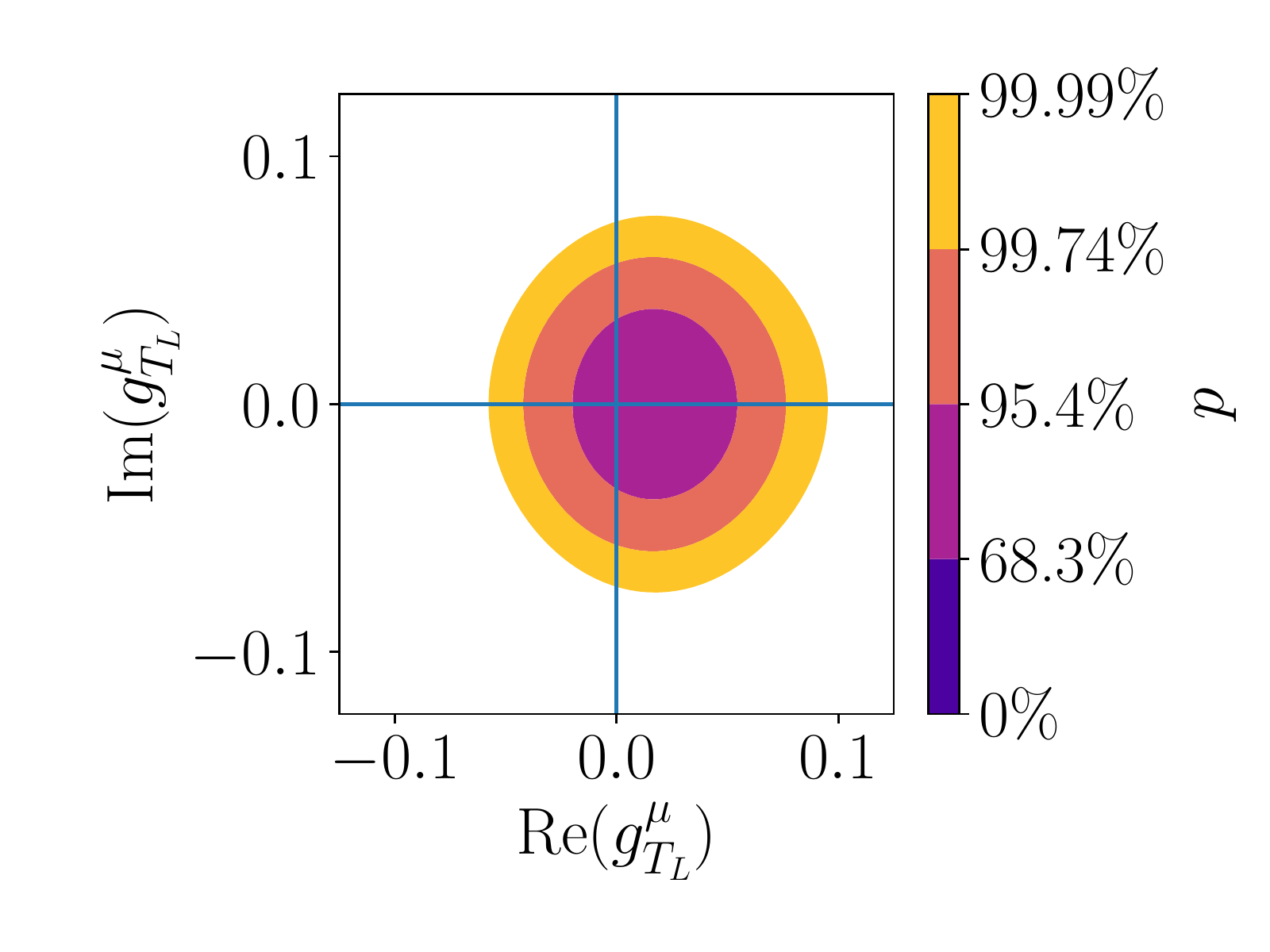}
\caption{\label{VcbRegT_L_mu_sigma_plot} Tension between our theory-only differential decay rate and the data for $B^0\to D^{*-} \mu^+{\nu}_\mu$  from Belle~\cite{Belle:2018ezy} for different combinations of $\mathrm{Re}(g^\mu_{T_L})$ and $V_{cb}$. The top~(bottom) plot corresponds to the $g_P\approx +(-)4 g_{T_L}$ case described in the text. The different shaded regions correspond to the confidence level to which those values are excluded with intervals of $p=0\%,~68.3\%,~95.4\%,~99.7\%,~99.99\%$. The unshaded regions of the plot have $p>99.99\%$. The vertical and horizontal blue line corresponds to the SM value. }
\end{figure}
Of the models considered in~\cite{Jung:2018lfu} only $S_1$ and $R_2$ produced a nonzero tensor coupling. These models also produced a correlated nonzero left-handed scalar operator, with $g_{S_L} = \pm 4 g_T$. Based on the expectation that Renormalisation Group mixing effects will leave the relations between left-handed scalar and tensor couplings approximately intact~\cite{Jung:2018lfu}, we include the pseudoscalar as a Gaussian random variable with central value $\pm 4 g_T$ and uncertainty $\pm 10\%$.

For the $\ell=e$ case the terms proportional to $m_\ell^2/q^2$ and $\sqrt{m_\ell^2/q^2}$ may be neglected. Then the only relevant combinations of helicity amplitudes are those appearing in~\cref{tab:mellzero_0}. These do not mix the tensor or pseudoscalar helicity amplitudes with the helicity amplitudes for the SM currents. For the $\ell=e$ mode we can then only determine constraints on the relative phase of $g_P$ and $g_{T_L}$. Since we fix $g_P=\pm 4 g_{T_L}$, we may look at just the real part of $g_{T_L}$, together with the value of $|V_{cb}|$. The constraints for both $g_P=\pm 4 g_{T_L}$ cases for $\ell=e$ are shown in~\cref{VcbRegT_L_e_sigma_plot}, where we see that the constraints in the $g_P= -4 g_{T_L}$ case are similar to the $g_P= +4 g_{T_L}$ case.

The situation for $\ell=\mu$ is more complicated. In the SM the lepton-mass-suppressed terms have factors $m_\ell^2/q^2$ but in NP scenarios combinations of the SM and NP helicity amplitudes appear at order $\sqrt{m_\ell^2/q^2}$. This contribution can be significant for $\ell=\mu$, depending on the size of $g^\mu_{T_L}$, and so we cannot remove the overall phase and must consider both the real and imaginary parts of $g^\mu_{T_L}$. The resulting constraint, using only the normalised differential rate, which is insensitive to $V_{cb}$, is shown in~\cref{VcbRegT_L_mu_sigma_plot}, where we see that the Belle $B^0\to D^{*-} \mu^+{\nu}_\mu$ data is consistent with $g^\mu_{T_L}=0$ at the level of $\approx 1\sigma$.

\subsubsection{Right-handed Vector Operator}
\begin{figure}
\includegraphics[scale=0.5]{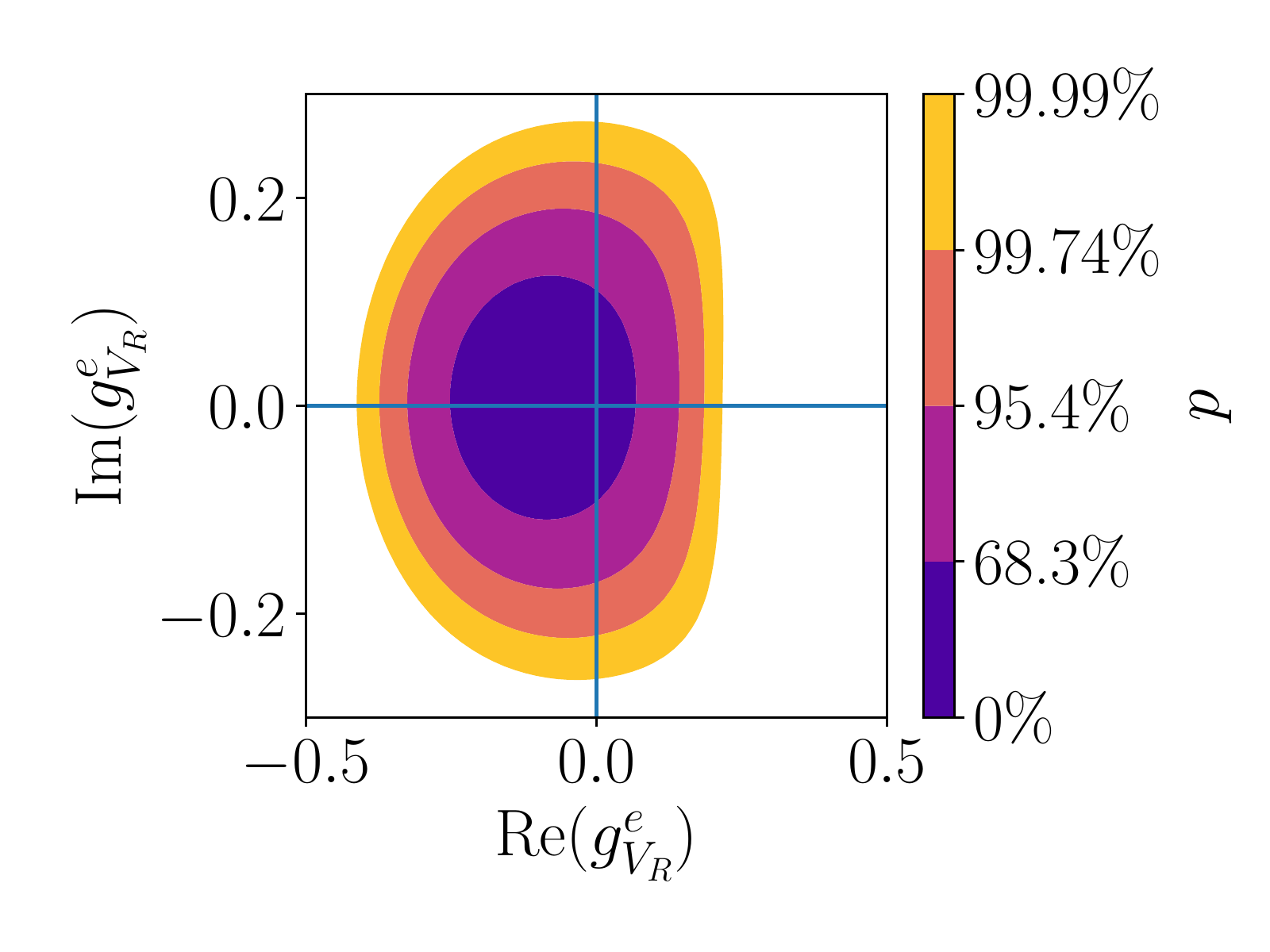}\\
\includegraphics[scale=0.5]{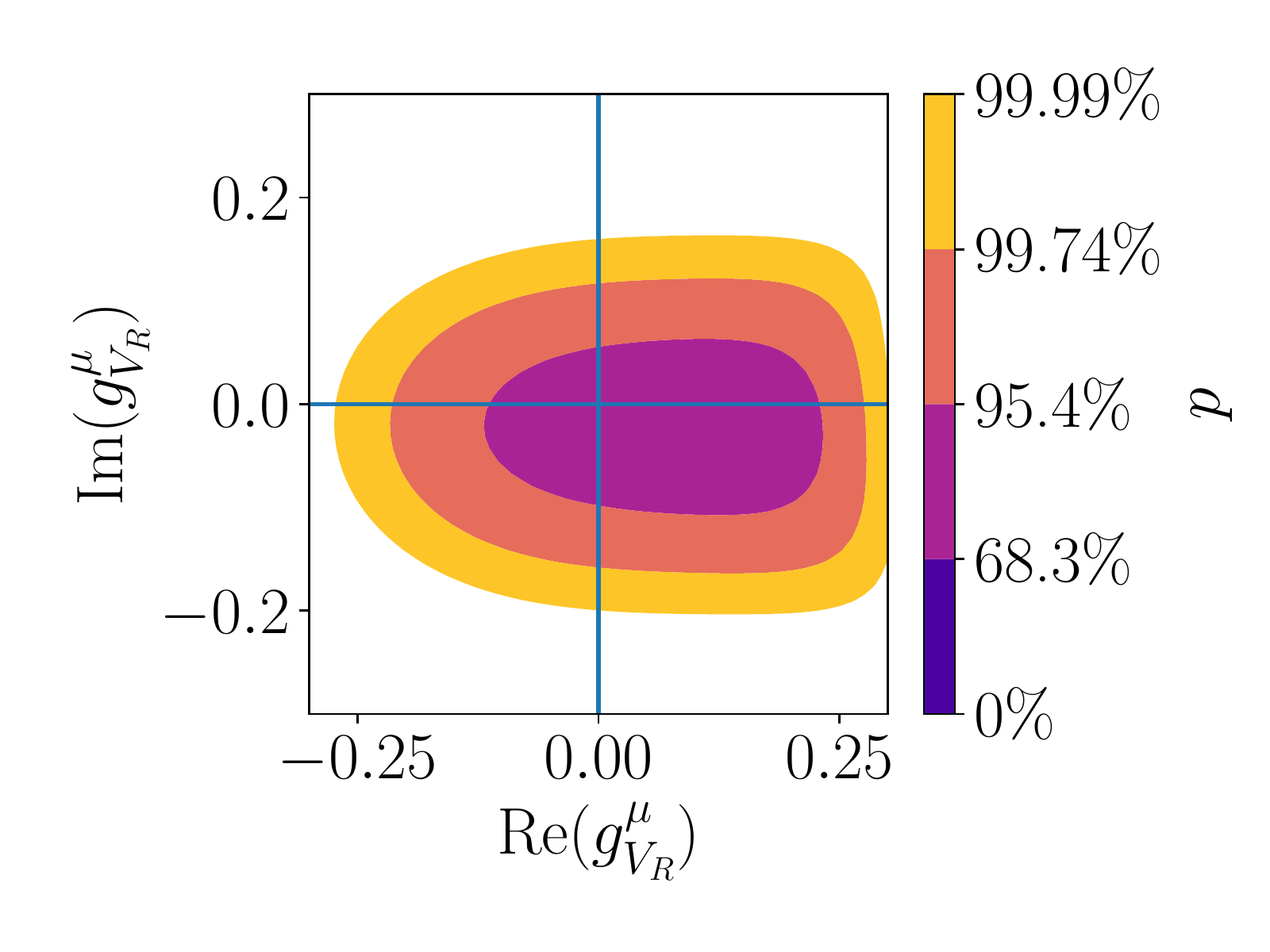}
\caption{\label{VcbRegV_R_sigma_plot} The top~(bottom) plot shows the constraints on $g_P^{e(\mu)}$ using our theory-only differential decay rate and the data for $B^0\to D^{*-} e^+(\mu^+){\nu}_{e(\mu)}$ from Belle~\cite{Belle:2018ezy} for different combinations of $g^{e(\mu)}_{V_R}$. The different shaded regions correspond to the confidence level to which those values are excluded with intervals of $p=0\%,~68.3\%,~95.4\%,~99.7\%,~99.99\%$. The unshaded regions of the plot have $p>99.99\%$. The vertical blue line corresponds to the SM value. }
\end{figure}

The constraints on the right-handed vector coupling, $g^\ell_{V_R}$, computed using our lattice results and the Belle data are shown in~\cref{VcbRegV_R_sigma_plot} for the $\ell=e$ and $\ell=\mu$ cases, where we also see no strong preference for a nonzero value.

\subsection{$V_{cb}^{e(\mu)}$, $\Delta A_{FB}$}

A clear feature of the constraints produced is that the $\ell=\mu$ data does not agree well with our SM predictions, compared to the $\ell=e$ case, for any values of the couplings considered. Having reconstructed the full $80\times 80$ Belle covariance matrix as described in~\cref{shapetolattbelle}, we may compute a value of $V_{cb}$ using $\ell=e$ and $\ell=\mu$ separately and calculate their difference including correlations. We find
\begin{align}
V_{cb}^e  = \VCBE,\nonumber\\
V_{cb}^\mu  = \VCBMU,
\end{align}
and
\begin{align}
V_{cb}^e/V_{cb}^\mu  = \RATIOVCBEMU,
\end{align}
consistent with $V_{cb}^e/V_{cb}^\mu  =1$ as we would expect in the absence of NP. This is a similar level of consistency as was seen in~\cite{Jung:2018lfu}.

In~\cite{Bobeth:2021lya} it was found that the 2018 Belle dataset was inconsistent with the SM prediction for $\Delta A_{FB} =  \langle A_{FB}^\mu \rangle - \langle A_{FB}^e \rangle$, using a combination of HQE, light-cone sum rules and lattice QCD results for the zero recoil $B\to D^*$ form factor, $h_{A_1}$, and for the SM $B\to D$ form factors across the full $q^2$ range. Here, using our lattice-only results, we find
\begin{align}
\langle A_{FB}^\mu \rangle =\AFBMU,\nonumber\\
\langle A_{FB}^e \rangle   = \AFBE.
\end{align}
This is in tension at the level of $\approx 2.5\sigma$ with the SM results for both $\ell=e$ and $\ell=\mu$ in~\cite{Bobeth:2021lya}. We also find
\begin{align}
\Delta A_{FB} = \DELTAAFBMUE.
\end{align}
This result is in tension with the SM results in~\cite{Bobeth:2021lya} at the level of $\approx 2\sigma$, although note that it is still much smaller than and of opposite sign to the corresponding result using fits to experimental data~\cite{Bobeth:2021lya}, in tension at the level of $3.5\sigma$.

\section{Conclusion}
\label{Conclusion}

We have computed the $B_{(s)}\to D^*_{(s)}$ form factors for the complete set of vector, axial-vector and tensor currents needed to describe both SM physics and potential new physics appearing in the effective Hamiltonian,~\cref{EffectiveHamiltonian}. These form factors include a fully relativistic treatment of both charm and bottom quarks in lattice QCD and span the full kinematic range of the decay. Our calculation includes two sets of gauge configurations with physical up/down quarks, which we use to constrain the chiral dependence in our chiral-continuum extrapolation using the full rooted staggered chiral perturbation theory (see~\cref{chilogs}). 

We have used our form factors to perform the first combined fit to both $B\to D^*$ data, from Belle~\cite{Belle:2018ezy}, as well $B_s\to D_s^*$ data, from LHCb~\cite{Aaij:2020xjy}. This gives a value of $V_{cb}$:
\begin{equation}
|V_{cb}| = \vcb.
\end{equation}
This is in good agreement with other exclusive determinations and confirms the tension seen with inclusive determinations~\cite{HFLAV:2022pwe}. For our result this tension is at the level of $\approx 3.6\sigma$ with the most recent inclusive value $|V_{cb}|=42.16(51)\times 10^{-3}$~\cite{Bordone:2021oof}. We have also determined a less precise value of \totalVCB, computed using only the total decay rate. This value is in good agreement with the inclusive result. The significant upward shift can be understood as a result of the observed tension between our results for the shape of the differential decay rate and the experimental data from Belle. This tension is similar to the tension seen by the Fermilab-MILC collaboration using their lattice QCD form factors, determined using a different formalism for $b$ and $c$ quarks~\cite{FermilabLattice:2021cdg}. A tension in the shape of the differential decay rate to light leptons is difficult to explain, since new physics is only expected to appear in the semitauonic mode. We have also computed the slope of $\mathcal{F}$, defined in~\cref{Fdefeq}, and plotted $|\mathcal{F}V_{cb}\eta_\mathrm{EW}|^2$ in~\cref{FsquareBelle}, where the difference in shape is visible.

We have also used our form factors to compute the phenomenologically important quantities, $R^{(\mathrm{imp})}(D_{(s)}^*)$, $\langle \mathcal{A}^{(s)}_{\lambda_\tau}\rangle$, $\langle F_{L}^{D^*_{(s)}}\rangle$ and $\langle \mathcal{A}^{(s)}_\mathrm{FB}\rangle$, given in~\cref{RDstartable,angobs,angobs_s} respectively. We find our value of $R(D^*)=\RDstar$ is in good agreement with the latest experimental measurements from Belle and LHCb~\cite{Belle:2019rba,LHCb:2017rln}, and with the most recent HFLAV average~\cite{HFLAV:2022pwe}. However, our value of the semitauonic $D^*$ longitudinal polarisation fraction is in tension with the recent Belle measurement~\cite{Belle:2019ewo} at the level of $2.2\sigma$. We have also determined a `lattice+experiment' value of $R(D^*)=\RDstarexp$, computed using the form factors resulting from the fit to both our lattice results and the experimental data from Belle, described in~\cref{shapetolattbelle}. The downward shift of the value of $R(D^*)$ when including the experimental differential rate data means that this lower value is in tension at the level of $3\sigma$ with the HFLAV average for $R(D^*)$. So we see that the `$R(D^*)$ anomaly', like the `$V_{cb}$ puzzle', arises from the inclusion of the experimental differential rate data.

The inclusion of $B_s\to D_s^*$ data in our chiral-continuum extrapolation has allowed us to also provide an update on these form factors from our previous calculation~\cite{Harrison:2021tol}. We find the additional data and chiral information, as well as changes to correlator fitting procedures and extrapolation method, result in $B_s\to D_s^*$ form factors which are compatible with our previous results but significantly more precise, particularly close to $w=1$ where we see an improvement in precision by a factor of $\approx 2$. We have used our updated $B_s\to D_s^*$ form factors to investigate $SU(3)_\mathrm{flav}$ symmetry breaking effects appearing in $R^{(\mathrm{imp})}(D_{(s)}^*)$, $\langle \mathcal{A}^{(s)}_{\lambda_\tau}\rangle$, $\langle F_{L}^{D^*_{(s)}}\rangle$ and $\langle \mathcal{A}^{(s)}_\mathrm{FB}\rangle$. In each observable we find that the $B\to D^*$ differs by $+0.6(1.4)\%$, $+4.0(2.9)\%$, $-5.8(4.6)\%$ and $+19.0(23.0)\%$ respectively from the $B_s\to D_s^*$ value.

We have used our form factors to generate synthetic data points which we fit using the popular BGL parameterisation, which we found gave a good fit with the unitarity bounds far from saturation. Our fitted BGL parameters agree well with those in~\cite{FermilabLattice:2021cdg} for $f$, $g$ and $F_1$ but for the form factor $F_2$, corresponding to the pseudoscalar current, are in significant disagreement. Note that in decay rates $F_2$ is suppressed by the square of the lepton mass and so only contributes to the semitauonic mode.

Finally we examined the constraints on the NP couplings for the $\ell=\mu/e$ modes resulting from combining our lattice results with the 2018 untagged Belle dataset. We found that none of the couplings, when varied in the combinations described in~\cref{muNP}, strongly preferred values different from the SM ones. We used our results to compute SM values for $\langle \mathcal{A}_\mathrm{FB}\rangle$ for both $\ell=\mu/e$ modes. Our values differ from the SM predictions given in~\cite{Bobeth:2021lya} using a combination of light-cone sum rules, HQE and lattice QCD results, at the level of $2.5\sigma$. We also computed the difference, $\Delta A_{FB} = \langle A_{FB}^\mu \rangle - \langle A_{FB}^e \rangle$, and found a value different to that given in~\cite{Bobeth:2021lya} by $\approx 2\sigma$, and in tension with fits to the 2018 Belle data at the level of $3.5\sigma$. This result confirms the need for further investigation of LFUV effects in the $\ell=\mu/e$ modes of the decay.

This work demonstrates the feasibility of computing a complete set of fully correlated SM and NP form factors for pseudoscalar to vector semileptonic decays using the heavy-HISQ approach, across different chiral regimes. Our calculation has allowed us to perform the first simultaneous analysis of data for $B\to D^*$ together with data for $B_s\to D^*_s$, paving the way for the analysis of more precise experimental data that is expected from LHCb and Belle~II in these channels in the near future.

\subsection*{\bf{Acknowledgements}}

We are grateful to the MILC Collaboration for the use
of their configurations and code. We thank C. Bouchard,
B. Colquhoun, J. Koponen, P. Lepage, E. McLean,
C. McNeile and A. Vaquero for useful discussions. Computing was done on
the Cambridge service for Data Driven Discovery (CSD3),
part of which is operated by the University of Cambridge
Research Computing on behalf of the DIRAC HPC Facility
of the Science and Technology Facilities Council (STFC).
The DIRAC component of CSD3 was funded by BEIS
capital funding via STFC capital Grants No. ST/P002307/1
and No. ST/R002452/1 and by STFC operations Grant
No. ST/R00689X/1. DiRAC is part of the national
e-infrastructure. We are grateful to the CSD3 support staff
for assistance. Funding for this work came from UK Science
and Technology Facilities Council Grants No. ST/L000466/1 and No. ST/P000746/1 and Engineering and Physical
Sciences Research Council Project No. EP/W005395/1.

\begin{appendix}

\section{Full Differential Decay Rate Including Tensor Operators}\label{fulldiff}
The matrix element $\mathcal{M}$ is given by
\begin{align}
 \mathcal{M}^{\lambda_{D^*}\lambda_\ell}=C\langle D\pi |D^*(\lambda_{D^*})\rangle \langle D^*(\lambda_{D^*})|J^\mathrm{had}_\alpha|B\rangle\nonumber\\
  \times\langle \ell(\lambda_\ell)\bar{\nu}|J^{\mathrm{lep}~\alpha}|0\rangle 
\end{align}
where $C^{-1}=g_{D^*D\pi}|\vec{p}_\pi|$ is a constant normalisation, with $\vec{p}_\pi$ the pion spatial momentum in the $D^*$ rest frame, such that integrating over $D\pi$ phase space yields the  rate for our choice of overall normalisation $N(q^2)$ including the $D^*\to D\pi$ branching fraction in~\cref{diffrateM}. The sum on $\alpha$ includes scalar, vector and tensor like currents
\begin{align}
\sum_\alpha J^\mathrm{had}_\alpha J^{\mathrm{lep}~\alpha} =&g_{S(P)}J^{\mathrm{had}_{S(P)}} J^{{\mathrm{lep}_{S(P)}}}\nonumber\\
+ &g_{V(A)}J^{\mathrm{had}_{V(A)}}_\mu J^{{\mathrm{lep}_{V(A)}}~\mu}\nonumber\\
+ &g_{T(T5)}J^{\mathrm{had}_{T(T5)}}_{\mu\nu} J^{{\mathrm{lep}_{T(T5)}}~\mu\nu}.
\end{align}
It is conventional to insert off-shell vector boson polarisation vectors in order to define \textit{helicity amplitudes}. These polarisation vectors, $\bar{\epsilon}(\lambda)$, possess the property that
\begin{equation}
\sum_{\lambda}\bar{\epsilon}_\mu(\lambda)^*\bar{\epsilon}_\nu(\lambda) \delta_\lambda = g_{\mu\nu}
\end{equation}
with $\delta_{0,\pm}=-1 $ and $\delta_{t}=1 $. We use vector boson polarisation vectors in the $W$ rest frame
\begin{align}
\bar{\epsilon}^\mu(\lambda=t)=\begin{pmatrix}1\\0\\0\\0\end{pmatrix},&~~\bar{\epsilon}^\mu(\lambda=0)=\begin{pmatrix}0\\0\\0\\-1\end{pmatrix},\nonumber\\
\bar{\epsilon}^\mu(\lambda=\pm)=&\pm\frac{1}{\sqrt{2}}\begin{pmatrix}0\\-1\\\pm i\\0\end{pmatrix}
\end{align}
and $D^*$ polarisation vectors in the $D^*$ rest frame
\begin{align}
{\epsilon}^\mu(\lambda=\pm)=\pm\frac{1}{\sqrt{2}}\begin{pmatrix}0\\-1\\\mp i\\0\end{pmatrix},~~{\epsilon}^\mu(\lambda=0)=\begin{pmatrix}0\\0\\0\\1\end{pmatrix}.
\end{align}
In the $B$ rest frame the polarisation vectors are
\begin{align}
\bar{\epsilon}^\mu(\lambda=t)=\frac{1}{\sqrt{q^2}}\begin{pmatrix}q_0\\0\\0\\-|\vec{q}|\end{pmatrix},&~~\bar{\epsilon}^\mu(\lambda=0)=\frac{1}{\sqrt{q^2}}\begin{pmatrix}|\vec{q}|\\0\\0\\-q_0\end{pmatrix},\nonumber\\
\bar{\epsilon}^\mu(\lambda=\pm)=&\pm\frac{1}{\sqrt{2}}\begin{pmatrix}0\\-1\\\pm i\\0\end{pmatrix},
\end{align}
and
\begin{align}
{\epsilon}^\mu(\lambda=\pm)=\pm\frac{1}{\sqrt{2}}\begin{pmatrix}0\\-1\\\mp i\\0\end{pmatrix},~~{\epsilon}^\mu(\lambda=0)=\frac{1}{M_{D^*}^2}\begin{pmatrix}|\vec{q}|\\0\\0\\E_{D^*}\end{pmatrix}.
\end{align}
We take $D$, $\pi$, $\ell$ and $\nu$ momenta
\begin{align}
p^\mu_D=&\begin{pmatrix}E_D\\k\cos(\chi)\sin(\theta_{D^*})\\-k\sin(\chi)\sin(\theta_{D^*})\\k\cos(\chi)\end{pmatrix},\nonumber\\
p^\mu_\pi=&\begin{pmatrix}k\\-k\cos(\chi)\sin(\theta_{D^*})\\k\sin(\chi)\sin(\theta_{D^*})\\-k\cos(\chi)\end{pmatrix},\\
p^\mu_\ell=&\begin{pmatrix}E_\ell\\k^\prime\sin(\theta_{W})\\0\\-k^\prime\cos(\theta_W)\end{pmatrix},
p^\mu_\nu=\begin{pmatrix}k^\prime\\-k^\prime\sin(\theta_{W})\\0\\k^\prime\cos(\theta_W)\end{pmatrix}.
\end{align}
The leptonic and hadronic helicity amplitudes are defined by
\begin{align}
g_{V(A)}\langle D^*(\lambda_{D^*})|&J^{\mathrm{had}_{V(A)}}_\mu|B\rangle \langle \ell(\lambda_\ell)\bar{\nu}|J^{\mathrm{lep}_{V-A}~\mu}|0\rangle\nonumber\\
=&\sum_\lambda  \delta_\lambda g_{V(A)}\langle D^*(\lambda_{D^*})|J^{\mathrm{had}_{V(A)}}_\mu|B\rangle\bar{\epsilon}^\mu(\lambda)^*  \nonumber\\
& \times\bar{\epsilon}^\nu(\lambda)\langle \ell(\lambda_\ell)\bar{\nu}|J^{\mathrm{lep}_{V-A}}_\nu|0\rangle\nonumber\\
=&\sum_\lambda \delta_\lambda H^{\lambda_{D^*},\lambda}_{V(A)} L_{V-A}^{\lambda_\ell,\lambda}.
\end{align}
The expressions for the vector and axial-vector helicity amplitudes, $H^\pm=H_V^{\pm,\pm}+H_A^{\pm,\pm}$, $H_0=-H_A^{0,0}$ and $H_t=-H_A^{0,t}$, are given in~\cref{smhelicityampls}. Note that for the $\lambda=0,t$ cases, it is conventional to define the helicity amplitude with an additional factor of $-1$.
For the tensor currents, it is conventional to also insert a factor of $i(-i)$
\begin{align}
ig_{T(T5)}\langle D^*(\lambda_{D^*})|&J^{\mathrm{had}_{T(T5)}}_{\mu\nu}|B\rangle \left(-i\langle \ell(\lambda_\ell)\bar{\nu}|J^{\mathrm{lep}_{T-T5}~{\mu\nu}}|0\rangle\right)\nonumber\\
=\sum_{\lambda\bar{\lambda} }& \delta_\lambda \delta_{\bar{\lambda}} H^{\lambda_{D^*},\lambda{\bar{\lambda}}}_{T(T5)} L_{T-T5}^{\lambda_\ell,\lambda{\bar{\lambda}}}.
\end{align}
Note that the tensor current $J^{\mathrm{had}_{T(T5)}}$ includes a continuum $\overline{\mathrm{MS}}$ renormalisation, defined at a scale $\mu$ which we take to be $\mu=4.8~\mathrm{GeV}$.

The nonzero tensor helicity amplitudes are, using~\cref{formfactors},
\begin{align}\label{tensorhelicityamplitudes}
H^{0,+-}_{T}=-H^{0,-+}_{T}=&-g_T \sqrt{M_BM_{D^*}}\Big((1+w)h_{T_1}\nonumber\\&+(w-1)h_{T_2}-h_{T_3}(w^2-1)\Big)\nonumber\\
H^{\pm,\pm 0}_{T}=-H^{\pm,0\pm}_{T}=&\pm g_T M_B\frac{\sqrt{M_BM_{D^*}}}{\sqrt{q^2}}\nonumber\\\times\Big(&h_{T_1}(1-r)(1+w) \nonumber\\
&-h_{T_2}(1+r)(w-1) \Big)\nonumber\\
H^{\pm,\pm t}_{T}=-H^{\pm,t\pm}_{T}=&\pm g_T M_B\frac{\sqrt{M_BM_{D^*}(w^2-1)}}{\sqrt{q^2}}\nonumber\\\times&\left(h_{T_1}(1+r)-h_{T_2}(1-r)\right)
\end{align}
For the axial-tensor current, we may use the fact that $\gamma^5 \sigma^{\mu\nu}=\frac{i}{2}\varepsilon^{\mu\nu\sigma\rho}\sigma_{\sigma\rho}$ to relate $\langle D^*|\bar{c} \gamma^5\sigma^{\mu\nu}b|B \rangle =\frac{i}{2}\varepsilon^{\mu\nu\sigma\rho}\langle D^*|\bar{c} \sigma_{\sigma\rho}b|B \rangle$. Inserting this into the definition of the helicity amplitudes allows us to relate the axial-tensor helicity amplitudes to the tensor helicity amplitudes.
\begin{align}
H^{\pm,\pm 0}_{T5} =-H^{\pm,0\pm}_{T5} =&\mp\frac{g_{T5}}{g_T}H^{\pm,\pm t}_{T}\nonumber\\
H^{\pm,\pm t}_{T5} =-H^{\pm,t\pm}_{T5} =&\mp \frac{g_{T5}}{g_T}H^{\pm,\pm 0}_{T}\nonumber\\	
H^{0,0t}_{T5} =-H^{0,t0}_{T5} =&- \frac{g_{T5}}{g_T}H^{0,+-}_{T}.\label{HT5defs}
\end{align}

The pseudoscalar current is straightforwardly
\begin{align}
g_{P}\langle D^*(\lambda_{D^*})|&J^{\mathrm{had}_{P}}|B\rangle \langle \ell(\lambda_\ell)\bar{\nu}|J^{\mathrm{lep}_{P}}|0\rangle=H^{\lambda_{D^*}}_P L^{\lambda_\ell}_{S-P}
\end{align}
We can use the PCAC relation, $\langle D^* | q_\mu \bar{c}\gamma^\mu\gamma^5 b | \bar{B}\rangle = -(m_b+m_c)\langle D^*| \bar{c}\gamma^5 b |\bar{B} \rangle$, to write
\begin{align}
H_P=H^{\lambda_{D^*}=0}_P = \frac{\sqrt{q^2}}{(m_b+m_c)}\frac{g_P}{g_A}H_t.
\end{align}
Together with the parameterisation of the amplitude
\begin{align}
\langle D\pi |D^*(\lambda_{D^*})\rangle=g_{D^*D\pi}\epsilon_\mu(\lambda_{D^*}) p_{D}^\mu
\end{align}
this gives
\begin{align}
\mathcal{M}^{\lambda_{D^*}\lambda_\ell}=&Cg_{D^*D\pi} \epsilon_\mu(\lambda_{D^*}) p_{D}^\mu \Big[H^{\lambda_{D^*}}_P L^{\lambda_\ell}_{S-P} \nonumber\\
+& \sum_\lambda \delta_\lambda \Big(H^{\lambda_{D^*},\lambda}_{V}  + H^{\lambda_{D^*},\lambda}_{A} \Big)L_{V-A}^{\lambda_\ell,\lambda} \nonumber\\
+&\sum_{\lambda\bar{\lambda}} \delta_\lambda\delta_{\bar{\lambda}}\Big( H^{\lambda_{D^*},\lambda{\bar{\lambda}}}_{T} + H^{\lambda_{D^*},\lambda{\bar{\lambda}}}_{T5}\Big)  L_{T-T5}^{\lambda_\ell,\lambda{\bar{\lambda}}}   \Big]
\end{align}
For the charge conjugate mode, we have 
\begin{align}
g_{V(A)}^*&\langle \overline{D^*}(\lambda_{D^*})|J^{\mathrm{had}_{V(A)}\dagger}_\mu|\overline{B}\rangle \langle \bar{\ell}(\lambda_\ell){\nu}|J^{\mathrm{lep}_{V-A}~\mu\dagger}|0\rangle\nonumber\\
=&\sum_\lambda \delta_\lambda \eta \frac{g_{V(A)}^*}{g_{V(A)}} H^{\lambda_{D^*},\lambda}_{V(A)} L_{V-A}^{\prime\lambda_\ell,\lambda},
\end{align}
\begin{align}
ig_{T(T5)}^*\langle \overline{D^*}(\lambda_{D^*})|&J^{\mathrm{had}_{T(T5)}\dagger}_{\mu\nu}|\overline{B}\rangle \nonumber\\
&\times\left(-i\langle \bar{\ell}(\lambda_\ell){\nu}|J^{\mathrm{lep}_{T-T5}~{\mu\nu}\dagger}|0\rangle\right)\nonumber\\
=\sum_{\lambda\bar{\lambda} }& \delta_\lambda \delta_{\bar{\lambda}} \frac{\eta g_{T(T5)}^*}{g_{T(T5)}} H^{\lambda_{D^*},\lambda{\bar{\lambda}}}_{T(T5)} L_{T-T5}^{\prime\lambda_\ell,\lambda{\bar{\lambda}}}.
\end{align}
and
\begin{align}
g_P^*\langle \overline{D^*}(\lambda_{D^*})|&J^{\mathrm{had}_{P}\dagger}|\overline{B}\rangle \langle \bar{\ell}(\lambda_\ell){\nu}|J^{\mathrm{lep}_{P}\dagger}|0\rangle\nonumber\\
&=\eta\frac{g_{P}^*}{g_{P}}H^{\lambda_{D^*}}_P L^{\prime\lambda_\ell}_{S-P}.
\end{align}
$\eta=\pm 1$ is a phase dependent upon the sign of the current under combined Hermitian conjugation and charge conjugation $\mathcal{C}J^{\mathrm{had}\dagger}_\alpha\mathcal{C}^{-1}=\eta J^\mathrm{had}_\alpha$. Specifically for the currents defined in~\cref{EffectiveHamiltonian}, $P,V,A,T,T5$ this phase is $-1,-1,1,-1,1$ respectively.  

Inserting these expressions, either for the normal or conjugate mode, into~\cref{diffrateM} gives the corresponding differential decay rate in terms of lepton tensors and helicity amplitudes.
The lepton tensor combinations may be evaluated straightforwardly using standard spinor identities when summing over polarisations,
\begin{align}
\sum_{\lambda_\ell} &L_{\gamma}^{\lambda_\ell} \left(L_{\Gamma}^{\lambda_\ell} \right)^*=\sum_{ss^\prime}  \langle \ell(s)\bar{\nu}({s^\prime})|\bar{\ell}\gamma \nu|0\rangle\nonumber\\
&\times \left( \langle \ell(s)\bar{\nu}(s^\prime)|\bar{\ell}\Gamma \nu|0\rangle \right)^\dagger\nonumber\\
=&\sum_{ss^\prime}  \bar{u}^s(p_\ell)\gamma v^{s^\prime}(p_{\bar{\nu}}) \left(  \bar{u}^s(p_\ell)\Gamma v^{s^\prime}(p_{\bar{\nu}}) \right)^\dagger\nonumber\\
=&\sum_{ss^\prime}  \bar{u}^s(p_\ell)\gamma v^{s^\prime}(p_{\bar{\nu}}) \bar{v}^{s^\prime}(p_{\bar{\nu}}) \gamma^0 \Gamma^\dagger \gamma^0 u^{s}(p_{\ell})\nonumber\\
=&\mathrm{Tr}\left[   \left(\slashed{p}_{\ell}+m_\ell  \right)           \gamma      \slashed{p}_{\bar{\nu}}      \gamma^0 \Gamma^\dagger \gamma^0      \right].
\end{align}
For the conjugate mode lepton tensors
\begin{align}
\sum_{\lambda_\ell}& L_{\gamma}^{\prime\lambda_\ell} \left(L_{\Gamma}^{\prime\lambda_\ell} \right)^*=\sum_{ss^\prime}  \langle \bar{\ell}(s){\nu}({s^\prime})|\bar{\nu}\gamma^0\gamma^\dagger\gamma^0 \ell|0\rangle\nonumber\\&\times \left( \langle \bar{\ell}(s){\nu}(s^\prime)|\bar{\nu}\gamma^0\Gamma^\dagger \gamma^0\ell|0\rangle \right)^\dagger\nonumber\\
=&\sum_{ss^\prime}  \bar{u}^{s^\prime}(p_\nu)\gamma^0\gamma^\dagger\gamma^0 v^{s}(p_{{\ell}}) \left(  \bar{u}^{s^\prime}(p_\nu)\gamma^0\Gamma^\dagger \gamma^0 v^{s}(p_{\ell}) \right)^\dagger\nonumber\\
=&\sum_{ss^\prime}  \bar{u}^{s^\prime}(p_\nu)\gamma^0\gamma^\dagger\gamma^0 v^{s}(p_{{\ell}}) \bar{v}^{s}(p_{\ell}) \Gamma {u}^{s^\prime}(p_\nu) \nonumber\\
=&\mathrm{Tr}\left[    \slashed{p}_{\nu}   \gamma^0\gamma^\dagger\gamma^0               \left(\slashed{p}_{\ell}-m_\ell  \right)                 \Gamma              \right].
\end{align}
The combinations of helicity amplitudes and $k_i$ factors entering the squared matrix element~\cref{eq:helicitytoMsq}, for general complex choices of $g_X$ in~\cref{EffectiveHamiltonian}, are given in~\cref{tab:mellzero_0,tab:mellsqrt_0,tab:mellfirst_0} for the conjugate mode ${B}^0\to D^{*-}\ell^+\nu$. Note that these include the factor $\eta$, and so should be used with the helicity amplitudes for ${B}^0\to D^{*+}\ell^-\bar{\nu}$, just taking $g_X\to g_X^*$. We have checked explicitly that our method of constructing the differential decay rate reproduces the results of~\cite{Becirevic:2019tpx}. Note, however, that in~\cite{Becirevic:2019tpx} $H_t$ and $H_0$ are defined without the additional $(-1)$, and also that our angular conventions for the lepton angle are related by $\theta_W\to \pi-\theta_W$. The full differential decay rate is also available in a slightly more compact notation in~\cite{Duraisamy:2014sna}, though the notation here makes clear which helicity amplitudes are suppressed by the lepton mass. 

Because 
\begin{align}
\left(\bar{c}\gamma^5\sigma_{\mu\nu}b\right)\left(\bar{\ell}_R\sigma^{\mu\nu}\nu_L\right)=-\left(\bar{c}\sigma_{\mu\nu}b\right)\left(\bar{\ell}_R\sigma^{\mu\nu}\nu_L\right),
\end{align}
the $g_{T5}$ term is redundant, and it is typical to identify $(g_{T}-g_{T5}) \equiv 2 g_{T_L}$. As such, we may replace $g_{T}\to 2 g_{T_L}$ and $g_{T5}\to 0$ in~\cref{tensorhelicityamplitudes,HT5defs} respectively and omit the helicity combinations including $H_{T5}$.

\begin{table}[H]
\centering
\caption{\label{tab:mellzero_0} The helicity amplitude combinations and coefficients for them that appear in~\cref{eq:helicitytoMsq} at order $(m_\ell^2/q^2)^0$ for ${B}^0\to D^{*-}\ell^+\nu$.}
\begin{tabular}{ c | c  }
\hline
$\mathcal{H}_i$&$k_i(\theta_W,\theta_{D^*},\chi)$\\\hline
$H_{T}^{0,+-}(H_{T}^{0,+-})^* $&$ 16 \cos ^2({\theta_W}) \cos ^2({\theta_{D^*}})$\\
$H_{T}^{0,+-}(H_{T}^{+,+0})^* $&$ 2 e^{i {\chi}} \sin (2 {\theta_W}) \sin (2 {\theta_{D^*}})$\\
$H_{T}^{0,+-}(H_{T}^{+,+t})^* $&$ -2 e^{i {\chi}} \sin (2 {\theta_W}) \sin (2 {\theta_{D^*}})$\\
$H_{T}^{0,+-}(H_{T}^{-,-0})^* $&$ -2 e^{-i {\chi}} \sin (2 {\theta_W}) \sin (2 {\theta_{D^*}})$\\
$H_{T}^{0,+-}(H_{T}^{-,-t})^* $&$ -2 e^{-i {\chi}} \sin (2 {\theta_W}) \sin (2 {\theta_{D^*}})$\\
$H_{T}^{0,+-}(H_P)^* $&$ 8 \cos ({\theta_W}) \cos ^2({\theta_{D^*}})$\\
$H_{T}^{+,+0}(H_{T}^{0,+-})^* $&$ 2 e^{-i {\chi}} \sin (2 {\theta_W}) \sin (2 {\theta_{D^*}})$\\
$H_{T}^{+,+0}(H_{T}^{+,+0})^* $&$ 4 \sin ^2({\theta_W}) \sin ^2({\theta_{D^*}})$\\
$H_{T}^{+,+0}(H_{T}^{+,+t})^* $&$ -4 \sin ^2({\theta_W}) \sin ^2({\theta_{D^*}})$\\
$H_{T}^{+,+0}(H_{T}^{-,-0})^* $&$ -4 e^{-2 i {\chi}} \sin ^2({\theta_W}) \sin ^2({\theta_{D^*}})$\\
$H_{T}^{+,+0}(H_{T}^{-,-t})^* $&$ -4 e^{-2 i {\chi}} \sin ^2({\theta_W}) \sin ^2({\theta_{D^*}})$\\
$H_{T}^{+,+0}(H_P)^* $&$ 2 e^{-i {\chi}} \sin ({\theta_W}) \sin (2 {\theta_{D^*}})$\\
$H_{T}^{+,+t}(H_{T}^{0,+-})^* $&$ -2 e^{-i {\chi}} \sin (2 {\theta_W}) \sin (2 {\theta_{D^*}})$\\
$H_{T}^{+,+t}(H_{T}^{+,+0})^* $&$ -4 \sin ^2({\theta_W}) \sin ^2({\theta_{D^*}})$\\
$H_{T}^{+,+t}(H_{T}^{+,+t})^* $&$ 4 \sin ^2({\theta_W}) \sin ^2({\theta_{D^*}})$\\
$H_{T}^{+,+t}(H_{T}^{-,-0})^* $&$ 4 e^{-2 i {\chi}} \sin ^2({\theta_W}) \sin ^2({\theta_{D^*}})$\\
$H_{T}^{+,+t}(H_{T}^{-,-t})^* $&$ 4 e^{-2 i {\chi}} \sin ^2({\theta_W}) \sin ^2({\theta_{D^*}})$\\
$H_{T}^{+,+t}(H_P)^* $&$ -2 e^{-i {\chi}} \sin ({\theta_W}) \sin (2 {\theta_{D^*}})$\\
$H_{T}^{-,-0}(H_{T}^{0,+-})^* $&$ -2 e^{i {\chi}} \sin (2 {\theta_W}) \sin (2 {\theta_{D^*}})$\\
$H_{T}^{-,-0}(H_{T}^{+,+0})^* $&$ -4 e^{2 i {\chi}} \sin ^2({\theta_W}) \sin ^2({\theta_{D^*}})$\\
$H_{T}^{-,-0}(H_{T}^{+,+t})^* $&$ 4 e^{2 i {\chi}} \sin ^2({\theta_W}) \sin ^2({\theta_{D^*}})$\\
$H_{T}^{-,-0}(H_{T}^{-,-0})^* $&$ 4 \sin ^2({\theta_W}) \sin ^2({\theta_{D^*}})$\\
$H_{T}^{-,-0}(H_{T}^{-,-t})^* $&$ 4 \sin ^2({\theta_W}) \sin ^2({\theta_{D^*}})$\\
$H_{T}^{-,-0}(H_P)^* $&$ -2 e^{i {\chi}} \sin ({\theta_W}) \sin (2 {\theta_{D^*}})$\\
$H_{T}^{-,-t}(H_{T}^{0,+-})^* $&$ -2 e^{i {\chi}} \sin (2 {\theta_W}) \sin (2 {\theta_{D^*}})$\\
$H_{T}^{-,-t}(H_{T}^{+,+0})^* $&$ -4 e^{2 i {\chi}} \sin ^2({\theta_W}) \sin ^2({\theta_{D^*}})$\\
$H_{T}^{-,-t}(H_{T}^{+,+t})^* $&$ 4 e^{2 i {\chi}} \sin ^2({\theta_W}) \sin ^2({\theta_{D^*}})$\\
$H_{T}^{-,-t}(H_{T}^{-,-0})^* $&$ 4 \sin ^2({\theta_W}) \sin ^2({\theta_{D^*}})$\\
$H_{T}^{-,-t}(H_{T}^{-,-t})^* $&$ 4 \sin ^2({\theta_W}) \sin ^2({\theta_{D^*}})$\\
$H_{T}^{-,-t}(H_P)^* $&$ -2 e^{i {\chi}} \sin ({\theta_W}) \sin (2 {\theta_{D^*}})$\\
$H_+(H_+)^* $&$ 4 \sin ^4\left(\frac{{\theta_W}}{2}\right) \sin ^2({\theta_{D^*}})$\\
$H_+(H_-)^* $&$ -e^{2 i {\chi}} \sin ^2({\theta_W}) \sin ^2({\theta_{D^*}})$\\
$H_+(H_0)^* $&$ -2 e^{i {\chi}} \sin ^2\left(\frac{{\theta_W}}{2}\right) \sin ({\theta_W}) \sin (2 {\theta_{D^*}})$\\
$H_-(H_+)^* $&$ -e^{-2 i {\chi}} \sin ^2({\theta_W}) \sin ^2({\theta_{D^*}})$\\
$H_-(H_-)^* $&$ 4 \cos ^4\left(\frac{{\theta_W}}{2}\right) \sin ^2({\theta_{D^*}})$\\
$H_-(H_0)^* $&$ e^{-i {\chi}} \sin ({\theta_W}) (\cos ({\theta_W})+1) \sin (2 {\theta_{D^*}})$\\
$H_0(H_+)^* $&$ -2 e^{-i {\chi}} \sin ^2\left(\frac{{\theta_W}}{2}\right) \sin ({\theta_W}) \sin (2 {\theta_{D^*}})$\\
$H_0(H_-)^* $&$ e^{i {\chi}} \sin ({\theta_W}) (\cos ({\theta_W})+1) \sin (2 {\theta_{D^*}})$\\
$H_0(H_0)^* $&$ 4 \sin ^2({\theta_W}) \cos ^2({\theta_{D^*}})$\\
$H_P(H_{T}^{0,+-})^* $&$ 8 \cos ({\theta_W}) \cos ^2({\theta_{D^*}})$\\
$H_P(H_{T}^{+,+0})^* $&$ 2 e^{i {\chi}} \sin ({\theta_W}) \sin (2 {\theta_{D^*}})$\\
$H_P(H_{T}^{+,+t})^* $&$ -2 e^{i {\chi}} \sin ({\theta_W}) \sin (2 {\theta_{D^*}})$\\
$H_P(H_{T}^{-,-0})^* $&$ -2 e^{-i {\chi}} \sin ({\theta_W}) \sin (2 {\theta_{D^*}})$\\
$H_P(H_{T}^{-,-t})^* $&$ -2 e^{-i {\chi}} \sin ({\theta_W}) \sin (2 {\theta_{D^*}})$\\
$H_P(H_P)^* $&$ 4 \cos ^2({\theta_{D^*}})$\\
\end{tabular}
\end{table}

\begin{table}[H]
\centering
\caption{\label{tab:mellsqrt_0} The helicity amplitude combinations and coefficients for them that appear in~\cref{eq:helicitytoMsq} at order $(m_\ell^2/q^2)^\frac{1}{2}$ for ${B}^0\to D^{*-}\ell^+\nu$.}
\begin{tabular}{ c | c  }
\hline
$\mathcal{H}_i$&$k_i(\theta_W,\theta_{D^*},\chi)$\\\hline
$\sqrt{\frac{m_\ell^2}{q^2}}H_{T}^{0,+-}(H_+)^* $&$ -2 e^{-i {\chi}} \sin ({\theta_W}) \sin (2 {\theta_{D^*}})$\\
$\sqrt{\frac{m_\ell^2}{q^2}}H_{T}^{0,+-}(H_-)^* $&$ 2 e^{i {\chi}} \sin ({\theta_W}) \sin (2 {\theta_{D^*}})$\\
$\sqrt{\frac{m_\ell^2}{q^2}}H_{T}^{0,+-}(H_0)^* $&$ 8 \cos ^2({\theta_{D^*}})$\\
$\sqrt{\frac{m_\ell^2}{q^2}}H_{T}^{0,+-}(H_t)^* $&$ -8 \cos ({\theta_W}) \cos ^2({\theta_{D^*}})$\\
$\sqrt{\frac{m_\ell^2}{q^2}}H_{T}^{+,+0}(H_-)^* $&$ -8 \cos ^2\left(\frac{{\theta_W}}{2}\right) \sin ^2({\theta_{D^*}})$\\
$\sqrt{\frac{m_\ell^2}{q^2}}H_{T}^{+,+0}(H_0)^* $&$ -2 e^{-i {\chi}} \sin ({\theta_W}) \sin (2 {\theta_{D^*}})$\\
$\sqrt{\frac{m_\ell^2}{q^2}}H_{T}^{+,+0}(H_t)^* $&$ -2 e^{-i {\chi}} \sin ({\theta_W}) \sin (2 {\theta_{D^*}})$\\
$\sqrt{\frac{m_\ell^2}{q^2}}H_{T}^{+,+t}(H_-)^* $&$ 8 \cos ^2\left(\frac{{\theta_W}}{2}\right) \sin ^2({\theta_{D^*}})$\\
$\sqrt{\frac{m_\ell^2}{q^2}}H_{T}^{+,+t}(H_0)^* $&$ 2 e^{-i {\chi}} \sin ({\theta_W}) \sin (2 {\theta_{D^*}})$\\
$\sqrt{\frac{m_\ell^2}{q^2}}H_{T}^{+,+t}(H_t)^* $&$ 2 e^{-i {\chi}} \sin ({\theta_W}) \sin (2 {\theta_{D^*}})$\\
$\sqrt{\frac{m_\ell^2}{q^2}}H_{T}^{-,-0}(H_+)^* $&$ 8 \sin ^2\left(\frac{{\theta_W}}{2}\right) \sin ^2({\theta_{D^*}})$\\
$\sqrt{\frac{m_\ell^2}{q^2}}H_{T}^{-,-0}(H_0)^* $&$ -2 e^{i {\chi}} \sin ({\theta_W}) \sin (2 {\theta_{D^*}})$\\
$\sqrt{\frac{m_\ell^2}{q^2}}H_{T}^{-,-0}(H_t)^* $&$ 2 e^{i {\chi}} \sin ({\theta_W}) \sin (2 {\theta_{D^*}})$\\
$\sqrt{\frac{m_\ell^2}{q^2}}H_{T}^{-,-t}(H_+)^* $&$ 8 \sin ^2\left(\frac{{\theta_W}}{2}\right) \sin ^2({\theta_{D^*}})$\\
$\sqrt{\frac{m_\ell^2}{q^2}}H_{T}^{-,-t}(H_0)^* $&$ -2 e^{i {\chi}} \sin ({\theta_W}) \sin (2 {\theta_{D^*}})$\\
$\sqrt{\frac{m_\ell^2}{q^2}}H_{T}^{-,-t}(H_t)^* $&$ 2 e^{i {\chi}} \sin ({\theta_W}) \sin (2 {\theta_{D^*}})$\\
$\sqrt{\frac{m_\ell^2}{q^2}}H_+(H_{T}^{0,+-})^* $&$ -2 e^{i {\chi}} \sin ({\theta_W}) \sin (2 {\theta_{D^*}})$\\
$\sqrt{\frac{m_\ell^2}{q^2}}H_+(H_{T}^{-,-0})^* $&$ 8 \sin ^2\left(\frac{{\theta_W}}{2}\right) \sin ^2({\theta_{D^*}})$\\
$\sqrt{\frac{m_\ell^2}{q^2}}H_+(H_{T}^{-,-t})^* $&$ 8 \sin ^2\left(\frac{{\theta_W}}{2}\right) \sin ^2({\theta_{D^*}})$\\
$\sqrt{\frac{m_\ell^2}{q^2}}H_+(H_P)^* $&$ -e^{i {\chi}} \sin ({\theta_W}) \sin (2 {\theta_{D^*}})$\\
$\sqrt{\frac{m_\ell^2}{q^2}}H_-(H_{T}^{0,+-})^* $&$ 2 e^{-i {\chi}} \sin ({\theta_W}) \sin (2 {\theta_{D^*}})$\\
$\sqrt{\frac{m_\ell^2}{q^2}}H_-(H_{T}^{+,+0})^* $&$ -8 \cos ^2\left(\frac{{\theta_W}}{2}\right) \sin ^2({\theta_{D^*}})$\\
$\sqrt{\frac{m_\ell^2}{q^2}}H_-(H_{T}^{+,+t})^* $&$ 8 \cos ^2\left(\frac{{\theta_W}}{2}\right) \sin ^2({\theta_{D^*}})$\\
$\sqrt{\frac{m_\ell^2}{q^2}}H_-(H_P)^* $&$ -e^{-i {\chi}} \sin ({\theta_W}) \sin (2 {\theta_{D^*}})$\\
$\sqrt{\frac{m_\ell^2}{q^2}}H_0(H_{T}^{0,+-})^* $&$ 8 \cos ^2({\theta_{D^*}})$\\
$\sqrt{\frac{m_\ell^2}{q^2}}H_0(H_{T}^{+,+0})^* $&$ -2 e^{i {\chi}} \sin ({\theta_W}) \sin (2 {\theta_{D^*}})$\\
$\sqrt{\frac{m_\ell^2}{q^2}}H_0(H_{T}^{+,+t})^* $&$ 2 e^{i {\chi}} \sin ({\theta_W}) \sin (2 {\theta_{D^*}})$\\
$\sqrt{\frac{m_\ell^2}{q^2}}H_0(H_{T}^{-,-0})^* $&$ -2 e^{-i {\chi}} \sin ({\theta_W}) \sin (2 {\theta_{D^*}})$\\
$\sqrt{\frac{m_\ell^2}{q^2}}H_0(H_{T}^{-,-t})^* $&$ -2 e^{-i {\chi}} \sin ({\theta_W}) \sin (2 {\theta_{D^*}})$\\
$\sqrt{\frac{m_\ell^2}{q^2}}H_0(H_P)^* $&$ 4 \cos ({\theta_W}) \cos ^2({\theta_{D^*}})$\\
$\sqrt{\frac{m_\ell^2}{q^2}}H_t(H_{T}^{0,+-})^* $&$ -8 \cos ({\theta_W}) \cos ^2({\theta_{D^*}})$\\
$\sqrt{\frac{m_\ell^2}{q^2}}H_t(H_{T}^{+,+0})^* $&$ -2 e^{i {\chi}} \sin ({\theta_W}) \sin (2 {\theta_{D^*}})$\\
$\sqrt{\frac{m_\ell^2}{q^2}}H_t(H_{T}^{+,+t})^* $&$ 2 e^{i {\chi}} \sin ({\theta_W}) \sin (2 {\theta_{D^*}})$\\
$\sqrt{\frac{m_\ell^2}{q^2}}H_t(H_{T}^{-,-0})^* $&$ 2 e^{-i {\chi}} \sin ({\theta_W}) \sin (2 {\theta_{D^*}})$\\
$\sqrt{\frac{m_\ell^2}{q^2}}H_t(H_{T}^{-,-t})^* $&$ 2 e^{-i {\chi}} \sin ({\theta_W}) \sin (2 {\theta_{D^*}})$\\
$\sqrt{\frac{m_\ell^2}{q^2}}H_t(H_P)^* $&$ -4 \cos ^2({\theta_{D^*}})$\\
$\sqrt{\frac{m_\ell^2}{q^2}}H_P(H_+)^* $&$ -e^{-i {\chi}} \sin ({\theta_W}) \sin (2 {\theta_{D^*}})$\\
$\sqrt{\frac{m_\ell^2}{q^2}}H_P(H_-)^* $&$ -e^{i {\chi}} \sin ({\theta_W}) \sin (2 {\theta_{D^*}})$\\
$\sqrt{\frac{m_\ell^2}{q^2}}H_P(H_0)^* $&$ 4 \cos ({\theta_W}) \cos ^2({\theta_{D^*}})$\\
$\sqrt{\frac{m_\ell^2}{q^2}}H_P(H_t)^* $&$ -4 \cos ^2({\theta_{D^*}})$\\
\end{tabular}
\end{table}

\begin{table}[H]
\centering
\caption{\label{tab:mellfirst_0} The helicity amplitude combinations and coefficients for them that appear in~\cref{eq:helicitytoMsq} at order $m_\ell^2/q^2$ for ${B}^0\to D^{*-}\ell^+\nu$.}
\begin{tabular}{ c | c  }
\hline
$\mathcal{H}_i$&$k_i(\theta_W,\theta_{D^*},\chi)$\\\hline
$\frac{m_\ell^2}{q^2}H_{T}^{0,+-}(H_{T}^{0,+-})^* $&$ 16 \sin ^2({\theta_W}) \cos ^2({\theta_{D^*}})$\\
$\frac{m_\ell^2}{q^2}H_{T}^{0,+-}(H_{T}^{+,+0})^* $&$ -4 e^{i {\chi}} \sin ({\theta_W}) (\cos ({\theta_W})+1) \sin (2 {\theta_{D^*}})$\\
$\frac{m_\ell^2}{q^2}H_{T}^{0,+-}(H_{T}^{+,+t})^* $&$ 4 e^{i {\chi}} \sin ({\theta_W}) (\cos ({\theta_W})+1) \sin (2 {\theta_{D^*}})$\\
$\frac{m_\ell^2}{q^2}H_{T}^{0,+-}(H_{T}^{-,-0})^* $&$ -8 e^{-i {\chi}} \sin ^2\left(\frac{{\theta_W}}{2}\right) \sin ({\theta_W}) \sin (2 {\theta_{D^*}})$\\
$\frac{m_\ell^2}{q^2}H_{T}^{0,+-}(H_{T}^{-,-t})^* $&$ -8 e^{-i {\chi}} \sin ^2\left(\frac{{\theta_W}}{2}\right) \sin ({\theta_W}) \sin (2 {\theta_{D^*}})$\\
$\frac{m_\ell^2}{q^2}H_{T}^{+,+0}(H_{T}^{0,+-})^* $&$ -4 e^{-i {\chi}} \sin ({\theta_W}) (\cos ({\theta_W})+1) \sin (2 {\theta_{D^*}})$\\
$\frac{m_\ell^2}{q^2}H_{T}^{+,+0}(H_{T}^{+,+0})^* $&$ 16 \cos ^4\left(\frac{{\theta_W}}{2}\right) \sin ^2({\theta_{D^*}})$\\
$\frac{m_\ell^2}{q^2}H_{T}^{+,+0}(H_{T}^{+,+t})^* $&$ -16 \cos ^4\left(\frac{{\theta_W}}{2}\right) \sin ^2({\theta_{D^*}})$\\
$\frac{m_\ell^2}{q^2}H_{T}^{+,+0}(H_{T}^{-,-0})^* $&$ 4 e^{-2 i {\chi}} \sin ^2({\theta_W}) \sin ^2({\theta_{D^*}})$\\
$\frac{m_\ell^2}{q^2}H_{T}^{+,+0}(H_{T}^{-,-t})^* $&$ 4 e^{-2 i {\chi}} \sin ^2({\theta_W}) \sin ^2({\theta_{D^*}})$\\
$\frac{m_\ell^2}{q^2}H_{T}^{+,+t}(H_{T}^{0,+-})^* $&$ 4 e^{-i {\chi}} \sin ({\theta_W}) (\cos ({\theta_W})+1) \sin (2 {\theta_{D^*}})$\\
$\frac{m_\ell^2}{q^2}H_{T}^{+,+t}(H_{T}^{+,+0})^* $&$ -16 \cos ^4\left(\frac{{\theta_W}}{2}\right) \sin ^2({\theta_{D^*}})$\\
$\frac{m_\ell^2}{q^2}H_{T}^{+,+t}(H_{T}^{+,+t})^* $&$ 16 \cos ^4\left(\frac{{\theta_W}}{2}\right) \sin ^2({\theta_{D^*}})$\\
$\frac{m_\ell^2}{q^2}H_{T}^{+,+t}(H_{T}^{-,-0})^* $&$ -4 e^{-2 i {\chi}} \sin ^2({\theta_W}) \sin ^2({\theta_{D^*}})$\\
$\frac{m_\ell^2}{q^2}H_{T}^{+,+t}(H_{T}^{-,-t})^* $&$ -4 e^{-2 i {\chi}} \sin ^2({\theta_W}) \sin ^2({\theta_{D^*}})$\\
$\frac{m_\ell^2}{q^2}H_{T}^{-,-0}(H_{T}^{0,+-})^* $&$ -8 e^{i {\chi}} \sin ^2\left(\frac{{\theta_W}}{2}\right) \sin ({\theta_W}) \sin (2 {\theta_{D^*}})$\\
$\frac{m_\ell^2}{q^2}H_{T}^{-,-0}(H_{T}^{+,+0})^* $&$ 4 e^{2 i {\chi}} \sin ^2({\theta_W}) \sin ^2({\theta_{D^*}})$\\
$\frac{m_\ell^2}{q^2}H_{T}^{-,-0}(H_{T}^{+,+t})^* $&$ -4 e^{2 i {\chi}} \sin ^2({\theta_W}) \sin ^2({\theta_{D^*}})$\\
$\frac{m_\ell^2}{q^2}H_{T}^{-,-0}(H_{T}^{-,-0})^* $&$ 16 \sin ^4\left(\frac{{\theta_W}}{2}\right) \sin ^2({\theta_{D^*}})$\\
$\frac{m_\ell^2}{q^2}H_{T}^{-,-0}(H_{T}^{-,-t})^* $&$ 16 \sin ^4\left(\frac{{\theta_W}}{2}\right) \sin ^2({\theta_{D^*}})$\\
$\frac{m_\ell^2}{q^2}H_{T}^{-,-t}(H_{T}^{0,+-})^* $&$ -8 e^{i {\chi}} \sin ^2\left(\frac{{\theta_W}}{2}\right) \sin ({\theta_W}) \sin (2 {\theta_{D^*}})$\\
$\frac{m_\ell^2}{q^2}H_{T}^{-,-t}(H_{T}^{+,+0})^* $&$ 4 e^{2 i {\chi}} \sin ^2({\theta_W}) \sin ^2({\theta_{D^*}})$\\
$\frac{m_\ell^2}{q^2}H_{T}^{-,-t}(H_{T}^{+,+t})^* $&$ -4 e^{2 i {\chi}} \sin ^2({\theta_W}) \sin ^2({\theta_{D^*}})$\\
$\frac{m_\ell^2}{q^2}H_{T}^{-,-t}(H_{T}^{-,-0})^* $&$ 16 \sin ^4\left(\frac{{\theta_W}}{2}\right) \sin ^2({\theta_{D^*}})$\\
$\frac{m_\ell^2}{q^2}H_{T}^{-,-t}(H_{T}^{-,-t})^* $&$ 16 \sin ^4\left(\frac{{\theta_W}}{2}\right) \sin ^2({\theta_{D^*}})$\\
$\frac{m_\ell^2}{q^2}H_+(H_+)^* $&$ \sin ^2({\theta_W}) \sin ^2({\theta_{D^*}})$\\
$\frac{m_\ell^2}{q^2}H_+(H_-)^* $&$ e^{2 i {\chi}} \sin ^2({\theta_W}) \sin ^2({\theta_{D^*}})$\\
$\frac{m_\ell^2}{q^2}H_+(H_0)^* $&$ -\frac{1}{2} e^{i {\chi}} \sin (2 {\theta_W}) \sin (2 {\theta_{D^*}})$\\
$\frac{m_\ell^2}{q^2}H_+(H_t)^* $&$ e^{i {\chi}} \sin ({\theta_W}) \sin (2 {\theta_{D^*}})$\\
$\frac{m_\ell^2}{q^2}H_-(H_+)^* $&$ e^{-2 i {\chi}} \sin ^2({\theta_W}) \sin ^2({\theta_{D^*}})$\\
$\frac{m_\ell^2}{q^2}H_-(H_-)^* $&$ \sin ^2({\theta_W}) \sin ^2({\theta_{D^*}})$\\
$\frac{m_\ell^2}{q^2}H_-(H_0)^* $&$ -\frac{1}{2} e^{-i {\chi}} \sin (2 {\theta_W}) \sin (2 {\theta_{D^*}})$\\
$\frac{m_\ell^2}{q^2}H_-(H_t)^* $&$ e^{-i {\chi}} \sin ({\theta_W}) \sin (2 {\theta_{D^*}})$\\
$\frac{m_\ell^2}{q^2}H_0(H_+)^* $&$ -\frac{1}{2} e^{-i {\chi}} \sin (2 {\theta_W}) \sin (2 {\theta_{D^*}})$\\
$\frac{m_\ell^2}{q^2}H_0(H_-)^* $&$ -\frac{1}{2} e^{i {\chi}} \sin (2 {\theta_W}) \sin (2 {\theta_{D^*}})$\\
$\frac{m_\ell^2}{q^2}H_0(H_0)^* $&$ 4 \cos ^2({\theta_W}) \cos ^2({\theta_{D^*}})$\\
$\frac{m_\ell^2}{q^2}H_0(H_t)^* $&$ -4 \cos ({\theta_W}) \cos ^2({\theta_{D^*}})$\\
$\frac{m_\ell^2}{q^2}H_t(H_+)^* $&$ e^{-i {\chi}} \sin ({\theta_W}) \sin (2 {\theta_{D^*}})$\\
$\frac{m_\ell^2}{q^2}H_t(H_-)^* $&$ e^{i {\chi}} \sin ({\theta_W}) \sin (2 {\theta_{D^*}})$\\
$\frac{m_\ell^2}{q^2}H_t(H_0)^* $&$ -4 \cos ({\theta_W}) \cos ^2({\theta_{D^*}})$\\
$\frac{m_\ell^2}{q^2}H_t(H_t)^* $&$ 4 \cos ^2({\theta_{D^*}})$\\
\end{tabular}
\end{table}

\section{$n_t$ Binning Strategy}\label{ntbinning}
\begin{figure}[H]
\includegraphics[scale=0.5]{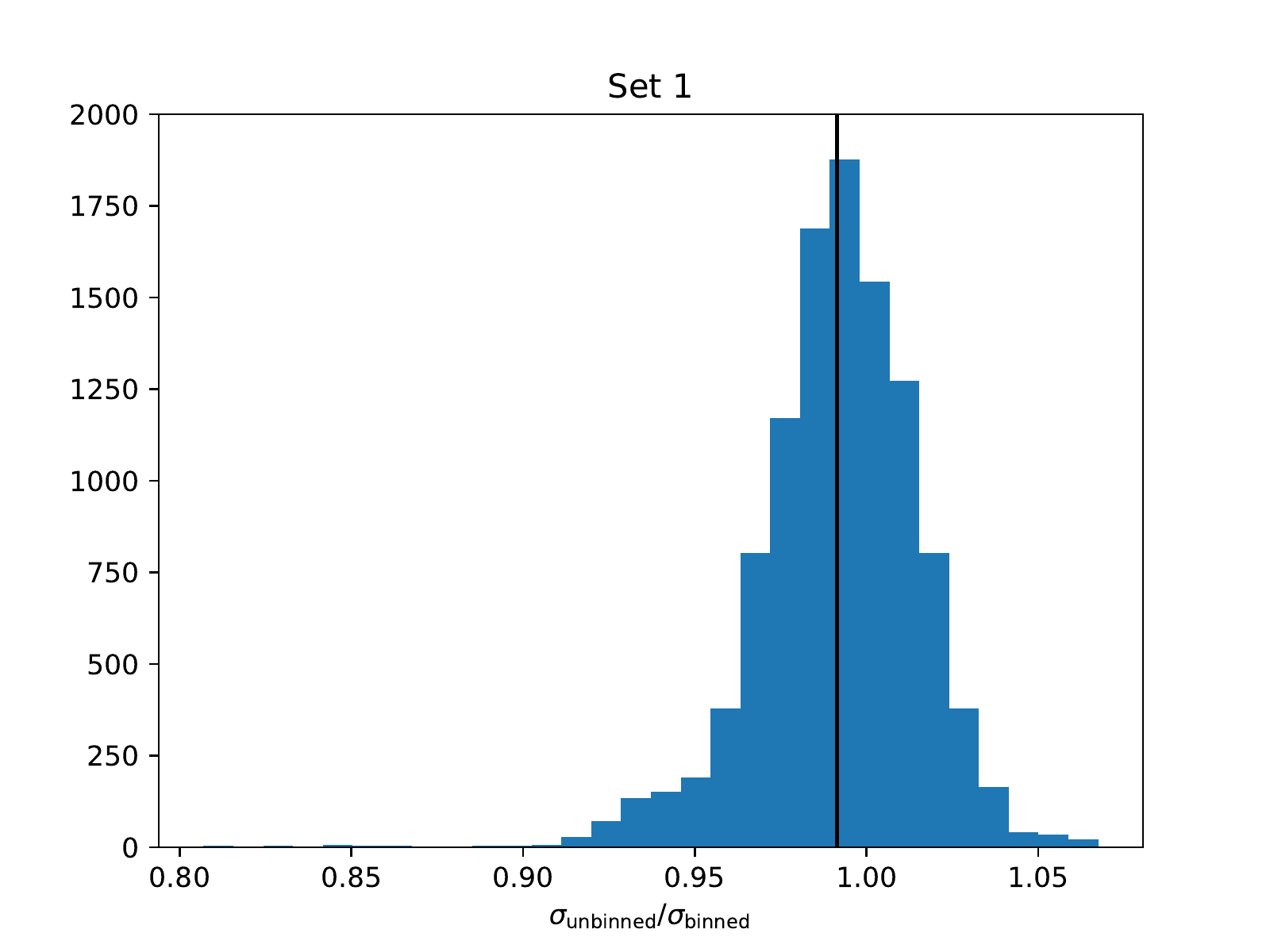}\\
\includegraphics[scale=0.5]{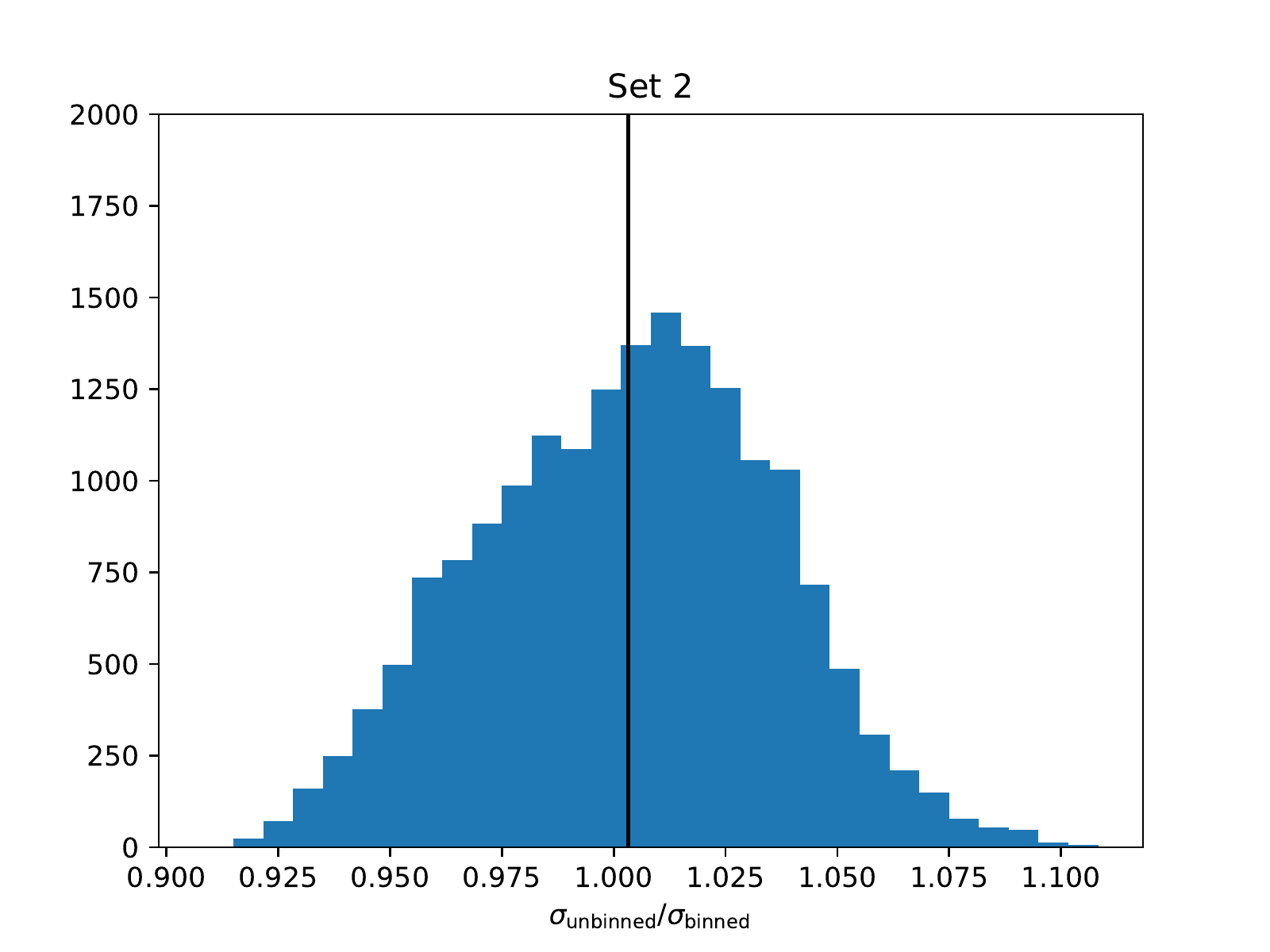}\\
\includegraphics[scale=0.5]{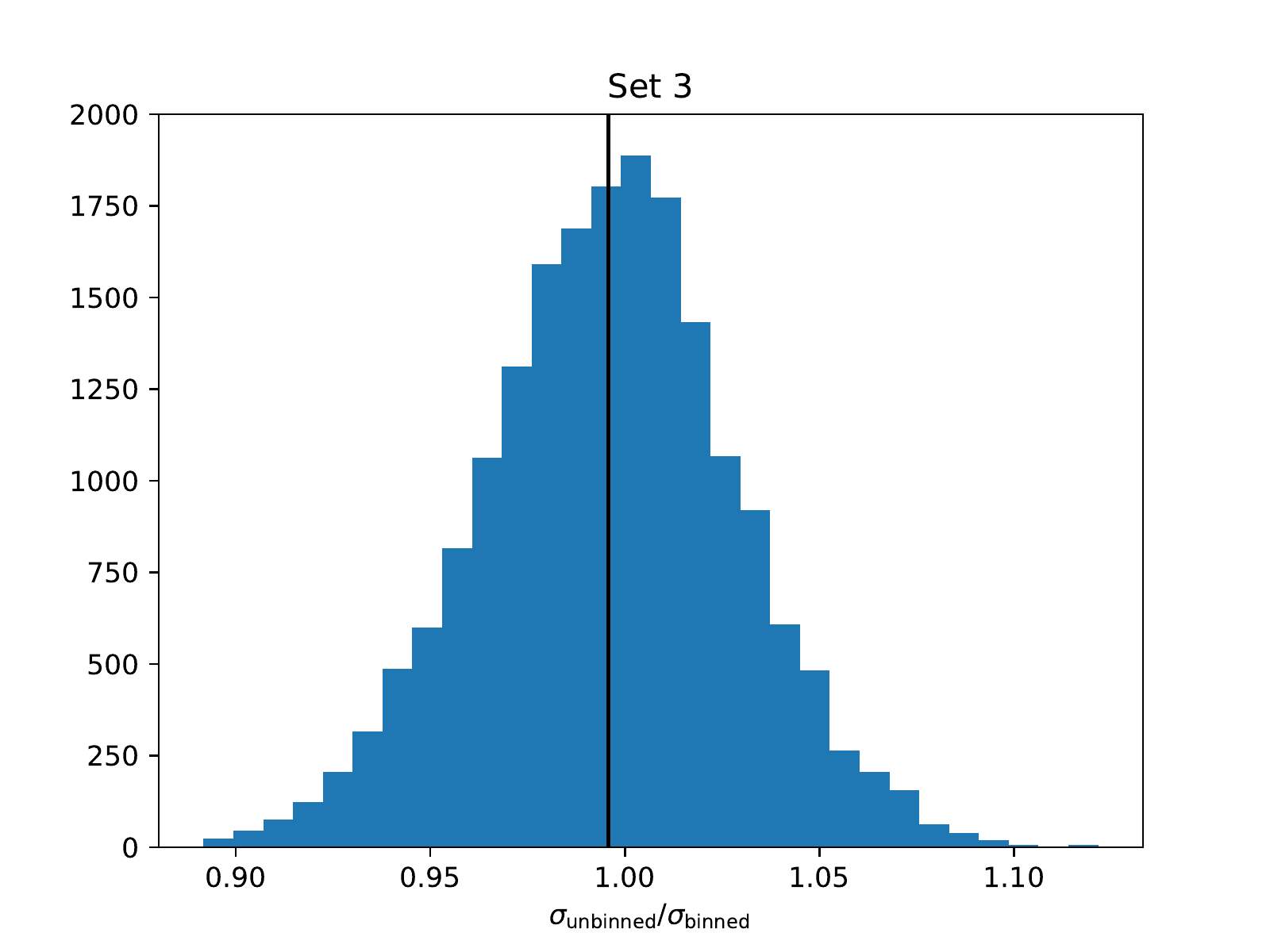}
\caption{\label{binnedvsunbinned1}Histogram plots showing the ratio of standard deviations, $\sigma_\mathrm{unbinned}/\sigma_\mathrm{binned}$, on each set for correlator data that has been only partially binned, or fully binned, as described in the text. The vertical black line corresponds to the mean. We only include data points in the range $t_\mathrm{min}\leq t \leq t_\mathrm{max}$.}
\end{figure}
\begin{figure}[H]
\includegraphics[scale=0.5]{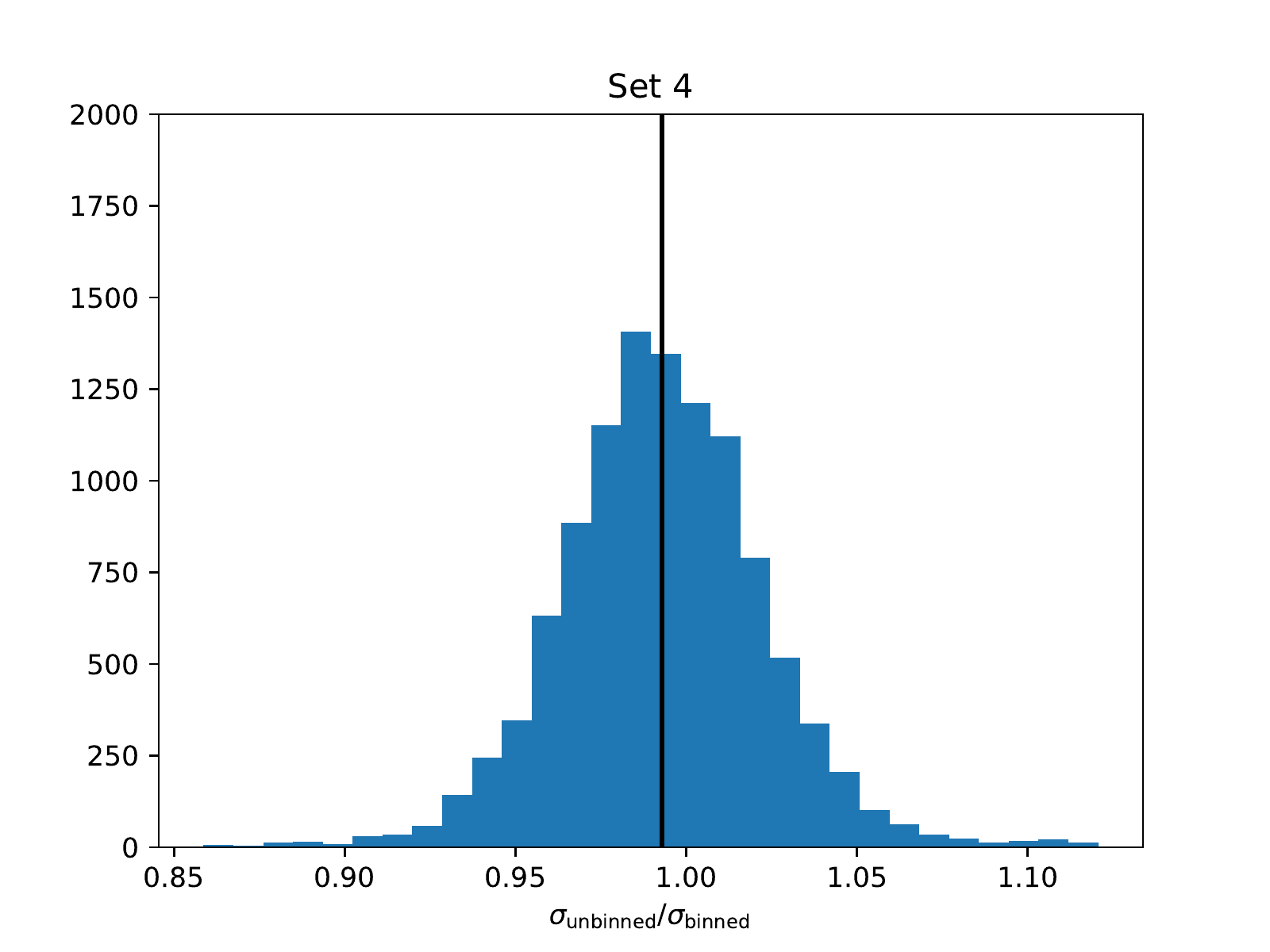}\\
\includegraphics[scale=0.5]{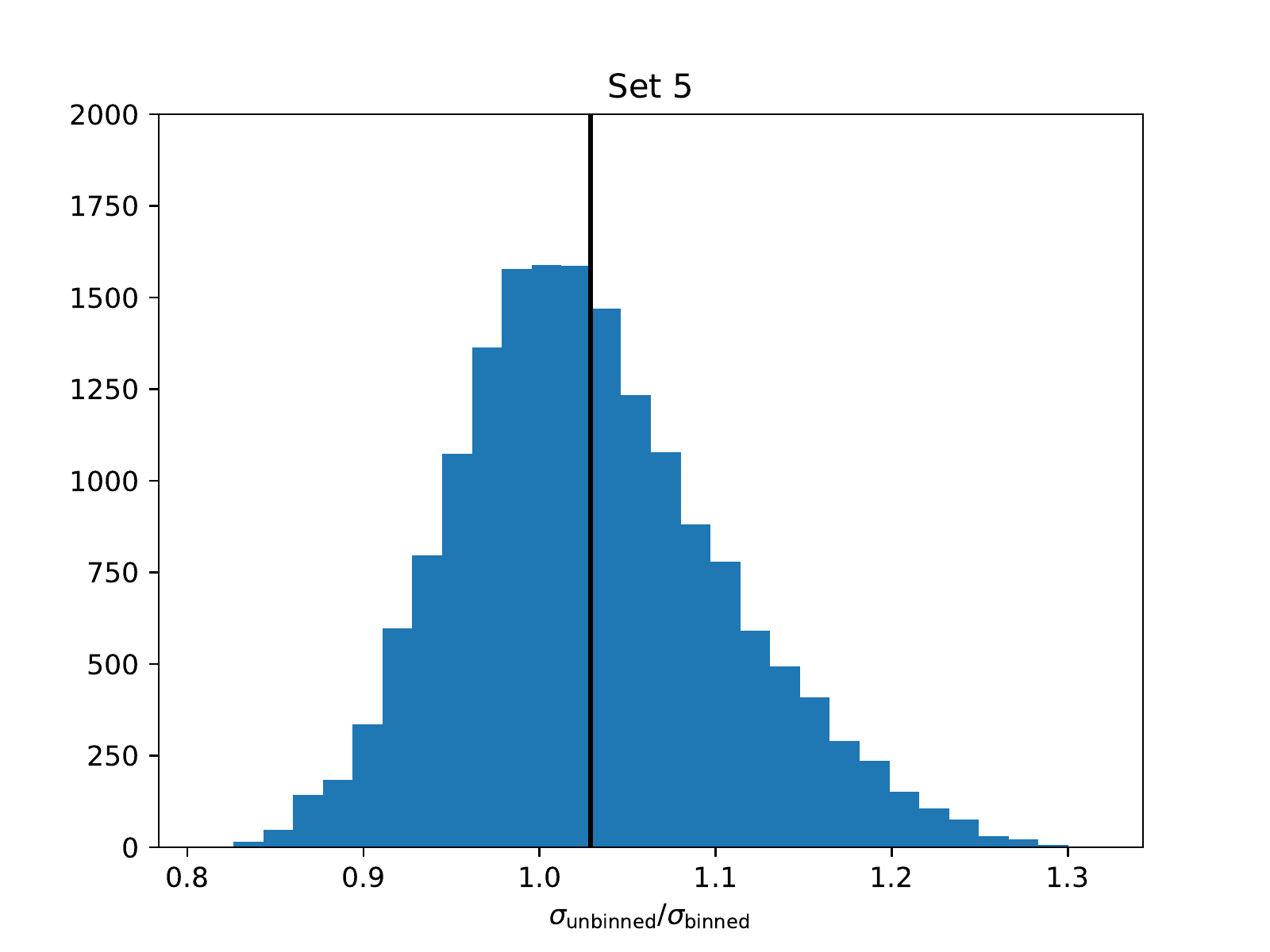}
\caption{\label{binnedvsunbinned2}Histogram plots showing the ratio of standard deviations, $\sigma_\mathrm{unbinned}/\sigma_\mathrm{binned}$, on each set for correlator data that has been only partially binned, or fully binned, as described in the text. The vertical black line corresponds to the mean. We only include data points in the range $t_\mathrm{min}\leq t \leq t_\mathrm{max}$.}
\end{figure}
On each gauge configuration we compute multiple instances of each correlation function, with $n_t$ sources placed at different values of $t_\mathrm{src}$, spaced equally across the time extent of a given configuration. The values of $n_t$ used on each configuration are given in~\cref{tab:gaugeinfo}. In previous calculations~\cite{Harrison:2021tol,Harrison:2020nrv} the correlation functions with different $t_\mathrm{src}$ on a given configuration were binned. Both calculations included states whose correlation functions exhibited significant correlation between the $n_t$ source times, such as the $\eta_h$. However, correlation functions that are sufficiently far apart in time are expected to be only weakly correlated and it is preferable not to bin the multiple $n_t$ in order to improve the resolution of the covariance matrix of our data points, particularly on set 3 and set 5 where $n_\mathrm{cfg}$ is relatively small. On set 1, we have computed the correlations between data generated from different values of $t_\mathrm{src}$. We find that excluding $\eta_h$ and $\eta_c$, the maximum correlation between any two data points using different $t_\mathrm{src}$, using 4 equally spaced values of $t_\mathrm{src}$, is $\approx 0.2$. On set 4, the maximum correlation between data from 4 different, equally spaced $t_\mathrm{src}$ is also $\approx 0.2$. With these findings in mind, we choose to use the $B_s$ and $D_s^*$ masses, instead of those of the $\eta_h$ and $\eta_c$ as was done in~\cite{Harrison:2021tol,Harrison:2020nrv}, to parameterise the physical heavy and charm quark masses. We do not then bin our data on set 2, 3 or 5, and on set 1 and 4 we only bin every 4 and 2 adjacent source times respectively. We have investigated the effect this has on the uncertainty of the raw correlator data points, expecting that for fully uncorrelated data the standard deviation will not change. Histogram plots of $\sigma_\mathrm{unbinned}/\sigma_\mathrm{binned}$ are shown in~\cref{binnedvsunbinned1,binnedvsunbinned2}, for data points in the range $t_\mathrm{min}\leq t \leq t_\mathrm{max}$, where we see that on each set binning results in very similar uncertainties consistent with the different time sources being uncorrelated. This allows us to use a smaller SVD cut when fitting our correlation functions and results in numerically more stable fits.

\section{Lattice Data}\label{lattdat}
\subsection{Lattice Form Factor Results}
Here we give our lattice results for the form factors extracted from correlator fits. The SM form factors for $B\to D^*$ are given in~\cref{l_SM_set1,l_SM_set2,l_SM_set3,l_SM_set4,l_SM_set5} and in~\cref{s_SM_set1,s_SM_set2,s_SM_set3,s_SM_set4,s_SM_set5} for $B_s\to D_s^*$. The tensor form factors for $B\to D^*$ are also given in~\cref{l_T_set1,l_T_set2,l_T_set3,l_T_set4,l_T_set5} and in~\cref{s_T_set1,s_T_set2,s_T_set3,s_T_set4,s_T_set5} for $B_s\to D_s^*$. These numbers include the renormalisation factors given in~\cref{Zfactors,ZTfactors}.
\begin{table}[H]
\centering
\caption{Lattice form factor results for set 1. $ak$ here is the value of the $x$ and $y$ components of the lattice momentum for the $D^*$. $ak$ is calculated from the corresponding twist in \cref{twists}.  \label{l_SM_set1}}
\begin{tabular}{ c c c c c c }
\hline
$am_h$& $ak$ & $h_{A_1}$& $h_{A_2}$& $h_{A_3}$& $h_{V}$\\\hline
0.65&	0.0	&0.933(15)	&$-$	&$-$	&$-$	\\
&	0.0378853	&0.932(14)	&5(22)	&-4(21)	&1.19(38)	\\
&	0.0757705	&0.926(15)	&2.0(6.1)	&-0.7(5.6)	&1.24(28)	\\
&	0.113656	&0.916(18)	&1.2(3.4)	&0.05(2.78)	&1.24(23)	\\
&	0.151541	&0.907(18)	&0.8(2.3)	&0.4(1.6)	&1.24(20)	\\
&	0.189426	&0.893(20)	&0.7(2.0)	&0.6(1.1)	&1.22(19)	\\
\hline
0.725&	0.0	&0.939(16)	&$-$	&$-$	&$-$	\\
&	0.0378853	&0.938(15)	&5(22)	&-3(22)	&1.19(38)	\\
&	0.0757705	&0.932(16)	&1.7(6.2)	&-0.6(5.8)	&1.24(28)	\\
&	0.113656	&0.922(18)	&1.0(3.3)	&0.08(2.89)	&1.25(24)	\\
&	0.151541	&0.913(18)	&0.7(2.1)	&0.4(1.7)	&1.24(21)	\\
&	0.189426	&0.898(21)	&0.5(1.7)	&0.6(1.2)	&1.23(19)	\\
\hline
0.8&	0.0	&0.946(16)	&$-$	&$-$	&$-$	\\
&	0.0378853	&0.945(15)	&4(23)	&-3(23)	&1.20(39)	\\
&	0.0757705	&0.939(16)	&1.6(6.3)	&-0.5(6.0)	&1.25(28)	\\
&	0.113656	&0.929(19)	&0.9(3.3)	&0.1(3.0)	&1.26(24)	\\
&	0.151541	&0.920(19)	&0.6(2.1)	&0.4(1.7)	&1.25(21)	\\
&	0.189426	&0.905(21)	&0.4(1.6)	&0.6(1.2)	&1.24(20)	\\
\hline
\end{tabular}
\end{table}
\begin{table}[H]
\centering
\caption{Lattice form factor results for set 2. $ak$ here is the value of the $x$ and $y$ components of the lattice momentum for the $D^*$. $ak$ is calculated from the corresponding twist in \cref{twists}.  \label{l_SM_set2}}
\begin{tabular}{ c c c c c c }
\hline
$am_h$& $ak$ & $h_{A_1}$& $h_{A_2}$& $h_{A_3}$& $h_{V}$\\\hline
0.427&	0.0	&0.916(37)	&$-$	&$-$	&$-$	\\
&	0.055399	&0.916(42)	&-9(15)	&10(14)	&1.11(31)	\\
&	0.110798	&0.900(44)	&-3.0(4.9)	&3.7(3.8)	&1.13(24)	\\
&	0.166197	&0.872(53)	&-1.9(3.7)	&2.4(2.0)	&1.10(22)	\\
&	0.221596	&0.832(66)	&-1.6(4.4)	&1.9(1.4)	&1.04(22)	\\
&	0.276995	&0.780(84)	&-2(10)	&1.6(1.1)	&0.98(25)	\\
\hline
0.525&	0.0	&0.921(40)	&$-$	&$-$	&$-$	\\
&	0.055399	&0.921(44)	&-10(16)	&11(15)	&1.11(32)	\\
&	0.110798	&0.904(47)	&-3.2(4.7)	&3.9(4.0)	&1.13(24)	\\
&	0.166197	&0.876(57)	&-1.9(3.0)	&2.5(2.1)	&1.10(22)	\\
&	0.221596	&0.836(71)	&-1.4(2.7)	&1.9(1.5)	&1.05(23)	\\
&	0.276995	&0.785(90)	&-1.2(3.1)	&1.6(1.2)	&0.99(26)	\\
\hline
0.65&	0.0	&0.930(43)	&$-$	&$-$	&$-$	\\
&	0.055399	&0.930(48)	&-11(17)	&12(16)	&1.12(32)	\\
&	0.110798	&0.913(51)	&-3.4(4.8)	&4.1(4.3)	&1.14(25)	\\
&	0.166197	&0.884(61)	&-1.9(2.9)	&2.6(2.3)	&1.11(23)	\\
&	0.221596	&0.845(77)	&-1.4(2.4)	&2.0(1.6)	&1.06(24)	\\
&	0.276995	&0.794(97)	&-1.2(2.4)	&1.7(1.3)	&1.00(27)	\\
\hline
0.8&	0.0	&0.943(46)	&$-$	&$-$	&$-$	\\
&	0.055399	&0.943(51)	&-11(18)	&12(17)	&1.13(33)	\\
&	0.110798	&0.927(55)	&-3.5(5.0)	&4.2(4.7)	&1.15(26)	\\
&	0.166197	&0.898(67)	&-2.0(3.0)	&2.7(2.5)	&1.13(24)	\\
&	0.221596	&0.859(83)	&-1.5(2.4)	&2.1(1.8)	&1.08(24)	\\
&	0.276995	&0.81(10)	&-1.2(2.2)	&1.7(1.5)	&1.02(28)	\\
\hline
\end{tabular}
\end{table}
\begin{table}[H]
\centering
\caption{Lattice form factor results for set 3. $ak$ here is the value of the $x$ and $y$ components of the lattice momentum for the $D^*$. $ak$ is calculated from the corresponding twist in \cref{twists}.  \label{l_SM_set3}}
\begin{tabular}{ c c c c c c }
\hline
$am_h$& $ak$ & $h_{A_1}$& $h_{A_2}$& $h_{A_3}$& $h_{V}$\\\hline
0.5&	0.0	&0.916(22)	&$-$	&$-$	&$-$	\\
&	0.061831	&0.902(25)	&0.05(3.91)	&0.9(3.7)	&1.21(14)	\\
&	0.123662	&0.850(39)	&-0.2(1.7)	&1.1(1.4)	&1.11(12)	\\
&	0.185493	&0.774(58)	&0.3(1.4)	&0.89(93)	&0.97(14)	\\
&	0.247324	&0.689(91)	&1.1(1.7)	&0.67(85)	&0.84(18)	\\
&	0.309155	&0.60(10)	&0.3(1.9)	&0.70(68)	&0.77(15)	\\
\hline
0.65&	0.0	&0.932(24)	&$-$	&$-$	&$-$	\\
&	0.061831	&0.917(28)	&-0.3(4.2)	&1.2(4.1)	&1.26(14)	\\
&	0.123662	&0.863(43)	&-0.4(1.8)	&1.3(1.6)	&1.15(12)	\\
&	0.185493	&0.789(63)	&-0.002(1.375)	&1.0(1.0)	&1.00(14)	\\
&	0.247324	&0.711(98)	&0.4(1.5)	&0.85(93)	&0.86(19)	\\
&	0.309155	&0.62(11)	&-0.3(1.4)	&0.88(73)	&0.77(16)	\\
\hline
0.8&	0.0	&0.950(26)	&$-$	&$-$	&$-$	\\
&	0.061831	&0.934(31)	&-0.3(4.6)	&1.2(4.5)	&1.30(15)	\\
&	0.123662	&0.880(47)	&-0.5(2.0)	&1.4(1.8)	&1.19(13)	\\
&	0.185493	&0.809(70)	&-0.2(1.5)	&1.2(1.2)	&1.03(15)	\\
&	0.247324	&0.74(11)	&0.02(1.52)	&1.0(1.0)	&0.87(20)	\\
&	0.309155	&0.64(12)	&-0.6(1.3)	&1.04(80)	&0.77(18)	\\
\hline
\end{tabular}
\end{table}
\begin{table}[H]
\centering
\caption{Lattice form factor results for set 4. $ak$ here is the value of the $x$ and $y$ components of the lattice momentum for the $D^*$. $ak$ is calculated from the corresponding twist in \cref{twists}.  \label{l_SM_set4}}
\begin{tabular}{ c c c c c c }
\hline
$am_h$& $ak$ & $h_{A_1}$& $h_{A_2}$& $h_{A_3}$& $h_{V}$\\\hline
0.65&	0.0	&0.935(29)	&$-$	&$-$	&$-$	\\
&	0.0376581	&0.937(27)	&-5(40)	&7(40)	&1.20(50)	\\
&	0.0753162	&0.934(27)	&-2(11)	&3(10)	&1.23(35)	\\
&	0.112974	&0.927(28)	&-0.9(5.6)	&2.1(4.7)	&1.23(29)	\\
&	0.150632	&0.919(31)	&-0.7(4.0)	&1.8(2.9)	&1.21(26)	\\
&	0.188291	&0.907(34)	&-0.7(3.3)	&1.7(2.1)	&1.19(24)	\\
\hline
0.725&	0.0	&0.941(30)	&$-$	&$-$	&$-$	\\
&	0.0376581	&0.943(28)	&-5(42)	&7(42)	&1.21(51)	\\
&	0.0753162	&0.939(28)	&-2(11)	&3(11)	&1.24(35)	\\
&	0.112974	&0.933(30)	&-1.1(5.6)	&2.2(5.0)	&1.24(29)	\\
&	0.150632	&0.924(32)	&-0.9(3.7)	&1.9(3.0)	&1.23(26)	\\
&	0.188291	&0.913(35)	&-0.8(3.0)	&1.8(2.2)	&1.21(25)	\\
\hline
0.8&	0.0	&0.948(31)	&$-$	&$-$	&$-$	\\
&	0.0376581	&0.950(29)	&-6(44)	&7(43)	&1.23(52)	\\
&	0.0753162	&0.947(30)	&-2(12)	&3(11)	&1.25(36)	\\
&	0.112974	&0.940(31)	&-1.2(5.7)	&2.2(5.2)	&1.25(30)	\\
&	0.150632	&0.931(33)	&-1.0(3.7)	&1.9(3.2)	&1.24(27)	\\
&	0.188291	&0.920(37)	&-0.9(2.9)	&1.8(2.3)	&1.22(25)	\\
\hline
\end{tabular}
\end{table}
\begin{table}[H]
\centering
\caption{Lattice form factor results for set 5. $ak$ here is the value of the $x$ and $y$ components of the lattice momentum for the $D^*$. $ak$ is calculated from the corresponding twist in \cref{twists}.  \label{l_SM_set5}}
\begin{tabular}{ c c c c c c }
\hline
$am_h$& $ak$ & $h_{A_1}$& $h_{A_2}$& $h_{A_3}$& $h_{V}$\\\hline
0.427&	0.0	&0.886(77)	&$-$	&$-$	&$-$	\\
&	0.055399	&0.882(73)	&7(24)	&-4(23)	&0.88(51)	\\
&	0.110798	&0.854(79)	&2.7(7.7)	&-0.5(6.1)	&1.03(38)	\\
&	0.166197	&0.804(95)	&2.1(5.5)	&0.2(3.2)	&1.02(35)	\\
&	0.221596	&0.74(12)	&2.2(6.0)	&0.4(2.2)	&0.95(37)	\\
&	0.276995	&0.67(14)	&4(10)	&0.4(1.7)	&0.83(38)	\\
\hline
0.525&	0.0	&0.888(81)	&$-$	&$-$	&$-$	\\
&	0.055399	&0.884(77)	&6(25)	&-4(24)	&0.88(51)	\\
&	0.110798	&0.857(83)	&2.1(7.3)	&-0.5(6.4)	&1.02(38)	\\
&	0.166197	&0.807(99)	&1.4(4.7)	&0.2(3.4)	&1.02(36)	\\
&	0.221596	&0.74(12)	&1.2(4.1)	&0.4(2.3)	&0.95(37)	\\
&	0.276995	&0.67(14)	&1.4(4.2)	&0.3(1.8)	&0.83(39)	\\
\hline
0.65&	0.0	&0.892(86)	&$-$	&$-$	&$-$	\\
&	0.055399	&0.889(81)	&6(26)	&-5(25)	&0.88(52)	\\
&	0.110798	&0.862(87)	&1.9(7.4)	&-0.5(6.7)	&1.03(38)	\\
&	0.166197	&0.81(10)	&1.1(4.4)	&0.2(3.6)	&1.02(36)	\\
&	0.221596	&0.75(13)	&0.9(3.6)	&0.4(2.5)	&0.95(38)	\\
&	0.276995	&0.67(15)	&1.0(3.3)	&0.3(1.9)	&0.83(40)	\\
\hline
0.8&	0.0	&0.899(90)	&$-$	&$-$	&$-$	\\
&	0.055399	&0.897(85)	&6(27)	&-5(26)	&0.88(52)	\\
&	0.110798	&0.870(91)	&1.8(7.6)	&-0.6(7.0)	&1.03(39)	\\
&	0.166197	&0.82(11)	&1.0(4.4)	&0.2(3.7)	&1.03(37)	\\
&	0.221596	&0.75(13)	&0.8(3.4)	&0.3(2.6)	&0.96(39)	\\
&	0.276995	&0.67(15)	&0.9(3.0)	&0.3(2.0)	&0.84(41)	\\
\hline
\end{tabular}
\end{table}
\begin{table}[H]
\centering
\caption{Lattice tensor form factor results for set 1. $ak$ here is the value of the $x$ and $y$ components of the lattice momentum for the $D^*$. $ak$ is calculated from the corresponding twist in \cref{twists}.  \label{l_T_set1}}
\begin{tabular}{ c c c c c c }
\hline
$am_h$& $ak$ & $h_{T_1}$& $h_{T_2}$& $h_{T_3}$&\\\hline
0.65&	0.0	&$-$	&$-$	&$-$	\\
&	0.0378853	&0.873(24)	&-0.07(27)	&0.06(38.61)	\\
&	0.0757705	&0.867(25)	&-0.08(20)	&-1(10)	\\
&	0.113656	&0.858(28)	&-0.08(17)	&-0.9(5.1)	\\
&	0.151541	&0.845(28)	&-0.07(15)	&-0.8(2.8)	\\
&	0.189426	&0.831(31)	&-0.06(14)	&-0.7(2.0)	\\
\hline
0.725&	0.0	&$-$	&$-$	&$-$	\\
&	0.0378853	&0.878(24)	&-0.08(28)	&-0.2(38.9)	\\
&	0.0757705	&0.873(26)	&-0.09(20)	&-1(10)	\\
&	0.113656	&0.863(29)	&-0.09(17)	&-1.0(5.1)	\\
&	0.151541	&0.850(28)	&-0.08(15)	&-0.9(2.8)	\\
&	0.189426	&0.836(31)	&-0.07(14)	&-0.7(2.0)	\\
\hline
0.8&	0.0	&$-$	&$-$	&$-$	\\
&	0.0378853	&0.885(24)	&-0.08(28)	&-0.5(39.2)	\\
&	0.0757705	&0.880(26)	&-0.10(21)	&-1(10)	\\
&	0.113656	&0.870(29)	&-0.10(18)	&-1.0(5.1)	\\
&	0.151541	&0.857(29)	&-0.09(16)	&-0.9(2.8)	\\
&	0.189426	&0.843(32)	&-0.08(15)	&-0.7(2.0)	\\
\hline
\end{tabular}
\end{table}
\begin{table}[H]
\centering
\caption{Lattice tensor form factor results for set 2. $ak$ here is the value of the $x$ and $y$ components of the lattice momentum for the $D^*$. $ak$ is calculated from the corresponding twist in \cref{twists}.  \label{l_T_set2}}
\begin{tabular}{ c c c c c c }
\hline
$am_h$& $ak$ & $h_{T_1}$& $h_{T_2}$& $h_{T_3}$&\\\hline
0.427&	0.0	&$-$	&$-$	&$-$	\\
&	0.055399	&0.878(59)	&-0.05(26)	&6(20)	\\
&	0.110798	&0.860(63)	&-0.07(20)	&1.7(5.3)	\\
&	0.166197	&0.831(76)	&-0.07(19)	&0.9(2.8)	\\
&	0.221596	&0.791(94)	&-0.06(20)	&0.7(1.9)	\\
&	0.276995	&0.74(12)	&-0.07(22)	&0.6(1.5)	\\
\hline
0.525&	0.0	&$-$	&$-$	&$-$	\\
&	0.055399	&0.887(59)	&-0.06(27)	&7(20)	\\
&	0.110798	&0.869(64)	&-0.08(21)	&2.1(5.3)	\\
&	0.166197	&0.840(76)	&-0.08(20)	&1.1(2.8)	\\
&	0.221596	&0.799(95)	&-0.07(21)	&0.8(1.9)	\\
&	0.276995	&0.75(12)	&-0.08(23)	&0.7(1.6)	\\
\hline
0.65&	0.0	&$-$	&$-$	&$-$	\\
&	0.055399	&0.899(61)	&-0.08(29)	&8(21)	\\
&	0.110798	&0.881(65)	&-0.10(22)	&2.4(5.4)	\\
&	0.166197	&0.851(78)	&-0.10(21)	&1.3(2.9)	\\
&	0.221596	&0.810(98)	&-0.09(23)	&0.9(2.0)	\\
&	0.276995	&0.76(12)	&-0.08(24)	&0.8(1.6)	\\
\hline
0.8&	0.0	&$-$	&$-$	&$-$	\\
&	0.055399	&0.915(63)	&-0.10(30)	&9(21)	\\
&	0.110798	&0.898(68)	&-0.12(23)	&2.7(5.6)	\\
&	0.166197	&0.867(81)	&-0.11(23)	&1.4(3.0)	\\
&	0.221596	&0.83(10)	&-0.10(24)	&1.0(2.1)	\\
&	0.276995	&0.77(13)	&-0.09(26)	&0.8(1.7)	\\
\hline
\end{tabular}
\end{table}
\begin{table}[H]
\centering
\caption{Lattice tensor form factor results for set 3. $ak$ here is the value of the $x$ and $y$ components of the lattice momentum for the $D^*$. $ak$ is calculated from the corresponding twist in \cref{twists}.  \label{l_T_set3}}
\begin{tabular}{ c c c c c c }
\hline
$am_h$& $ak$ & $h_{T_1}$& $h_{T_2}$& $h_{T_3}$&\\\hline
0.5&	0.0	&$-$	&$-$	&$-$	\\
&	0.061831	&0.877(32)	&-0.04(14)	&-1.8(4.6)	\\
&	0.123662	&0.820(50)	&-0.03(13)	&-0.6(1.8)	\\
&	0.185493	&0.748(73)	&0.001(168)	&-0.4(1.1)	\\
&	0.247324	&0.66(12)	&0.02(24)	&-0.5(1.1)	\\
&	0.309155	&0.65(16)	&0.45(48)	&0.23(86)	\\
\hline
0.65&	0.0	&$-$	&$-$	&$-$	\\
&	0.061831	&0.898(33)	&-0.04(15)	&-1.2(4.7)	\\
&	0.123662	&0.843(52)	&-0.04(14)	&-0.2(1.8)	\\
&	0.185493	&0.772(76)	&0.007(177)	&-0.1(1.2)	\\
&	0.247324	&0.69(13)	&0.05(25)	&-0.2(1.1)	\\
&	0.309155	&0.68(16)	&0.47(51)	&0.41(89)	\\
\hline
0.8&	0.0	&$-$	&$-$	&$-$	\\
&	0.061831	&0.920(35)	&-0.05(16)	&-0.9(5.1)	\\
&	0.123662	&0.866(56)	&-0.04(15)	&0.0004(1.9847)	\\
&	0.185493	&0.798(82)	&0.01(19)	&0.08(1.25)	\\
&	0.247324	&0.73(13)	&0.07(27)	&0.06(1.14)	\\
&	0.309155	&0.71(17)	&0.50(54)	&0.61(94)	\\
\hline
\end{tabular}
\end{table}
\begin{table}[H]
\centering
\caption{Lattice tensor form factor results for set 4. $ak$ here is the value of the $x$ and $y$ components of the lattice momentum for the $D^*$. $ak$ is calculated from the corresponding twist in \cref{twists}.  \label{l_T_set4}}
\begin{tabular}{ c c c c c c }
\hline
$am_h$& $ak$ & $h_{T_1}$& $h_{T_2}$& $h_{T_3}$&\\\hline
0.65&	0.0	&$-$	&$-$	&$-$	\\
&	0.0376581	&0.875(44)	&-0.19(40)	&4(67)	\\
&	0.0753162	&0.872(44)	&-0.18(29)	&1(17)	\\
&	0.112974	&0.867(47)	&-0.17(24)	&0.8(7.9)	\\
&	0.150632	&0.859(51)	&-0.16(22)	&0.5(4.8)	\\
&	0.188291	&0.849(56)	&-0.15(21)	&0.4(3.4)	\\
\hline
0.725&	0.0	&$-$	&$-$	&$-$	\\
&	0.0376581	&0.881(44)	&-0.19(41)	&3(68)	\\
&	0.0753162	&0.878(45)	&-0.18(30)	&1(17)	\\
&	0.112974	&0.873(48)	&-0.17(25)	&0.6(8.1)	\\
&	0.150632	&0.865(52)	&-0.16(23)	&0.5(4.9)	\\
&	0.188291	&0.855(57)	&-0.15(22)	&0.4(3.4)	\\
\hline
0.8&	0.0	&$-$	&$-$	&$-$	\\
&	0.0376581	&0.888(45)	&-0.19(42)	&2(70)	\\
&	0.0753162	&0.885(46)	&-0.18(30)	&0.8(17.7)	\\
&	0.112974	&0.880(49)	&-0.17(26)	&0.5(8.2)	\\
&	0.150632	&0.872(53)	&-0.17(23)	&0.4(5.0)	\\
&	0.188291	&0.861(58)	&-0.16(22)	&0.4(3.5)	\\
\hline
\end{tabular}
\end{table}
\begin{table}[H]
\centering
\caption{Lattice tensor form factor results for set 5. $ak$ here is the value of the $x$ and $y$ components of the lattice momentum for the $D^*$. $ak$ is calculated from the corresponding twist in \cref{twists}.  \label{l_T_set5}}
\begin{tabular}{ c c c c c c }
\hline
$am_h$& $ak$ & $h_{T_1}$& $h_{T_2}$& $h_{T_3}$&\\\hline
0.427&	0.0	&$-$	&$-$	&$-$	\\
&	0.055399	&0.840(92)	&-0.07(42)	&-3(28)	\\
&	0.110798	&0.81(10)	&-0.12(33)	&-0.8(7.6)	\\
&	0.166197	&0.75(12)	&-0.15(32)	&-0.3(4.1)	\\
&	0.221596	&0.69(16)	&-0.18(35)	&-0.2(2.9)	\\
&	0.276995	&0.61(19)	&-0.22(39)	&-0.3(2.2)	\\
\hline
0.525&	0.0	&$-$	&$-$	&$-$	\\
&	0.055399	&0.845(91)	&-0.06(43)	&-3(28)	\\
&	0.110798	&0.81(10)	&-0.12(34)	&-0.8(7.6)	\\
&	0.166197	&0.76(12)	&-0.15(34)	&-0.3(4.1)	\\
&	0.221596	&0.69(16)	&-0.18(37)	&-0.2(2.9)	\\
&	0.276995	&0.61(19)	&-0.22(41)	&-0.3(2.2)	\\
\hline
0.65&	0.0	&$-$	&$-$	&$-$	\\
&	0.055399	&0.851(92)	&-0.03(45)	&-4(28)	\\
&	0.110798	&0.82(10)	&-0.11(35)	&-0.9(7.6)	\\
&	0.166197	&0.76(12)	&-0.15(35)	&-0.4(4.1)	\\
&	0.221596	&0.69(16)	&-0.18(39)	&-0.2(2.9)	\\
&	0.276995	&0.61(19)	&-0.22(43)	&-0.3(2.2)	\\
\hline
0.8&	0.0	&$-$	&$-$	&$-$	\\
&	0.055399	&0.861(92)	&-0.01(47)	&-4(28)	\\
&	0.110798	&0.83(10)	&-0.10(37)	&-1.0(7.6)	\\
&	0.166197	&0.77(12)	&-0.15(37)	&-0.4(4.1)	\\
&	0.221596	&0.70(16)	&-0.18(41)	&-0.2(2.9)	\\
&	0.276995	&0.62(20)	&-0.22(46)	&-0.3(2.3)	\\
\hline
\end{tabular}
\end{table}
\begin{table}[H]
\centering
\caption{Lattice form factor results for set 1. $ak$ here is the value of the $x$ and $y$ components of the lattice momentum for the $D_s^*$. $ak$ is calculated from the corresponding twist in \cref{twists}.  \label{s_SM_set1}}
\begin{tabular}{ c c c c c c }
\hline
$am_h$& $ak$ & $h^s_{A_1}$& $h^s_{A_2}$& $h^s_{A_3}$& $h^s_{V}$\\\hline
0.65&	0.0	&0.9293(50)	&$-$	&$-$	&$-$	\\
&	0.0378853	&0.9281(48)	&0.5(7.1)	&0.8(6.9)	&1.27(16)	\\
&	0.0757705	&0.9231(53)	&0.3(2.1)	&0.9(1.9)	&1.27(11)	\\
&	0.113656	&0.9152(58)	&0.2(1.1)	&0.99(93)	&1.258(87)	\\
&	0.151541	&0.9040(64)	&0.25(82)	&1.02(58)	&1.241(74)	\\
&	0.189426	&0.8900(72)	&0.30(69)	&1.02(41)	&1.219(67)	\\
\hline
0.725&	0.0	&0.9342(51)	&$-$	&$-$	&$-$	\\
&	0.0378853	&0.9330(50)	&0.3(7.2)	&0.8(7.1)	&1.27(17)	\\
&	0.0757705	&0.9280(55)	&0.2(2.1)	&0.9(2.0)	&1.27(11)	\\
&	0.113656	&0.9200(59)	&0.1(1.1)	&0.97(96)	&1.264(89)	\\
&	0.151541	&0.9088(66)	&0.08(74)	&1.00(60)	&1.248(76)	\\
&	0.189426	&0.8947(74)	&0.08(59)	&1.01(42)	&1.226(69)	\\
\hline
0.8&	0.0	&0.9404(52)	&$-$	&$-$	&$-$	\\
&	0.0378853	&0.9392(51)	&0.2(7.4)	&0.9(7.3)	&1.28(17)	\\
&	0.0757705	&0.9341(56)	&0.1(2.1)	&0.9(2.0)	&1.28(11)	\\
&	0.113656	&0.9261(61)	&0.02(1.09)	&0.97(99)	&1.274(91)	\\
&	0.151541	&0.9148(68)	&-0.005(722)	&1.00(61)	&1.257(78)	\\
&	0.189426	&0.9006(76)	&-0.01(56)	&1.01(44)	&1.235(70)	\\
\hline
\end{tabular}
\end{table}
\begin{table}[H]
\centering
\caption{Lattice form factor results for set 2. $ak$ here is the value of the $x$ and $y$ components of the lattice momentum for the $D_s^*$. $ak$ is calculated from the corresponding twist in \cref{twists}.  \label{s_SM_set2}}
\begin{tabular}{ c c c c c c }
\hline
$am_h$& $ak$ & $h^s_{A_1}$& $h^s_{A_2}$& $h^s_{A_3}$& $h^s_{V}$\\\hline
0.427&	0.0	&0.908(15)	&$-$	&$-$	&$-$	\\
&	0.055399	&0.902(16)	&-3.3(6.0)	&4.2(5.6)	&1.23(16)	\\
&	0.110798	&0.879(17)	&-1.1(2.0)	&2.0(1.5)	&1.21(12)	\\
&	0.166197	&0.844(21)	&-0.7(1.5)	&1.50(81)	&1.15(11)	\\
&	0.221596	&0.801(26)	&-0.6(1.9)	&1.30(57)	&1.08(12)	\\
&	0.276995	&0.753(34)	&-0.2(4.2)	&1.15(48)	&1.00(13)	\\
\hline
0.525&	0.0	&0.913(15)	&$-$	&$-$	&$-$	\\
&	0.055399	&0.907(17)	&-3.5(6.1)	&4.3(5.8)	&1.23(16)	\\
&	0.110798	&0.884(18)	&-1.2(1.9)	&2.0(1.6)	&1.21(12)	\\
&	0.166197	&0.849(22)	&-0.8(1.2)	&1.50(86)	&1.15(11)	\\
&	0.221596	&0.804(27)	&-0.6(1.1)	&1.29(61)	&1.08(12)	\\
&	0.276995	&0.756(36)	&-0.5(1.3)	&1.15(51)	&1.00(13)	\\
\hline
0.65&	0.0	&0.922(16)	&$-$	&$-$	&$-$	\\
&	0.055399	&0.916(18)	&-3.6(6.3)	&4.3(6.1)	&1.24(16)	\\
&	0.110798	&0.893(19)	&-1.2(1.9)	&2.0(1.7)	&1.21(12)	\\
&	0.166197	&0.856(23)	&-0.8(1.1)	&1.50(91)	&1.16(11)	\\
&	0.221596	&0.812(29)	&-0.65(96)	&1.30(65)	&1.09(12)	\\
&	0.276995	&0.764(38)	&-0.53(99)	&1.17(55)	&1.01(13)	\\
\hline
0.8&	0.0	&0.936(17)	&$-$	&$-$	&$-$	\\
&	0.055399	&0.930(19)	&-3.6(6.6)	&4.3(6.5)	&1.26(16)	\\
&	0.110798	&0.906(21)	&-1.3(2.0)	&2.0(1.8)	&1.23(13)	\\
&	0.166197	&0.869(25)	&-0.9(1.2)	&1.51(97)	&1.18(12)	\\
&	0.221596	&0.824(31)	&-0.69(93)	&1.32(70)	&1.11(12)	\\
&	0.276995	&0.776(41)	&-0.57(92)	&1.19(60)	&1.02(14)	\\
\hline
\end{tabular}
\end{table}
\begin{table}[H]
\centering
\caption{Lattice form factor results for set 3. $ak$ here is the value of the $x$ and $y$ components of the lattice momentum for the $D_s^*$. $ak$ is calculated from the corresponding twist in \cref{twists}.  \label{s_SM_set3}}
\begin{tabular}{ c c c c c c }
\hline
$am_h$& $ak$ & $h^s_{A_1}$& $h^s_{A_2}$& $h^s_{A_3}$& $h^s_{V}$\\\hline
0.5&	0.0	&0.9169(92)	&$-$	&$-$	&$-$	\\
&	0.061831	&0.8984(90)	&-0.4(1.4)	&1.2(1.3)	&1.247(58)	\\
&	0.123662	&0.845(12)	&-0.45(53)	&1.18(44)	&1.142(38)	\\
&	0.185493	&0.772(17)	&-0.48(42)	&1.15(28)	&1.021(41)	\\
&	0.247324	&0.688(30)	&-0.33(58)	&1.02(28)	&0.901(92)	\\
&	0.309155	&0.597(54)	&-0.25(97)	&0.87(33)	&0.86(12)	\\
\hline
0.65&	0.0	&0.931(10)	&$-$	&$-$	&$-$	\\
&	0.061831	&0.9117(99)	&-0.4(1.5)	&1.2(1.4)	&1.272(64)	\\
&	0.123662	&0.857(13)	&-0.48(56)	&1.19(49)	&1.162(43)	\\
&	0.185493	&0.785(19)	&-0.54(42)	&1.18(31)	&1.035(46)	\\
&	0.247324	&0.705(33)	&-0.46(52)	&1.08(32)	&0.902(99)	\\
&	0.309155	&0.620(58)	&-0.50(76)	&0.98(38)	&0.84(13)	\\
\hline
0.8&	0.0	&0.948(11)	&$-$	&$-$	&$-$	\\
&	0.061831	&0.928(11)	&-0.4(1.6)	&1.2(1.6)	&1.300(69)	\\
&	0.123662	&0.872(14)	&-0.52(60)	&1.21(54)	&1.186(48)	\\
&	0.185493	&0.801(20)	&-0.60(44)	&1.23(35)	&1.054(51)	\\
&	0.247324	&0.726(36)	&-0.58(52)	&1.17(36)	&0.91(11)	\\
&	0.309155	&0.651(64)	&-0.70(74)	&1.12(43)	&0.82(14)	\\
\hline
\end{tabular}
\end{table}
\begin{table}[H]
\centering
\caption{Lattice form factor results for set 4. $ak$ here is the value of the $x$ and $y$ components of the lattice momentum for the $D_s^*$. $ak$ is calculated from the corresponding twist in \cref{twists}.  \label{s_SM_set4}}
\begin{tabular}{ c c c c c c }
\hline
$am_h$& $ak$ & $h^s_{A_1}$& $h^s_{A_2}$& $h^s_{A_3}$& $h^s_{V}$\\\hline
0.65&	0.0	&0.9271(42)	&$-$	&$-$	&$-$	\\
&	0.0376581	&0.9269(41)	&-1.8(5.7)	&2.9(5.6)	&1.31(14)	\\
&	0.0753162	&0.9222(46)	&-0.7(1.8)	&1.8(1.6)	&1.284(95)	\\
&	0.112974	&0.9148(51)	&-0.39(96)	&1.51(80)	&1.262(77)	\\
&	0.150632	&0.9043(56)	&-0.23(68)	&1.38(50)	&1.237(68)	\\
&	0.188291	&0.8911(62)	&-0.12(57)	&1.31(35)	&1.207(64)	\\
\hline
0.725&	0.0	&0.9323(43)	&$-$	&$-$	&$-$	\\
&	0.0376581	&0.9321(42)	&-1.9(5.9)	&2.9(5.8)	&1.31(14)	\\
&	0.0753162	&0.9274(48)	&-0.8(1.8)	&1.9(1.7)	&1.290(97)	\\
&	0.112974	&0.9199(52)	&-0.51(94)	&1.51(83)	&1.268(79)	\\
&	0.150632	&0.9093(58)	&-0.36(64)	&1.38(52)	&1.242(70)	\\
&	0.188291	&0.8960(64)	&-0.28(50)	&1.31(37)	&1.212(65)	\\
\hline
0.8&	0.0	&0.9387(44)	&$-$	&$-$	&$-$	\\
&	0.0376581	&0.9384(43)	&-2.0(6.0)	&3.0(6.0)	&1.32(15)	\\
&	0.0753162	&0.9337(49)	&-0.9(1.8)	&1.9(1.7)	&1.30(10)	\\
&	0.112974	&0.9261(54)	&-0.58(94)	&1.52(86)	&1.276(81)	\\
&	0.150632	&0.9155(60)	&-0.44(63)	&1.39(54)	&1.250(72)	\\
&	0.188291	&0.9021(66)	&-0.36(49)	&1.31(38)	&1.220(67)	\\
\hline
\end{tabular}
\end{table}
\begin{table}[H]
\centering
\caption{Lattice form factor results for set 5. $ak$ here is the value of the $x$ and $y$ components of the lattice momentum for the $D_s^*$. $ak$ is calculated from the corresponding twist in \cref{twists}.  \label{s_SM_set5}}
\begin{tabular}{ c c c c c c }
\hline
$am_h$& $ak$ & $h^s_{A_1}$& $h^s_{A_2}$& $h^s_{A_3}$& $h^s_{V}$\\\hline
0.427&	0.0	&0.898(22)	&$-$	&$-$	&$-$	\\
&	0.055399	&0.894(21)	&1.4(6.3)	&-0.3(6.0)	&1.24(15)	\\
&	0.110798	&0.874(21)	&0.1(1.9)	&0.9(1.5)	&1.22(11)	\\
&	0.166197	&0.838(24)	&0.002(1.409)	&1.01(80)	&1.16(11)	\\
&	0.221596	&0.790(31)	&0.05(1.65)	&0.97(61)	&1.09(11)	\\
&	0.276995	&0.734(42)	&0.2(3.0)	&0.90(53)	&1.01(13)	\\
\hline
0.525&	0.0	&0.903(23)	&$-$	&$-$	&$-$	\\
&	0.055399	&0.900(22)	&1.5(6.5)	&-0.5(6.3)	&1.24(15)	\\
&	0.110798	&0.880(22)	&0.06(1.86)	&0.8(1.6)	&1.22(11)	\\
&	0.166197	&0.843(25)	&-0.1(1.2)	&0.97(86)	&1.17(11)	\\
&	0.221596	&0.794(33)	&-0.1(1.1)	&0.94(65)	&1.10(12)	\\
&	0.276995	&0.737(44)	&-0.06(1.37)	&0.87(58)	&1.02(13)	\\
\hline
0.65&	0.0	&0.912(24)	&$-$	&$-$	&$-$	\\
&	0.055399	&0.909(23)	&1.8(6.9)	&-0.9(6.7)	&1.25(16)	\\
&	0.110798	&0.889(23)	&0.08(1.92)	&0.8(1.7)	&1.23(12)	\\
&	0.166197	&0.851(27)	&-0.1(1.2)	&0.93(94)	&1.18(11)	\\
&	0.221596	&0.801(35)	&-0.1(1.0)	&0.91(71)	&1.11(12)	\\
&	0.276995	&0.743(48)	&-0.1(1.1)	&0.84(63)	&1.03(14)	\\
\hline
0.8&	0.0	&0.926(26)	&$-$	&$-$	&$-$	\\
&	0.055399	&0.922(25)	&2.1(7.4)	&-1.3(7.3)	&1.27(16)	\\
&	0.110798	&0.901(25)	&0.1(2.0)	&0.7(1.9)	&1.25(12)	\\
&	0.166197	&0.862(29)	&-0.1(1.2)	&0.9(1.0)	&1.19(12)	\\
&	0.221596	&0.811(38)	&-0.1(1.0)	&0.88(78)	&1.12(13)	\\
&	0.276995	&0.752(52)	&-0.1(1.0)	&0.81(69)	&1.04(14)	\\
\hline
\end{tabular}
\end{table}
\begin{table}[H]
\centering
\caption{Lattice tensor form factor results for set 1. $ak$ here is the value of the $x$ and $y$ components of the lattice momentum for the $D_s^*$. $ak$ is calculated from the corresponding twist in \cref{twists}.  \label{s_T_set1}}
\begin{tabular}{ c c c c c c }
\hline
$am_h$& $ak$ & $h^s_{T_1}$& $h^s_{T_2}$& $h^s_{T_3}$&\\\hline
0.65&	0.0	&$-$	&$-$	&$-$	\\
&	0.0378853	&0.8654(76)	&-0.08(13)	&0.06(10.90)	\\
&	0.0757705	&0.8597(75)	&-0.083(82)	&-0.4(2.7)	\\
&	0.113656	&0.8506(79)	&-0.079(64)	&-0.3(1.3)	\\
&	0.151541	&0.8383(84)	&-0.075(56)	&-0.29(76)	\\
&	0.189426	&0.8231(93)	&-0.070(53)	&-0.25(53)	\\
\hline
0.725&	0.0	&$-$	&$-$	&$-$	\\
&	0.0378853	&0.8712(77)	&-0.09(13)	&0.08(10.96)	\\
&	0.0757705	&0.8654(76)	&-0.095(84)	&-0.4(2.7)	\\
&	0.113656	&0.8563(80)	&-0.091(65)	&-0.4(1.3)	\\
&	0.151541	&0.8440(85)	&-0.086(57)	&-0.30(77)	\\
&	0.189426	&0.8286(94)	&-0.081(54)	&-0.26(54)	\\
\hline
0.8&	0.0	&$-$	&$-$	&$-$	\\
&	0.0378853	&0.8778(78)	&-0.10(14)	&0.1(11.1)	\\
&	0.0757705	&0.8720(77)	&-0.106(86)	&-0.4(2.7)	\\
&	0.113656	&0.8629(81)	&-0.101(67)	&-0.4(1.3)	\\
&	0.151541	&0.8504(87)	&-0.097(58)	&-0.32(78)	\\
&	0.189426	&0.8349(95)	&-0.091(55)	&-0.27(55)	\\
\hline
\end{tabular}
\end{table}
\begin{table}[H]
\centering
\caption{Lattice tensor form factor results for set 2. $ak$ here is the value of the $x$ and $y$ components of the lattice momentum for the $D_s^*$. $ak$ is calculated from the corresponding twist in \cref{twists}.  \label{s_T_set2}}
\begin{tabular}{ c c c c c c }
\hline
$am_h$& $ak$ & $h^s_{T_1}$& $h^s_{T_2}$& $h^s_{T_3}$&\\\hline
0.427&	0.0	&$-$	&$-$	&$-$	\\
&	0.055399	&0.870(22)	&-0.09(14)	&2.3(7.7)	\\
&	0.110798	&0.845(25)	&-0.09(11)	&0.5(2.2)	\\
&	0.166197	&0.807(30)	&-0.07(11)	&0.2(1.2)	\\
&	0.221596	&0.763(37)	&-0.05(12)	&0.20(80)	\\
&	0.276995	&0.714(48)	&-0.05(14)	&0.19(65)	\\
\hline
0.525&	0.0	&$-$	&$-$	&$-$	\\
&	0.055399	&0.878(22)	&-0.11(14)	&2.5(7.7)	\\
&	0.110798	&0.852(25)	&-0.11(12)	&0.5(2.2)	\\
&	0.166197	&0.814(30)	&-0.09(12)	&0.3(1.2)	\\
&	0.221596	&0.769(38)	&-0.07(13)	&0.21(82)	\\
&	0.276995	&0.720(49)	&-0.06(14)	&0.20(67)	\\
\hline
0.65&	0.0	&$-$	&$-$	&$-$	\\
&	0.055399	&0.889(22)	&-0.13(15)	&2.5(7.7)	\\
&	0.110798	&0.863(25)	&-0.13(12)	&0.6(2.2)	\\
&	0.166197	&0.824(31)	&-0.11(12)	&0.3(1.2)	\\
&	0.221596	&0.778(39)	&-0.09(14)	&0.22(84)	\\
&	0.276995	&0.729(51)	&-0.08(15)	&0.21(69)	\\
\hline
0.8&	0.0	&$-$	&$-$	&$-$	\\
&	0.055399	&0.904(23)	&-0.16(16)	&2.6(7.9)	\\
&	0.110798	&0.877(26)	&-0.15(13)	&0.6(2.3)	\\
&	0.166197	&0.838(32)	&-0.13(13)	&0.3(1.2)	\\
&	0.221596	&0.791(41)	&-0.11(14)	&0.23(88)	\\
&	0.276995	&0.742(54)	&-0.10(16)	&0.22(73)	\\
\hline
\end{tabular}
\end{table}
\begin{table}[H]
\centering
\caption{Lattice tensor form factor results for set 3. $ak$ here is the value of the $x$ and $y$ components of the lattice momentum for the $D_s^*$. $ak$ is calculated from the corresponding twist in \cref{twists}.  \label{s_T_set3}}
\begin{tabular}{ c c c c c c }
\hline
$am_h$& $ak$ & $h^s_{T_1}$& $h^s_{T_2}$& $h^s_{T_3}$&\\\hline
0.5&	0.0	&$-$	&$-$	&$-$	\\
&	0.061831	&0.883(12)	&-0.135(70)	&-0.04(1.74)	\\
&	0.123662	&0.823(16)	&-0.120(56)	&-0.10(59)	\\
&	0.185493	&0.747(23)	&-0.072(79)	&0.02(36)	\\
&	0.247324	&0.670(41)	&0.04(15)	&0.11(36)	\\
&	0.309155	&0.628(92)	&0.42(39)	&0.36(51)	\\
\hline
0.65&	0.0	&$-$	&$-$	&$-$	\\
&	0.061831	&0.902(13)	&-0.159(75)	&-0.04(1.90)	\\
&	0.123662	&0.841(17)	&-0.140(60)	&-0.06(65)	\\
&	0.185493	&0.764(25)	&-0.091(84)	&0.06(41)	\\
&	0.247324	&0.688(45)	&0.02(16)	&0.16(40)	\\
&	0.309155	&0.645(97)	&0.39(40)	&0.42(55)	\\
\hline
0.8&	0.0	&$-$	&$-$	&$-$	\\
&	0.061831	&0.921(14)	&-0.176(81)	&-0.03(2.05)	\\
&	0.123662	&0.859(19)	&-0.154(65)	&-0.009(705)	\\
&	0.185493	&0.782(27)	&-0.102(89)	&0.11(45)	\\
&	0.247324	&0.710(49)	&0.008(167)	&0.24(45)	\\
&	0.309155	&0.67(10)	&0.38(42)	&0.53(59)	\\
\hline
\end{tabular}
\end{table}
\begin{table}[H]
\centering
\caption{Lattice tensor form factor results for set 4. $ak$ here is the value of the $x$ and $y$ components of the lattice momentum for the $D_s^*$. $ak$ is calculated from the corresponding twist in \cref{twists}.  \label{s_T_set4}}
\begin{tabular}{ c c c c c c }
\hline
$am_h$& $ak$ & $h^s_{T_1}$& $h^s_{T_2}$& $h^s_{T_3}$&\\\hline
0.65&	0.0	&$-$	&$-$	&$-$	\\
&	0.0376581	&0.8644(69)	&-0.10(12)	&2.9(9.2)	\\
&	0.0753162	&0.8586(68)	&-0.102(72)	&0.7(2.2)	\\
&	0.112974	&0.8498(70)	&-0.099(55)	&0.3(1.1)	\\
&	0.150632	&0.8377(74)	&-0.096(48)	&0.10(63)	\\
&	0.188291	&0.8226(79)	&-0.093(45)	&0.02(44)	\\
\hline
0.725&	0.0	&$-$	&$-$	&$-$	\\
&	0.0376581	&0.8703(70)	&-0.11(12)	&2.9(9.3)	\\
&	0.0753162	&0.8644(69)	&-0.112(73)	&0.8(2.3)	\\
&	0.112974	&0.8556(71)	&-0.109(57)	&0.3(1.1)	\\
&	0.150632	&0.8434(75)	&-0.107(50)	&0.10(64)	\\
&	0.188291	&0.8282(80)	&-0.104(46)	&0.03(44)	\\
\hline
0.8&	0.0	&$-$	&$-$	&$-$	\\
&	0.0376581	&0.8770(71)	&-0.12(12)	&2.9(9.5)	\\
&	0.0753162	&0.8711(70)	&-0.122(75)	&0.8(2.3)	\\
&	0.112974	&0.8622(73)	&-0.119(58)	&0.3(1.1)	\\
&	0.150632	&0.8499(76)	&-0.116(51)	&0.11(65)	\\
&	0.188291	&0.8346(82)	&-0.113(48)	&0.03(45)	\\
\hline
\end{tabular}
\end{table}
\begin{table}[H]
\centering
\caption{Lattice tensor form factor results for set 5. $ak$ here is the value of the $x$ and $y$ components of the lattice momentum for the $D_s^*$. $ak$ is calculated from the corresponding twist in \cref{twists}.  \label{s_T_set5}}
\begin{tabular}{ c c c c c c }
\hline
$am_h$& $ak$ & $h^s_{T_1}$& $h^s_{T_2}$& $h^s_{T_3}$&\\\hline
0.427&	0.0	&$-$	&$-$	&$-$	\\
&	0.055399	&0.858(24)	&-0.08(16)	&-1.7(7.6)	\\
&	0.110798	&0.835(25)	&-0.09(13)	&-0.3(2.0)	\\
&	0.166197	&0.794(30)	&-0.09(12)	&-0.2(1.0)	\\
&	0.221596	&0.743(39)	&-0.08(13)	&-0.13(76)	\\
&	0.276995	&0.686(53)	&-0.08(15)	&-0.10(64)	\\
\hline
0.525&	0.0	&$-$	&$-$	&$-$	\\
&	0.055399	&0.866(25)	&-0.09(17)	&-1.9(7.7)	\\
&	0.110798	&0.843(26)	&-0.10(13)	&-0.3(2.1)	\\
&	0.166197	&0.802(31)	&-0.10(12)	&-0.2(1.1)	\\
&	0.221596	&0.751(40)	&-0.09(13)	&-0.12(79)	\\
&	0.276995	&0.693(55)	&-0.09(15)	&-0.10(67)	\\
\hline
0.65&	0.0	&$-$	&$-$	&$-$	\\
&	0.055399	&0.878(26)	&-0.11(17)	&-2.1(8.0)	\\
&	0.110798	&0.854(27)	&-0.12(14)	&-0.4(2.1)	\\
&	0.166197	&0.813(32)	&-0.12(13)	&-0.2(1.1)	\\
&	0.221596	&0.761(42)	&-0.11(14)	&-0.12(83)	\\
&	0.276995	&0.702(57)	&-0.10(16)	&-0.09(70)	\\
\hline
0.8&	0.0	&$-$	&$-$	&$-$	\\
&	0.055399	&0.893(27)	&-0.14(18)	&-2.4(8.3)	\\
&	0.110798	&0.869(28)	&-0.14(15)	&-0.4(2.2)	\\
&	0.166197	&0.827(34)	&-0.13(14)	&-0.2(1.2)	\\
&	0.221596	&0.774(45)	&-0.12(15)	&-0.11(89)	\\
&	0.276995	&0.714(61)	&-0.11(17)	&-0.09(76)	\\
\hline
\end{tabular}
\end{table}

\subsection{Error Band Plots}
\begin{figure}[H]
\includegraphics[scale=0.5]{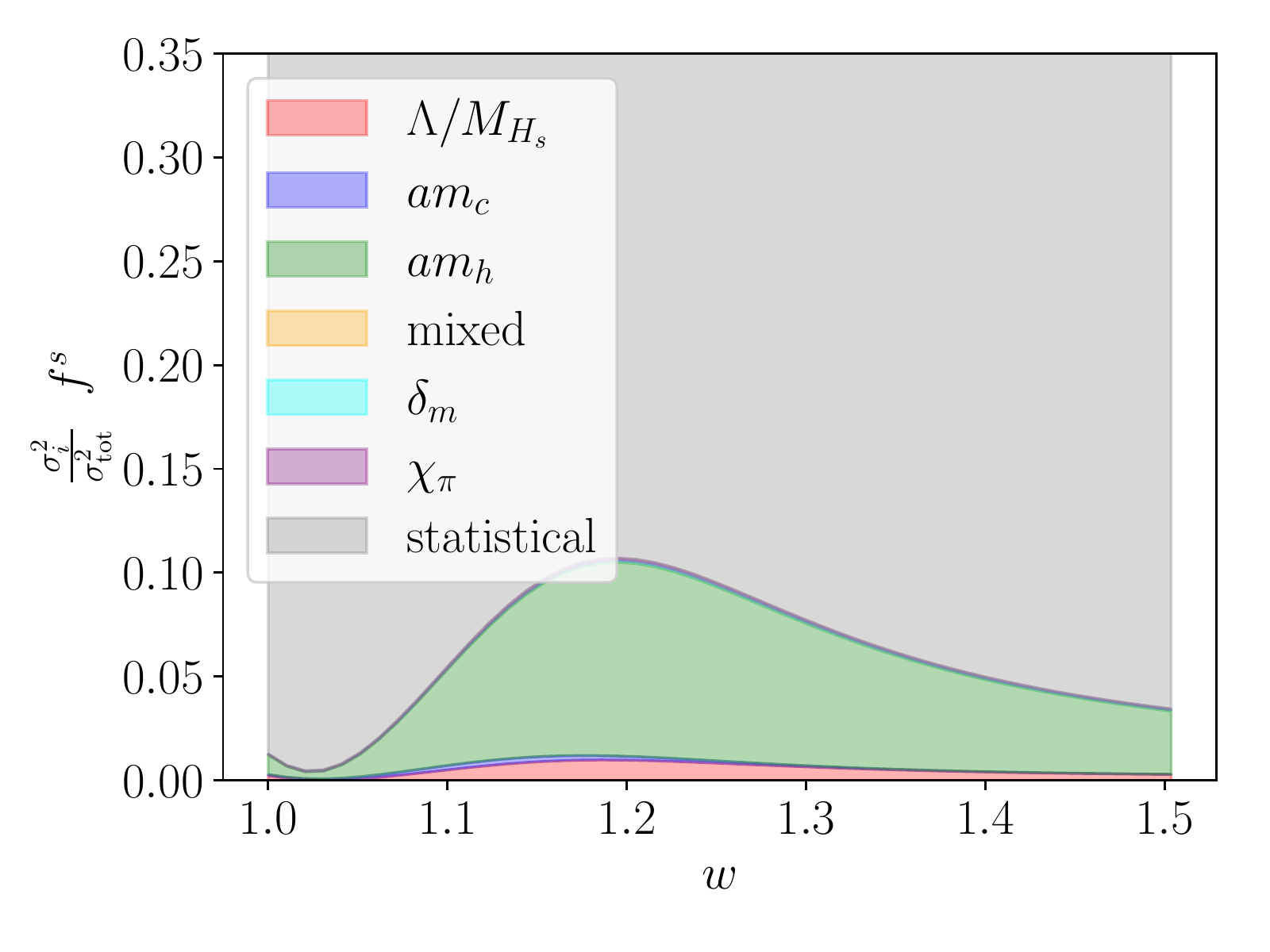}\\
\includegraphics[scale=0.5]{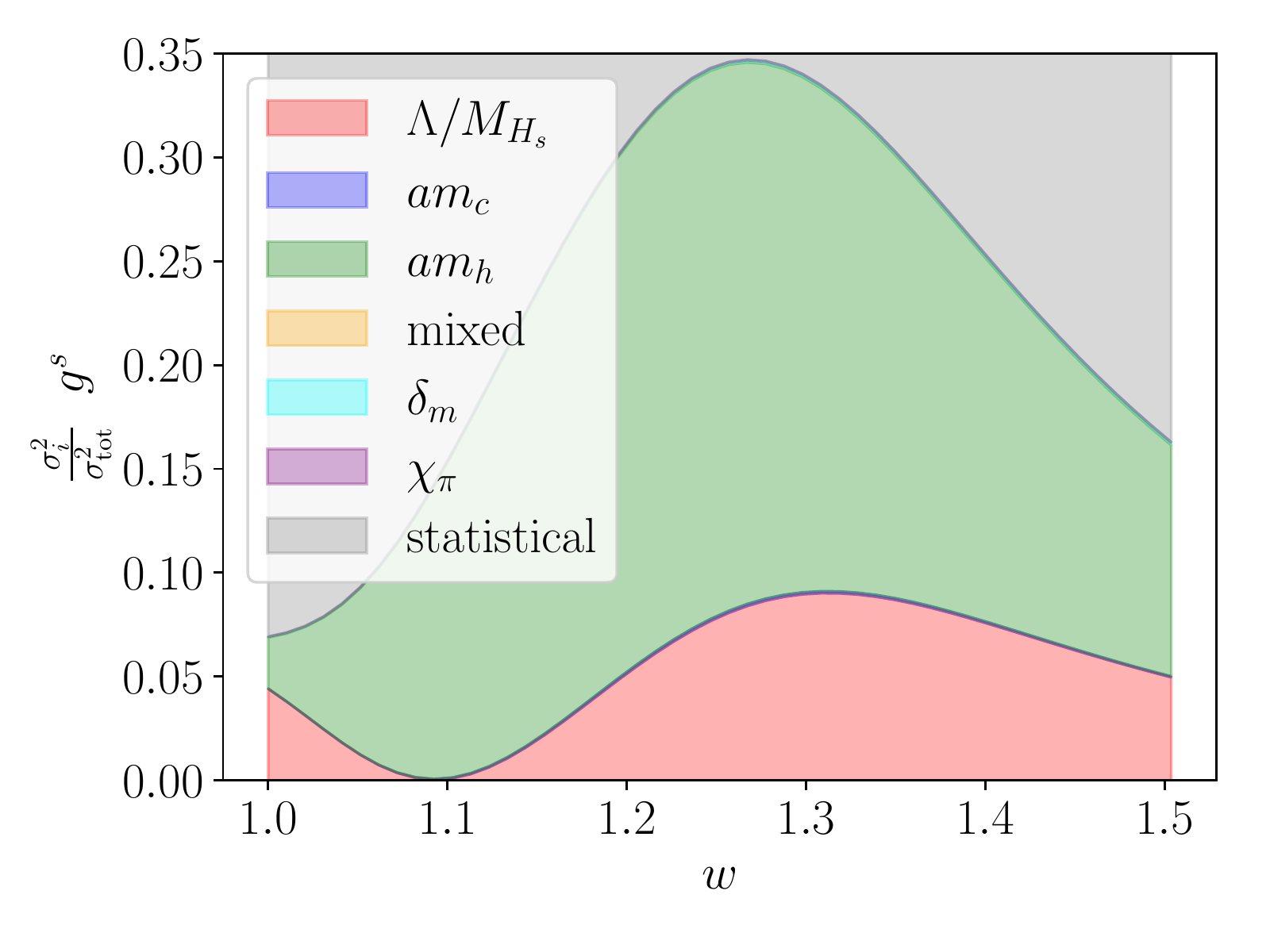}
\caption{\label{errorbandsfgs}Plots showing the fractional contribution of each source of uncertainty to the total variance for the $B_s\to D_s^*$ form factors $f^s$ and $g^s$ across the full kinematic range. The vertical axis is truncated at $0.35$ for clarity, with the remaining variance between $0.35$ and $1$ attributable to statistics.}
\end{figure}

\begin{figure}[H]
\includegraphics[scale=0.5]{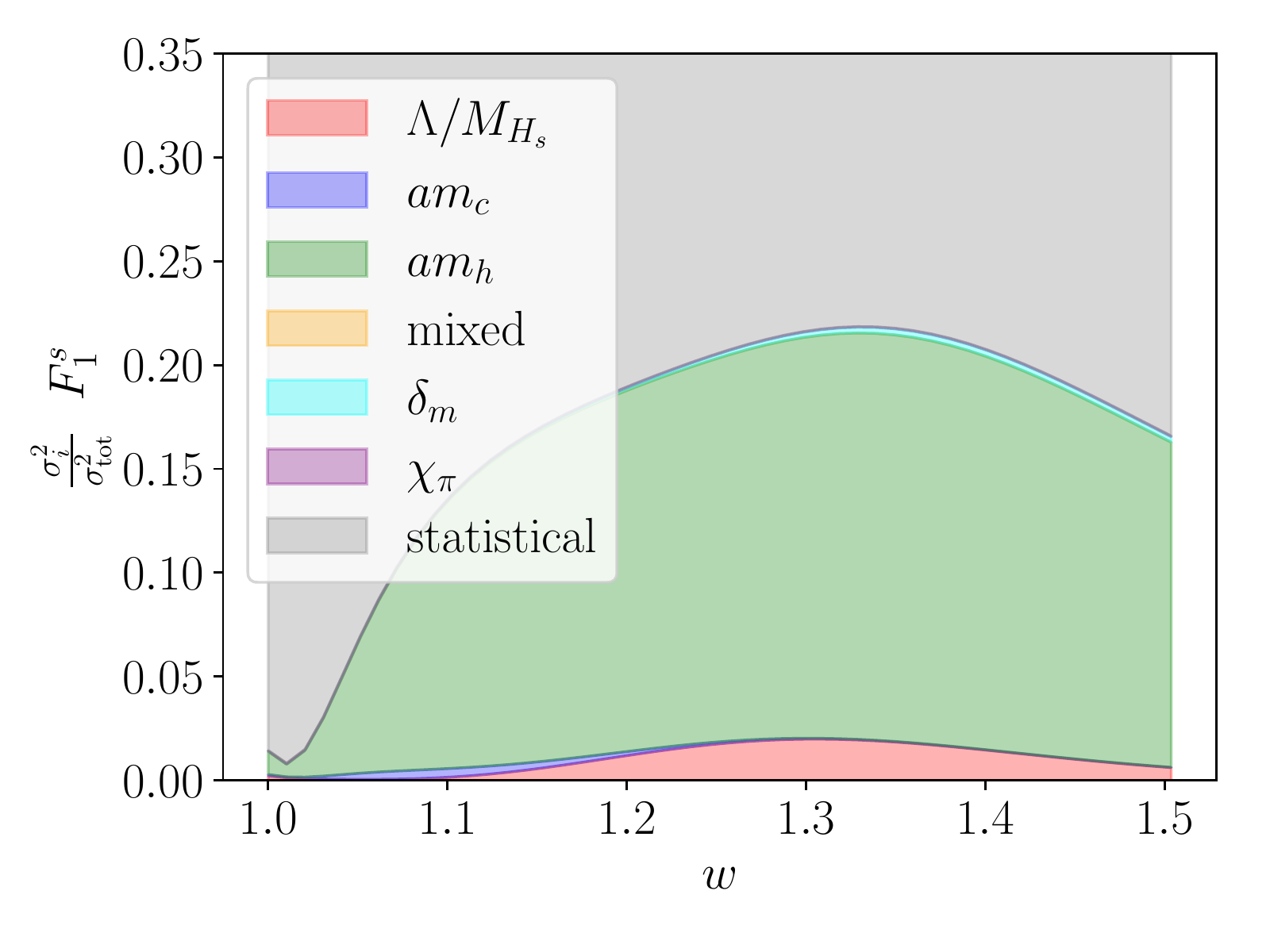}\\
\includegraphics[scale=0.5]{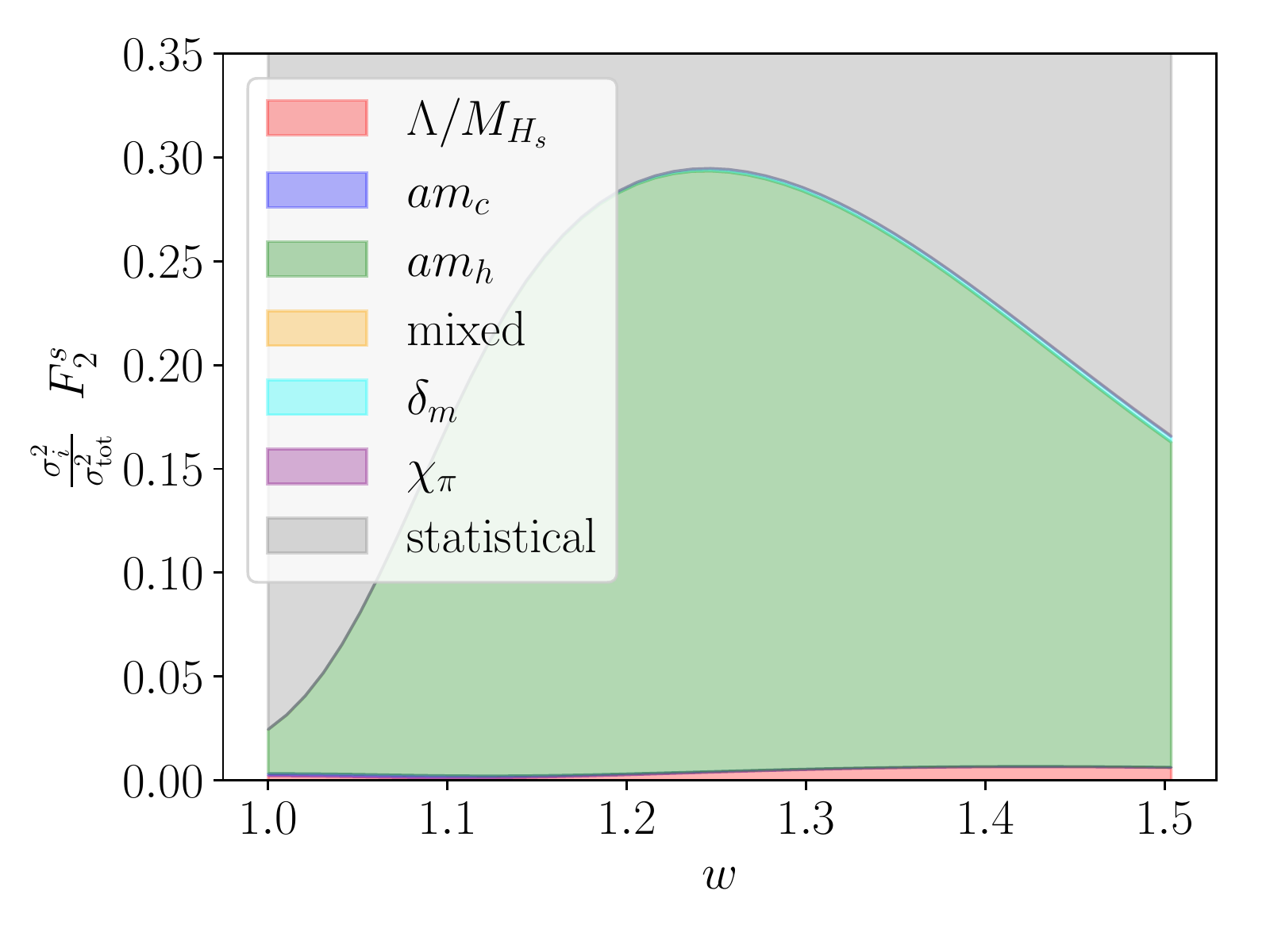}
\caption{\label{errorbandsF1F2s}Plots showing the fractional contribution of each source of uncertainty to the total variance for the $B_s\to D^*_s$ form factors $F_1^s$ and $F_2^s$ across the full kinematic range. The vertical axis is truncated at $0.35$ for clarity, with the remaining variance between $0.35$ and $1$ attributable to statistics.}
\end{figure}

\begin{figure}[H]
\includegraphics[scale=0.5]{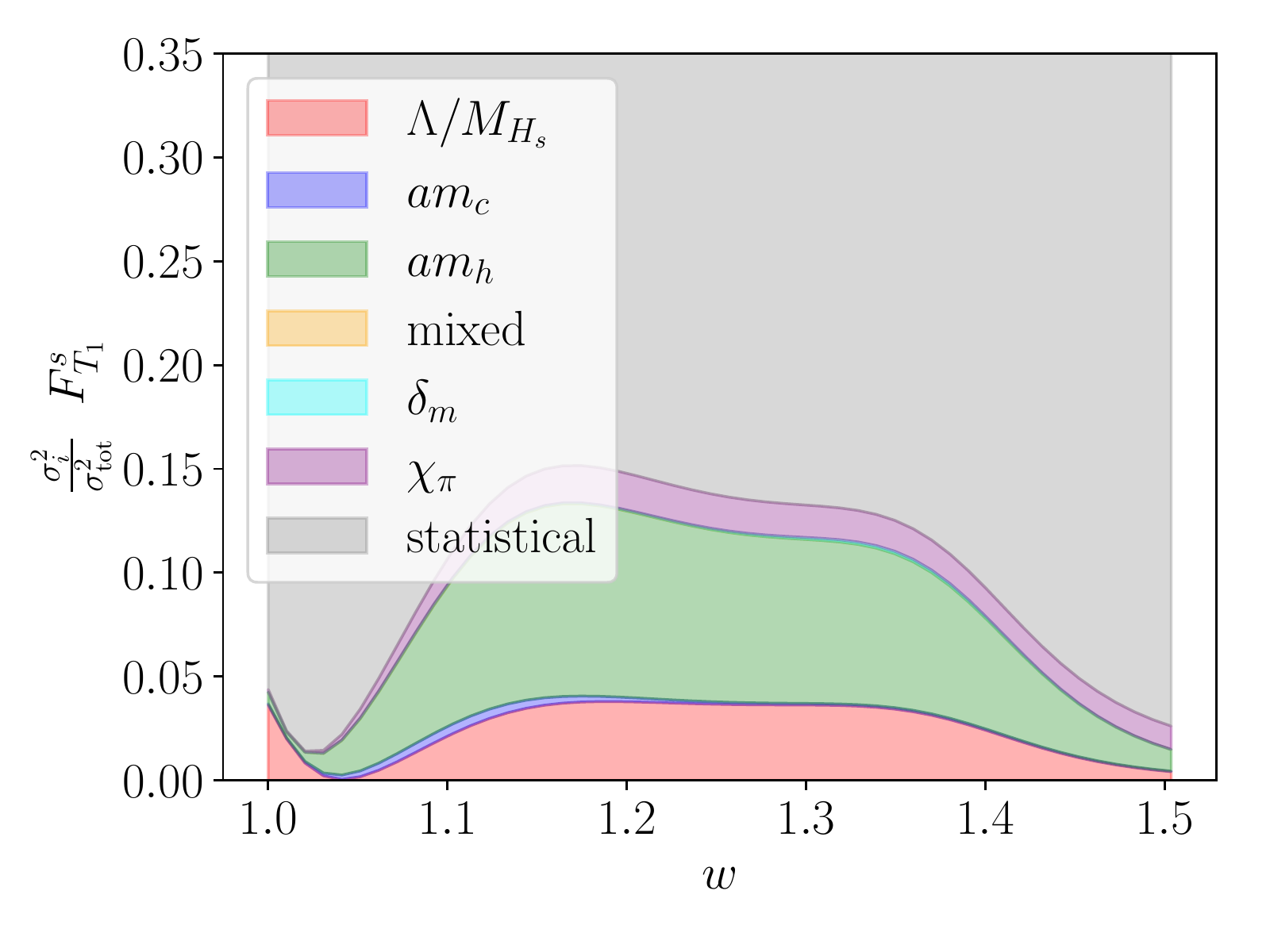}\\
\includegraphics[scale=0.5]{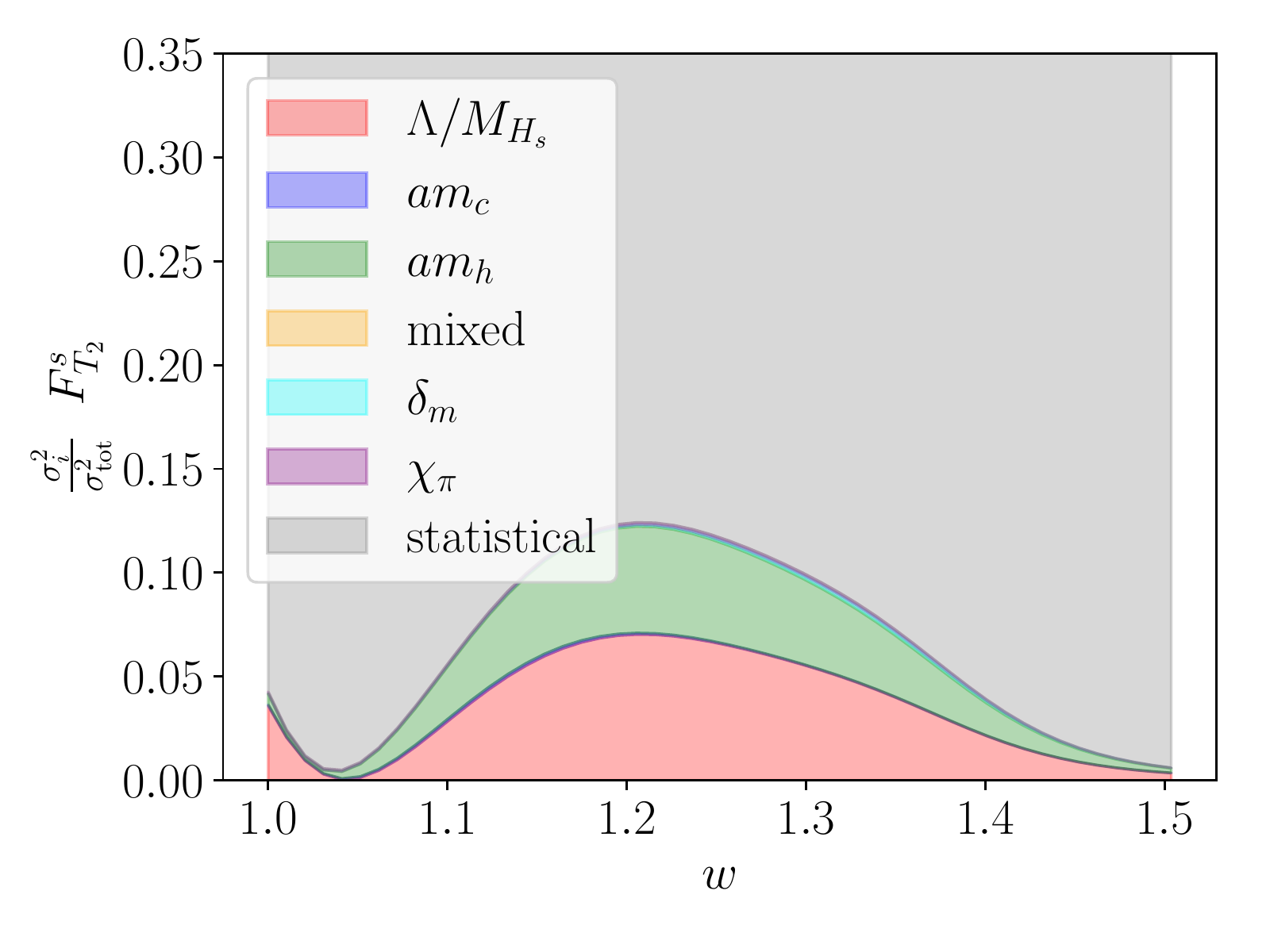}\\
\includegraphics[scale=0.5]{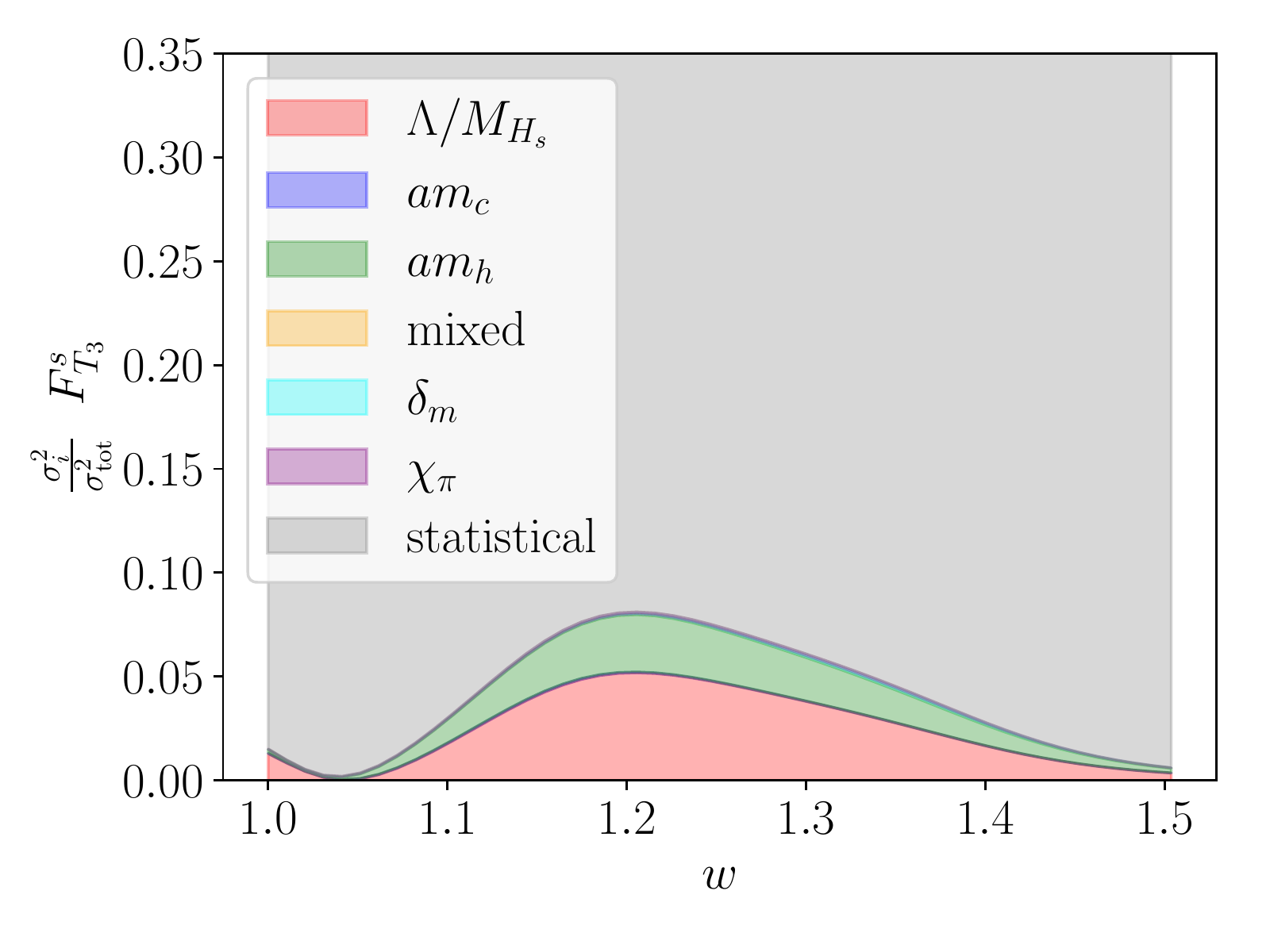}
\caption{\label{errorbandsFT1FT2FT3s}Plots showing the fractional contribution of each source of uncertainty to the total variance for the $B_s\to D^*_s$ tensor form factors in the helicity basis defined in~\cref{helicitybasisT}, across the full kinematic range. The vertical axis is truncated at $0.35$ for clarity, with the remaining variance between $0.35$ and $1$ attributable to statistics. Note the large contribution of the unconstrained chiral dependence entering $F_{T_1}$ that originates from $h_{T_3}$.}
\end{figure}

Here we show plots for the fractional contribution of each source of uncertainty to the total variance for the $B\to D^*$ form factors $F_1$ and $F_2$, as well as the tensor form factors in the helicity basis defined in~\cref{helicitybasisT}~(plots for $g$ and $f$ for $B\to D^*$ are given in~\cref{errorbandsfg} in the main text). Plots for the full set of $B_s\to D_s^*$ form factors are given in~\cref{errorbandsfgs,errorbandsF1F2s,errorbandsFT1FT2FT3s}.
\begin{figure}[H]
\includegraphics[scale=0.5]{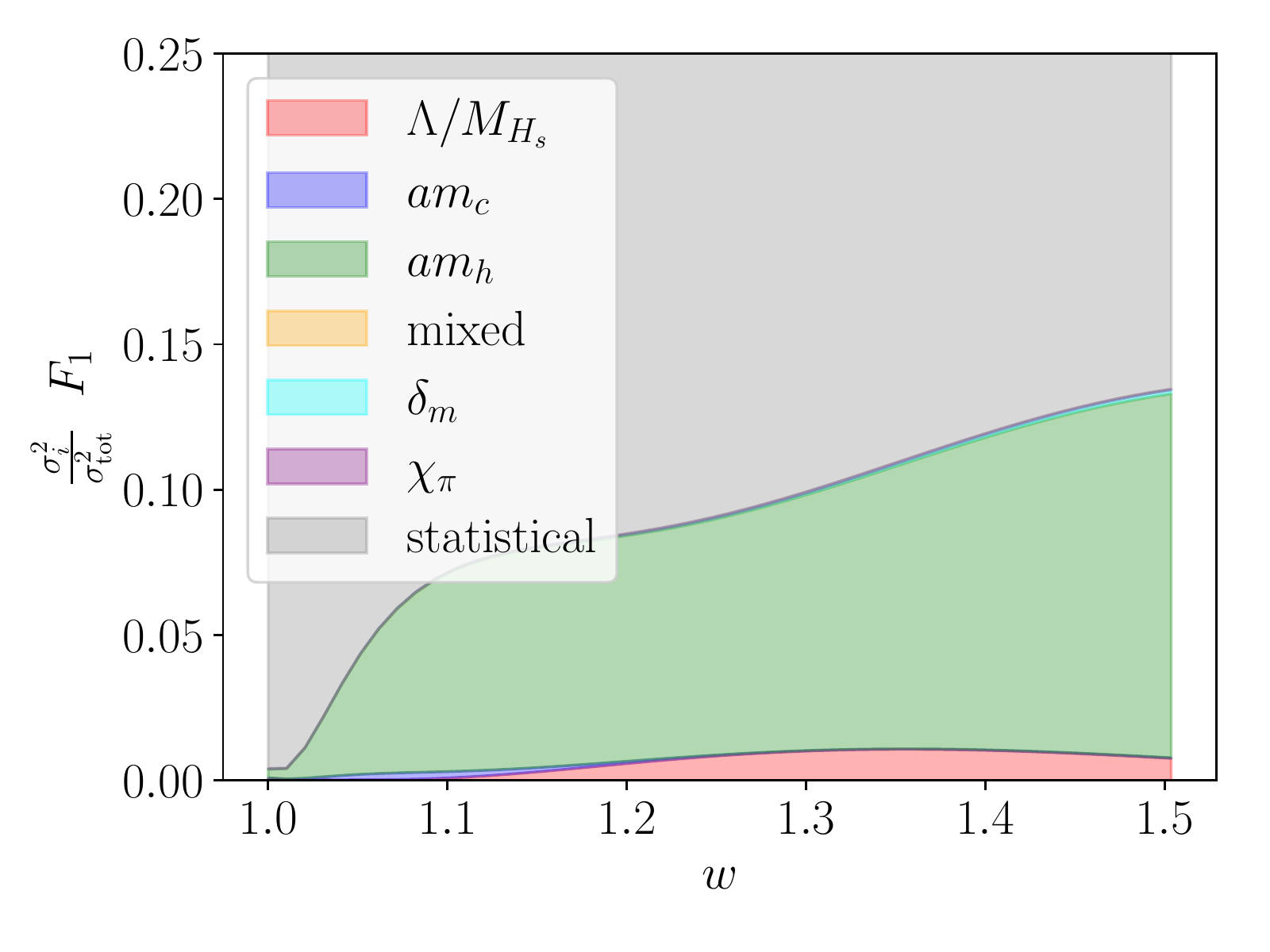}\\
\includegraphics[scale=0.5]{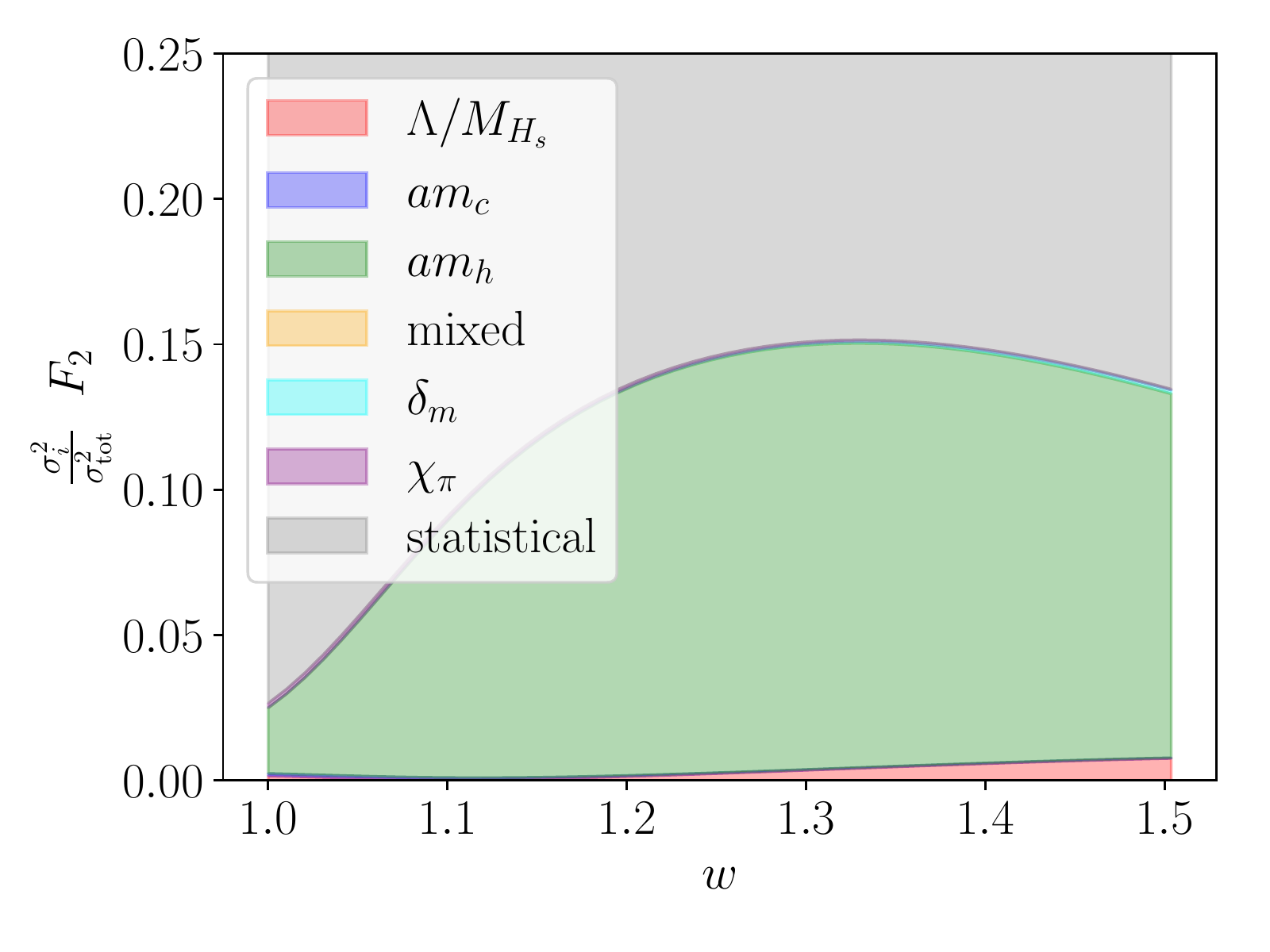}
\caption{\label{errorbandsF1F2}Plots showing the fractional contribution of each source of uncertainty to the total variance for the $B\to D^*$ form factors $F_1$ and $F_2$ across the full kinematic range. The vertical axis is truncated at $0.25$ for clarity, with the remaining variance between $0.25$ and $1$ attributable to statistics.}
\end{figure}

\begin{figure}[H]
\includegraphics[scale=0.5]{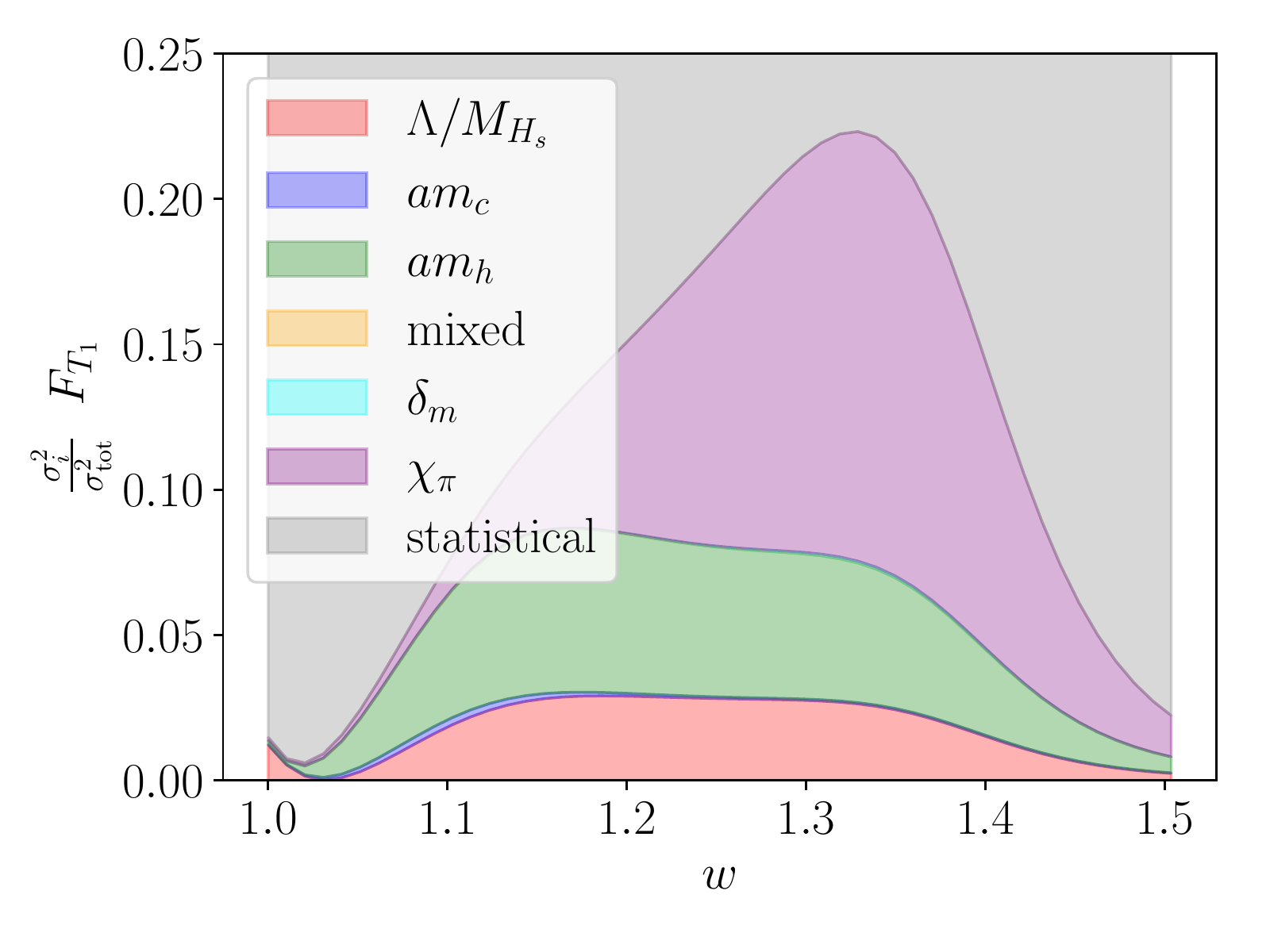}\\
\includegraphics[scale=0.5]{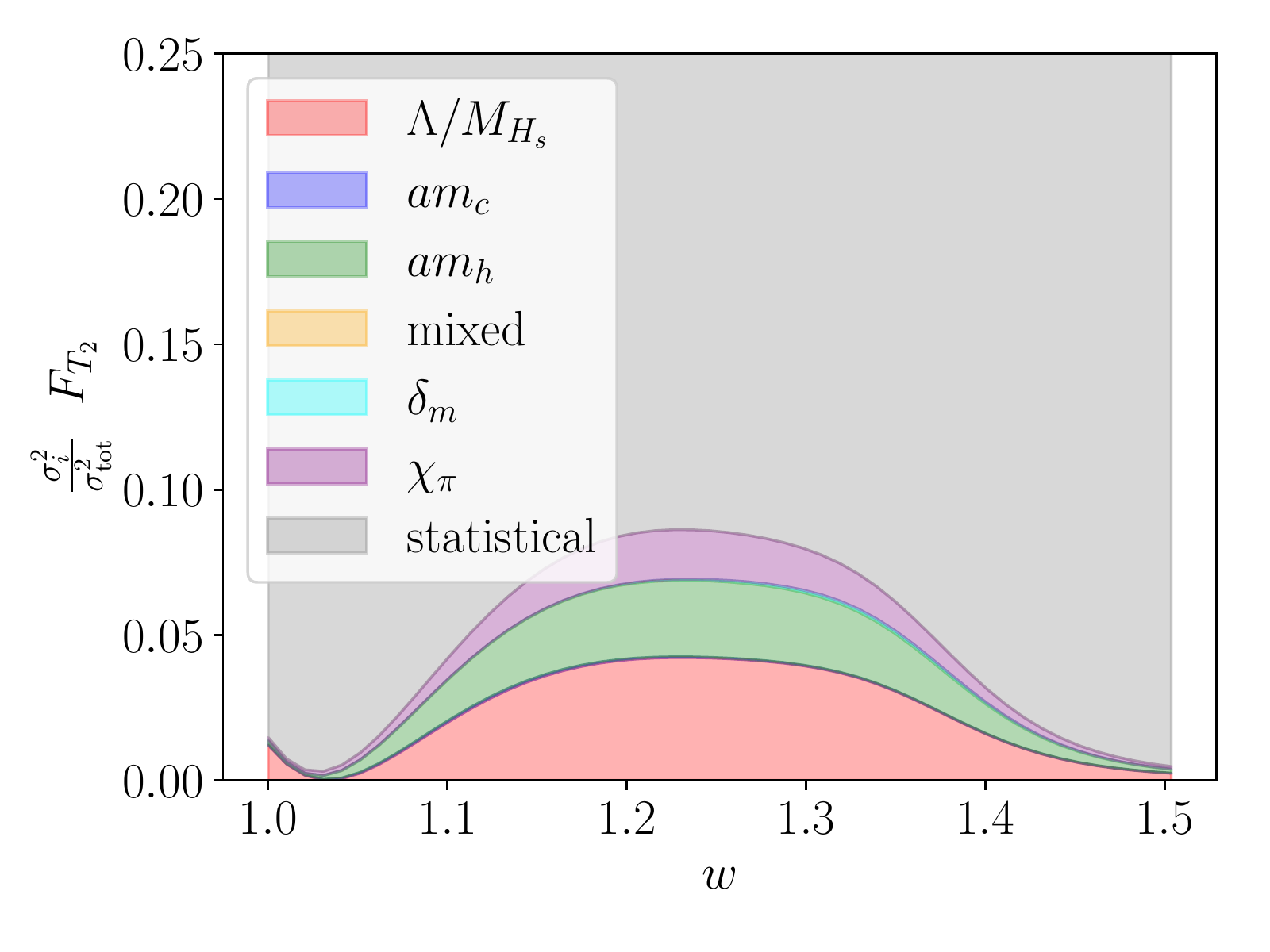}\\
\includegraphics[scale=0.5]{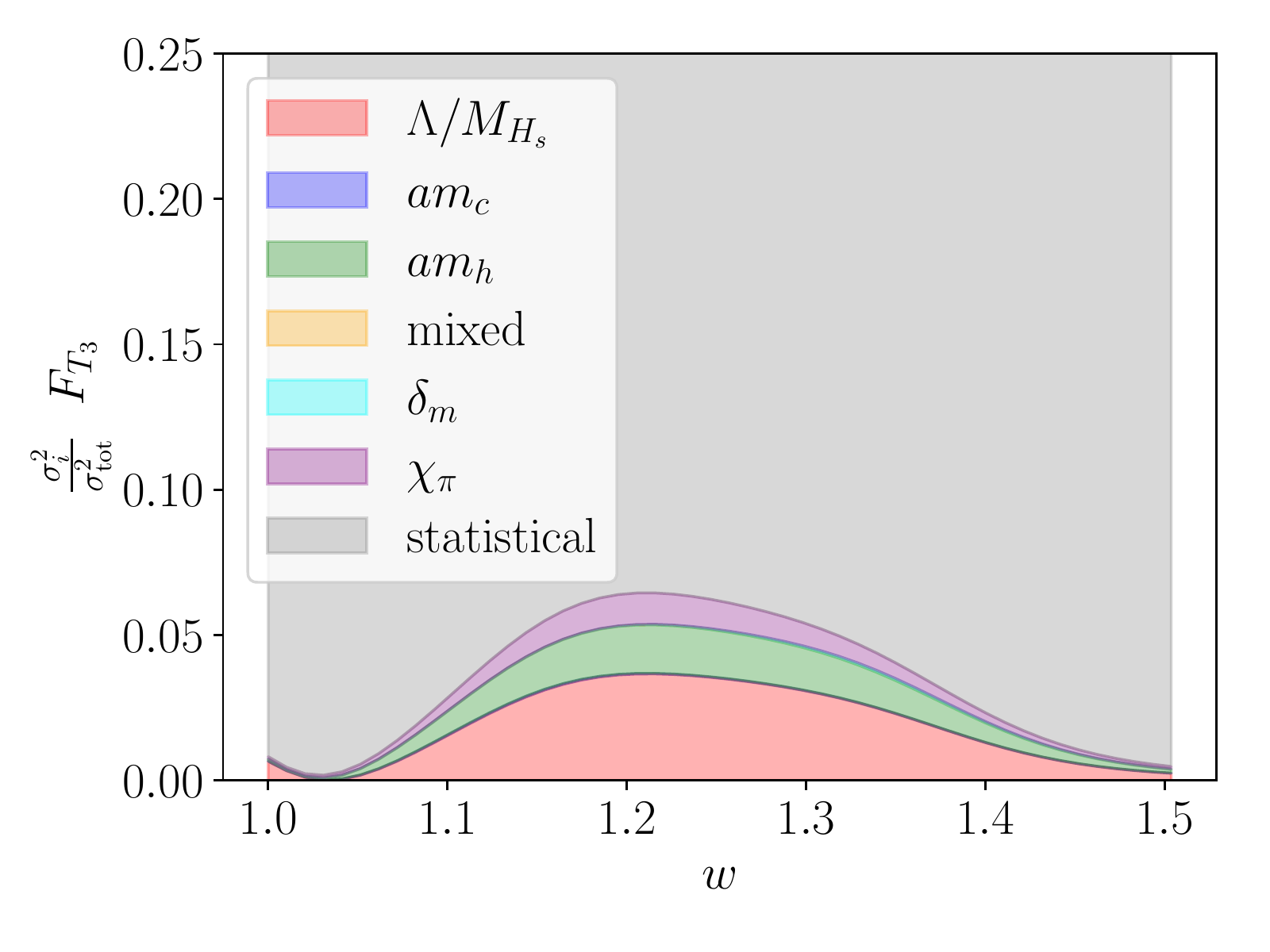}
\caption{\label{errorbandsFT1FT2FT3}Plots showing the fractional contribution of each source of uncertainty to the total variance for the $B\to D^*$ tensor form factors in the helicity basis defined in~\cref{helicitybasisT}, across the full kinematic range. The vertical axis is truncated at $0.25$ for clarity, with the remaining variance between $0.25$ and $1$ attributable to statistics. Note the large contribution of the unconstrained chiral dependence entering $F_{T_1}$ that originates from $h_{T_3}$.}
\end{figure}

\section{Staggered Chiral Perturbation Theory}
\label{chilogs}
To compute the chiral logarithms for the $B\to D^{(*})$ tensor form factors we use heavy-meson chiral perturbation theory, modified to account for the multiple tastes present when using staggered quarks~\cite{Aubin:2005aq}. The heavy meson fields are given by
\begin{align}
H_a=\frac{1+\slashed{v}}{2}\left[\gamma^\mu B^*_{a\mu} +i\gamma^5 B_a\right]\nonumber\\
\overline{H}_a=\gamma_0 H_a^\dagger \gamma_0 = \left[\gamma^\mu B^{*\dagger}_{a\mu} +i\gamma^5 B_a^\dagger\right]\frac{1+\slashed{v}}{2}
\end{align}
where $a$ labels taste and flavour. The pion fields are
\begin{align}
\Sigma = \mathrm{exp}\left(i\Phi/f\right),
\end{align}
where
\begin{align}
\Phi_{ab} = \Phi_{i\alpha,j\beta} = \pi_{ij,\Xi}\tilde{T}^\Xi_{\alpha\beta},
\end{align}
with $\Xi$ labelling the taste of the pion and the $SU(4)$ taste generators $\tilde{T}^\Xi=\{\xi_5,i\xi_{\mu 5},i\xi_{\mu\nu},\xi_\mu,\xi_I\}$. $\xi_\mu$ are the Euclidean gamma matrices, with $\xi_I = 1, \xi_{\mu\nu} = \frac{1}{2}[\xi_\mu,\xi_\nu]$ and $\xi_{\mu 5}=\xi_\mu\xi_5$.

The leading order Minkowski staggered chiral Lagrangian for 3 flavours of light quarks is given by~\cite{Aubin:2005aq}, including heavy quarks,
\begin{align}\label{lagrangian}
\mathcal{L}_\Sigma =& \frac{f^2}{8}\mathrm{STr}\left[\partial_\mu \Sigma \partial^\mu \Sigma^\dagger\right] + \frac{1}{4}\mu f^2 \mathrm{STr}\left[\mathcal{M}\Sigma + \mathcal{M}\Sigma^\dagger\right] \nonumber\\
&- \frac{2m_0}{3}(U_I+D_I+S_I)^2-a^2\mathcal{V}\nonumber\\
&-i\mathrm{tr}\left[\overline{H}_av^\mu \partial_\mu H_a\right]+\mathrm{tr}\left[\overline{H}_a H_b\right]v^\mu V^{ba}_\mu\nonumber\\
&+g_\pi \mathrm{tr}\left[\overline{H}_aH_b\gamma^\nu\gamma^5\right]A_{\nu}^{ba}\nonumber\\
&+ \frac{\lambda_2}{m_Q}\mathrm{tr}\left[\overline{H}_a \sigma^{\mu\nu} H_a \sigma_{\mu\nu}\right]
\end{align}
where $U_I$, $D_I$ and $S_I$ are the diagonal elements of $\Phi$. We use these rather than the physical basis in order to simplify the quark flow analysis. We will take $m_0\to \infty$ at the end. We use `tr' to indicate a trace over dirac indices, and `STr' to indicate a trace over SU(4n) indices. The final term generates a mass splitting for the $D^*$ and $D$, $\Delta^c = (m_{D^*}-m_D)=-\lambda_2/{8m_c}$. $\mathcal{V}$ contains operators that generate the taste splittings, as well as operators that mix the taste-(axial-)vector, flavour neutral mesons. In~\cref{lagrangian}, $A_\mu$ and $V_\mu$ are constructed from the pion fields and couple to the heavy-meson fields. They are given by
\begin{align}
V_\mu = \frac{i}{2}\left[\sigma^\dagger\partial_\mu\sigma + \sigma\partial_\mu\sigma^\dagger\right],\nonumber\\
A_\mu = \frac{i}{2}\left[\sigma^\dagger\partial_\mu\sigma - \sigma\partial_\mu\sigma^\dagger\right],
\end{align}
where $\sigma=\sqrt{\Sigma}$. At first order in the pion fields, these are
\begin{align}
V^{i\alpha,j\beta}_\mu &= 0 +\mathcal{O}(\pi^2),\nonumber\\
A^{i\alpha,j\beta}_\mu &= -\frac{1}{2f}\partial_\mu \pi_{ij,\Xi}\tilde{T}^\Xi_{\alpha\beta}.\nonumber.
\end{align}
Expanding~\cref{lagrangian} to first order in the pion fields we find
\begin{align}\label{lagrangian1st}
\mathcal{L}^1_\Sigma =& \frac{1}{2}\partial_\mu \pi_{ij,\Xi} \partial^\mu \pi_{ji,\Xi} + \frac{1}{2}M^2_{ij,\Xi}\pi_{ij,\Xi}\pi_{ji,\Xi} \nonumber\\
&- \frac{2m_0}{3}(U_I+D_I+S_I)^2-a^2\mathcal{V}'\nonumber\\
&-i\mathrm{tr}\left[\overline{H}_av^\mu \partial_\mu H_a\right]\nonumber\\
&-g_\pi \mathrm{tr}\left[\overline{H}_{j\beta}H_{i\alpha}\gamma^\nu\gamma^5\right]\frac{1}{2f}\partial_\mu \pi_{ij,\Xi}\tilde{T}^\Xi_{\alpha\beta}\nonumber\\
&+ \frac{\lambda_2}{m_Q}\mathrm{tr}\left[\overline{H}_a \sigma^{\mu\nu} H_a \sigma_{\mu\nu}\right]
\end{align}
where $M^2_{ij,\Xi} = \mu(m_i+m_j) + a^2 \Delta_\Xi$ and $a^2\mathcal{V}'$ contains the remaining hairpin vertices mixing flavor neutral taste vector and axial-vector pions. The pion propagator for flavour non-neutral pions is then:
\begin{align}
\{\pi_{ij,\Xi}\pi_{j'i',\Xi'}\}_\mathrm{con} = \frac{i\delta_{ii'}\delta_{jj'}\delta_{\Xi\Xi'}}{p^2-M^2_{ij,\Xi}+i\varepsilon}.
\end{align}
For vector, axial-vector and singlet taste, flavour-neutral pions there is an additional disconnected hairpin contribution. In Minkowski space this is given by~\cite{Aubin:2005aq} 
\begin{align}
\{\pi_{ij,\Xi}\pi_{j'i',\Xi'}\}_\mathrm{disc} = {\delta_{ij}\delta_{j'i'}\delta_{\Xi\Xi'}}\mathcal{D}^\Xi_{ii,i'i'}
\end{align}
where
\begin{widetext}
\begin{align}
\mathcal{D}^\Xi_{ii,i'i'}=ia^2\delta'_\Xi\frac{(p^2-m^2_{U{\Xi}})(p^2-m^2_{D{\Xi}})(p^2-m^2_{S{\Xi}})}{(p^2-m^2_{ii{\Xi}})(p^2-m^2_{i'i'{\Xi}})(p^2-m^2_{\pi^0{\Xi}})(p^2-m^2_{\eta{\Xi}})(p^2-m^2_{\eta'{\Xi}})}
\end{align}
\end{widetext}
such that
\begin{align}
\{\pi_{ij,\Xi}\pi_{j'i',\Xi'}\} = \frac{i\delta_{ii'}\delta_{jj'}\delta_{\Xi\Xi'}}{p^2-M^2_{ij,\Xi}+i\varepsilon}+{\delta_{ij}\delta_{j'i'}\delta_{\Xi\Xi'}}\mathcal{D}^\Xi_{ii,i'i'}.
\end{align}
For $m_u=m_d$ relevant to 2+1+1 simulations for $B\to D^*$ where the spectator quark is either a $u$ or a $d$, the flavour neutral disconnected propagator is
\begin{widetext}
\begin{align}
\mathcal{D}^\Xi_{ii,i'i'}&=ia^2\delta'_\Xi\frac{(p^2-m^2_{S{\Xi}})}{(p^2-m^2_{\pi^0{\Xi}})(p^2-m^2_{\eta{\Xi}})(p^2-m^2_{\eta'{\Xi}})}\nonumber\\
&=ia^2\delta'_\Xi\frac{m^2_{\pi^0{\Xi}}-m^2_{S{\Xi}}}{(m^2_{\pi^0{\Xi}}-m^2_{\eta{\Xi}})(m^2_{\pi^0{\Xi}}-m^2_{\eta'{\Xi}})}\frac{1}{p^2-m^2_{\pi^0{\Xi}}}\nonumber\\
&+ia^2\delta'_\Xi\frac{m^2_{\eta{\Xi}}-m^2_{S{\Xi}}}{(m^2_{\eta{\Xi}}-m^2_{\pi^0{\Xi}})(m^2_{\eta{\Xi}}-m^2_{\eta'{\Xi}})}\frac{1}{p^2-m^2_{\eta{\Xi}}}\nonumber\\
&+ia^2\delta'_\Xi\frac{m^2_{\eta'{\Xi}}-m^2_{S{\Xi}}}{(m^2_{\eta'{\Xi}}-m^2_{\pi^0{\Xi}})(m^2_{\eta'{\Xi}}-m^2_{\eta{\Xi}})}\frac{1}{p^2-m^2_{\eta'{\Xi}}}\nonumber\\
&=ia^2\delta'_\Xi\left[A_\Xi\frac{1}{p^2-m^2_{\pi^0{\Xi}}+i\varepsilon}+B_\Xi\frac{1}{p^2-m^2_{\eta{\Xi}}+i\varepsilon}+C_\Xi\frac{1}{p^2-m^2_{\eta'{\Xi}}+i\varepsilon}\right]\label{hairpinprop}
\end{align}
\end{widetext}
for $i,i'=u,d$. For $B_s\to D_s^*$ we are interested in the case $i,i'=s$. In this case, using the fact that $M_{\pi_\Xi}=M_{U_\Xi}=M_{D_\Xi}$, we just swap $M_{\pi_\Xi}\leftrightarrow M_{S_\Xi}$ in~\cref{hairpinprop}. We write the pion propagator as
\begin{align}\label{staggeredpiprop}
\{\pi_{ij,\Xi}\pi_{j'i',\Xi'}\} = \delta_{\Xi\Xi'}\sum_n P_{ii'jj'}^{\Xi,n}\frac{i}{p^2-M^2_{ij,\Xi, n}+i\varepsilon}
\end{align}
where 
\begin{align}
M^2_{ij,\Xi, n}=
\begin{pmatrix}
M^2_{ij,\Xi}\\
m^2_{\pi^0{\Xi}}\\
m^2_{\eta{\Xi}}\\
m^2_{\eta'{\Xi}}
\end{pmatrix}_n
\end{align}
and 
\begin{align}
P_{ii'jj'}^{\Xi,n} =
\begin{pmatrix}
\delta_{ii'}\delta_{jj'}\\
a^2\delta'_\Xi A_\Xi \delta_{ij}\delta_{j'i'}\\
a^2\delta'_\Xi B_\Xi \delta_{ij}\delta_{j'i'}\\
a^2\delta'_\Xi C_\Xi \delta_{ij}\delta_{j'i'}
\end{pmatrix}_n.
\end{align}
Here $\delta'_V,\delta'_A$ are the parameters determined from $\mathcal{V}'$, $a^2\delta'_I=4m_0^2/3$ and $\delta'_5=\delta'_T=0$. 

The heavy meson propagators for the $B$, $B^*$, $D$ and $D^*$ are given by
\begin{align}
\{B_aB_b^\dagger\} &=&\{B_{i\alpha}B_{j\beta}^\dagger\} &=& \frac{i\delta_{ij}\delta_{\alpha\beta}}{2(v\cdot k +i\varepsilon)}\nonumber\\
\{B^*_{a,\mu}B_{b,\nu}^{*\dagger}\}&=&\{B^*_{i\alpha,\mu}B_{{j\beta},\nu}^{*\dagger}\} &=& -\frac{i\delta_{ij}\delta_{\alpha\beta}(g_{\mu\nu}-v_\mu v_\nu)}{2(v\cdot k -\Delta^b+i\varepsilon)}\nonumber\\
\{D_aD_b^\dagger\}&=&\{D_{i\alpha}D_{j\beta}^\dagger\} &=& \frac{i\delta_{ij}\delta_{\alpha\beta}}{2(v\cdot k +i\varepsilon)}\nonumber\\
\{D^*_{a,\mu}D_{b,\mu}^{*\dagger}\}&=&\{D^*_{i\alpha,\mu}D_{j\beta,\nu}^{*\dagger}\} &=& -\frac{i\delta_{ij}\delta_{\alpha\beta}(g_{\mu\nu}-v_\mu v_\nu)}{2(v\cdot k -\Delta^c+i\varepsilon)}.\nonumber\\
\end{align}
Throughout the remainder of this section we will assume $\Delta^b=0$ and write $\Delta^c=\Delta$. We can expand the $\overline{H}H\pi$ interaction from~\cref{lagrangian}
\begin{align}
g_\pi \mathrm{tr}&\left[\overline{H}_aH_b\gamma^\nu\gamma^5\right]A_{\nu}^{ba}\nonumber\\
=&i\frac{g_\pi}{f}\varepsilon^{\mu\kappa\lambda\nu}B_{i\alpha,\mu}^{*\dagger}B_{j\beta,\lambda}^*v_\kappa T^\Xi_{\beta\alpha} \partial_\nu\pi_{ji,\Xi} \nonumber\\
+&i\frac{g_\pi}{f}(B^{*\dagger}_{i\alpha,\lambda}B_{j\beta}-B^\dagger_{i\alpha}B^*_{j\beta,\lambda})T^\Xi_{\beta\alpha} \partial^\lambda\pi_{ji,\Xi}
\end{align}
Finally, we must add terms corresponding to the electroweak $b\to c$ current whose matrix elements we are interested in computing. These will take the form:
\begin{align}
-\epsilon(w)\mathrm{tr}\left[\overline{H}^{(c,v')}_a \Gamma H^{(b,v)}_a\right]
\end{align}
where $w=v'\cdot v$ and $\epsilon(w)$ is the Isgur-Wise function. We define
\begin{align}
-\epsilon(w)\mathrm{tr}\left[\overline{H}^{(c,v')}_a \Gamma H^{(b,v)}_a\right]=&\nonumber\\
&D^{*\dagger}_{a,\mu} \mathcal{J}^{\Gamma,\mu} B_{a} - D^{\dagger}_{a} \mathcal{J}^{\Gamma,\mu} B^{*}_{a,\mu} \nonumber\\
+&D^{*\dagger}_{a,\mu} {\mathcal{K}}^{\Gamma,\mu\nu} B^*_{a,\nu} +D^{\dagger}_{a} {\mathcal{P}}^{\Gamma} B_{a} 
\end{align}
where $\mathcal{J}$, $\mathcal{P}$ and $\mathcal{K}$ depend on $v$, $v'$ and $\Gamma$. Note that $\mathcal{P}$ does not contribute at 1-loop to the current corrections, since there is no $BB\pi$ coupling. $\mathcal{P}$ will only enter for $B\to D$ at tree level and multiplied by the wavefunction and current renormalisation. $\mathcal{J}$, $\mathcal{P}$ and $\mathcal{K}$ maybe be computed straightforwardly for the currents of interest from standard $\gamma$-matrix trace methods.

\subsection{Current renormalisation}

We will follow the conventions in Manohar and Wise~\cite{manohar_wise_2000} and write the renormalised operator of which we wish to compute the matrix elements as
\begin{align}
\mathcal{O}^R_\Gamma = \frac{\sqrt{Z^B Z^{D^*}}}{Z_\mathcal{O}} \mathcal{O}_\Gamma.
\end{align}
Here $\mathcal{O}_\Gamma$ is the local composite operator built from renormalised fields $\sqrt{Z^q} H^q_R = H^q_0$.
We write this as
\begin{align}
\mathcal{O}^R_{[\bar{H}^c\Gamma H^b]}=&-\epsilon(w)\left(1+\frac{1}{2}(\delta_{Z^B}+\delta_{Z^{D^*}}) - \delta_{Z_\mathcal{O}} \right)\nonumber\\
&\times \mathrm{tr}\left[\bar{H}^{c,v'}_a\Gamma H^{b,v}_a\right].
\end{align}
Here we have defined $\delta_{Z_\mathcal{O}}=Z_\mathcal{O}-1$ and, for the charm and bottom fields, $\delta_{Z^q}=Z^q-1$.

The wavefunction renormalisation of the $B$ may be computed from the self energy~\cite{Aubin:2005aq} $(1/2)\partial_{v\cdot p}\Sigma^B(v\cdot p)\Big{|}_{v\cdot p=0}=\delta_{Z^B}$, where $-i\Sigma(p\cdot v)$ is the 1PI diagram with two external lines and with the overall identity in taste and flavour space removed. For the $D^*$, $(1/2)\partial_{v\cdot p}\Sigma^{D^*}(v\cdot p)\Big{|}_{v\cdot p=\Delta}=\delta_{Z^{D^*}}$. Evaluating the Feynman diagram for the $B$ self energy we have
\begin{align}
\Sigma^B =& -i\left[\frac{g_\pi}{f}\right]^2\int\frac{d^4k}{(2\pi)^4}\sum_{\Xi,in}P^{\Xi,n}_{uu,ii}\frac{1}{k^2-M^2_{ui,\Xi,n}+i\varepsilon}\nonumber\\
&\times \frac{ k^\nu (g_{\mu\nu}-v_\mu v_\mu) k^\mu}{2(v\cdot(k+p)+i\varepsilon)}
\end{align}
We look just at the contribution of a single mass of pion, as the sum can be reinserted straightforwardly.
\begin{align}
\Sigma^B_m =& -i\left[\frac{g_\pi}{f}\right]^2\int\frac{d^4k}{(2\pi)^4}\frac{1}{k^2-m^2+i\varepsilon} \frac{ k^\nu (g_{\mu\nu}-v_\mu v_\mu) k^\mu}{2(v\cdot(k+p)+i\varepsilon)}.
\end{align}
Following the notation of~\cite{Chow:1993hr}, we denote
\begin{align}
\mathcal{I}_3(w,m,\Delta)=&\nonumber\\
\int_0^\infty d\alpha \int \frac{d^4k}{(2\pi)^4}&\frac{\alpha k^2}{\left[k^2-(\alpha^2+2\alpha\Delta+m^2)+i\varepsilon\right]^3},
\end{align}
and find $\delta_{Z^B}^m=\frac{3}{2}i({g_\pi}/{f})^2 \mathcal{I}_3(w,m,0)$. A similar calculation yields $\delta_{Z^{D^*}}^m=\frac{i}{2}({g_\pi}/{f})^2 [\mathcal{I}_3(w,m,-\Delta)+2\mathcal{I}_3(w,m,0)]$. The combination appearing in the current renormalisation is then given by
\begin{align}
\frac{1}{2}(\delta_{Z^B}^m+\delta_{Z^{D^*}}^m) =& i\frac{1}{2}\left[\frac{g_\pi}{f}\right]^2\nonumber\\
\times &\left(\frac{1}{2}\mathcal{I}_3(w,m,-\Delta)+\frac{5}{2}\mathcal{I}_3(w,m,0)\right)
\end{align}

\subsection{1-loop matrix element contribution}

The 1-loop contribution to the matrix elements of $-\epsilon(w)\mathrm{tr}[\bar{H\Gamma H}]$ are given by the amputated on-shell two point correlation functions in momentum space, contracted with the appropriate $D^*$ polarisation vector, $\epsilon_{\lambda'}(v')$, for which $\epsilon_{\lambda'}(v')v'^{\lambda'}=0$. For $B\to D^*$ the current correction is given by
\begin{align}
\left[\frac{g_\pi}{f}\right]^2\int\frac{d^4k}{(2\pi)^4}\sum_{\Xi,in}P^{\Xi,n}_{uu,ii}\frac{i}{k^2-M^2_{ui,\Xi,n}+i\varepsilon} \epsilon^*_{\lambda'}\nonumber\\
\times\Big[\frac{i\epsilon^{\lambda'}_{~\kappa\gamma\nu}v'^\kappa  k^\nu}{2(v'\cdot k +i\varepsilon)} \mathcal{K}^{\Gamma,\gamma\rho} \frac{i(k_\rho-(k \cdot v) v_\rho )}{2(v\cdot k +i\varepsilon)} \nonumber\\
+\frac{i k^{\lambda'}}{2(v'\cdot k+\Delta +i\varepsilon)}  \mathcal{J}^{\Gamma,\rho} \frac{i(k_\rho-(k \cdot v) v_\rho )}{2(v\cdot k+i\varepsilon)}\Big].
 \nonumber\\
\end{align}
Here we have left implicit that we will divide sea quark loops by a factor of 4 to reduce the number of tastes from 4 to 1. Noting that the sum over tastes and hairpin terms may be straightforwardly reinserted, we evaluate the contribution of a single mass,
\begin{align}
\left[\frac{g_\pi}{f}\right]^2\int\frac{d^4k}{(2\pi)^4}\frac{i}{k^2-m^2+i\varepsilon} \epsilon^*_{\lambda'}\nonumber\\
\times\Big[\frac{i\epsilon^{\lambda'}_{~\kappa\gamma\nu}v'^\kappa  k^\nu}{2(v'\cdot k +i\varepsilon)} \mathcal{K}^{\Gamma,\gamma\rho} \frac{i(k_\rho-(k \cdot v) v_\rho )}{2(v\cdot k +i\varepsilon)} \nonumber\\
+\frac{i k^{\lambda'}}{2(v'\cdot k+\Delta +i\varepsilon)}  \mathcal{J}^{\Gamma,\rho} \frac{i(k_\rho-(k \cdot v) v_\rho )}{2(v\cdot k+i\varepsilon)}\Big].
 \nonumber\\
\end{align}
For the currents considered here, these may all be expressed in terms of the integral
\begin{align}
\int\frac{d^4k}{(2\pi)^4}\frac{1  }{2(v'\cdot k -\Delta+i\varepsilon)}  \frac{1}{2(v\cdot k +i\varepsilon)}
\frac{ k^\delta k^{\nu}}{k^2-m^2+i\varepsilon}\nonumber\\
=\frac{1}{2}\mathcal{I}_1(w,m,\Delta) g^{\delta\nu}+\frac{1}{2}\mathcal{I}_2(w,m,\Delta) v'^\delta v^{\nu}+...
\end{align}
where only the $\mathcal{I}_1$ contains a UV divergence and the $...$ indicates terms which give zero when summing over lorentz indices. These definitions for the integrals match those given in Chow et al~\cite{Chow:1993hr}. They include a factor of $i/16\pi^2$ that has been removed from the definitions used in Appendix~A of~\cite{FermilabLattice:2021cdg} that has ${I}=\mathcal{I}^\mathrm{FNAL} = i16\pi^2\times\mathcal{I}^\mathrm{Chow}$. We will adopt the conventions in Appendix~A of~\cite{FermilabLattice:2021cdg}. With these conventions we have
\begin{align}
\frac{1}{2}(\delta_{Z^B}^m+\delta_{Z^{D^*}}^m) =& \frac{1}{2}\left[\frac{g_\pi}{4\pi f}\right]^2\nonumber\\
\times &\left(\frac{1}{2}I_3(w,m,-\Delta/m)+\frac{5}{2}I_3(w,m,0)\right)
\end{align}

This yields a total 1-loop current correction for $B\to D^*$, for a single pion mass, 
\begin{align}
\frac{-1}{2}&\left[\frac{g_\pi}{4\pi f}\right]^2 \epsilon^*_{\lambda'}\times\nonumber\\
\Big[&{I}_1(w,m,0){\epsilon^{\lambda'}_{~\kappa\gamma\nu}v'^\kappa  }\big(\mathcal{K}^{\Gamma,\gamma\nu} - \mathcal{K}^{\Gamma,\gamma\rho} v_\rho v^\nu\big) \nonumber\\
+ &{I}_1(w,m,-\Delta/m) \big(\mathcal{J}^{\Gamma,\lambda'} - \mathcal{J}^{\Gamma,\rho}v_\rho v^{\lambda'}\big)   \nonumber\\
+ &{I}_2(w,m,0) {\epsilon^{\lambda'}_{~\kappa\gamma\nu}v'^\kappa  }v^\nu\big(\mathcal{K}^{\Gamma,\gamma\rho}  v'_{\rho} -w\mathcal{K}^{\Gamma,\gamma\rho}  { v_\rho }\big)\nonumber\\
+&{I}_2(w,m,-\Delta/m)   v^{\lambda'} \big(\mathcal{J}^{\Gamma,\rho}  v'_{\rho} -w\mathcal{J}^{\Gamma,\rho}  { v_\rho }\big)   \Big].
\end{align}
For $B\to D$ this procedure gives the 1-loop current correction for a single pion mass
\begin{align}
\frac{1}{2}\left[\frac{g_\pi}{4\pi f}\right]^2\Big[& {I}_1(w,m,\Delta)\mathcal{K}^{\Gamma,\gamma\rho} \left(g_{\gamma\rho}/2 + w v_\rho v'_\gamma\right)\nonumber\\
+&{I}_2(w,m,\Delta)\mathcal{K}^{\Gamma,\gamma\rho} \left( v'_\rho-wv_\rho \right)\left( v_\gamma-wv'_\gamma \right)\Big].\nonumber\\
\end{align}

We evaluate $\mathcal{K}$, $\mathcal{J}$ and $\mathcal{P}$ for $\Gamma=\gamma^i, \gamma^5\gamma^i$ and $\sigma^{\alpha\beta}$ corresponding to the vector, axial-vector and tensor currents respectively. Defining $\delta^m _{h_X}$ as the deviation from the tree level value resulting from loops including a pion $m$, we find, dropping $\delta_{Z_\mathcal{O}}$ and including only the finite parts of the integrals regularised using dimensional regularisation,

\begin{align}
\delta^m _{h_{A_1}}/\varepsilon(w)=& \frac{1}{2}\left(\delta_{Z^c}^m+\delta_{Z^b}^m\right) - \frac{1}{2}\left[\frac{g_\pi}{4\pi f}\right]^2\nonumber\\
\times&\Big((1+w)I_1(w,m,0) + I_1(w,m,-\Delta/m)\nonumber\\
&+(w^2-1)I_2(w,m,0)  \Big)    \nonumber\\
 =&-\frac{1}{2}\left[\frac{g_\pi}{4\pi f}\right]^2\Big( -\frac{1}{2}{I}_3(w,m,-\Delta/m)\nonumber\\
 &-\frac{5}{2}{I}_3(w,m,0)+ (1+w)I_1(w,m,0) \nonumber\\
 &+ I_1(w,m,-\Delta/m)+(w^2-1)I_2(w,m,0)  \Big)  \nonumber\\
=&\frac{1}{4}\frac{g_\pi^2}{16\pi^2 f^2} F^{h_{A_1}}(w,m,-\Delta/m),
\end{align}
\begin{align}
\delta^m _{h_{A_2}}/\varepsilon(w)=&  -\frac{1}{2}\left[\frac{g_\pi}{4\pi f}\right]^2\Big(I_1(w,m,-\Delta/m)\nonumber\\
&-I_1(w,m,0) +(w+1)I_2(w,m,-\Delta/m) \nonumber\\
&-(1+w)I_2(w,m,0)  \Big)\nonumber\\
=&\frac{1}{4}\frac{g_\pi^2}{16\pi^2 f^2} F^{h_{A_2}}(w,m,-\Delta/m)\nonumber\\
\delta^m _{h_{A_3}}=&\delta^m _{h_{A_1}}-\delta^m _{h_{A_2}}\nonumber\\
\delta^m _{h_{V}}=&\delta^m _{h_{A_1}}
\end{align}
and for the tensor current
\begin{align}
\delta^m _{h_{T_1}}/\varepsilon(w)=& \frac{1}{2}\left(\delta_{Z^c}^m+\delta_{Z^b}^m\right) -\frac{1}{2}\left[\frac{g_\pi}{4\pi f}\right]^2\nonumber\\
&\times\Big( (1+w)I_1(w,m,0) +I_1(w,m,-\Delta/m) \nonumber\\
&+(w^2-1)I_2(w,m,0) \Big)\nonumber\\
 =-\frac{1}{2}\left[\frac{g_\pi}{4\pi f}\right]^2&\Big( -\frac{1}{2}{I}_3(w,m,-\Delta/m)\nonumber\\
 &-\frac{5}{2}{I}_3(w,m,0)+ (1+w)I_1(w,m,0) \nonumber\\
 &+ I_1(w,m,-\Delta/m)+(w^2-1)I_2(w,m,0)  \Big)  \nonumber\\
=&\delta^m _{h_{A_1}}/\varepsilon(w)\nonumber\\
\delta^m _{h_{T_2}}/\varepsilon(w)=&0 \nonumber\\
\delta^m _{h_{T_3}}/\varepsilon(w)=&-\frac{1}{2}\left[\frac{g_\pi}{4\pi f}\right]^2 \nonumber\\
&\times\Big(\left(I_1(w,m,-\Delta/m)-I_1(w,m,0)\right) \nonumber\\
&-(w+1)I_2(w,m,0)\nonumber\\
&+(1+w)I_2(w,m,-\Delta/m)\Big)\nonumber\\
=&\delta^m _{h_{A_2}}/\varepsilon(w).\nonumber\\
\end{align}

Here we have defined the quantities $F^{h_{A_i}}(w,m,-\Delta/m)$ to match those given in~\cite{FermilabLattice:2021cdg}:
\begin{align}
F^{h_{A_1}}(w,m,x)&=-2\Big[ I_1(w,m,x)-\frac{1}{2}I_3(w,m,x) \nonumber\\
&+(w+1)I_1(w,m,0)+(w^2-1)I_2(w,m,0) \nonumber\\
&-\frac{5}{2}I_3(w,m,0) \Big],\nonumber\\
F^{h_{A_2}}(w,m,x)&=-2\Big[ I_1(w,m,x)+(w+1)I_2(w,m,x)\nonumber\\
&-I_1(w,m,0)-(w+1)I_2(w,m,0)  \Big],\nonumber\\
F^{h_{A_3}}(w,m,x)&=F^{h_{A_1}}(w,m,x)-F^{h_{A_2}}(w,m,x).\nonumber\\
\end{align}
We also have the tree level values $h_{A_1}^\mathrm{tree}=h_{A_3}^\mathrm{tree}=h_{V}^\mathrm{tree}=h_{T_1}^\mathrm{tree}=\varepsilon(w)\nonumber$ and 
$h_{A_2}^\mathrm{tree}=h_{T_2}^\mathrm{tree}=h_{T_3}^\mathrm{tree}=0$.
With these definitions we have 
\begin{align}\label{sumovertastesfinal}
h_X= \left(h_X^\mathrm{tree} +\sum_{\Xi,in}P^{\Xi,n}_{uu,ii}\delta^{M_{ui,\Xi,n}}_{h_X} \right).
\end{align}
Since the sum over tastes acts in the same way for different form factors, we find
\begin{align}
h_{T_1}&=h_{A_1}\nonumber\\
h_{T_2}&=0\nonumber\\
h_{T_3}&=h_{A_2}
\end{align}
We also find for $B\to D$ that $f_T=f_+$ and confirm the 1-loop relation $h_{A_1}=h_{V}$ and $f_-=0$.

\subsection{Chiral Logarithms}
Denoting $F^{h_{Y}}(w,m_j,-\Delta/m_j)=\bar{F}^Y_j$ as in~\cite{FermilabLattice:2021cdg}, the sum in~\cref{sumovertastesfinal} over $n$ and $i$ gives

\begin{align}\label{sumovertastesexpanded}
\delta_{h_X}=&\frac{g_\pi^2}{16\pi^2 f^2}\sum_{\Xi,in}P^{\Xi,n}_{uu,ii} \frac{1}{4}\bar{F}^{Y}_{{ui,\Xi,n}}\nonumber\\
=&\frac{g_\pi^2}{16\pi^2 f^2}\times\Big( \frac{1}{4}\sum_{\Xi} \left( 2 \bar{F}^{Y}_{\pi_\Xi} +   \bar{F}^{Y}_{K_\Xi} \right) \Big)\nonumber\\
&+\frac{1}{4} \sum_{\Xi} a^2\delta'_\Xi\left[A_\Xi \bar{F}^{Y}_{\pi^0_\Xi} +B_\Xi \bar{F}^{Y}_{\eta_\Xi}+C_\Xi \bar{F}^{Y}_{\eta'_\Xi}\right]\Big)\nonumber\\
=&\frac{g_\pi^2}{16\pi^2 f^2}\times\Big( \frac{1}{4}\sum_{\Xi} \left( 2 \bar{F}^{Y}_{\pi_\Xi} +   \bar{F}^{Y}_{K_\Xi} \right) \Big)\nonumber\\
&+\sum_{\Xi=A,V} a^2\delta'_\Xi\left[A_\Xi \bar{F}^{Y}_{\pi^0_\Xi} +B_\Xi \bar{F}^{Y}_{\eta_\Xi}+C_\Xi \bar{F}^{Y}_{\eta'_\Xi}\right]\nonumber\\
&+\frac{m_0^2}{3}\left[A_I \bar{F}^{Y}_{\pi^0_I} +B_I \bar{F}^{Y}_{\eta_I}+C_I \bar{F}^{Y}_{\eta'_I}\right]
\end{align}
where $A$, $B$ and $C$ are defined as in~\cref{hairpinprop}. Following~\cite{Aubin:2003mg} to move from 4+4+4 to 1+1+1 flavor-tastes of light quark, 
we find after taking $m_0\to \infty$,
\begin{align}
\delta_{h_X}=\left(\frac{g_\pi^2}{16\pi^2 f^2}\right)\Big[\frac{1}{16}\sum_\Xi \left(2\bar{F}^Y_{\pi_\Xi}+\bar{F}^Y_{K_\Xi}\right)-\frac{1}{2}\bar{F}^Y_{\pi_I}+\frac{1}{6}\bar{F}^Y_{\eta_I}\nonumber\\
+\sum_{\Xi=V,A}a^2\delta_\Xi\Big(\frac{m^2_{S_{\Xi}}-m^2_{\pi^0_{\Xi}}}{(m^2_{\eta_{\Xi}}-m^2_{\pi^0_{\Xi}})(m^2_{\pi^0_{\Xi}}-m^2_{\eta'_{\Xi}})} \bar{F}^Y_{\pi_\Xi} \nonumber\\
+\frac{m^2_{\eta_{\Xi}}-m^2_{S_{\Xi}}}{(m^2_{\eta_{\Xi}}-m^2_{\pi^0_{\Xi}})(m^2_{\eta_{\Xi}}-m^2_{\eta'_{\Xi}})} \bar{F}^Y_{\eta_\Xi}\nonumber\\
+\frac{m^2_{S_{\Xi}}-m^2_{\eta'_{\Xi}}}{(m^2_{\pi^0_{\Xi}}-m^2_{\eta'_{\Xi}})(m^2_{\eta'_{\Xi}}-m^2_{\eta_{\Xi}})}\bar{F}^Y_{\eta'_\Xi}\Big)\Big],\nonumber\\
\end{align}
where we have used the relations given in~\cite{Laiho:2005ue} for the flavour-neutral pion mass eigenstates. Note that our results here differ from those given in~\cite{FermilabLattice:2021cdg} by an overall factor of 3. We have checked that our results match those in~\cite{Chow:1993hr} for $h_{A_1},~h_{A_2},~h_{A_3}$, and we have also checked that our zero recoil results match those in~\cite{Laiho:2005ue}. The chiral logarithms, $\mathrm{logs}_{SU(3)}^Y$, in~\cref{fitfunctionequation}, for $B\to D^*$ thus take the form
\begin{align}\label{su3logschipt}
\mathrm{logs}_{SU(3)}^Y=\frac{1}{16}\sum_\Xi \left(2\bar{F}^Y_{\pi_\Xi}+\bar{F}^Y_{K_\Xi}\right)-\frac{1}{2}\bar{F}^Y_{\pi_I}+\frac{1}{6}\bar{F}^Y_{\eta_I}\nonumber\\
+\sum_{\Xi=V,A}a^2\delta_\Xi\Big(\frac{m^2_{S_{\Xi}}-m^2_{\pi^0_{\Xi}}}{(m^2_{\eta_{\Xi}}-m^2_{\pi^0_{\Xi}})(m^2_{\pi^0_{\Xi}}-m^2_{\eta'_{\Xi}})} \bar{F}^Y_{\pi_\Xi} \nonumber\\
+\frac{m^2_{\eta_{\Xi}}-m^2_{S_{\Xi}}}{(m^2_{\eta_{\Xi}}-m^2_{\pi^0_{\Xi}})(m^2_{\eta_{\Xi}}-m^2_{\eta'_{\Xi}})} \bar{F}^Y_{\eta_\Xi}\nonumber\\
+\frac{m^2_{S_{\Xi}}-m^2_{\eta'_{\Xi}}}{(m^2_{\pi^0_{\Xi}}-m^2_{\eta'_{\Xi}})(m^2_{\eta'_{\Xi}}-m^2_{\eta_{\Xi}})}\bar{F}^Y_{\eta'_\Xi}\Big)\nonumber\\
\end{align}
and for $B_s \to D_s^*$ a similar calculation gives the logs
\begin{align}\label{su3logschipt_s}
\mathrm{logs}_{SU(3)}^{Y^s}=\frac{1}{16}\sum_\Xi \left(2\bar{F}^{Y}_{K_\Xi}+\bar{F}^{Y}_{S_\Xi}\right)-\bar{F}^{Y}_{S_I}+\frac{2}{3}\bar{F}^{Y}_{\eta_I}\nonumber\\
+\sum_{\Xi=V,A}a^2\delta_\Xi\Big(\frac{m^2_{S_{\Xi}}-m^2_{\pi^0_{\Xi}}}{(m^2_{S_{\Xi}}-m^2_{\eta_{\Xi}})(m^2_{S_{\Xi}}-m^2_{\eta'_{\Xi}})}\bar{F}^{Y}_{S_\Xi}\nonumber\\
+\frac{m^2_{\eta_{\Xi}}-m^2_{\pi_{\Xi}}}{(m^2_{\eta_{\Xi}}-m^2_{S_{\Xi}})(m^2_{\eta_{\Xi}}-m^2_{\eta'_{\Xi}})}\bar{F}^{Y}_{\eta_\Xi}\nonumber\\
+\frac{m^2_{\eta'_{\Xi}}-m^2_{\pi_{\Xi}}}{(m^2_{\eta'_{\Xi}}-m^2_{S_{\Xi}})(m^2_{\eta'_{\Xi}}-m^2_{\eta_{\Xi}})}\bar{F}^{Y}_{\eta'_\Xi}\Big).\nonumber\\
\end{align}
The chiral logarithms are most sensitive to variation of the pion and $\eta_{V,A}$ masses, and the effect of the taste splittings and hairpin contributions is most pronounced near the `cusp' $m_\pi\approx \Delta$, which roughly coincides with the physical pion mass. The sum over tastes appearing in~\cref{su3logschipt} acts to wash out the cusp, averaging over the masses of the different tastes~\cite{Laiho:2005ue}. This effect is somewhat mitigated for HISQ quarks by the fact that the taste splittings are all approximately proportional~\cite{PhysRevD.87.054505}, with $M_{\pi_\xi}^2-M_{\pi_5}^2\approx n_\xi a^2\delta_t$, where $n_A=1$, $n_T=2$, $n_V=3$ and $n_I=4$. 

We can analyse the effect of taste-splittings by expanding the pion log terms to first order in $a^2$, and dropping terms proportional to $a^2\bar{F}_{\pi_5}^Y$ that produce only normal discretisation effects. This gives
\begin{align}
\mathrm{logs}_{SU(3)}^{Y} \approx a^2\left(2\delta_t+\delta'_V+\delta'_A\right)\frac{\partial\bar{F}^Y_{\pi_5}}{\partial m_{\pi_5}^2}.
\end{align}
Together with the approximate relation, $\delta'_V+ \delta'_A\approx -2\delta_t$, for  HISQ~\cite{PhysRevD.88.074504,Colquhoun:2015mfa}, this leading order correction is suppressed so there is no non-analytic behaviour in $a$. This matches what was seen in~\cite{Colquhoun:2015mfa}, where a similar approximate cancellation of leading order taste-splitting and hairpin terms was seen. Note, however, that in our fits we use the full expressions given in~\cref{su3logschipt,su3logschipt_s}.


\section{Comparison to Previous HPQCD $B_s\to D_s^*$ Form Factors}\label{comptoprev}
\begin{figure}
\includegraphics[scale=0.25]{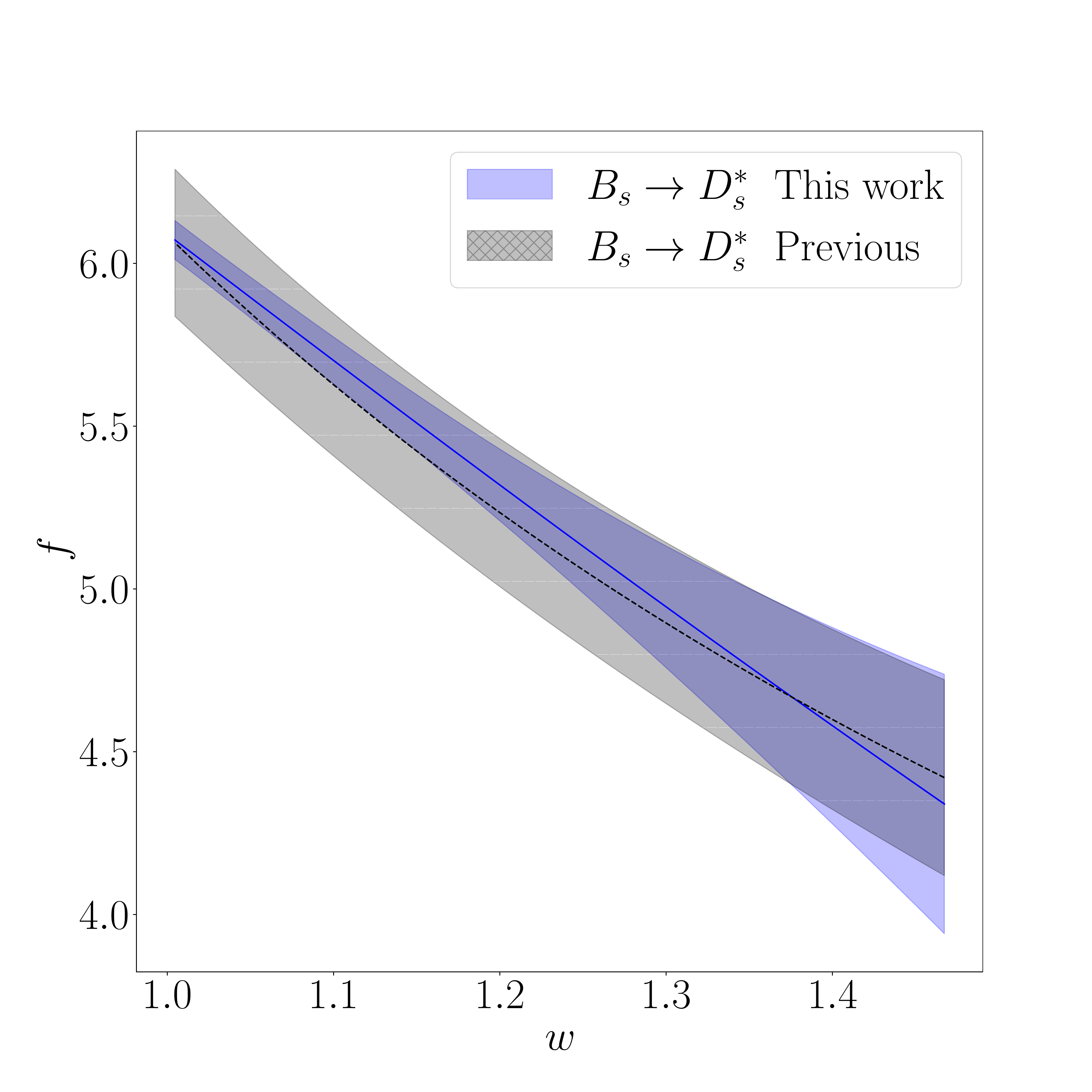}\\
\includegraphics[scale=0.25]{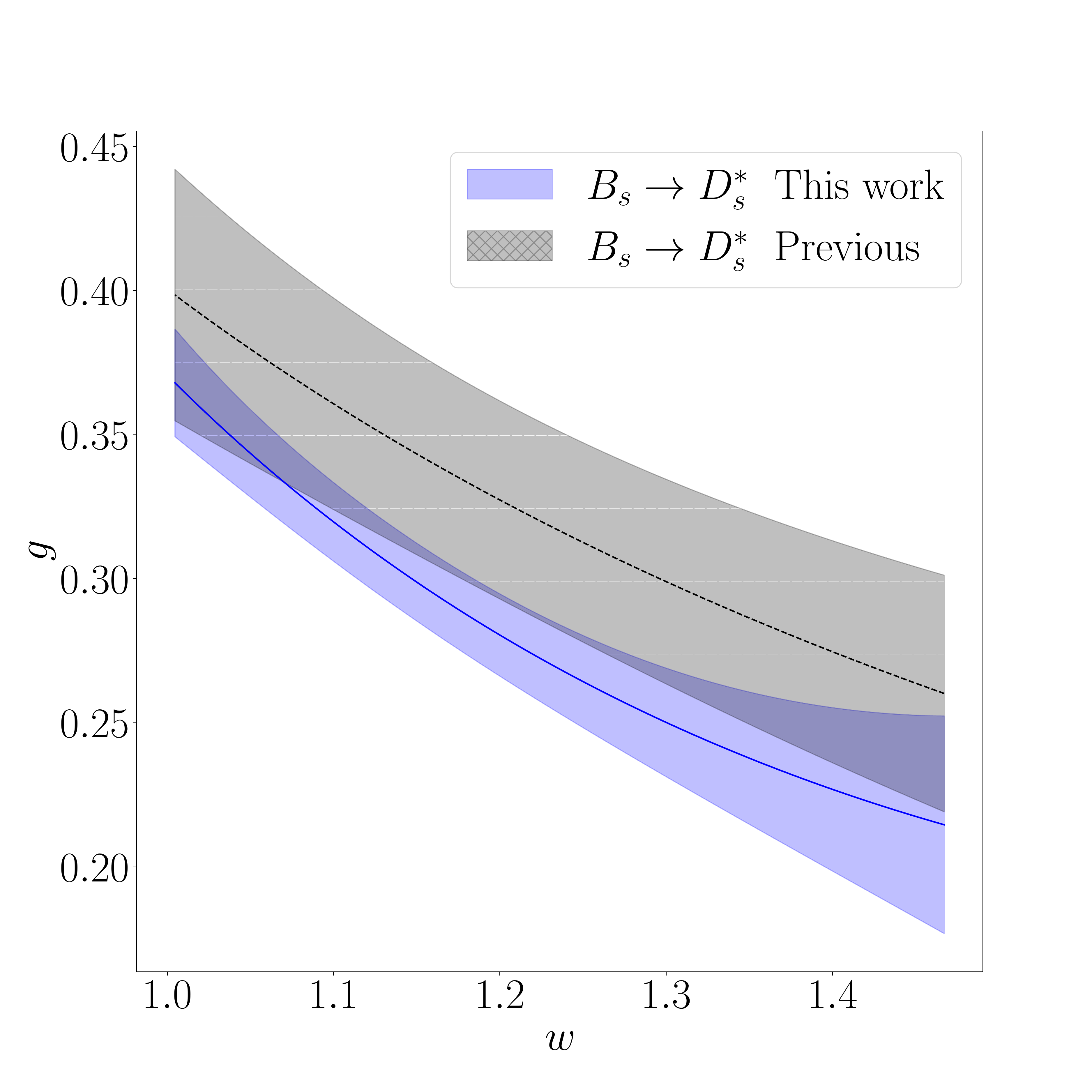}
\caption{\label{HPQCD-HQET_bsdss_old_comparison1}$B_s\to D_s^*$ helicity basis form factors $f$ and $g$, defined in~\cref{helicitybasis}.  We show the results of this work as a blue band, compared to the results of~\cite{Harrison:2021tol} given as a grey band.}
\end{figure}

\begin{figure}
\includegraphics[scale=0.25]{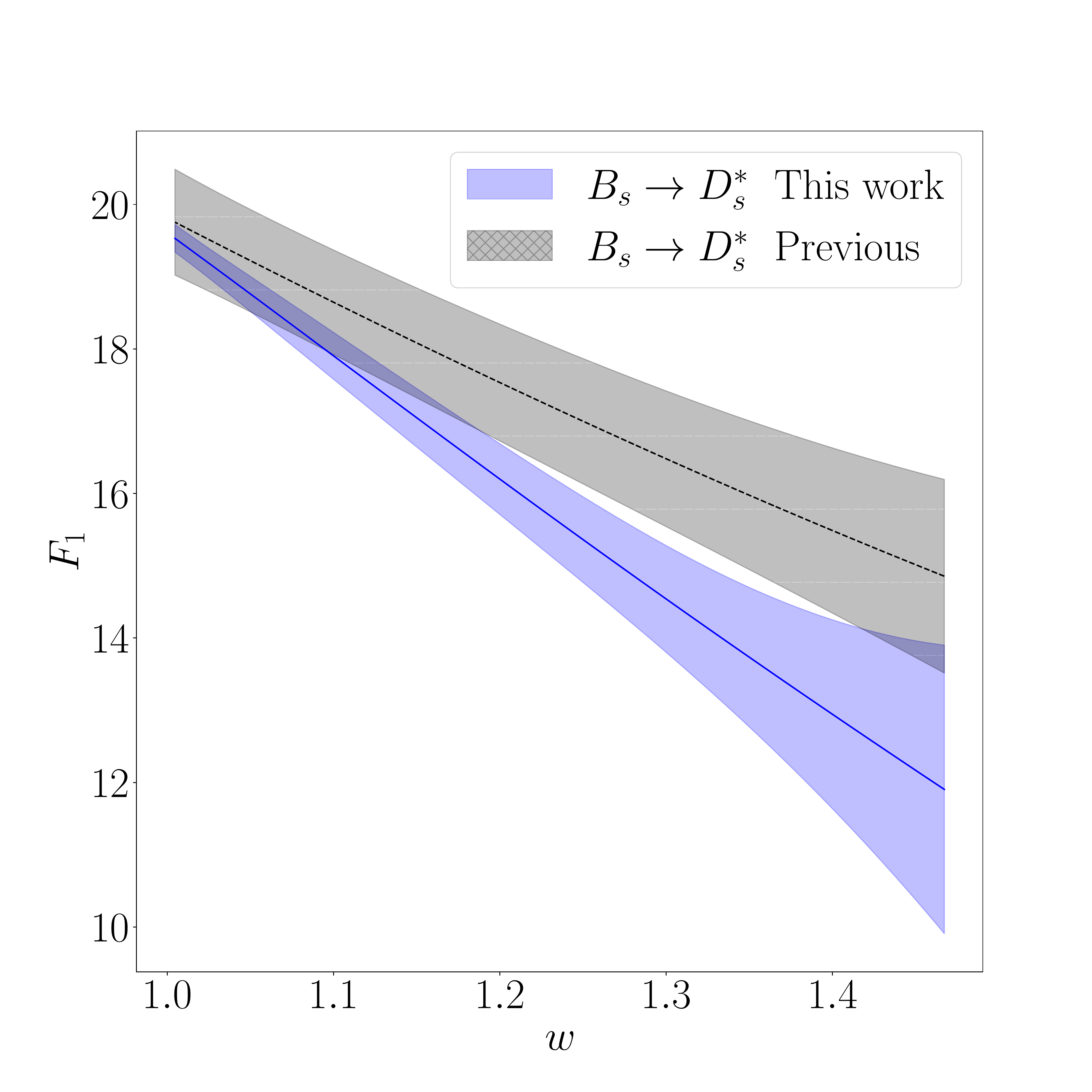}\\
\includegraphics[scale=0.25]{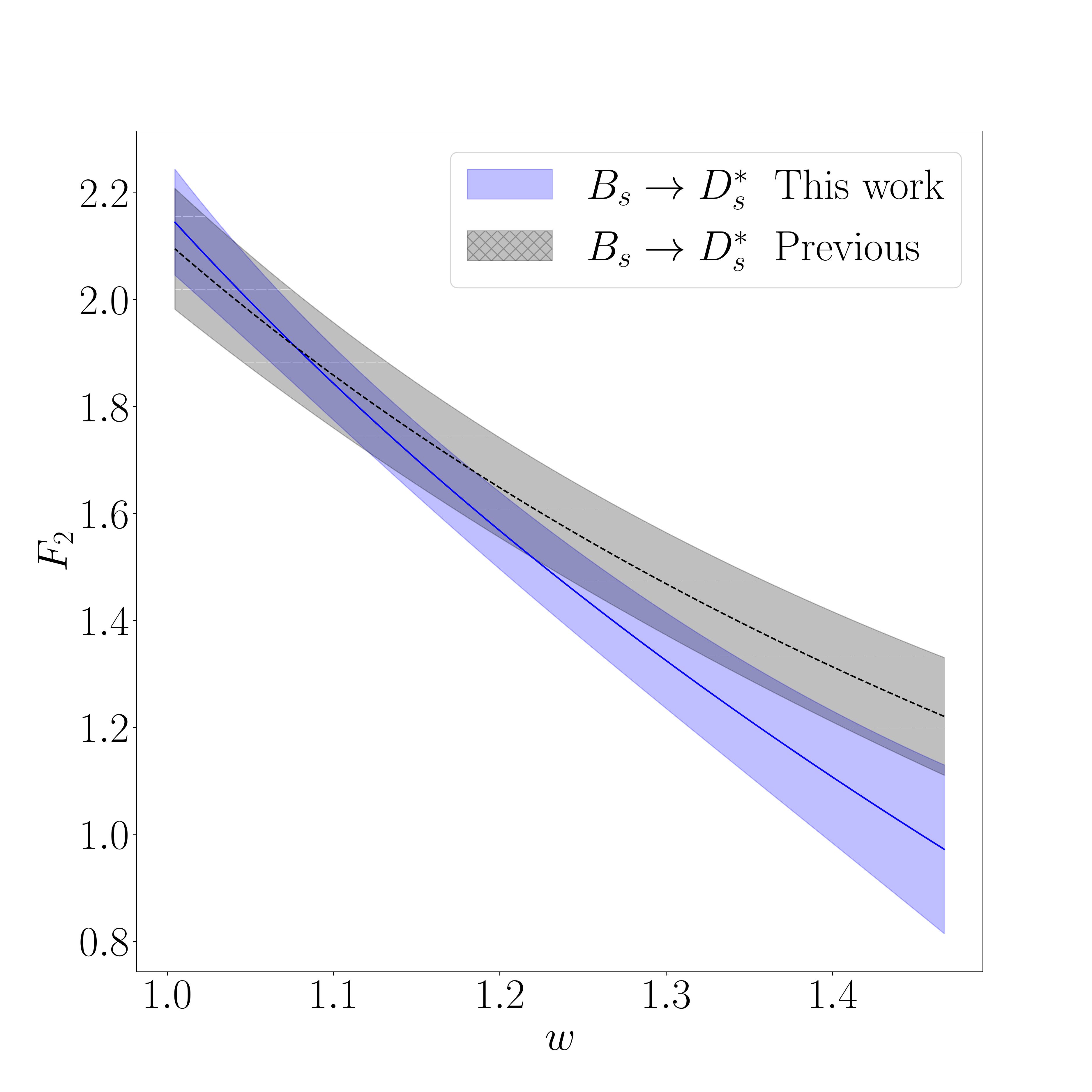}
\caption{\label{HPQCD-HQET_bsdss_old_comparison2}$B_s\to D_s^*$ helicity basis form factors $F_1$ and $F_2$, defined in~\cref{helicitybasis}. We show the results of this work as a blue band, compared to the results of~\cite{Harrison:2021tol} given as a grey band.}
\end{figure}

In~\cref{HPQCD-HQET_bsdss_old_comparison1,HPQCD-HQET_bsdss_old_comparison2} we plot our previous results for $B_s\to D_s^*$ from~\cite{Harrison:2021tol} together with the updated form factors given in this work. The improved calculation presented here has the additon of a physical ensemble with $a\approx 0.06\mathrm{fm}$, Set 5, and includes the additional $B\to D^*$ correlator data that informs the $B_s \to D_s^*$ form factors through our chiral extrapolation. Additionally, in this work we adopt the time source binning strategy described in~\cref{ntbinning}, providing improved resolution of the correlator covariance matrices. \cite{Harrison:2021tol}~also used a BGL-like parameterisation to describe the kinematic dependence of the form factors, compared to the simpler expression in powers of $(w-1)$ used here. Despite these methodological differences, reasonable agreement is seen between form factors in the physically important helicity basis and good agreement is seen near $w=1$. We have also verified that fitting $B\to D^*$ and $B_s\to D_s^*$ separately produces essentially identical form factor results to the simultaneous fit described in~\cref{physextrap}.

Further investigation of the differences shows that in~\cite{Harrison:2021tol} the choice of $t_\mathrm{min}$ ($\Delta T_\mathrm{2pt}$ in that work) for correlator fits on set 3 was too small, and that this resulted in excited state contamination which shifted the extracted matrix elements on set 3 upwards by $\approx 1\sigma$. This shift resulted in a change in the result in the chiral-continuum limit that was most pronounced near $w_\mathrm{max}$ where lattice data was only available on set 3. As a cross check, we have applied the chiral-continuum extrapolation described in~\cite{Harrison:2021tol} to the $B_s \to D_s^*$ dataset used in this work, excluding set 5. This results in similar form factors to those given in this work, except near $w_\mathrm{max}$ where $\approx 1\sigma$ differences are seen in $F_1$ and $F_2$. 

We conclude that the $\approx 1-2\sigma$ differences seen between this work and~\cite{Harrison:2021tol}, particularly near $w_\mathrm{max}$, are due to a combination of the excited state contamination present in set 3 of~\cite{Harrison:2021tol} the much more conservative description of kinematic dependence of the form factors used in this work~(\cref{physextrap}) and the use of a $z$-expansion in~\cite{Harrison:2021tol} whose form led to a bias in the shape of the continuum form factors and the underestimation of uncertainties near $w_\mathrm{max}$.

\end{appendix}

\bibliographystyle{apsrev4-1}
\bibliography{BsDsstar}

\end{document}